\documentclass[a4paper,12pt]{report}
\usepackage{amsmath, amsthm, amsfonts, amssymb}
\usepackage[svgnames]{xcolor}
\colorlet{color1}{NavyBlue}
\usepackage[colorlinks=true,allcolors=color1]{hyperref}
\usepackage{physics}
\usepackage{dsfont}
\usepackage{graphicx}
\usepackage{cleveref} 
\usepackage{xfrac}
\usepackage{cite}
\usepackage[margin=1in]{geometry}
\usepackage{mathrsfs}
\usepackage{bm}
\usepackage{pdfpages}

\usepackage{fancyhdr}
\usepackage[font=small,labelfont=bf,margin=10pt]{caption}

\newcommand{\xbb}{\bar{\bm x}}

\newcommand{\vbb}{\bar{\bm v}}

\AtBeginDocument{\addtocontents{toc}{\protect\thispagestyle{empty}}}

\begin{document}

\begin{titlepage}
\centering
\Large
\textsc{Universidad Complutense de Madrid}\par
\vspace*{0.1cm}
\textsc{Facultad de Ciencias Físicas}\par
\vspace*{3cm}
\LARGE
\hrule
\vspace*{0.3cm}
\textbf{Vacuum polarisation and regular gravitational collapse}\par
\vspace*{0.2cm}
\hrule
\vspace*{0.3cm}
\textbf{Polarización de vacío y colapso gravitatorio regular}\par
\vspace*{0.2cm}
\hrule
\vspace*{2cm}
\large
Valentin Boyanov Savov\par
\vspace*{2cm}
\textsc{Supervisors:}\par
\vspace*{0.2cm}
Luis Javier Garay Elizondo,\\
Carlos Barceló Serón,\\
Raúl Carballo Rubio\par
\vspace*{3cm}
\includegraphics[scale=1]{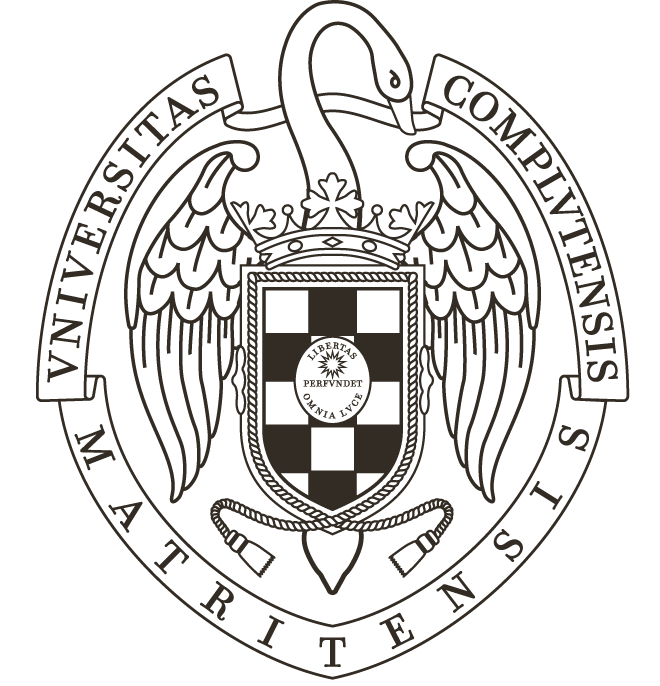}\par

\newpage
\thispagestyle{empty}
\null

\newpage
\thispagestyle{empty}

\centering
\Large
\textsc{Universidad Complutense de Madrid}\par
\vspace*{0.5cm}
\textsc{Facultad de Ciencias Físicas}\par
\vspace*{1.2cm}
\includegraphics[scale=1]{escudo.png}\par
\vspace*{1cm}
\textsc{Tesis Doctoral}\par
\vspace*{2cm}
\LARGE
Vacuum polarisation and regular gravitational collapse\par
\vspace*{0.1cm}
Polarización de vacío y colapso gravitatorio regular\par
\vspace*{1.5cm}
\large
\textsc{Memoria para optar al grado de Doctor en Física}\par
\vspace*{0.1cm}
\textsc{presentada por}\par
\vspace*{0.2cm}
Valentin Boyanov Savov\footnote{email: vboyanov@ucm.es; valentinboyanov@tecnico.ulisboa.pt}\par
\vspace*{1.5cm}
\textsc{Directores:}\par
\vspace*{0.2cm}
Luis Javier Garay Elizondo,\\
Carlos Barceló Serón,\\
Raúl Carballo Rubio

\newpage
\thispagestyle{empty}
\null

%

\newpage
\thispagestyle{empty}

\vspace*{4cm}
\begin{flushleft}
	``First, the \textit{matter} of thought exists in itself, while \textit{though} in itself is empty [\dots] Second, the matter object in itself is something complete, which does not require thought at all, while thought is something to be completed by it [\dots] Third, the object and thought are each in a sphere separate from the other [\dots] thought does not leave itself when receiving and adapting to matter, but rather just modifies itself [\dots] These are the fallacies which bar the gateway into philosophy, and must be overcome before entry."
\end{flushleft}
\begin{flushright}
	- G. W. F. Hegel, \textit{Science of Logic} (1812).
\end{flushright}

\end{titlepage}

\thispagestyle{empty}

\hrule
\begin{abstract}
	\pagestyle{empty}
	\noindent
	It is the goal of this work to revisit and revise the problem of black hole (BH) formation and evolution in semiclassical gravity---a theory in which spacetime is treated classically, while matter admits a quantum description, coupling to gravity through an expectation value of a stress-energy tensor operator. Particularly, we analyse the vacuum expectation value of this operator for a test scalar field in spherically symmetric spacetimes in which trapped regions either form or are close to forming. First, we look at the magnitude of potential corrections to the spacetime evolution in the vicinity of outer horizons formed by collapsing matter in different dynamical regimes. We find that when the matter approaches an adiabatic collapse regime while close to forming a trapped region (i.e. close to crossing its Schwarzschild radius), the vacuum energy tends to grow unboundedly. This relates to the Boulware state divergence at horizons, which in turn can be related to the existence of static horizonless BH mimicker solutions to the semiclassical Einstein equations. This suggests that the growing vacuum energy in slow collapse regimes may stabilise the matter into a final horizonless configuration.
	
	We then look at how the dynamics of such horizonless ultracompact objects can lead to the emission of Hawking radiation (without the formation of trapped surfaces). We find that an oscillatory movement of its surface results in the emission of bursts of radiation, while a slow collapse which asymptotically approaches the crossing of the Schwarzschild radius can result in a thermal spectrum akin to that of BHs, but with a modified temperature. We also analyse the causal structure of the latter family of spacetimes (in which the formation of a trapped surface is approached asymptotically), finding that they posses an event horizon and either a Cauchy horizon, or two separate future (null and timelike) infinity regions in the interior and exterior.
	
	Finally, we look at the vacuum energy content in the vicinity of a BH inner horizon, a region of spacetime which is classically known to amplify perturbations in a highly non-linear manner. On a classical level, the evolution of the inner horizon in the presence of generic perturbations present in the astrophysical medium leads to the so-called ``mass inflation instability", wherein curvature around and below the initial position of the inner horizon grows exponentially, while the inner horizon itself tends to approach the origin. On a semiclassical level, we find a negative ingoing flux of energy, akin to the one which drives Hawking evaporation at the outer horizon. However, backreaction from this flux seems to indeed be amplified, growing exponentially and quickly overcoming of the Planck scale suppression suffered by semiclassical dynamics. Indeed, classical mass inflation and the semiclassical effect (which we dub inner horizon inflation) act in opposite ways on the inner horizon: one pushing it inwards and the other outwards. Analyses comparing the two effects suggest that the semiclassical one may dominate at late times, making it possible for the trapped region to disappear from the inside out, on a time scale much shorter than the Hawking evaporation time. As an aside, we find that the techniques used for analysing the semiclassical stability of the inner horizon can also be applied to other geometries with similar causal properties. In particular, we look at the geometry of a warp drive spacetime, and use its causal structure to argue that an instability previously found in 2-dimensional models can be avoided in higher dimensions.
	
	To complete the picture of BH objects in semiclassical gravity, we propose that, if trapped regions indeed disappear on short timescales, then the ultracompact objects observed astrophysically may indeed be horizonless BH mimickers, formed from slowly collapsing matter, the initial conditions for which are obtained after the dissipation from one or several iteration of trapped region formation, inner horizon inflation and recollapse.
	
	\newpage
	\begin{center}
		\textbf{Resumen}
	\end{center}
	
	\noindent
	El objetivo de este trabajo es reexaminar y revisar el problema de la formación y evolución de agujeros negros en gravedad semiclásica, una teoría en la que el espaciotiempo se trata de forma clásica, mientras que la materia admite una descripción cuántica, acoplándose a la gravedad a través de un valor esperado de un operador tensor de energía-momento. En particular, analizamos el valor esperado en vacío de este operador para un campo escalar de prueba en espaciotiempos esféricamente simétricos en los que se forman o están a punto de formarse regiones atrapadas. En primer lugar, examinamos la magnitud de las potenciales correcciones a la evolución del espaciotiempo en las proximidades de horizontes externos formados por materia en colapso en diferentes regímenes dinámicos. Encontramos que cuando la materia se aproxima a un régimen de colapso adiabático mientras está cerca de formar una región atrapada (es decir, cerca de cruzar su radio de Schwarzschild), la energía de vacío tiende a crecer ilimitadamente. Esto se relaciona con la divergencia presente en el estado de Boulware en horizontes, que a su vez se puede relacionar con la existencia de soluciones de las ecuaciones semiclásicas de Einstein de objetos estáticos sin horizonte capaces de imitar observacionalmente a los agujeros negros. Esto sugiere que la creciente energía de vacío en los regímenes de colapso lento puede estabilizar la materia en una configuración final sin horizonte.
	
	Después, estudiamos cómo la dinámica de tales objetos ultracompactos sin horizonte puede conducir a la emisión de radiación de Hawking (sin la formación de superficies atrapadas). Encontramos que un movimiento oscilatorio de su superficie da lugar a la emisión de ráfagas de radiación, mientras que un colapso lento que se aproxime asintóticamente al cruce del radio de Schwarzschild puede dar lugar a un espectro térmico similar al de los agujeros negros, pero con una temperatura modificada. También analizamos la estructura causal de esta última familia de espaciotiempos (en los que la formación de una superficie atrapada se aproxima asintóticamente), encontrando que poseen un horizonte de sucesos y, o bien un horizonte de Cauchy, o bien una separación en dos de la región asintótica futura (de género tiempo y nulo) entre el interior y el exterior.
	
	Por último, examinamos el contenido de energía de vacío en las proximidades del horizonte interno de un agujero negro, una región del espaciotiempo clásicamente conocida por su amplificación no lineal de perturbaciones. A nivel clásico, la evolución del horizonte interno en presencia de perturbaciones genéricas presentes en el medio astrofísico conduce a la inestabilidad llamada "inflación de masa", debido a la que la curvatura alrededor y por debajo de la posición inicial del horizonte interno crece exponencialmente, mientras que el propio horizonte interno tiende a acercarse al origen. A nivel semiclásico, encontramos un flujo de energía entrante negativo, similar al responsable de la evaporación de Hawking del horizonte externo. Sin embargo, el efecto que tiene este flujo sobre la geometría parece ser amplificado, creciendo exponencialmente y superando rápidamente la supresión Planckiana que sufre la dinámica semiclásica. De hecho, la inflación de masa clásica y el efecto semiclásico (que denominamos inflación del horizonte interno) actúan de forma opuesta sobre el horizonte interno: uno lo empuja hacia dentro y el otro hacia fuera. Los análisis que comparan ambos efectos sugieren que el semiclásico puede dominar en tiempos tardíos, haciendo posible que la región atrapada desaparezca desde dentro hacia fuera, en una escala temporal mucho más corta que el tiempo de evaporación de Hawking. Adicionalmente, observamos que las técnicas utilizadas para analizar la estabilidad semiclásica del horizonte interno también pueden aplicarse a otras geometrías con propiedades causales similares. En particular, examinamos la geometría de un espaciotiempo de warp drive y utilizamos su estructura causal para argumentar que una inestabilidad encontrada previamente en modelos bidimensionales puede evitarse en dimensiones más altas.
	
	Para completar la imagen de los agujeros negros en gravedad semiclásica, proponemos que, si las regiones atrapadas desaparecen en escalas de tiempo cortas, los objetos ultracompactos observados astrofísicamente pueden ser imitadores de agujeros negros que no tienen horizonte, formados a partir de materia que colapsa lentamente, cuyas condiciones iniciales se obtienen tras la disipación de una o varias iteraciones de formación de regiones atrapadas, inflación del horizonte interno y recolapso.
\end{abstract}

\newpage
\null
\thispagestyle{empty}

\noindent
\begin{center}
	\textbf{Acknowledgements}
\end{center}

\vspace*{0.2cm}

\noindent
I would like to express my gratitude to my supervisors, Carlos, Raúl and Luis, for teaching me most of what I know about black holes and quantum fields, and for all their work in our joint research endeavours. Also, I would like to thank Vitor Cardoso and Niayesh Afshordi for the hospitality and guidance they provided during my research visits.

I also thank my friends and colleagues, both in Madrid and in Granada, who have made my work and everyday life a lot more enjoyable during these past few years. Also, I thank my family, as well as my sister's cats (\textit{Fluff no1} and \textit{Fluff no2}), for continuing to provide a home whenever I need it. Also, shout-out to the ``Empanadas Margarita" shop near the Genil river in Granada, who make the best Argentinian empanadas (and possibly the best food) I have ever had. Last but not least, I would like to thank Kentaro Miura for the inspiration he has given to so many people, including myself.

I acknowledge financial support from the Spanish Government through the fellowship FPU17/04471, as well as the projects PID2020-118159GB-C43 and PID-2020-118159GB-C44.

\newpage
\thispagestyle{empty}
\tableofcontents
\thispagestyle{empty}

\newpage
\thispagestyle{empty}
\null

\setcounter{page}{0}


\fancyhead[R]{}
\fancyhead[L]{$\mathscr{I}$ntroduction}

\setcounter{secnumdepth}{-1}
\chapter{Introduction}\label{chI}

\addtocontents{toc}{\protect\thispagestyle{empty}}

\renewcommand{\theequation}{$\mathscr{I}$.\arabic{equation}}

A black hole (BH) in classical general relativity (GR) is a finite region of spacetime in which the future of all causal trajectories is disconnected from that of the rest of the universe. The geometric description of its inner region reveals a complex structure even in the classical theory~\cite{Wald1984,PoissonIsrael89,Belinsky1970,Marolf2012}, though this is all hidden behind an event horizon from which nothing can escape. Incidentally, no information to confirm the existence of this inner region as described by GR can escape either, making its study within this theory a purely mathematical exercise.

This is no longer the case when quantum corrections are considered. As Hawking showed \cite{Hawking1975}, when one takes into account the presence of quantum fields in a curved spacetime \cite{BD,Wald1995} with an event horizon, the mass of the BH will slowly be reduced, while an equivalent amount of energy will be emitted at infinity---a process known as Hawking evaporation. As the mass is depleted, the BH outer horizon would slowly shrink and reveal (i.e. bring into causal contact with the outside universe) the innermost regions of the object. This is the first indication that a complete description of BHs is physically necessary, including an appropriate regularisation or dynamical avoidance of classical singularities.

On the classical side, the GR description of the evolution of a BH interior has gone a long way since the first model of spherical collapse \cite{Oppenheimer1939}. It is now known that the region of large curvature begins much further out than what might be expected from looking at the Schwarzschild solution \cite{Marolf2012}; that the central singularity may well have a chaotic oscillatory character \cite{Belinsky1970}, and that another oscillatory singularity (albeit a milder one) develops at finite radii over a Cauchy horizon \cite{Ori1992}.

Although many works on semiclassical BH physics still treat the interior of these objects as essentially the Schwarzschild solution (particularly, many of those which analyse the issue of information loss, e.g.~\cite{Susskind2008,Giddings2019}), there have been analyses which partially incorporate a more realistic version of a classical BH interior \cite{Birrell1978,BalbinotPoisson93,Hollands2020a,Hollands2020b,Ori2019,Zilberman2022}. Still, a self-consistent semiclassical solution of gravitational collapse which incorporates all the inherent complexity of this problem does not yet exist. While it may well be the case that only a full quantum description of gravity can provide us with a consistent picture of BH spacetimes as a whole (see e.g. \cite{Ashtekar2005,Mathur2005,Ashtekar2018}), the semiclassical approach used by Hawking still has a wide range of potential applications, especially pertaining to BHs, which have not yet been fully explored.

It is the goal of this thesis to bring together and expand upon the different aspects of semiclassical BH analysis: from effects around the formation of trapped regions, to their short- and long-term evolution under backreaction, as well as the relation of this evolution to semiclassically-sustained static horizonless BH mimickers~\cite{Carballo-Rubio2017,Arrechea2021}. We work in asymptotically flat spacetimes, and test these effects with a quantum massless scalar field in the ``in" vacuum state, defined as the Minkowski vacuum at past null infinity. We also work in spherical symmetry, and our main tool for probing semiclassical corrections is the renormalised stress-energy tensor (RSET) of this massless field, calculated in the Polyakov approximation~\cite{Fabbri2005}, which makes use of dimensional reduction in the angular variables and a quantisation in a 1+1 dimensional spacetime to approximate the radial and temporal components of the RSET, much like an $s$-wave approximation.

We begin by studying the magnitude of semiclassical corrections in several spacetimes which represent a collapse of matter in different dynamical regimes. Building on previous work in this direction~\cite{Barcelo2008}, we find that this magnitude is large enough for backreaction to be relevant to the evolution when the collapse is slowed down or reversed just before a trapped region is formed~\cite{Barcelo2019}. This can be understood in terms of the fact that if the background dynamics is stopped before a trapped region forms, the ``in" vacuum state quickly relaxes to the static Boulware state, which approaches a divergent energy contribution at the horizon~\cite{Boulware1975,Arrechea2019,Fabbri2005b}. Particularly, we use a toy model with a spherical thin shell of matter to explore three different dynamical regimes: a small oscillation of the matter surface just above the Schwarzschild radius, an asymptotic approach toward this radius, and a crossing of this radius (and trapped region formation) at arbitrarily low speeds. In the oscillating case, we find a series of bursts of outgoing radiation emitted when the shell is closest to crossing its gravitational radius but bounces back, separated by periods of approximately thermal emission during the rest of each oscillation cycle. In the case of horizon crossing at low speeds, we find large values in all components of the RSET, tending to a divergence in the zero limit of the speed parameter, suggesting backreaction would become extremely large when matter collapses slowly, possibly leading to the formation of horizonless static ultracompact objects~\cite{Carballo-Rubio2017,Arrechea2021}.

In the case of asymptotic approach toward the formation of an apparent horizon, we find an exact thermal emission of Hawking-like radiation, although at a temperature lower than that of Hawking radiation of a Schwarzschild BH of the same mass. In light of this intriguing result, we subsequently analysed the causal structure of these spacetimes~\cite{Barcelo2020}, finding that an asymptotic tendency in time toward the formation of a trapped region, if continued indefinitely, is in fact sufficient to generate an event horizon, since the expansion of outgoing radial light rays can tend to zero sufficiently quickly for their overall radial displacement to be finite in infinite time. The temperature of the quantum emission of these objects depends not only on the surface gravity of the asymptotically formed apparent horizon, but also on the details of their dynamical approach toward this formation. This discrepancy leads to an exponential growth of the RSET components in the vicinity of the would-be horizon, suggesting that these periods of thermal BH behaviour could only last for a short time in a full self-consistent semiclassical evolution of objects without trapped surfaces.

We then turn our attention to semiclassical effects in standard scenarios of BH formation, where matter collapses quickly and the RSET is initially very small in comparison to the classical energy in the system. We perturbatively analyse backreaction, focusing in particular on the dynamics of the trapped region. Classically, when a trapped region forms, it usually does not extend all the way to the origin (at least initially). This is the case when a BH has electric charge or angular momentum~\cite{Wald1984}, or when an effectively classical regular central region is present~\cite{Ansoldi,Hayward}. The lower boundary of the trapped region is known as the \textit{inner apparent horizon}, and it has some unique geometric properties. On the one hand, geodesic observers which approach it experience a slow-down of their proper time parameter with respect to the outside universe, making their trajectories extendable past the bounds of the initial universe through what is known as a Cauchy horizon (as beyond it, Cauchy initial data is no longer sufficient to determine the evolution). On the other hand, small energy perturbations in its vicinity have a disproportionately large effect on the growth of spacetime curvature. This latter property in particular has lead to the discovery of the \textit{mass inflation instability}, wherein generic decaying classical perturbations present in the astrophysical medium lead to an exponential growth of curvature (and, in spherical symmetry, of the locally defined Misner-Sharp mass)~\cite{PoissonIsrael89,Ori1991,Ori1992}.

We first look at the effect that backreaction from the RSET has on a classically static inner horizon, as well as on an inner horizon which moves to the origin prior to the formation of a spacelike singularity (as e.g. happens in the Oppenheimer-Snyder model~\cite{Oppenheimer1939}). We find that the amplifying effect the inner horizon has on perturbations makes it so the seemingly negligible RSET can lead to drastic changes in the evolution of the trapped region~\cite{Barcelo2021}. Particularly, since the RSET violates energy positivity conditions~\cite{Curiel2014}, it actually pushes the inner horizon outwards, tending to reduce the size of the trapped region from the inside. Our perturbative analysis reveals that the initial tendency of this inner horizon semiclassical correction is exponential in time, making it clear that, on the one hand, a quasi-stationary approximation such as the one used to analyse Hawking evaporation would not be adequate, and on the other, that the time scale involved in inner-horizon-related dynamics is much shorter than the Hawking evaporation time. This latter conclusion in particular leads to the possibility that the whole semiclassical picture of BHs should be revised, and full attention should be placed on how the inner horizon evolve in realistic scenarios which involve both classical and semiclassical perturbations.

In an attempt to address this issue, we construct a toy model for mass-inflation-inducing classical perturbations based on spherical thin shells interacting with a generic (spherical) BH with an inner horizon. We analyse when and how mass inflation is triggered from this interaction, and then use these backgrounds to once again perform a perturbative semiclassical backreaction analysis~\cite{Barcelo2022}. The classical and semiclassical perturbations basically have opposite effects: the former pushing the inner horizon inwards, while the latter pushing it outwards; we find that initially, the classical evolution continues, but after a short time (linear in the mass) the semiclassical outward push tends to dominate. This is further corroborated by the asymptotic analyses performed in~\cite{Hollands2020a,Hollands2020b,Ori2019,Zilberman2022}, where it is found that semiclassical backreaction at the Cauchy horizon, if such a horizon were to form, would dominate over classical mass inflation, forming a stronger curvature singularity. Extrapolating from our analysis, we argue that even in the presence of mass inflation, the inner horizon may still end up having an outward movement, extinguishing the trapped region from the inside and changing the lifetime of BHs to an extremely short one. If a full numerical analysis of the semiclassical Einstein equations were to indeed lead to this outcome, then the extremely compact dark astrophysical objects observed~\cite{LIGO,EHT} may likely turn out to be horizonless BH mimickers, such as those obtained semiclassically~\cite{Carballo-Rubio2017,Arrechea2021}.

Under this hypothesis, we conjecture that the full picture of BHs in semiclassical gravity consists in the following. Initially, a trapped region is formed by quickly collapsing classical matter, followed by an equally quick outward inflation of the inner horizons and disappearance of this trapped region. After some energy is dissipated from this process, the dispersed classical matter recollapses under its own gravity. With possibly several iterations of this process, enough energy is dissipated for the initial recollapse conditions to be those of ``slow collapse" in which the Boulware-like terms in the RSET are ignited, leading to a final relaxation into a semiclassical ultracompact horizonless BH mimicker.

This thesis is based on several publications~\cite{Barcelo2019,Barcelo2020,Arrechea2020,Barcelo2021,Barcelo2022,Boyanov2022,Barcelo2022w}. The structure is as follows. The remainder of the \hyperref[chI]{Introduction} chapter presents a brief overview of the theory of quantum fields in curved spacetimes, explaining the ambiguities in quantising on non-trivial backgrounds and how they can translate into different possible values for the RSET, and thus to different self-consistent evolutions of spacetime. We briefly show how quantisation is performed and how the RSET is calculated in a 1+1 dimensional spacetime, on the one hand, because the simplicity allows for analytical results, and on the other, because we then use the result to approximate the RSET for spherically-symmetric 3+1 dimensional spacetimes. A short review of how this theory has been applied to the study of BHs thus far is also presented.

In \hyperref[pt1]{part I} we study the magnitude of semiclassical effects in the vicinity of horizon formation for collapse scenarios with different classical dynamics. In \hyperref[ch1]{Chapter 1} we start by looking at the model of an oscillating thin shell which periodically approaches crossing its own Schwarzschild radius only to bounce back outwards each time. We find periodic emission of nearly thermal Hawking-like radiation, separated by non-thermal bursts, along with large values of the RSET. In \hyperref[ch2]{Chapter 2} we use the same model, but allow the matter surface to cross the Schwarzschild radius, and do so at arbitrarily low speeds. We find that nearly static classical matter close to crossing this radius results in large values of the RSET. In \hyperref[ch3]{Chapter 3} we analyse a matter distribution which collapses so slowly that it only approaches the formation of a trapped surface asymptotically in time. We find an emission of thermal Hawking-like radiation along with a growing value of the RSET, as well as some interesting causal behaviours.

In \hyperref[pt2]{part II} we study semiclassical corrections to BHs with an inner apparent horizon. In \hyperref[ch4]{Chapter 4} we begin by reviewing the classical behaviour of such a horizon, focusing in particular on the instability under perturbations known as ``mass inflation". We present a simple model which contains the essential characteristics necessary to reproduce this instability, while also being analytically solvable. On the other hand, in \hyperref[ch5]{Chapter 5} we focus on the semiclassically sourced evolution of an inner horizon in the absence of classical perturbations, obtaining the result that trapped regions have a tendency to evaporate not only form the outside, but also from the inside. In \hyperref[ch6]{Chapter 6} we take a brief detour, wherein we apply the conceptual results of the inner horizon analysis to determine the semiclassical stability of a different class of spacetimes: the warp drive. Finally, in \hyperref[ch7]{Chapter 7} we return to BHs and put together the classical and semiclassical perturbations in order to see what the evolution of a generic inner horizon may be. We find a tendency for classical evolution to dominate initially, but for semiclassical effects to become relevant after a very short timescale (of the order of the BH mass in geometric units).

In the \hyperref[chC]{Conclusions} we summarise our results and link them together to list the possibilities of what the ultimate fate of BHs in semiclassical gravity could be.

We use the metric signature $(-,+,+,+)$. Greek letters $\mu,\nu$, etc. are used for spacetime indices in 3+1 dimensions, and Latin letters $a,b$, etc. for indices in 1+1 dimensions, unless otherwise stated. In general we use natural units, where $c=1$, $G=1$ and $\hbar=1$, except when writing the semiclassical Einstein equations, where having $\hbar$ written explicitly is useful for comparison between the classical and semiclassical terms (to make the dimensionality of the constant clear we in fact use the square of the Planck length, $l_{\rm p}^2$, rather than $\hbar$).

\section{$\mathscr{I}$-1. Preliminaries to quantisation in curved spacetimes}

The quantisation of a field in Minkowski spacetime follows a well established procedure (see e.g.~\cite{Peskin1995}), the physical validity of which is shown by a myriad of experimental results. However, this procedure relies heavily on the symmetries present in flat spacetime (i.e. Poincaré invariance), making its generalisation to curved spacetimes a complicated task. We begin by briefly explaining the method of this generalisation and its inherent ambiguities, following in part the discussion presented in~\cite{BD}.

Here and throughout the thesis, we make use of a massless real scalar $\phi$ minimally coupled to gravity. Given that we will work in spherical symmetry, the quantisation of this field is sufficient to probe generic effects from the presence of quantised matter. Its action is
\begin{equation}
S_{\phi}=\int d^4x\sqrt{-g}\,\mathcal{L}_\phi=-\frac{1}{2}\int d^4x\sqrt{-g}\,\nabla^\mu\phi\nabla_\mu\phi,
\end{equation}
where $g$ is the determinant of the metric tensor $g_{\mu\nu}$, and $\mathcal{L}_\phi$ is the Lagrangian density. The equation of motion which determines the dynamics of this field, known as the Klein-Gordon equation, is simply
\begin{equation}\label{KG}
\nabla^\mu\nabla_\mu\phi=0,
\end{equation}
where $\nabla^\mu$ is the covariant derivative compatible with $g_{\mu\nu}$.

\subsection{$\mathscr{I}$-1.1. Flat spacetime quantisation}

In flat spacetime, using Cartesian coordinates adapted to an inertial observer we have $\nabla_\mu=\partial_\mu$. We define the Klein-Gordon product,
\begin{equation}\label{KGprod}
(\phi_1,\phi_2)_{\rm KG}=-i\int_{\mathbf{\Sigma}_t}\phi_1\overset{\text{\tiny$\leftrightarrow$}}{\partial}_t\phi_2^*\,d\mathbf{\Sigma}_t,
\end{equation}
where $\mathbf{\Sigma}_t$ is a $t=$ const.~hypersurface, $t$ being $x^0$, and $f\overset{\text{\tiny$\leftrightarrow$}}{\partial}_t g=f\partial_tg-(\partial_tf)g$. For pair of solutions to the Klein-Gordon equation, this constitutes a pseudo-scalar product (as the norm it generates is not positive-definite), the values of which are independent of the particular choice of $t=$ const.~slice. A basis of solutions to the Klein-Gordon equation which is orthonormal with respect to the product \eqref{KGprod} are the normalised plane waves
\begin{equation}\label{Mmode}
\begin{split}
u_\mathbf{k}&=\frac{1}{\sqrt{2\omega(2\pi)^3}}e^{-i\omega t+i\mathbf{k}\cdot\mathbf{x}}\quad\text{with norm }\delta(\mathbf{k}-\mathbf{k}'),\\
v_\mathbf{k}=u_\mathbf{k}^*&=\frac{1}{\sqrt{2\omega(2\pi)^3}}e^{i\omega t-i\mathbf{k}\cdot\mathbf{x}}\quad\text{with norm }-\delta(\mathbf{k}-\mathbf{k}'),
\end{split}
\end{equation}
where $\omega=|\mathbf{k}|$ is the frequency and $\mathbf{k}$ is the wave number, $\mathbf{x}$ being the 3-vector of the spatial Minkowski coordinates. A generic solution can be expanded in terms of this basis as
\begin{equation}
\phi=\sum_\mathbf{k}(a_\mathbf{k}u_\mathbf{k}+a_\mathbf{k}^* u_\mathbf{k}^*).
\end{equation}
Formally, if the domain in $\mathbf{x}$ is infinite, then $\mathbf{k}$ becomes continuous and should be integrated over rather than summed, here and in the expressions below.

The basis \eqref{Mmode} allows us to divide solutions into two subspaces: one comprised of solutions which can be obtained from an expansion solely in $u_{\mathbf{k}}$'s, which we will denote by $S_\oplus$, and one of solutions which are obtained from $v_{\mathbf{k}}$'s, which will be denoted by $S_\ominus$. The $S_\oplus$ functions are referred to as \textit{positive-frequency} solutions, and the $S_\ominus$ ones as \textit{negative-frequency} solutions. Any solution can be expressed as $s=s_++s_-$, where $s_+\in S_\oplus$ and $s_-\in S_\ominus$. Note that while all positive-frequency solutions have positive norm, the reverse is not true: a solution with an overall positive norm can have negative-frequency components. In other words, the product \eqref{KGprod} is not sufficient to define a unique positive- and negative-frequency subspace separation. This will play a key role in the ambiguity of quantisation in curved spacetimes.

The quantisation of this field is performed by promoting $\phi$ to an operator $\hat{\phi}$ which satisfies the equal time commutation relations
\begin{equation}\label{Mcomm}
\begin{split}
[\hat{\phi}(t,\mathbf{x}),\hat{\phi}(t,\mathbf{x}')]&=0,\\
[\hat{\pi}(t,\mathbf{x}),\hat{\pi}(t,\mathbf{x}')]&=0,\\
[\hat{\phi}(t,\mathbf{x}),\hat{\pi}(t,\mathbf{x}')]&=i\delta^3(\mathbf{x}-\mathbf{x}'),
\end{split}
\end{equation}
where $\hat{\pi}=\partial_t\hat{\phi}$ is the canonical conjugate of the field. The field operator can now be expanded in terms of the same mode basis, but the coefficients in the expansion become operators themselves,
\begin{equation}\label{flatfield}
\hat{\phi}=\sum_\mathbf{k}(\hat{a}_\mathbf{k}u_\mathbf{k}+\hat{a}_\mathbf{k}^\dagger u_\mathbf{k}^*).
\end{equation}
The operators $\hat{a}_\mathbf{k}$ and $\hat{a}^\dagger_\mathbf{k}$ are known as particle annihilation and creation operators, which act on the Hilbert space which spans all quantum states. Thanks to the orthonormality of the modes \eqref{Mmode}, the relations \eqref{Mcomm} translate into the simple commutation relations for $\hat{a}_\mathbf{k}$ and $\hat{a}^\dagger_\mathbf{k}$,
\begin{equation}
\begin{split}
[\hat{a}_\mathbf{k},\hat{a}_{\mathbf{k}'}]&=0,\\
[\hat{a}_\mathbf{k}^\dagger,\hat{a}_{\mathbf{k}'}^\dagger]&=0,\\
[\hat{a}_\mathbf{k},\hat{a}_{\mathbf{k}'}^\dagger]&=\delta_{\mathbf{kk}'},
\end{split}
\end{equation}
which are akin to those of a harmonic oscillator. We can then define the vacuum state as the one annihilated by all the $a_\mathbf{k}$ operators,
\begin{equation}
\hat{a}_{\mathbf{k}}\ket{0}=0,\quad \forall \mathbf{k}.
\end{equation}
A state of $n$ particles with momentum $\mathbf{k}$ is then obtained by acting $n$ times on $\ket{0}$ with the creation operator $\hat{a}_{\mathbf{k}}^\dagger$. A quantum field can thus be thought of as an infinite assembly of quantum harmonic oscillators.

Ultimately, the plane-wave basis used for the flat spacetime quantisation is, in a sense, a natural choice, given the symmetry of the background: the total number of particles in this quantisation is Poincaré invariant, though their momentum is of course relative. In other words, if the whole procedure were carried out in any other set of inertial Minkowski coordinates, the resulting vacuum state would be the same, and the momentum of particle states would be related by a boost.

\subsection{$\mathscr{I}$-1.2. Curved spacetime quantisation}

If we attempt to repeat the same procedure in an arbitrary curved spacetime background, then we quickly run into problems. In a globally hyperbolic spacetime~\cite{Hawking1973}, we can define a covariant generalisation of the Klein-Gordon product,
\begin{equation}\label{KGprod2}
	(\phi_1,\phi_2)_{\rm KG}=-i\int_{\mathbf{\Sigma}_t}\phi_1\overset{\text{\tiny$\leftrightarrow$}}{\nabla}_\mu\phi_2^*\,n^\mu\sqrt{-h}\,d\mathbf{\Sigma}_t,
\end{equation}
where $t$ is any time function and $\mathbf{\Sigma}_t$ is again a $t=$ const.~surface, $n^\mu$ is a unit vector orthogonal to this surface, and $h$ is the determinant of the induced metric on the surface. For solutions to the Klein-Gordon equation, this product is invariant under changes of the time function $t$. Any solution $s$ therefore has a specific well-defined norm, and orthonormal bases, such as the one used for quantisation in flat spacetime, can be constructed. However, there are infinitely many such bases, which generally have a different separation into $s_+$ and $s_-$ sectors, (which determine the separation between creation and annihilation variables and thus define the vacuum state).

Quantising the field in a curved spacetime brings with it two difficulties. The first is a practical one: obtaining explicit expressions for any basis of solutions to \eqref{KG} is generally a very challenging task. Only in a small handful of spacetimes can any basis be written in terms of known functions. The second difficulty is a conceptual one: for generic spacetimes, there is no a priori indication as to which one of the infinitely many bases should be chosen for quantisation~\cite{Ashtekar1975,BD,Wald1995}, different choices generally leading to physically inequivalent results.

To see that this is the case, let us first write the field operator as
\begin{equation}\label{field}
\hat{\phi}=\sum_i(\hat{a}_iu_i+\hat{a}^\dagger_iu_i^*),
\end{equation}
where $u_j$ is a set of functions which form an orthonormal basis with respect to the Klein Gordon product, and $u_i$ and $u_i^*$ generalise the positive- and negative-frequency subspaces of solutions (the range of the $i$ index would then be half of the number of elements in the basis). The separation into these two subspaces is unique for each basis of solutions (as defined by the associated complex structure \cite{Ashtekar1975,Wald1995}) and, being directly related to the definition of the creation and annihilation operators, defines the vacuum and particle states. If we consider an expansion using an alternative basis,
\begin{equation}\label{field2}
\hat{\phi}=\sum_i(\hat{b}_i\tilde{u}_i+\hat{b}^\dagger_i\tilde{u}_i^*),
\end{equation}
where $\tilde{u}_j$ are also orthonormal, and $\hat{b}$ and $\hat{b}^\dagger$ are our new annihilation and creation operators, then the annihilation operator of the first quantisation can be written as
\begin{equation}
\hat{a}_i=\sum_k(\alpha_{ki}\hat{b}_k+\beta^*_{ki}\hat{b}^\dagger_k),
\end{equation}
where
\begin{equation}\label{beta}
\alpha_{ki}=(\tilde{u}_k,u_i)_{\rm KG},\quad \beta_{ki}=-(\tilde{u}_k,u_i^*)_{\rm KG}
\end{equation}
are known as the Bogolyubov coefficients. When $\beta_{ki}\neq 0$, the vacuum state of one quantisation can appear full of particles for another, and vice versa. Therefore, some additional criterion appears to be needed in order to single out a particular basis and vacuum state.

In flat spacetime, this criterion is derived from the symmetries of the background. The free quantisation which respects these symmetries lies at the base of the standard model of particle physics. The same criterion can be extended directly to stationary spacetimes, where the timelike Killing vector field can be used for a unique separation between the positive- and negative-frequency subspaces $S_\oplus$ and $S_\ominus$, and consequently to a set of equivalent quantisations~\cite{Ashtekar1975}. Particularly, when the temporal part of the Klein-Gordon equation is separable, specifying a basis of solutions for this part already allows for a separation between the two subspaces, and choosing different spatial components for the remainder of the basis function alone cannot mix them.

Another case which allows for a natural choice of quantisation is found in spacetimes which are stationary at least in a region large enough to define the initial conditions of a field solution. The quantisation is then defined as the preferred one in the region of symmetry, and then the corresponding modes are evolved through the subsequently dynamical spacetime. For instance, this can be done in cosmology if one considers that the scale factor of the Friedmann–Lemaître–Robertson–Walker geometry is initially constant for some duration of time~\cite{Cortez2015}. For  asymptotically flat spacetimes, a similar approach can be applied. Particularly, the spacetime need only be flat at past null infinity in order to provide preferred initial conditions for the modes of a massless field. We will refer to the vacuum state of this quantisation as the ``in" state~\cite{Hawking1975,Fabbri2005,Frolov2017}, and we will use it throughout this thesis for calculating the RSET probing semiclassical effects in spacetimes of gravitational collapse.

For generic dynamical spacetimes, however, the choice of modes and vacuum state remains ambiguous. This is why much of the techniques developed for working with quantum fields in curves spacetimes, such as the covariant renormalisation we will discuss below, are formulated in a way which can be applied for any choice of modes. The matter of how the ambiguity is fixed in our universe as a whole is, for the semiclassical theory, relegated to a choice of initial conditions, which can only be determined by the observation of cosmological deviations from classicality.

\section{$\mathscr{I}$-2. Semiclassical gravity}

The physical implications of the quantisation ambiguity presented above may not appear immediately clear, as the discussion so far has been focused on the dynamics of a single field free of any interaction. In order to endow this field with a physical meaning, one has to model its interactions, either with other fields or with some effective detector system. For a field in curved spacetimes, it turns out that interaction is in fact inevitable, as the covariant description of its dynamics already couples it (minimally) to gravity, and thus endows it with a physical role as a source of curvature. In other words, one may define a field with no interactions to other fields or detectors, but so long as it has energy it will gravitate, and thus have an observable effect.

This brings with it an issue of its own: namely, after quantising the field, it would still have to dictate how the (seemingly) classical spacetime curves. What would the gravitational field of a superposition of states look like? Consistently coupling these systems requires either the quantisation of gravity (allowing for superpositions of gravitational fields and causalities)~\cite{DeWitt1967,Aharony1999,Ashtekar2004}, or a ``classicalisation" of matter when it reaches the energy scale at which it interacts gravitationally~\cite{Diosi1989,Penrose1996,Barcelo2021b}. Despite the fact that, as of yet, no approach in either direction has been fully successful, probing the interface between the realms of gravity and quantum matter is still possible, albeit only approximately and in certain situations.

The theory of semiclassical gravity describes the evolution of a classical spacetime with a quantum stress-energy tensor source, which is brought to classicality by means of an expectation value, i.e. the RSET mentioned above. This theory can be argued to be the leading order contribution in a fully quantum system of gravity and matter~\cite{Birrell1978,Anderson1983}, and in general is intuitively expected to provide a good approximation when quantum matter is not highly delocalised (c.f.~\cite{Ford1982,Kuo1993,Diosi1998,Hu2008}). From a classical standpoint, it is merely a modified version of Einstein's theory in which source terms (the right-hand side of the field equations) have a dependence on the derivatives of the metric of order higher than one (unlike in GR). From a quantum standpoint, it reflects the ambiguity in the quantisation (different choices leading to different spacetime evolutions) and the change in the energy content of the vacuum and particle states as the background evolves, properly encoding these characteristics in the source term~\cite{BD,Wald1995}. In practice, the RSET is usually computed in the vacuum state of the chosen quantisation, and focus is placed on the counter-intuitive fact that this state has an energy content even after renormalisation.

Let us explicitly see how this theory is constructed in the case of the massless scalar seen above. Starting from the classical theory, when considering the dynamics of spacetime, the full action of the theory must include the Einstein-Hilbert term \cite{Misner1973}, becoming
\begin{equation}
S=\int\mathop{d^4x}\sqrt{-g}\left(\frac{R}{16\pi}-\frac{1}{2}\nabla^\sigma\phi\nabla_\sigma\phi\right),
\end{equation}
where $R$ is the Ricci scalar (the trace of the Ricci tensor $R_{\mu\nu}$). The variation of this action with respect to the metric gives the Einstein equations,
\begin{equation}
R_{\mu\nu}-\frac{1}{2}g_{\mu\nu}R=8\pi T_{\mu\nu},
\end{equation}
with the scalar field stress-energy tensor
\begin{equation}\label{SET}
T_{\mu\nu}=\nabla_\mu\phi\nabla_\nu\phi-\frac{1}{2}g_{\mu\nu}\nabla^\sigma\phi\nabla_\sigma\phi.
\end{equation}
Once the field is quantised, a stress-energy tensor operator can be constructed by substituting $\phi\to\hat{\phi}$ (if the field is real and the operator $\hat{\phi}$ is self-adjoint, then no additional symmetrisation is needed).\footnote{Alternatively, the operator expectation value can be constructed from a variational principle of an effective action~\cite{BD,Barcelo2012}, which provides alternative methods for renormalisation to the one presented below, giving equivalent results.} The resulting tensor operator, being quadratic in the field, has a divergence in its expectation value. In flat spacetime, this is none other than the divergence which is subtracted through ``normal ordering" of the creation and annihilation operators \cite{Peskin1995}, equivalent to simply subtracting the expectation value in vacuum from the ones in all other states. In curved spacetimes, rather than directly subtracting the vacuum state value, a more involved renormalisation procedure is followed, making use of the local geometric nature of the diverging terms~\cite{BD,Wald1995}, as we will see. Once the procedure is complete, we are left with a prescription for the RSET which sources the semiclassical Einstein equations.

\subsection{$\mathscr{I}$-2.1. RSET in spherically symmetric spacetimes}

To summarise, calculating the RSET of our scalar field involves finding a basis of solutions to eq.~\eqref{KG} corresponding to a physically reasonable quantisation (e.g. the ``in" quantisation in asymptotically flat spacetimes \cite{Fabbri2005,Frolov2017}, or adiabatic quantisation in cosmology \cite{Parker1968}), then using the field operator \eqref{field} to construct the stress-energy tensor operator through the functional expression \eqref{SET}, and computing its renormalised expectation value in the vacuum state of the chosen quantisation. This calculation can only be performed analytically in cases with either a high degree of symmetry, such as a conformally coupled field in a homogeneous and isotropic cosmological model \cite{Davies1977,Anderson1983}, or for a lower number of dimensions, such as the 1+1 case \cite{DF}, or a combination of the two, such as the 2+1 dimensional BH \cite{Casals2017}.

For BH spacetimes in 3+1 dimensions, one seems to be left with no other choice but to attempt to calculate the RSET numerically. Indeed, tremendous progress has been made in this direction over the past decades \cite{Howard1984,Anderson1995,Ottewill2012,Taylor2017,Ori2017,Taylor2021,Anderson2020}. However, the trade-off for the precision which these calculations provide lies in the fact that they can be performed only on certain fixed backgrounds, and only in certain quantum vacuum states. The full process of gravitational collapse and BH evolution in semiclassical gravity cannot as yet be treated self-consistently with this approach.

A trade-off in the opposite direction can be made by sacrificing the precision of the result (indeed, only preserving its qualitative character in certain regions) for a tool which can be applied to calculating the RSET in a variety of dynamical geometries which represent the formation and evolution of BHs and stellar objects: the Polyakov approximation \cite{Fabbri2005}. This approximation consists in dimensionally reducing a spherically symmetric spacetime by integrating out the angular variables, then quantising in the 1+1 dimensional radial-temporal sector and calculating the RSET there, and finally applying the result back to 3+1 dimensions.

The line element of a 3+1 dimensional spherically-symmetric geometry can be written as
\begin{equation}\label{46}
ds^2=h_{ab}(t,r)x^ax^b+r^2d\Omega^2,
\end{equation}
where the indices $a,b$ refer to the temporal and radial dimensions and $h_{ab}$ is a $2\times2$ tensor. We can consider a field $\phi$ propagating on this spacetime for which the $s$-wave contribution is dominant, i.e. which essentially has a dependence only on the radial and temporal coordinates, $\phi=\phi(t,r)$. As a functional of the field, the Lagrangian density $\mathcal{L}_\phi$ also loses its angular dependence, and the part of the action dictating the field dynamics can then be reduced to two dimensions by integrating out the angular variables
\begin{equation}\label{action}
S_\phi=\int\mathop{d^4x}\sqrt{-g}\,\mathcal{L}_\phi(t,r)=4\pi\int\mathop{d^2x}\sqrt{-h}\,r^2\,\mathcal{L}_\phi(t,r),
\end{equation}
where $h$ is the determinant of $h_{ab}$. The relation between the stress-energy tensor in the four-dimensional theory $T_{\mu\nu}^{(4)}$ and the stress-energy tensor corresponding to the dimensionally-reduced field dynamics $T_{ab}^{(2)}$ is then
\begin{equation}\label{dimred}
T_{\mu\nu}^{(4)}=\frac{\delta_\mu^a\delta_\nu^bT_{ab}^{(2)}}{4\pi r^2}.
\end{equation}
We note, however, that the two dimensional theory described by \eqref{action} is not that of a free field---a part of the four-dimensional kinetic term takes on the role of a potential after the dimensional reduction, namely the part responsible for backscattering in the $s$-wave sector.
	
The Polyakov approximation to the RSET is obtained with two simplifying assumptions. The first one is that this potential in the equation of motion of the dimensionally-reduced field can be disregarded. How well this assumption is justified varies depending on the spacetime in question, but it is worth noting in particular that the potential is in fact zero both at infinity and at the horizons of BHs~\cite{Nollert1999}. It therefore may be expected that horizon-related effects are captured well in this approximation; indeed, Hawking's result only changes slightly when one includes backscattering (correcting the black-body spectrum with grey-body factors)~\cite{BD}. The second assumption is that the two-dimensional theory retains at least a qualitative similarity to the full four-dimensional one even after quantisation and renormalisation. In other words, that eq.~\eqref{dimred} can be applied to relate the RSET of the two dimensional theory with an approximation to the four dimensional RSET. Of course, this approximation is far from exact, as dimensional reduction and renormalisation do not commute \cite{Frolov1999}. Additionally, the Polyakov approximation becomes less reliable the closer one gets to the origin of spherical coordinates $r=0$, as it tends to a non-physical divergence there. In spite of these issues, it works well in providing an analytical expression for a conserved RSET which captures some of the essential characteristics of quantum field theory in curved spacetimes, such as the ambiguity in the choice of vacuum and the violation of energy conditions. For the purposes of this work, it is important to note that it does capture some well-established horizon-related effects, such as Hawking evaporation, the Boulware state outer horizon divergence, and the Hadamard state Cauchy horizon divergence (all of which will be explained in more detail below)~\cite{DFU,Boulware1975,Birrell1978,Fabbri2005}. We will therefore work with this approximation to probe semiclassical effects near the edges of trapped regions or, more generally, in regions where the causal structure approaches a light-trapping behaviour.

\subsection{$\mathscr{I}$-2.2. Quantisation and renormalisation in 1+1 dimensions}

To construct the Polyakov RSET for spherically symmetric spacetimes, we need to quantise our test field in 1+1 dimensions, isolate the divergence in the expectation value of the stress-energy tensor operator, subtract this divergence in a covariant manner (i.e. with counterterms proportional to tensor quantities), and finally plug the result into the approximation \eqref{dimred}. Luckily, in 1+1 dimensions the system is simple enough for this whole procedure to be done analytically, even while keeping the spacetime and the choice of quantum modes arbitrary. This fact in particular allows not only the easy calculation of the RSET on fixed backgrounds, but also to determine the background at the same time as the Polyakov RSET in a semiclassically self-consistent manner.

In 1+1 dimensions, a spacetime metric has only one physical degree of freedom, and therefore takes on a conformally flat form in appropriate sets of coordinates. For the calculation of the RSET, it is convenient to work in double null coordinates,
\begin{equation}\label{metric2d}
	ds^2=-C(u,v)\,du\,dv,
\end{equation}
where $C$ is the conformal factor. The Klein-Gordon equation for the minimally coupled scalar \eqref{KG} in this coordinate system takes the form
\begin{equation}
	\phi_{,uv}=0,
\end{equation}
where the comma indicates partial differentiation. Note that this equation is conformally invariant, and therefore has the same form as in flat spacetime for any function $C$. Analogously to the flat spacetime case, a basis of solutions which is orthonormal with respect to the Klein-Gordon product \eqref{KGprod2} is given by the right- and left-moving plane waves
\begin{equation}\label{modes}
	\phi_\omega(u)=\frac{1}{\sqrt{4\pi\omega}}e^{-i\omega u},\quad\phi_\omega(v)=\frac{1}{\sqrt{4\pi\omega}}e^{-i\omega v},
\end{equation}
and their complex conjugates, with $\omega$ being a positive real number (the frequency of the wave). Quantization can be performed by defining the field operator
\begin{equation}\label{phi}
	\hat{\phi}=\sum_\omega\left[\hat{a}^R_\omega\phi_\omega(u)+(\hat{a}^R_\omega)^\dagger\phi_\omega^*(u)+\hat{a}^L_\omega\phi_\omega(v)+(\hat{a}^L_\omega)^\dagger\phi_\omega^*(v)\right],
\end{equation}
with the creation and annihilation operators for each type of mode satisfying the non-zero commutation relations
\begin{equation}\label{commut}
	[\hat{a}^R_\omega,(\hat{a}^R_{\omega'})^\dagger]=[\hat{a}^L_\omega,(\hat{b}^L_{\omega'})^\dagger]=\delta_{\omega\omega'}.
\end{equation}
The vacuum state $\ket{0}$ is defined by $\hat{a}^R_\omega\ket{0}=\hat{a}^L_\omega\ket{0}=0$. An orthonormal basis of the standard Fock space containing all particle states can then be obtained by applying the creation operators $(\hat{a}^R_\omega)^\dagger$ and $(\hat{a}^L_\omega)^\dagger$ on the vacuum state, with appropriate normalization.

Since the field operator in \cref{phi} is constructed to be self-adjoint, its Lagrangian density is the same as that of the real, massless scalar field, and its corresponding stress-energy tensor operator can be obtained directly from \cref{SET} with the substitution $\phi\to\hat{\phi}$. The vacuum expectation value of this tensor can be obtained by calculating the expectation value $\bra{0}\nabla_\mu\hat{\phi}\nabla_\nu\hat{\phi}\ket{0}$. With \cref{phi} and the commutation relations in \cref{commut}, we get
\begin{equation}\label{expv1}
\begin{split}
&\bra{0}\nabla_u\hat{\phi}\nabla_v\hat{\phi}\ket{0}=0,\\
&\bra{0}\nabla_u\hat{\phi}\nabla_u\hat{\phi}\ket{0}=\sum_\omega\omega^2\phi_\omega(u)\phi^*_\omega(u)=\frac{1}{4\pi}\sum_\omega\omega,\\
&\bra{0}\nabla_v\hat{\phi}\nabla_v\hat{\phi}\ket{0}=\sum_\omega\omega^2\phi_\omega(v)\phi^*_\omega(v)=\frac{1}{4\pi}\sum_\omega\omega.
\end{split}
\end{equation}
As the frequency $\omega$ is not bounded from above, the last two expressions are manifestly divergent, and consequently so is the expectation value of the stress-energy tensor.

The divergence of the expectation value of operators which are quadratic in the field is a well known problem from quantum field theory in flat spacetime. There, the standard procedure is to perform ``normal ordering" of the creation and annihilation operators or, equivalently, to subtract the infinity of the expectation value in the vacuum state off of the ``same type of infinities" in the expectation values in other states, the result being a finite value. This procedure works well due to the fact that, in the absence of coupling to spacetime, only the differences between energy states are measurable, making the absolute value of energy physically irrelevant. However, in curved spacetimes this is no longer the case: there is no reason to believe that zero-point energy would not gravitate. There have been arguments that instead of a full subtraction, an ultraviolet cutoff should be put on the sums and integrals of the form \eqref{expv1}, giving rise to a vacuum energy which would manifest itself globally in spacetime as a cosmological constant. However, putting the cutoff at the Planck scale, where the semiclassical approximation is expected to break down, makes this type of energy much larger than the observed cosmological constant~\cite{Weinberg1988}. The interpretation of this issue is still an open problem.

Performing a full subtraction of the divergence, rather than putting in an arbitrary cutoff, seems like the more reasonable choice. However, the construction of the counter-terms should be done carefully. While it is possible to directly subtract the whole vacuum expectation value in any given quantisation, resulting in zero vacuum energy and finite expectation values in particle states, there are strong indications that this (generally non-covariant) procedure is not the most reasonable option. Firstly, calculations of beta coefficients \eqref{beta} between the ``preferred" (in terms of symmetry) vacuum states in different regions of evolving spacetimes suggest the creation of particles by the gravitational field~\cite{Parker1968,Hawking1975}, which should be taken into account energetically. Secondly, the nature of the divergence turns out to be such that the counter-terms can be local in curvature tensors, making them the same for any choice of vacuum state (of Hadamard type, i.e. stemming from sufficiently regular mode solutions)~\cite{BD,Wald1995,Christensen1976}. After such a subtraction, the vacuum expectation value of the stress-energy tensor operator is generally non-zero (though finite) when spacetime is not flat.

One of the most used methods for isolating these divergent terms is known as \textit{covariant point-splitting}~\cite{Christensen1976}, and it consists of expressing the two-point correlation function $\bra{0}\nabla_\mu\hat{\phi}(x^\rho)\nabla'_{\nu'}\hat{\phi}(x'^\rho)\ket{0}$ as a Laurent series in a geodesic distance parameter which becomes zero when $x'^\rho\to x^\rho$. The precise method used in \cite{DF}, where the 1+1 dimensional calculation was originally performed, involves symmetrically separating the two terms in the two-point function in opposite directions along an arbitrary geodesic passing through the initial centre point. Both covariant derivatives are translated to the tangent spaces of the new points through parallel transport and evaluated there, leaving all terms as functions of the original point and the derivatives of the geodesic curve, as well as the small distance parameter. Note that this is slightly different from simply evaluating the two-point function directly at the two final points, as it gives a covariant prescription of how the coincidence limit is to be approached.

We have summarised the calculation of the 1+1 RSET, performed in~\cite{DF}, in \hyperref[RSETcalc]{Appendix A}. After the covariant subtraction of the divergence, the RSET of this theory becomes
\begin{equation}\label{e33}
\expval{T_{\mu\nu}}=\frac{1}{48\pi}Rg_{\mu\nu}+\Theta_{\mu\nu},
\end{equation}
where the tensor $\Theta_{\mu\nu}$ is defined through its value in the null coordinates of the mode basis as
\begin{equation}
\Theta_{\mu\nu}=-\frac{\sqrt{C}}{12\pi}\text{diag}\left(\partial_u^2\frac{1}{\sqrt{C}},\partial_v^2\frac{1}{\sqrt{C}}\right)_{\mu\nu}.
\end{equation}
The first term in the RSET \eqref{e33} is expressed in terms of curvature tensors, and is thus independent of the particular choice of modes and vacuum state of the quantisation. The second term, $\Theta_{\mu\nu}$, is the one which encodes the information about the chosen vacuum state, and is also the one which can become significant even in regions of low curvature. The RSET is conserved covariantly, making it an appropriate source term for the semiclassical Einstein equations. It also reduces to zero for the Minkowski quantisation of flat spacetime, making this spacetime a solution of the semiclassical equations as well.

\subsection{Vacuum states and thermal particle fluxes}

As we discussed above, the quantisation defined in \eqref{phi} depends on the choice of modes \eqref{modes}, which in this case is encoded in a pair of null coordinates $\{u,v\}$ in which these take the form of plane waves. This dependence can be seen now through the fact that with a different pair of null coordinates $\{\tilde{u},\tilde{v}\}$, the metric \eqref{metric2d} would take the form
\begin{equation}
ds^2=-C[u(\tilde{u}),v(\tilde{v})]g(\tilde{u})h(\tilde{v})\,d\tilde{u}\,d\tilde{v},\qquad\text{with}\quad g(\tilde{u})=\frac{du}{d\tilde{u}},\quad h(\tilde{v})=\frac{dv}{d\tilde{v}}.
\end{equation}
From here, all the above calculations follow in an analogous way, with the different conformal factor $C\to\tilde{C}=Cgh$. The part of \eqref{e33} which is written in terms of local curvature remains the same, but the non-local part $\Theta_{\mu\nu}$ changes. In terms of components, the relation between the RSETs in these two different vacua (written in the same coordinate basis) is
\begin{subequations}\label{25}
	\begin{align}
	\expval{T_{uu}}_{\tilde{0}}&=\frac{1}{24\pi}\left(\frac{g''}{g^3}-\frac{3}{2}\frac{g'^2}{g^4}\right)+\expval{T_{uu}}_0,\\
	\expval{T_{vv}}_{\tilde{0}}&=\frac{1}{24\pi}\left(\frac{h''}{h^3}-\frac{3}{2}\frac{h'^2}{h^4}\right)+\expval{T_{vv}}_0,\\
	\expval{T_{uv}}_{\tilde{0}}&=\expval{T_{uv}}_0,
	\end{align}
\end{subequations}
where $g'\equiv\partial_{\tilde{u}}g(\tilde{u})$ and $h'\equiv\partial_{\tilde{v}}h(\tilde{v})$, and the subscripts $0$ and $\tilde{0}$ refer to the quantisations with respect to the plane wave modes in the $\{u,v\}$ and $\{\tilde{u},\tilde{v}\}$ coordinates respectively. We can see that a change in the vacuum state translates into the addition of right- and left-moving radiation flux terms (which we will identify with ingoing and outgoing spherical fluxes through the Polyakov approximation in spherically-symmetric spacetimes).

The fact that the state-dependent terms in the RSET are the ones which are non-local in curvature brings to light an interesting observation: the difference between RSETs in two vacuum states eliminates the terms which are local in curvature, and thus can also remove the divergences present in the operator expectation value before renormalisation. In other words, in situations where the counter-terms for renormalisation have not been constructed explicitly, one can still obtain information about vacuum energy effects from differences between RSETs in different quantisations (see e.g.~\cite{Zilberman2022,BBGJ16}).

Apart from the RSET, a useful tool for measuring energy content in BH spacetimes is the effective temperature function (ETF) \cite{BLSV11,BBGJ16}, defined as
\begin{equation}\label{26}
\kappa_{\tilde{u}}^u\equiv-\left.\frac{d^2\tilde{u}}{du^2}\right/\frac{d\tilde{u}}{du}=\frac{g'}{g^2},\quad\kappa_{\tilde{v}}^v\equiv-\left.\frac{d^2\tilde{v}}{dv^2}\right/\frac{d\tilde{v}}{dv}=\frac{h'}{h^2}
\end{equation}
for the outgoing and ingoing (right- and left-moving) radiation sectors, respectively. In the case of a spacetime representing the formation of a BH, the usual Hawking effect is encoded in the constant value $\kappa_{u_{in}}^{u_{out}}=1/2=2\pi T_{\rm H}$ between the ``in" and ``out" vacuum states, where $T_{\rm H}$ is the Hawking temperature in natural units. In more general terms, if $\kappa^u_{\tilde{u}}$ or $\kappa^v_{\tilde{v}}$ remains constant for a sufficiently long period of time (defined by an adiabaticity condition), the vacuum state defined by the $\{\tilde{u},\tilde{v}\}$ coordinates [through the modes in~\eqref{modes}] will be seen by an observer with proper coordinates proportional to $\{u,v\}$ as a state of outgoing or ingoing thermal radiation respectively \cite{BLSV11}.

This function is also directly related to the outgoing and ingoing radiation fluxes which appear in the RSET after a change of vacuum state \cite{BBGJ16}. Specifically, equations~\eqref{25} can be written as
\begin{subequations}\label{27}
	\begin{align}
	\expval{T_{uu}}_{\tilde{0}}&=\frac{1}{24\pi}\left(\frac{d\kappa_{\tilde{u}}^u}{du}+\frac{1}{2}(\kappa_{\tilde{u}}^u)^2\right)+\expval{T_{uu}}_0,\\
	\expval{T_{vv}}_{\tilde{0}}&=\frac{1}{24\pi}\left(\frac{d\kappa_{\tilde{v}}^v}{dv}+\frac{1}{2}(\kappa_{\tilde{v}}^v)^2\right)+\expval{T_{vv}}_0,\\
	\expval{T_{uv}}_{\tilde{0}}&=\expval{T_{uv}}_0.
	\end{align}
\end{subequations}
In other words, the information about the difference between the RSETs in two different vacuum states in 1+1 dimensions is entirely contained in their relative ETFs (and first derivatives thereof).

For the 3+1 dimensional spherically-symmetric spacetimes we will study, we are interested in calculating these quantities for two special quantum vacuum states: the ``in" and the ``out" states. The ``in" (``out") state is the one defined by null coordinates proportional to the proper time of inertial observers at past (future) null infinity in asymptotically flat regions. In order to carry out calculations, we will want to extend these sets of coordinates throughout the whole spacetime, if possible, and obtain the relations between them. However, if there is a horizon present in the geometry, one or both of these extensions may cover the spacetime only partially. For example, in a collapse geometry which begins by being almost flat and ends up forming a BH, the ``in" state corresponds to the natural Minkowski vacuum at the asymptotic past which then evolves according to the dynamics of the system. On the other hand, the ``out" state corresponds to the Minkowski-like vacuum at future null infinity, the backwards extension of which is ambiguous (as it requires additional data at either the horizon or the singularity), but generally exhibits a singular behaviour at the horizon~\cite{Christensen1977}. This discrepancy leads to a variety of interesting horizon-related semiclassical effects, which are the subject of the first half of this work.

\subsection{$\mathscr{I}$-2.3. Semiclassical Einstein equations}

The semiclassical field equations we will work with are
\begin{equation}\label{semicl}
G_{\mu\nu}=8\pi (T_{\mu\nu}^{\rm class}+\expval{T_{\mu\nu}}).
\end{equation}
Aside from the RSET described above, $\expval{T_{\mu\nu}}$, we also include an effectively classical matter source, $T_{\mu\nu}^{\rm class}$, which takes into account the bulk of the macroscopic matter responsible for the curvature of spacetime (planets, stars, etc.). It is standard to assume that the two stress-energy tensors should be covariantly conserved independently from each other. In other words, the only interaction between vacuum energy and the effectively classical matter contemplated in this theory is the one mediated by gravity itself.

Most of the spacetimes we will work with are spherically symmetric. Their line element can be written as
\begin{equation}\label{geo}
ds^2=-C(u,v)dudv+r(u,v)^2d\Omega^2,
\end{equation}
where $u$ and $v$ are radial null coordinates, $r$ is the area-radius of the spherical slicing, and $d\Omega^2=d\theta^2+\sin^2\theta\,d\varphi$ is the line element of the unit sphere. The function $r$ is positive, and so is $C$ if the null coordinates are regular throughout the spacetime. For the RSET, we will use the operator expectation value derived from a massless scalar field in the Polyakov approximation, by combining \eqref{e33} and \eqref{dimred}. If we take $\{u,v\}$ to be the coordinates in which the modes of the chosen 1+1 dimensional quantisation are written as plane waves (or one with an equivalent vacuum state), then the components of the semiclassical equations can be written explicitly as
\begin{subequations}
\begin{align}
\frac{2C_{,u}r_{,u}}{Cr}-\frac{2 r_{,uu}}{r}&=8\pi T^{\rm class}_{uu}+\frac{l_{\rm p}^2}{12\pi r^2}\left(\frac{C_{,uu}}{C}-\frac{3}{2}\frac{C_{,u}^2}{C^2}\right),\\
\frac{2 C_{,v} r_{,v}}{Cr}-\frac{2 r_{,vv}}{r}&=8\pi T_{vv}^{\rm class}+\frac{l_{\rm p}^2}{12\pi r^2}\left(\frac{C_{,vv}}{C}-\frac{3}{2}\frac{C_{,v}^2}{C^2}\right),\\
\frac{C}{2r^2}+\frac{2 r_{,v} r_{,u}}{r^2}+\frac{2 r_{,uv}}{r}&=8\pi T_{uv}^{\rm class}+\frac{l_{\rm p}^2}{12\pi r^2}\left(\frac{C_{,u}C_{,v}}{C^2}-\frac{C_{,uv}}{C}\right),\\
\frac{2 C_{,v} C_{,u} r^2}{C^3}-\frac{2 C_{,uv} r^2}{C^2}-\frac{4 r_{,uv}r}{C}&=8\pi T_{\theta\theta}^{\rm class},
\end{align}
\end{subequations}
where $l_{\rm p}$ is the Planck length, which we will write explicitly in these equations in order to emphasise the suppression of the RSET with respect to the classical terms. Note that, unlike standard classical stress-energy tensors, the RSET contains second derivatives of the metric, changing the principal part of the evolution equations (with whichever evolution parameter one chooses). Though no explicit proof of the well-posedness of these equations exists as of yet, they have been used in numerical calculations with stable results~\cite{Parentani1994,Ayal1997}.

With exact calculations of the RSET in 3+1 dimensions (rather than the Polyakov approximation), the situation gets even more complicated. Derivatives of up to fourth order appear in the RSET, which, if taken at face value, can lead to the semiclassical destabilisation of well-established classical solutions, such as Minkowski spacetime, under arbitrary perturbations~\cite{HorowitzWald1978,Horowitz1980,Suen1989}. This issue has been subsequently formally remedied by considering the semiclassical approximation as an expansion in $\hbar$, which allows an effective reduction of the order of the derivatives through constraints obtained form the zeroth order of this expansion (the classical limit)~\cite{Simon1991,ParkerSimon1993,Halpern1993,Siemieniec1999}. 

In practice, the semiclassical analysis rarely gets far enough along for the order reduction procedure to come into play. Indeed, due to the great difficulty involved in obtaining the RSET in 3+1 dimensions, even in BH spacetimes with angular symmetries, the calculation is usually only performed on fixed backgrounds without complicated (or any) dynamics (see e.g.~\cite{Hollands2020a,Zilberman2022,Taylor2021,Anderson2020}). The effects of backreaction are then only inferred in certain regions of these spacetimes.

By contrast, the Polyakov approximation can be used for simple calculations of the RSET in a large variety of dynamical spacetimes (albeit, restricted to spherical symmetry), often allowing analytical perturbative analyses of backreaction, and even full self-consistent solutions when numerics are involved. In this thesis we mainly focus on analytical studies of the magnitude of the RSET and backreaction, bringing to light the dynamical scenarios in which semiclassical horizon-related corrections are relevant, and thus paving the way for future numerical computations of self-consistent solutions.

\section{$\mathscr{I}$-3. Black holes in classical and semiclassical gravity}

When one looks for a gravitational system in which deviations from classicality are expected to occur, the most natural candidate is the densest object in the observed universe: the astrophysical BH~\cite{LIGO,EHT}. BHs are a generic outcome of the gravitational collapse of classical matter~\cite{Penrose1965}. Their classical description involves the formation a trapped region and, ultimately, a singularity, which is a tell-tale sign of the breakdown of the classical theory. These objects therefore provide a natural testing ground for theories which model the interface between the gravitational and quantum realms~\cite{Hawking1975,Hawking1974,Mathur2005,Ashtekar2005,Ashtekar2018,Almheiri2013,Bekenstein1995,Modesto2006,Oshita2020,Haggard2015,Calmet2022,Ikeda2021,Dvali2013}. We will now conclude this introduction by presenting a brief overview of what the classical and semiclassical theories have told us so far about BH formation and evolution, comparing in particular what these two approaches (or partial admixtures of the two) have argued the ultimate fate of these objects might be.

\subsection{$\mathscr{I}$-3.1. Classical BH formation and evolution}

The first solution to the Einstein equations which described the formation of a BH, obtained by Oppenheimer and Snyder~\cite{Oppenheimer1939}, involved the collapse of a perfectly spherical distribution of a homogeneous and pressureless ideal fluid (or dust cloud). The end result of this collapse is (the future part of) a Schwarzschild BH~\cite{Schwarzschild1916,Kruskal1960}, i.e. a spacelike curvature singularity enclosed by a trapped region, as shown in fig.~\ref{fig23}. For details on the Oppenheimer-Snyder model, see \hyperref[OSmodel]{Appendix B}.

\begin{figure}
	\centering
	\includegraphics[scale=.78]{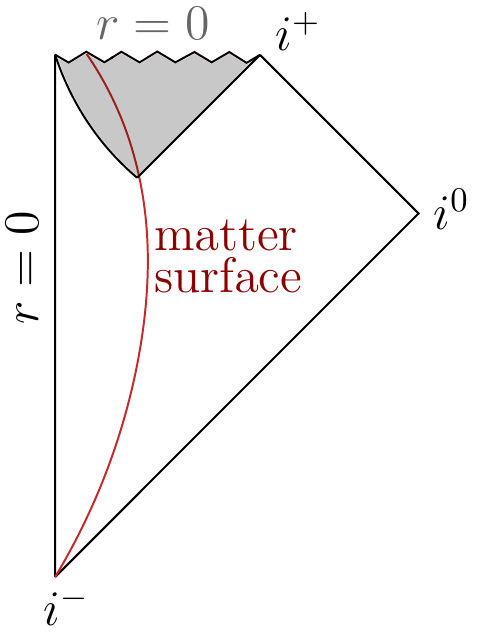}
	\hspace*{1cm}
	\includegraphics[scale=1.12]{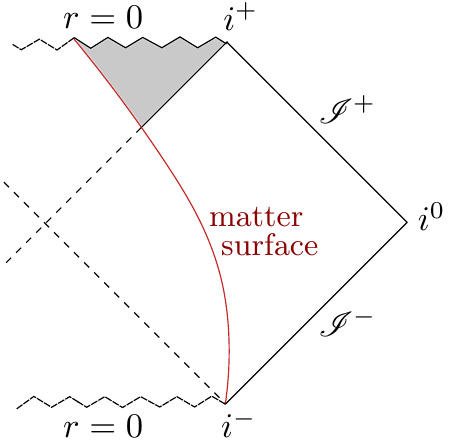}
	\caption{Left: Causal structure of the Oppenheimer-Snyder-type geometry. The red curve represents the surface of the collapsing sphere of matter, and the shaded part is the trapped region, delimited by the outer and inner horizons. Right: The part of the Oppenheimer-Snyder geometry external to the matter distribution is identified with a section of the Kruskal maximal extension of Schwarzschild spacetime.}
	\label{fig23}
\end{figure}

Initially it was not clear whether the BH singularity was a generic result, or rather a consequence of the imposed idealisation of spherical symmetry~\cite{Thorne1970}. The result was subsequently shown to indeed be generic with the \textit{singularity theorems}~\cite{Penrose1965,Hawking1973,Senovilla1998}: when a trapped region forms, if certain energy positivity and causality conditions are satisfied~\cite{Curiel2014}, the collapse process necessarily continues until a singularity (in the sense of inextendability of incomplete geodesics) forms. However, these theorems do not give indications as to the type of singularity, nor indeed the causal structure surrounding it.

As it turns out, the inner structure of a realistic BH is rather more complicated than that of the Oppenheimer-Snyder solution. On the one hand, it was shown that perturbations away from spherical symmetry can make spacelike singularities develop in a significantly different way from their symmetric counterparts~\cite{Belinsky1970,Misner1969}. Particularly, it was shown that at the singular end of a big-bang-like region of non-isotropic universes (akin to the interior of BHs), a so-called Belinski-Khalatnikov-Lifshitz (BKL) singularity develops, which has an unbounded oscillatory behaviour that is different in different spatial directions, and which essentially evolves independently form the matter content which generates it. A BKL singularity can also be found at the endpoint of gravitational collapse~\cite{Berger2002,Garfinkle2007,Saotome2010}.

On the other hand, it was discovered that the causal structure inside realistic (spinning and/or electrically charged) BHs is altogether quite different from that of the Schwarzschild solution, and that environmental perturbations play a key role in its evolution. Particularly, the Kerr-Newman solution~\cite{Kerr1963,Newman1965,Boyer1967} possesses not only an outer apparent horizon, but also an inner one. Given that the trapped region does not go all the way down to the singularity, the nature of this singularity becomes timelike.

\begin{figure}
	\centering
	\includegraphics[scale=.75]{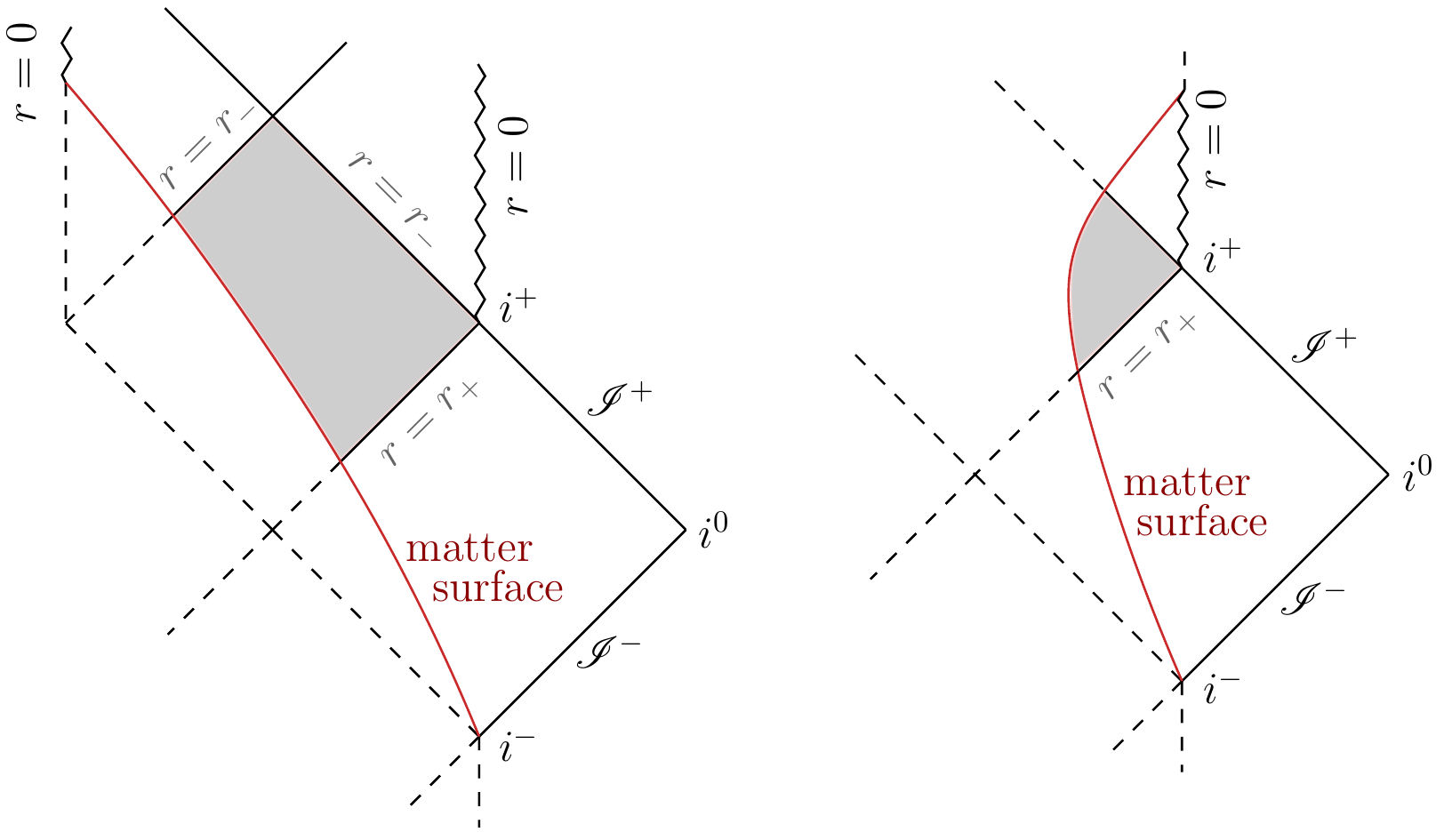}
	\caption{Causal diagrams for the formation of a charged BH, and their analytical extensions. On the left diagram matter collapses to form a singularity before crossing the Cauchy horizon. On the right, matter crosses the Cauchy horizon before forming a singularity. The shaded regions are the parts of the trapped regions external to the matter distribution in each case. $r_+$ indicates the stationary part of the outer horizon and $r_-$ of the inner and Cauchy horizons.}
	\label{f11}
\end{figure}

The interesting causal features do not end there. Observers which approach the inner horizon in an outgoing manner actually have their proper time slowed down with respect to the outside universe, to the point of making their trajectories extendable beyond this universe through a Cauchy horizon~\cite{Carter1968,Wald1984}. Even more strangely, the matter which forms the BH can itself cross the Cauchy horizon before collapsing all the way to form a singularity~\cite{Boulware1973,Wald1984}, as shown in the right causal diagram of fig.~\ref{f11} for a charged Reissner-Nordström BH. 

However, the picture changes yet again when taking into account the generic decaying energy perturbations typically present in BH spacetimes~\cite{Price1972}. The so-called \textit{mass inflation} instability is triggered~\cite{PoissonIsrael89,Poisson1990,Page1991}, wherein the small perturbation to the position of the inner horizon caused by the external energy fluxes leads to an exponential growth of the energy contained in the matter sector in the core of the BH, and consequently of the spacetime curvature. Though this does not affect how the BH is perceived from the outside (the core region of the BH being causally disconnected from the outside universe), its inner structure does change significantly: the inner horizon plummets to the origin (possibly forming a spacelike singularity at a finite time~\cite{Brady1995}), and the Cauchy horizon is substituted by a (weak) curvature singularity~\cite{Ori1991}, which in the case of a rotating BH has an oscillating structure~\cite{Ori1992} not unlike the BKL singularity. In \hyperref[pt2]{part II} of this work, we will present a particular geometric model which captures the essential characteristics of mass inflation, and we will discuss in more detail the properties of the internal BH region.

All this said, for observers outside the BH outer horizon the Oppenheimer-Snyder model, or indeed even the Schwarzschild solution itself, may be considered sufficient for modelling the causal properties of a BH: a region of spacetime from which nothing, not even light, can escape. One may then wonder whether this more precise study of the inner structure of BHs is worthwhile, given the lack of causal contact and the consequent impossibility of observational corroboration (at least, for those of us not willing to leap into BHs ourselves). However, as we will now see, this study becomes quite relevant when semiclassical effects are considered.

\subsection{$\mathscr{I}$-3.2. Semiclassical BHs: the standard picture and beyond}

\subsubsection{Vacuum states on static BH backgrounds}

One of the most frequently analysed problems in quantum field theory in curved spacetimes is that of the vacuum energy present in BH geometries. If we take the simplest BH solution, the Schwarzschild BH, the most direct way of quantising a field in the region exterior to the horizon would be to take advantage of the timelike Killing vector to separate the mode solutions into positive and negative frequencies. The resulting quantum vacuum, commonly referred to as the Boulware state~\cite{Boulware1975}, turns out to have non-regular properties at both the past and future horizons (of the Kruskal maximal extension, represented in fig.~\ref{fig23}). This is directly related to the fact that the time coordinate in the direction of symmetry becomes singular at the horizons, these being Killing horizons beyond which the BH is no longer static. The unbounded oscillatory behaviour of the modes (with respect to any physical time parameter) as they approach the horizons translates to singular energy expectation values for the field.

In fact, if one looks for a semiclassically self-consistent vacuum solution in spherical symmetry compatible with staticity, one finds something which looks like a Schwarzschild BH at large radii, where backreaction is small, but close to the would-be horizon one finds instead a wormhole throat, which opens up into a strange high-curvature region not hidden by horizons~\cite{Arrechea2019,Fabbri2005b}. The large backreaction around the would-be horizon has in fact sparked a search for static objects with classical (positive-energy) matter and semiclassical vacuum energy, with the latter compensating the tendency for collapse of the former, resulting in potentially stable ultra-compact configurations~\cite{Carballo-Rubio2017,Arrechea2021}. It is also the inspiration for the work presented in \hyperref[pt1]{part I} of this thesis, where effects related to how the dynamical ``in" vacuum state can approach the high-energy levels of the static Boulware state in the vicinity of would-be horizons are studied.

In objects with actual trapped regions, the Boulware state is of course considered non-physical. Eternal BHs are idealised objects, whereas realistic BHs are expected to form dynamically from regular initial conditions, both classically and semiclassically. The main argument against the physicality of the Boulware state is the fact that if a quantisation is initially renormalisable (i.e. divergences of quadratic operator expectation values are of Hadamard type), then it continues as such throughout the Cauchy evolution of the initial data~\cite{Fulling1978}. Thus, the ``in" state of gravitational collapse can never evolve into the Boulware state if a closed trapped region is formed.

If one attempts to change the vacuum state to one which gives a regular RSET at either the past or future horizons (or both), one finds an inevitable introduction of energy fluxes at infinity~\cite{Christensen1977}. In the case of regularising only the future horizon, the resulting flux is none other than that of thermal Hawking radiation. This vacuum is known as the Unruh state~\cite{Unruh1976}, and it reproduces the general characteristics of the ``in" state after BH formation, independently of the details of the collapse. It is therefore useful for studying the behaviour of quantum fields in evaporating BHs.

If we take a state which is regular at both past and future horizons, the presence of fluxes at both past and future null infinity translates into a time-invariant thermal bath in the bulk of the spacetime. This vacuum is known as the Hartle-Hawking state~\cite{HartleHawking1976,Jacobson1994}, and due to its time invariance it is useful for a variety of semiclassical calculations (see e.g.~\cite{Anderson1995,Frolov1989,Taylor2022}).

\subsubsection{BH formation and evaporation}

In order to study semiclassical backreaction on BHs, it is useful to work with the ``in" state of gravitational collapse. As mentioned earlier, this state is defined from the Minkowski quantisation in the past region of asymptotically flat spacetimes. When a BH forms, this state behaves much like the Unruh state, giving a flux of outgoing thermal radiation at future null infinity, and also a compensatory negative ingoing energy flux at the horizon~\cite{DFU}.

\begin{figure}
	\centering
	\includegraphics[scale=.75]{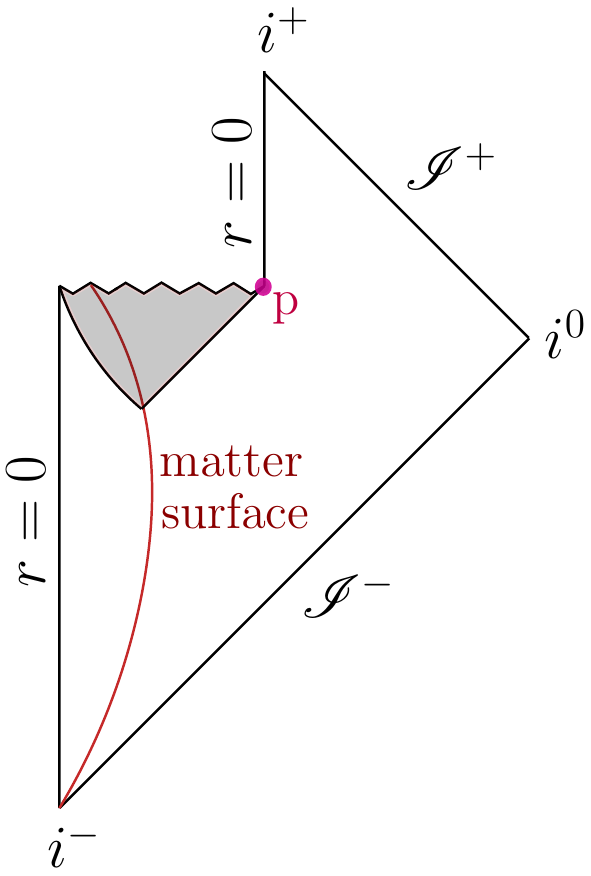}
	\caption{Causal diagram of BH formation and its subsequent Hawking evaporation. Unlike in the classical case, the outer horizon area decreases, until it reaches $r=0$. The point $p$ represents the so-called \textit{thunderbolt} event, where the singularity meets the outside universe, leading to an ambiguity in the future evolution.}
	\label{fig24}
\end{figure}

When backreaction on dynamically formed BH was studied, it lead to one of the most striking results of modern physics: the BH outer horizon, classically seen as a one-way barrier from which nothing escapes, actually tends to gradually reduce in size, and eventually lets out all that was trapped inside---a process known as Hawking evaporation \cite{Hawking1975,Hawking1974}. Though it is not clear what happens at the very end of this process, where the central singularity comes into contact with the outside universe at the same moment in which it disappears due to the depletion of its mass, the qualitative features of the typically expected causal structure of an evaporating BH are represented in fig.~\ref{fig24}.

Remarkable though this result may be, it actually opens the door to a plethora of problems for the consistency of BH solutions. For instance, it makes it hard to sweep singularities under the rug, as is usually done in the classical theory, formally through the well-known Cosmic Censorship Conjecture \cite{Penrose2002}. In the semiclassical theory, the issue of the high-curvature central region has to be dealt with eventually, but the question turns into how long this can be avoided for.

One may believe that this evaporation picture is perfectly valid and free of inconsistencies up until the outer horizon shrinks to a Planck size, from where a quantum theory of gravity would resolve the issue. However, one important observation bring this into question. The thermal nature of the Hawking radiation flux, if continued throughout the whole evaporation process, becomes incompatible with a unitary evolution of the field involved. However, as mentioned above, this flux is necessary for the regularity of the RSET at the outer horizon, and modifying it leads to a resurgence of an energetic firewall at this horizon reminiscent of the Boulware state divergence~\cite{Almheiri2013}. While a transient Hawking evaporation phase can be compatible with unitarity, if it is continued until the horizon reaches a region of Planckian curvature (as the Hawking picture suggests), one would be left with an extremely small region keeping the correlations which can purify the whole seemingly thermal external region of the vacuum state. In other words, a Planck-sized region would have to contain an amount of information regarding the field which is well beyond what can be intuitively expected no matter what the quantum gravity theory which describes this region might be. This issue is commonly referred to as the \textit{information paradox} of BHs, and it has been a central part of semiclassical gravity research for several decades~\cite{Page1993,Susskind2008,Stephens1993,Almheiri2013,Kaplan2018,Page2004,Giddings2019,Preskill1992}.

Note that the assumptions underlying this discussion have been that the classical BH formation scenario to be corrected is the one shown in fig.~\ref{fig23}, and that the RSET only produces notable backreaction at the outer horizon. However, neither of these assumptions is actually justified. As discussed above, the inner structure of a classical BH is notably more complex than this, and backreaction from the RSET in this region has only been studied partially.

\subsubsection{Backreaction at the inner horizon}

Studies of BHs are permeated by the idea that nothing can escape from these objects. Though this is justified when working with classical matter, the intuition has been extended even to semiclassical analyses. The Hawking effect is indeed best seen as a consequence of effectively negative energy falling into the BH, rather than something coming out, hence the information loss problem. However, the modification of the causal structure of the BH produced by this negative energy intake does imply that causal geodesics which were once inside the trapped region can subsequently escape to the outside universe. Although the amount of actual matter which could feasibly escape the BH in this manner in a scenario like the one represented in fig.~\ref{fig24} is miniscule (due to the time scale of Hawking evaporation), this need not be the case in scenarios with a more complex inner structure. Even more crucially, the idea that backreaction from the RSET can change the fate of classical matter by altering the causal properties of the spacetime has far-reaching implications. For instance, these modifications to the light-cone structure may not be limited to the vicinity of the outer horizon.

Particularly, there is an obvious place where curvature may still be small (making the semiclassical approximation valid), but where effects non-local in curvature may lead to significant backreaction from the RSET: the inner horizon. As discussed above, the inner horizon is classically believed to be unstable, making the Cauchy horizon it generates weakly singular. As an add-on to the study of this instability, semiclassical analyses have been performed in the vicinity of BH Cauchy horizons~\cite{Birrell1978,BalbinotPoisson93,Hollands2020a,Hollands2020b,Ori2019,Zilberman2022}. What is found is that the RSET diverges in all initially renormalisable (Hadamard-type) quantum states. This divergence is completely analogous to the one present in the Boulware state at the outer horizon: the modes of the quantum field attain a divergent behaviour with respect to the physical distance parameters at the Cauchy horizon. The conclusions drawn from this behaviour are typically that semiclassical effects would produce a stronger curvature singularity at the Cauchy horizon and definitively prevent the extendability of the geometry into another asymptotic region.

The fact that the RSET tends to be larger than the classical matter source in the vicinity of the Cauchy horizon, which is the asymptotic limit of the inner horizon, is indeed an interesting result. However, the conclusions drawn thus far regarding backreaction in this region are, at the very least, incompatible with the evaporation of the outer horizon, as a Cauchy horizon only tends to form in these geometries if a trapped region continues to exist indefinitely (as seen from the outside). Scenarios involving Cauchy horizons in the framework of semiclassical gravity can only be consistent either if the BH is extremal, or if Hawking evaporation ceases for some other reason (in which case the RSET would also tend to a divergence at the outer horizon~\cite{Christensen1977,Almheiri2013}).

A fully consistent semiclassical analysis of the evolution of a generic trapped region formed dynamically from gravitational collapse involves the study of backreaction at both the outer and inner horizon \textit{at finite times}. In \hyperref[pt2]{part II} of this thesis we perform this analysis using the RSET in the Polyakov approximation. We recover the result for Hawking evaporation of the outer horizon, and we find that the inner horizon has a tendency to move outward, reducing the size of the trapped region from the inside. Although our results do not describe the evolution of the trapped region in its entirety, they are highly suggestive of a process by which this region disappears from the inside-out on a fairly short timescale, as will be discussed in detail.


\setcounter{secnumdepth}{2}

\part{Semiclassical effects near the outer horizon}\label{pt1}

\fancyhead[R]{}
\fancyhead[L]{Part I}

\addtocontents{toc}{\protect\thispagestyle{empty}}

\renewcommand{\theequation}{\thechapter.\arabic{equation}}

When quantising a field on a spherically-symmetric Schwarzschild spacetime, there are three particular choices for a vacuum state which illustrate the relation between the quantisation and the symmetries of the spacetime. If one focuses on the staticity of the Kruskal maximal extension and quantises with respect to the timelike Killing vector, one obtains the Boulware state~\cite{Boulware1975}. This state is well behaved at infinity, having an energy content which goes to zero there. However, its energy becomes divergent at the past and future horizons, making it inconsistent with full semiclassical description of a BH. If one attempts to regularise the state at both horizons while retaining a stationary behaviour, one obtains the Hartle-Hawking state~\cite{HartleHawking1976}, in which there are fluxes of Hawking radiation at both past and future null infinity, resulting in an overall thermal bath in the bulk of the spacetime. However, these eternal fluxes at infinity make this state incompatible with asymptotic flatness when the semiclassical Einstein equations are considered. It thus appears that eternal BHs are disallowed in semiclassical gravity.

A more physically reasonable scenario is that of an asymptotically flat (ignoring cosmological backgrounds) spacetime with an initially dispersed distribution of matter, which eventually collapses to form a BH. In it, we can choose an asymptotically Minkowskian quantisation in the asymptotic past, which can be extended to the whole spacetime (i.e. the ``in" vacuum state). This is in fact the state for which the Hawking radiation result was obtained~\cite{Hawking1975}. In it, observables are regular at the horizon, and the future asymptotic structure is made consistent through the evaporation process, allowing for a (nearly) complete semiclassically self-consistent description.

When using the ``in" vacuum, the overall resulting picture is that any stellar-mass object which \emph{collapses rapidly toward the formation of a horizon} generates extremely small RSETs. It is important to stress that ``rapidly'' here corresponds precisely to the standard situation one would expect when working in the framework of general relativity (defined by the Einstein field equations coupled to matter satisfying the standard energy conditions \cite{Curiel2014}) and taking into account the forces that are known to play a role in stellar evolution. In these situations, semiclassical effects appear so small that the collapse can be expected to proceed in almost exactly the same manner as in classical general relativity, forming a trapped surface and thus a BH (see e.g. \cite{DFU,Parentani1994} for the first treatments of this problem and \cite{Barcelo2008,Unruh2018} for modern retakes). The crucial hypothesis of ``\emph{rapid approach toward the formation of a horizon}" is, therefore, perfectly sensible in most scenarios, even semiclassically. However, the possibility of modifications to the geometry which begin inside BHs and propagate outwards (see e.g.~\cite{Barcelo2008,Barcelo2014,Barcelo2014b,Haggard2015} and \hyperref[pt2]{part II} of this thesis) might lead to situations in which this hypothesis is questionable. For instance, the divergent behaviour of the Boulware vacuum may be taken as a hint of the possibility that even in a physical vacuum, the surroundings of a black-hole outer horizon may be a region where semiclassical corrections become large enough to be relevant to the evolution of the system. Indeed, as we will show, the hypothetical formation of ultracompact objects sustained very close to horizon formation (an alternative to BHs \cite{Visser2004,Visser2008,Cardoso2017,Carballo-Rubio2018,Arrechea2021}) appears to require at least a semiclassical treatment. Generally, if the RSET contribution overcomes its suppression by Planck's constant and becomes comparable to the classical stress-energy tensor, then a complete, non-perturbative semiclassical treatment of the problem is in order.

In this first part of the thesis we study the values of the RSET for the ``in" vacuum of a free massless scalar field in spherically symmetric geometries which approach the formation of a horizon in different ways. Previous works with the same motivation have checked some of the semiclassical effects produced by a collapse of matter which quickly decelerates just before reaching the formation of a horizon \cite{Pad2009,Harada2019}. Our present goal is to more generally identify the precise geometric characteristics of the dynamical situations which would cause large backreaction close to horizon formation. We will use the ETF introduced above, as well as the RSET, to measure these effects.

For our geometric setup, we will mainly use a thin shell model with a Schwarzschild exterior and a Minkowski interior, moving radially along some timelike curve; in chapter \ref{ch3}, in the interest of analysing a larger variety of causal structures and thermal effects, we will also generalise this setup to arbitrary exterior and interior geometries. We will use trajectories for the shell surface suitably chosen to represent scenarios in which the formation of a horizon is approached in different ways. We note that the use of the term ``horizon'' throughout this part of the thesis must be identified with the local notion of an outer apparent/trapping horizon~\cite{Ashtekar2004b,Visser2014}, and not with the global (and generally unobservable) event horizon~\cite{Hawking1973,Geroch1977}, unless explicitly stated. Let us briefly make some comments on this.

While in some of the geometries analysed below the position of apparent/trapping horizon and event horizon are coincident, this should not be taken as an indication that our results are tied in any way with the formation of event horizons. In fact, it is always possible to deform these geometries in a way that event horizons are removed completely, but the local geometric conditions that eventually lead to their formation in the undeformed geometries are maintained for arbitrarily long times (for geometries in which apparent/trapping horizons are formed in finite time, this would imply that they remain present for a large, but finite, amount of time), which would yield the same results except for arbitrarily small deviations. One of the shortcomings of these \mbox{(quasi-)}local definitions of the boundaries of BHs (with respect to the notion of event horizon) is their non-uniqueness \cite{Ashtekar2005b}. This issue disappears in practice when dealing with spherically-symmetric backgrounds, as one can focus on trapping horizons that are spherically-symmetric as well, the location of which turn out to be determined by the quasi-local Misner-Sharp mass \cite{Nielsen2008} that measures the overall energy enclosed in a given sphere \cite{Hayward1994}. When the external geometry to the shell is the Schwarzschild geometry, the location of the horizon defined this way is simply the Schwarzschild radius.

In this work, ``close to horizon formation'' will therefore mean that the shell has trajectories exploring the surroundings of the Schwarzschild radius. There, we expect to find interesting semiclassical effects, and we want to understand their dependence on the precise dynamical properties of the spacetime as it approaches this point. To this end, we have chosen three types of shell trajectories, the study of which will be sufficient to provide a general intuition for judging the magnitude of semiclassical effects in a much larger variety of geometries. The first type of situation, studied in chapter \ref{ch1}, is that of a shell oscillating between two radii, outside but near the Schwarzschild radius. This situation models in a simple way the effects of an ultracompact horizonless object undergoing small pulsations. Varying the characteristics of this oscillation will allow us to explore a wide range of short-time dynamical behaviours.

For the second type of dynamical situation (chapter \ref{ch2}), we look at the case in which a shell forms a horizon while moving at an arbitrarily low speed (relative e.g. to stationary observers in its interior). Our analysis here, which is an extension to \cite{Barcelo2008}, allows us to clearly see how the strength of semiclassical effects depends crucially on the collapsing velocity at horizon formation. This will provide a counterpoint to the already well-known results for a shell collapsing at high velocities or even light speed (see e.g. \cite{Fabbri2005,Unruh2018}).

The third type of situation (chapter \ref{ch3}) will explore the consequences of a long-term monotonous dynamical behaviour, particularly one which we expect (both \textit{a priori} and based on results of chapter \ref{ch1}) to present interesting semiclassical effects---a shell approaching the Schwarzschild radius asymptotically in a regular time coordinate. In this study, we will even go beyond the simple scenario of a thin shell in vacuum and attempt an exhaustive analysis of all possible spherically symmetric geometries which present an asymptotic tendency toward the trapping of outgoing light rays. This asymptotic approach can be stopped at any time, so that these configurations could model, for example, a relaxation phase towards an ultracompact object. On the other hand, if the asymptotic process is continued indefinitely, the spacetimes present the interesting feature of generating an event horizon without having any trapped surfaces formed at finite time.

It is worth mentioning at this point that our analysis will be purely geometrical, and thus goes beyond the Einstein equations. In other words, we will be exploring the effects of a geometry on semiclassical quantities without being concerned with how the geometry itself is generated. We will not require that the evolution of the geometry be governed by the Einstein equation with a stress-energy tensor which satisfies some energy conditions (with the exception of chapter \ref{ch3}, where we consider energy conditions in the generalisation of the thin shell model on either side of the object surface). The goal of our geometry-based analysis is to point the way toward the configurations which should be analysed in further detail in future works with fully self-consistent semiclassical evolution in mind.

\chapter{Oscillating shell model and radiation emission}\label{ch1}

\fancyhead[R]{}
\fancyhead[L]{Part I -- \chaptername\ \thechapter: \leftmark}

The first type of geometry we will look at is that of a spherical thin shell of mass $M$ oscillating between two radii at high frequencies. The lower of the radii will be close to (but above) the Schwarzschild radius $2M$, such that the effect the geometry has on light rays and modes of quantum fields is close to what a horizon would produce. Particularly, it is this closeness to a horizon-like behaviour combined with the wide range of dynamics produced during an oscillation that is worth a close examination through a semiclassical lens. Aside from gaining a general intuition on effects produced on backgrounds close to horizon formation, this model may also be useful for analysing the (potentially observable) bursts of particle creation in perturbations of semiclassically sustained horizonless BH mimickers~\cite{Carballo-Rubio2017,Arrechea2021}.

\section{Spherical thin-shell geometries}

The thin-shell geometries that we will analyse consist of an internal Minkowskian region matched to an external Schwarzschild region of mass $M$ through a moving timelike shell. In the interior region one can write the metric as
\begin{equation}\label{1}
ds_-^2=-du_-dv_-+r_-^2d\Omega^2,
\end{equation}
where the subscript ``$-$" refers to the interior region, and the radial null coordinates are related to the Minkowski time $t_-$ and radius $r_-$ through
\begin{equation}\label{2}
u_-=t_--r_-,\quad v_-=t_-+r_-.
\end{equation}
Equivalently we can construct natural null coordinates in the Schwarzschild region as
\begin{equation}\label{3}
ds_+^2=-|f(r_+)|du_+dv_++r_+^2d\Omega^2,
\end{equation}
where $f(r)=1-2M/r$ is the redshift function, and in this case the null coordinates are related to the Schwarzschild time $t_+$ and radius $r_+$ through
\begin{equation}\label{eq:u+def}
u_+=\text{sign}\left[f(r_+)\right](t_+-r_+^*),\quad v_+=t_++r_+^*.
\end{equation}
Here $r_+^*$ is the tortoise coordinate obtained by integrating $dr_+^*=dr_+/f(r_+)$. The $u_+$ coordinate goes from $-\infty$ to $+\infty$, that is, between past null infinity and the Schwarzschild radius (if the exterior region reaches that far in). Inside the Schwarzschild radius (but outside the shell) we must define a different coordinate $u^i_+$, given by the same relation to $t_+$ and $r_+^*$ as $u_+$ above, and which goes from $-\infty$ at the horizon, until it reaches some point of the spacelike singularity at some finite value. On the horizon itself, relations with this variable can only be obtained as a limit from either side. The sign of $f(r_+)$ ensures that $u_+$ and $u^i_+$ advance in the same direction as $u_-$, both outside and inside the horizon. Though the shells we will use in this chapter will not cross the horizon, we define these more general expressions to set up the formalism which will also be used in the next chapter.

The two geometries are connected by a thin spherical shell of mass $M$. In general, this matching is only possible if the shell's radial position follows a spacetime curve of the same causality type as seen from either side. In our case, we will require that this be a timelike trajectory, parametrised by $v_-=T_-(u_-)$ from the inside and by $v_+=T_+(u_+)$ from the outside. Of course, given one of these curves the other is also fixed. For convenience we will also define the velocity parameters
\begin{equation}\label{5}
\alpha_-\equiv\left.\frac{dv_-}{du_-}\right|_{\rm shell},\qquad \alpha_+\equiv\left.\frac{dv_+}{du_+}\right|_{\rm shell}
\end{equation}
(which are simply the derivatives of $T_\pm$), both of which take values in $(0,\infty)$ for a timelike trajectory. For an ingoing shell to approach the speed of light would imply approaching the limit $\alpha_\pm\to0$. On the other hand, for an outgoing shell reaching the speed of light $\alpha_{\pm}\to\infty$. A static shell has $\alpha_\pm=1$.

In order to complete the definition of this geometry, we must require that the metric be continuous at the shell. This will allow us to determine the trajectory of the shell as seen from one side if it is defined on the other. It will also allow us to extend the ``$+$" coordinates into the ``$-$" region and vice versa.

From matching the null part of the line elements we obtain the functions,
\begin{equation}\label{6}
g=\frac{du_+}{du_-}=\left.\sqrt{\frac{\alpha_-}{|f|\alpha_+}}\right|_{\rm shell},\qquad h=\frac{dv_+}{dv_-}=\left.\sqrt{\frac{\alpha_+}{|f|\alpha_-}}\right|_{\rm shell},
\end{equation}
which can be expressed in either the ``$+$" or ``$-$" variables. From matching the radial parts we get the relation between the velocity parameters of the shell from either side,
\begin{equation}\label{7}
\alpha_+=\text{sign}(f)+\frac{1}{2|f|}\frac{(1-\alpha_-)^2}{\alpha_-}-\frac{1}{2|f|}\frac{1-\alpha_-}{\alpha_-}\sqrt{4\alpha_- f+(1-\alpha_-)^2}.
\end{equation}
Thus if we define the trajectory in terms of $T_-$, we can obtain $T_+$ by integrating $\alpha_+$ from the same initial radial position. We can also obtain the relations $u_+(u_-)$ and $v_+(v_-)$ by integrating the functions $g$ and $h$.

From the square root in \eqref{7} we deduce a condition for the continuous matching of the geometries, namely that the $\alpha_-$ parameter which defines the movement of their separation surface must be such that $4\alpha_- f+(1-\alpha_-)^2$ remains positive. In other words, $\alpha_-$ must tend to zero (the infalling shell must approach light-speed) if the shell goes below the Schwarzschild radius, in such a way as to compensate the increasingly negative value of the redshift function. The parameter $\alpha_-$ which satisfies
\begin{equation}
4\alpha_- f|_{\rm shell}+(1-\alpha_-)^2=0
\end{equation}
defines the slowest possible collapse below the Schwarzschild radius as seen from the (rapidly disappearing) Minkowski region.

\subsection{Interpretation of the terms in $g$ and $h$}

Let us focus on the function $g$ outside the Schwarzschild radius,
\begin{equation}\label{8}
g=\frac{du_+}{du_-}=\frac{1}{\sqrt{f}}\sqrt{\frac{\alpha_-}{\alpha_+}}.
\end{equation}
The presence of the term $1/\sqrt{f}$ is to be expected, as it represents the redshift experienced by an outgoing light ray. This can be seen most clearly in the case of a static shell (which, of course, would sit outside the horizon), for which $\alpha_\pm=1$. There, this term is necessary for a rescaling of the coordinates compatible with a matching of the angular parts of the geometry.

\begin{figure}
	\centering
	\includegraphics[scale=.7]{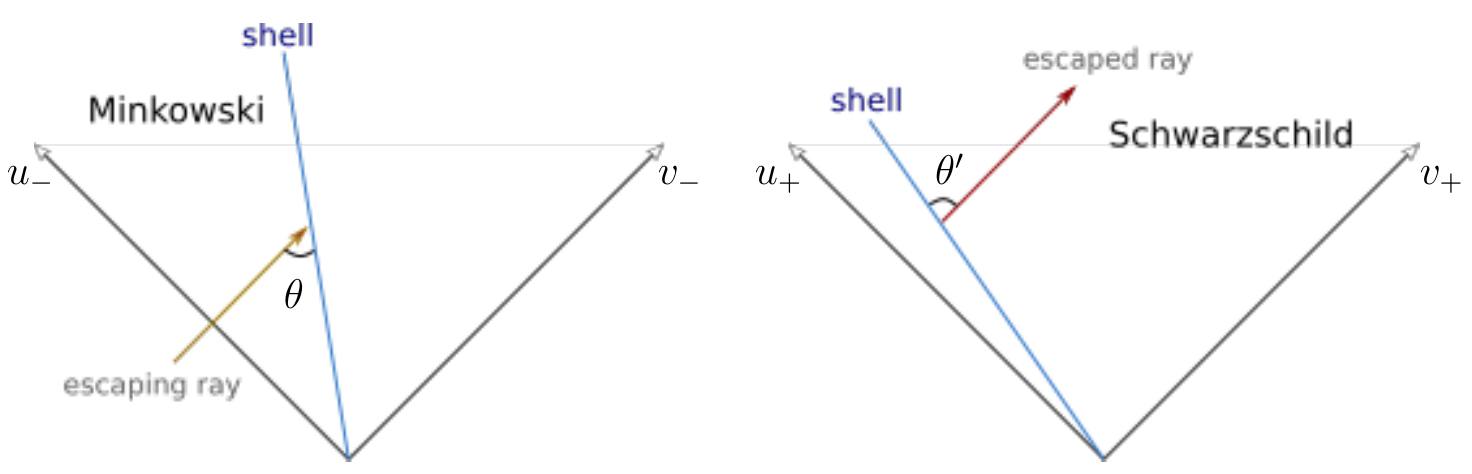}
	\caption{Change in angle with respect to the shell of on outgoing light ray, as measured by static observers on either side.}
	\label{f1}
\end{figure}

The $\alpha_-/\alpha_+$ term has a purely dynamical origin. The velocity of the shell seen by a static observer on one of its sides is different from the one seen by a static observer on the other. In their respective null coordinates this can be seen as a change in the slope of the line tangent to the shell trajectory, namely $\alpha_-\to\alpha_+$ (see fig.~\ref{f1}). From the perspective of the shell, which can use the appropriate coordinates for each side, this looks something like a spacetime refraction phenomenon. If a light ray incides with an angle $\theta$ with respect to the shell trajectory from the inside, it exits with an angle $\theta'$ related to the first by
\begin{equation}
\frac{\tan \theta'}{\tan \theta}=\frac{\alpha_-}{\alpha_+}.
\end{equation}
For the angles formed by an ingoing ray, the relation is the inverse of the above.

Another way to interpret the $\alpha_-/\alpha_+$ term is as a kind of Doppler effect. Even though technically there is no interaction between the matter in the shell and the light ray crossing it which could cause absorption and reemission, the similarity with the Doppler effect can be seen clearly with the following. If we define
\begin{equation}
R(t_\pm)\equiv r|_{\rm shell}(t_\pm)\quad\text{and}\quad \dot{R}=dR/dt_-,\quad R'=dR/dt_+,
\end{equation}
then
\begin{equation}\label{10}
\alpha_-=\frac{1+\dot{R}}{1-\dot{R}},\quad \alpha_+=\frac{1+R'/f(R)}{1-R'/f(R)}. 
\end{equation} 
That is, the quotient $\alpha_-/\alpha_+$ represents the Doppler shift for a ray that is ``absorbed" at one side by a shell moving at a velocity $\dot{R}$ and ``reemitted" on the other by a shell moving at a different velocity, $R'/f(R)$. If the geometry on both sides were the same, there would be no net effect, as these velocities would be the same.

It is worth mentioning that there may be a difficulty in interpreting the above expressions at the Schwarzschild radius, since the coordinate $t_+$ used for the derivative in the second equation in \eqref{10} is not regular there. To see the behaviour of $\alpha_+$ more clearly we can switch to a regular time coordinate, say the Painlevé-Gullstrand $\tau_+$ defined as the proper time of a free-falling observer from infinity in the Schwarzschild region \cite{Martel2001}, which satisfies
\begin{equation}
d\tau_+=dt_++\frac{\sqrt{1-f(r_+)}}{f(r_+)}dr_+.
\end{equation}
We can then define the radial velocity $R_{,\tau}\equiv dR/d\tau_+$, which is regular at the horizon. Then the second equation in \eqref{10} becomes
\begin{equation}\label{12}
\alpha_+=\frac{1+R_{,\tau}/(1+\sqrt{2M/R})}{1-R_{,\tau}/(1-\sqrt{2M/R})},
\end{equation}
from which we can see that at $r=2M$, $\alpha_+=0$ and the function $g$ in \eqref{8} diverges.

In light of these results, we will call the $1/\sqrt{|f|}$ terms in the functions $g$ and $h$ the ``redshift" terms, and the ones with a quotient of $\alpha$'s the ``Doppler" terms. Combining equations \eqref{6}, \eqref{10} and \eqref{12} we can write the functions $g$ and $h$ as
\begin{subequations}\label{13}
	\begin{align}
	&g(u)=\frac{1}{\sqrt{f}}\sqrt{\frac{1+\dot{R}}{1-\dot{R}}}\sqrt{\frac{1-R_{,\tau}/(1-\sqrt{2M/R})}{1+R_{,\tau}/(1+\sqrt{2M/R})}},\\
	&h(v)=\frac{1}{\sqrt{f}}\sqrt{\frac{1-\dot{R}}{1+\dot{R}}}\sqrt{\frac{1+R_{,\tau}/(1+\sqrt{2M/R})}{1-R_{,\tau}/(1-\sqrt{2M/R})}},
	\end{align}
\end{subequations}
in which all quantities are evaluated at the points where the lines $u=$ const.~and $v=$ const.~intersect the shell trajectory respectively. We could work directly with these expressions instead of \eqref{6} by defining the trajectory through the velocities $\dot{R}$ and $R_{,\tau}$, which must satisfy a relation similar to \eqref{7}. However, throughout this work we will keep using the $\alpha_\pm$ parameters, as they are more natural and simple when dealing with the relations between null coordinates needed for the calculation of semiclassical quantities.

\section{Oscillating shells and semiclassical effects}\label{s4}

In this section we will study the behaviour of the functions $g$ and $h$, which relate the ``$+$" and ``$-$" coordinates, when the shell gets near the formation of a horizon, but does not reach it. We will then use equations \eqref{25}-\eqref{27} and the properties of the static Boulware state to directly obtain conclusions regarding the nature and magnitude of semiclassical effects in geometries with this kind of light-bending behaviour.

\begin{figure}
	\centering
	\includegraphics[scale=.6]{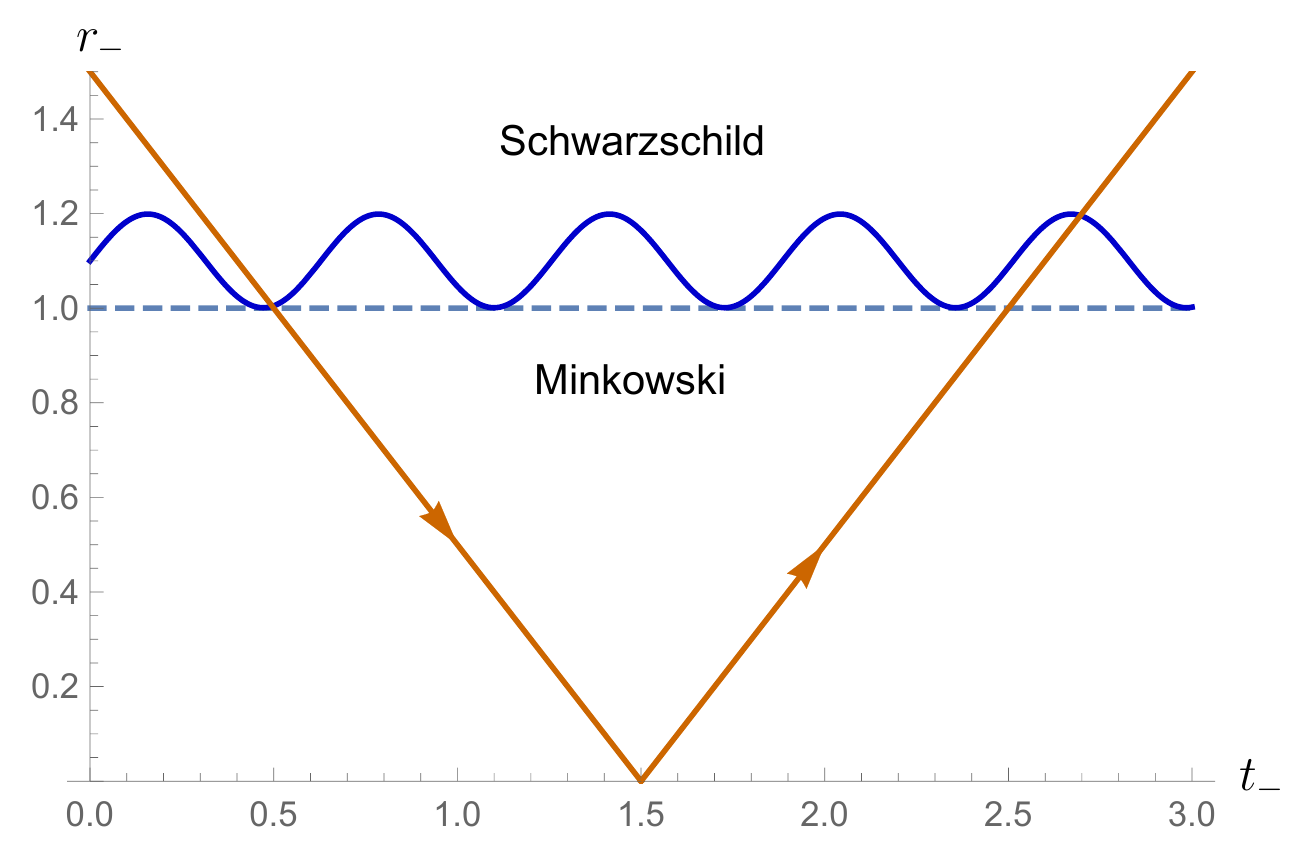}	
	\caption{Oscillatory radial trajectory of the shell (with parameters $d=0.1$, $A=0.099$ and $\omega=10$). The dashed line represents the $r=2M$ $(=1)$ surface, and the diagonal lines represent a light ray entering and exiting the interior region. Although not perceived in the figure, the thick oscillatory curve does not touch the $r=2M$ line.}
	\label{f2}
\end{figure}

The shell trajectories we will look at cover a wide range of dynamical configurations, in which both the redshift and Doppler effects have significant contributions to the values of these functions and their derivatives. Namely, we will consider a high-speed radial oscillation about a point just above the surface with radius $r_{\rm s}=2M$ (in the following, we will always take $r_{\rm s}=1$ for numerical evaluations). We will use three parameters to describe this movement: the distance $d$ of the centre of oscillation to the gravitational radius $r_{\rm s}$, the amplitude $A$ and the frequency $\omega$. Then, the radius at which the shell is located will follow the spacetime curve (see fig.~\ref{f2})
\begin{equation}
R(t_-)=r_{\rm s}+d+A\sin(\omega t_-).
\end{equation}
In order to avoid the formation of a horizon and maintain a timelike trajectory, the parameters must satisfy the relations
\begin{equation}
A<d\quad \text{and}\quad A\omega<1.
\end{equation}
We stress once again that the purpose of this study is to gain a better understanding of the relation between dynamical regimes close to horizon formation and the magnitude of semiclassical effects, and not to provide a self-consistent solution with a classical matter content which satisfies some energy conditions. Thus we only impose that the shell be causal, with no further restrictions to its trajectory.

Since the trajectory is described in terms of the interior coordinate system, we can obtain the simple expression for the interior velocity parameter
\begin{equation}
\alpha_-=\frac{r_{\rm s}+A\omega\cos(\omega t_-)}{r_{\rm s}-A\omega\cos(\omega t_-)},
\end{equation}
while for $\alpha_+$ we must use eq. \eqref{7}. To evaluate these quantities on the points where the shell trajectory intersects the lines of constant $u$ or $v$ we must solve a transcendental equation, which we will do numerically. First we will obtain the individual values of the functions $g$ and $h$, which represent the change in the coordinate description of outgoing and ingoing radial light rays respectively. Then we will calculate the quotient $g/h$ with $h$ evaluated at a point of entry $v_-$ of a light ray into the Minkowski region and $g$ evaluated at the point of exit $u_-$, which carries information of how light rays suffer a temporal dispersion by passing through this region. With the convention fixed in eq. \eqref{2} we can see that an ingoing ray $v_-$ connects with an outgoing ray through $u_-=v_-$, so the quotient we are looking for is $g(v_-)/h(v_-)$. This quantity will also describe the evolution of the ``in" quantum vacuum state, defined at the asymptotically flat region at past null infinity, and its comparison with the ``out" vacuum state, defined at future null infinity.

\begin{figure}
	\centering
	\includegraphics[scale=.5]{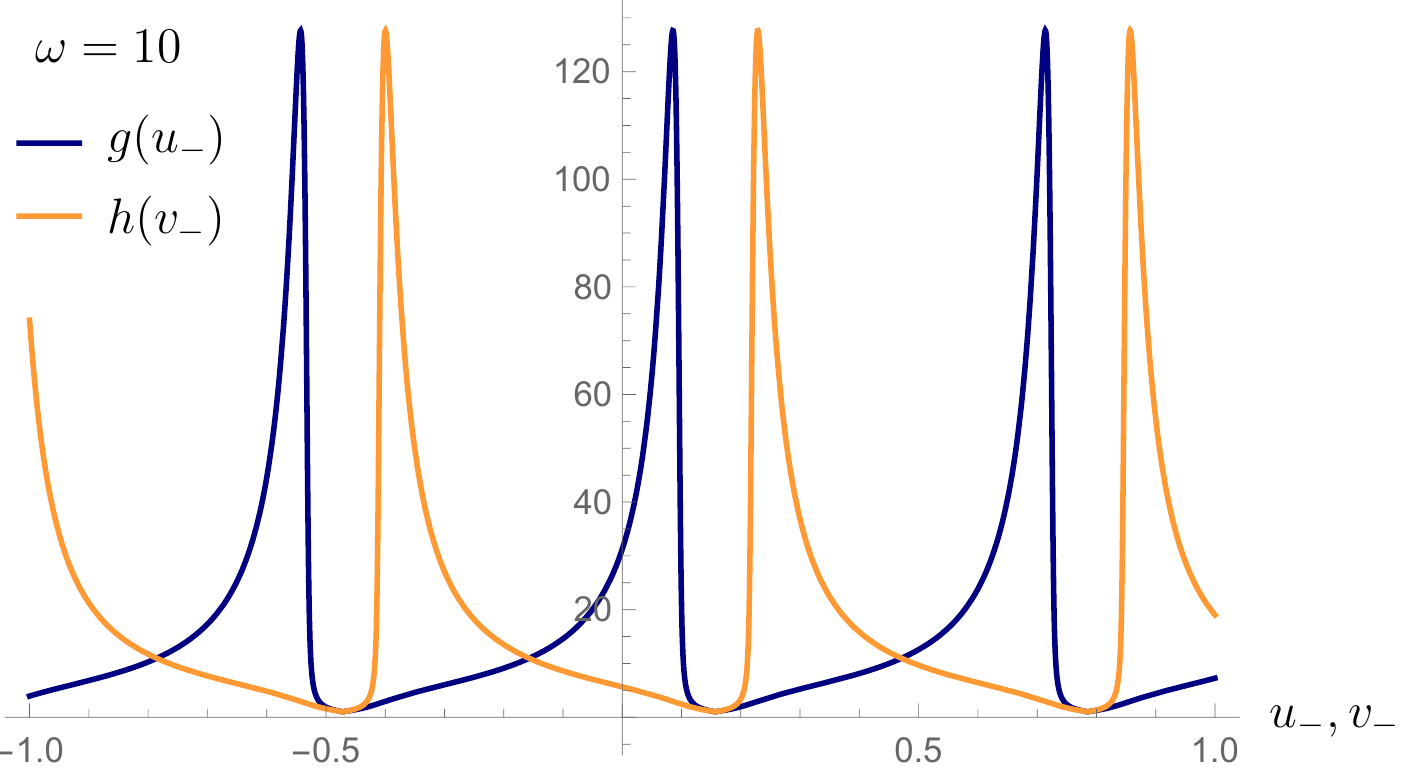}
	\includegraphics[scale=.5]{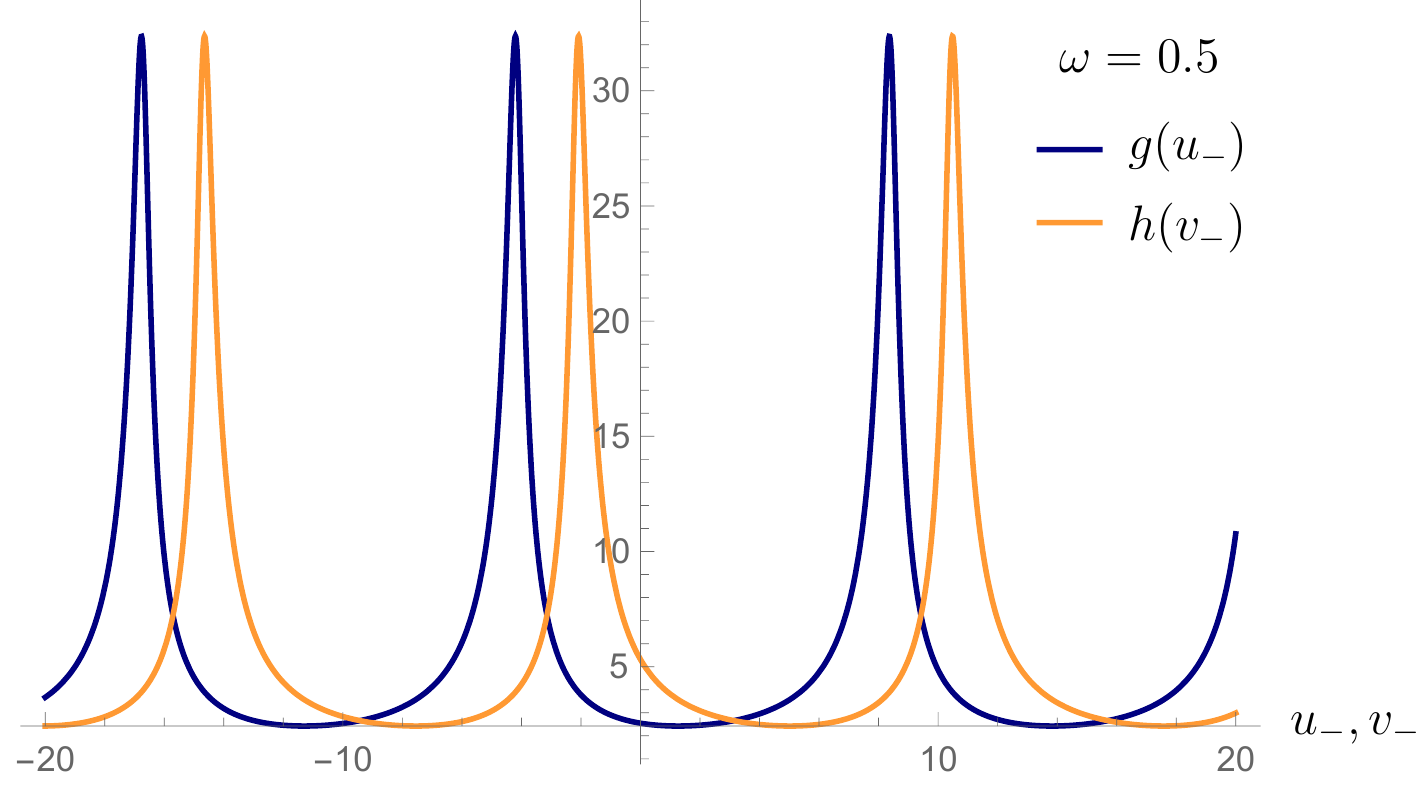}
	\includegraphics[scale=.5]{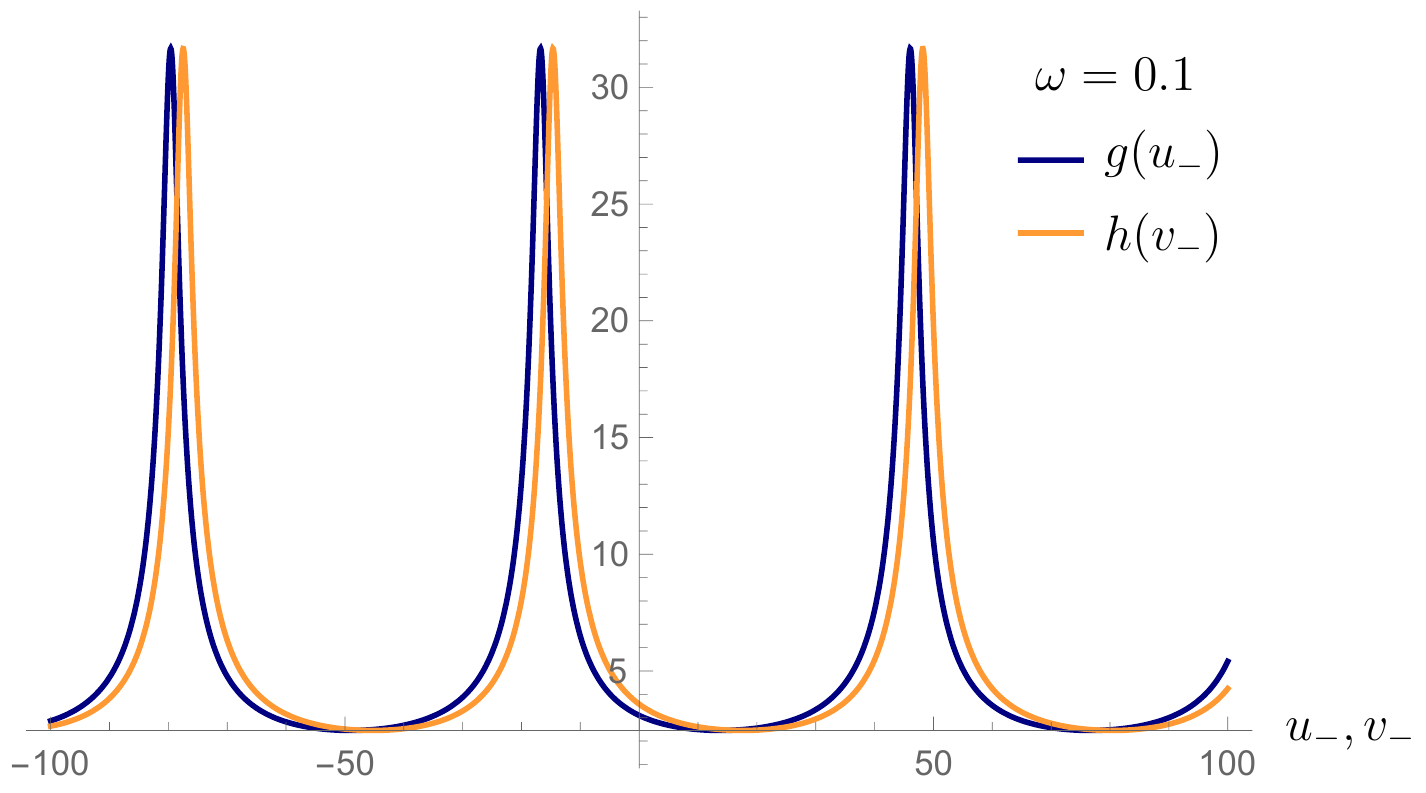}
	\caption{Functions $g$ and $h$ for an oscillation with parameters $d=0.1$, $A=0.099$ and three different frequencies: $\omega=10$, $\omega=0.5$ and $\omega=0.1$. The peaks are produced when the shell is nearly at the closest point to the horizon, as will be discussed below. We observe that at low frequencies the functions practically coincide since the light rays enter and leave the interior region in a time much smaller than $\omega^{-1}$, so the in-crossing and out-crossing dispersion effects would almost cancel out (i.e. $g(u_-)/h(v_-=u_-)\simeq1$). At somewhat larger frequencies the light rays enter and exit at appreciably different points of the oscillation and the functions attain a relative displacement. Finally, at frequencies which make the shell move at nearly light-speed the displacement is greater still, and the peaks become somewhat tilted to one side for each function, due to the fact that the peaks of the sine function in $t_-$ become tilted when seen in the $u_-$ and $v_-$ coordinates (in opposite directions).}
	\label{f3}
\end{figure}

In figure \ref{f3} we observe the values of the functions $g$ and $h$ evaluated at $u_-$ and $v_-=u_-$, representing the dispersion of a light ray when it is exiting and entering the interior region respectively. The net effect, given by $g/h$, reduces to nearly unity when $g(u_-)\simeq h(u_-)$, which occurs when the shell is oscillating very slowly (at low $\omega$) compared to the time it takes for light to cross it (in the static limit, $g/h=1$). At higher frequencies the light rays enter and exit at completely different points of the oscillation, as in the case represented in fig.~\ref{f2}, and the net effect becomes appreciable. It is easy to notice that there are some special cases for this net effect corresponding to different resonances between the oscillation frequency and the crossing time of the light ray: say, when it enters crossing a maximum and also exits crossing one, or crossing a minimum, and a few other such situations. These will be studied in more detail in the following subsection.

\subsection{Resonance between in-crossing and out-crossing effects}

From equations \eqref{6} we can obtain the expression for the total temporal dispersion suffered by a light ray entering the shell at a point ``in" and exiting at a point ``out",
\begin{equation}\label{17}
\frac{du_{+,{\rm out}}}{dv_{+,in}}=\frac{g|_{\rm out}}{h|_{\rm in}}=\frac{\sqrt{f}|_{\rm in}}{\sqrt{f}|_{\rm out}}\left.\sqrt{\frac{\alpha_-}{\alpha_+}}\right|_{\rm out}\left.\sqrt{\frac{\alpha_-}{\alpha_+}}\right|_{\rm in},
\end{equation}
where in the first step we have made use of the fact that for rays reflecting at the origin $dv_-|_{\rm in}/du_-|_{\rm out}=1$, as can be seen from eq.~\eqref{2}. We can see again that for a static shell, for which the surface redshift function would be constant and $\alpha_\pm=1$, this quotient reduces to unity. For a moving shell the effects can cancel out again only in one special case, which occurs when not only the $in$ and $out$ redshift functions are the same, but also when $\alpha_-=1$ (and therefore $\alpha_+=1$ as well, as can be seen from eq. \eqref{7}) at both points. For the case of an oscillating shell this can occur only when a light ray exists such that it both enters and exits at a minimum or at a maximum of $R(t_-)$. Then the effects cancel out locally, but they continue being non-trivial for the rest of the light rays. These local resonances are possible only when the frequency, amplitude and distance from the horizon satisfy the relations
\begin{equation}\label{18}
\omega=\frac{n\pi}{r_{\rm s}+d-A}, \quad\text{with }n\text{ integer less than }\quad\frac{r_{\rm s}+d-A}{a\pi},
\end{equation}
for a ray entering and exiting at a minimum, and likewise
\begin{equation}\label{19}
\omega=\frac{n\pi}{r_{\rm s}+d+A}, \quad\text{with }n\text{ integer less than }\quad \frac{r_{\rm s}+d+A}{a\pi},
\end{equation}
for a maximum. These expressions are obtained simply by comparing the ray crossing time and the oscillation periods in the coordinate $t_-$. The upper bound on the values of $n$ comes from the causal restriction $A \omega<1$. For a shell following an arbitrary (known) radial motion such cases can be found just as easily.

\begin{figure}
	\centering
	\includegraphics[scale=.59]{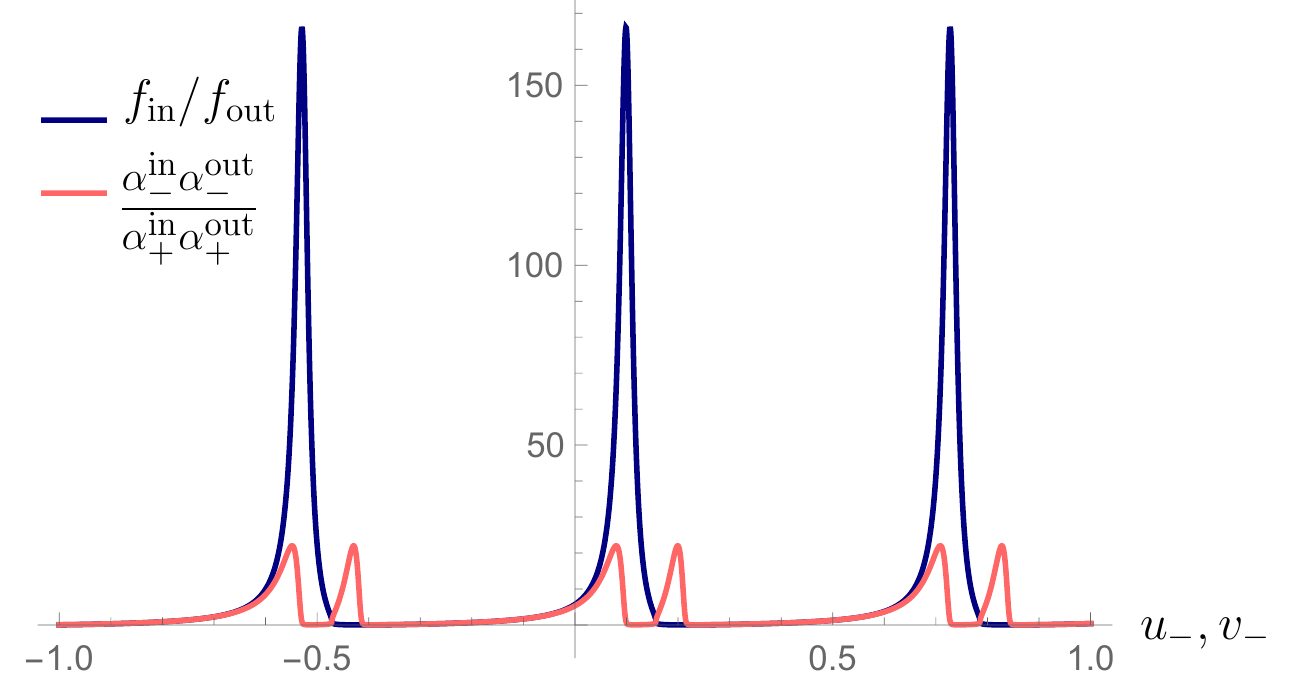}
	\includegraphics[scale=.59]{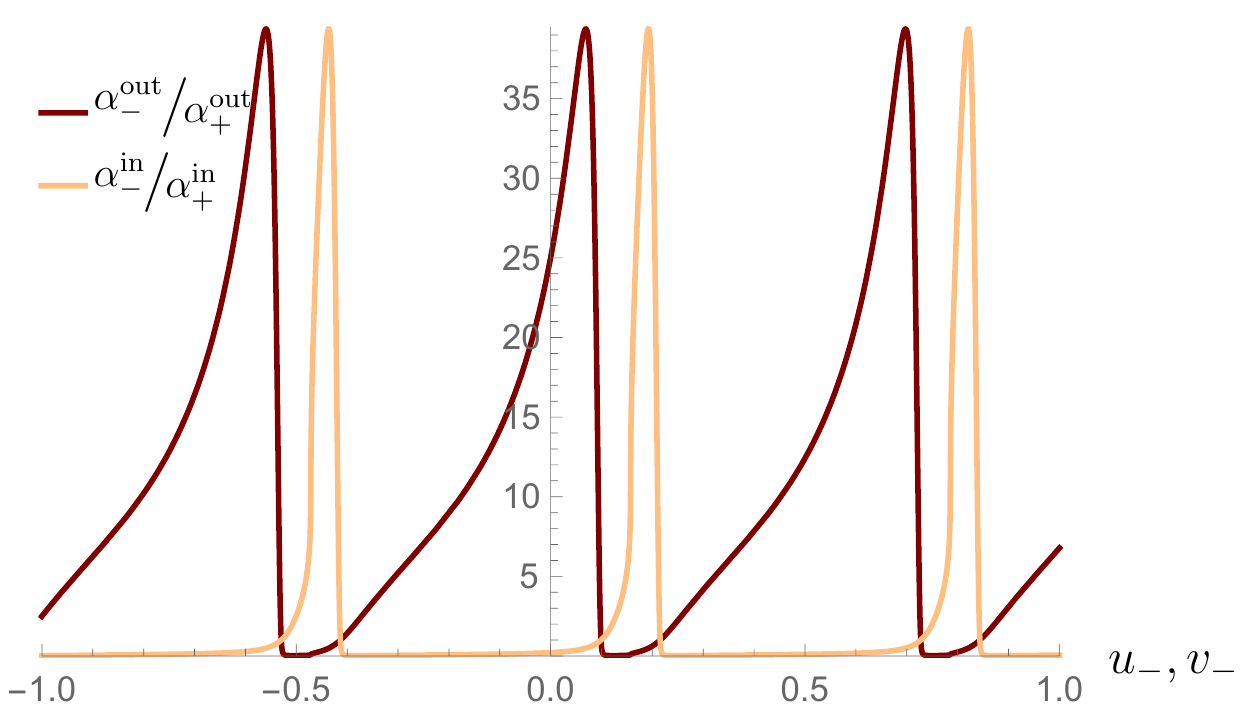}
	\caption{Left: total redshift and Doppler terms plotted separately (the square root of their product gives $du_{\rm out}/dv_{\rm in}$) for on oscillation with parameters $d=0.1$, $A=0.099$ and $\omega\simeq9.996$, given by \eqref{20} with $n=3$. We see that even though the shell reaches 99\% of the speed of light (as seen in the Minkowski coordinates) and the Doppler terms become quite large, the redshift term clearly gives the dominant contribution around its maxima. Right: in-crossing and out-crossing Doppler terms plotted separately. We observe that the peaks and valleys are completely out of phase between the two.}
	\label{f4}
\end{figure}

On the other hand, if we want to see when a maximisation of $du_{+,{\rm out}}/dv_{-,{\rm in}}$ in eq. \eqref{17} takes place (which will be related to the largest bursts of radiation in the RSET, as we will see), a more detailed analysis is necessary. First, we may notice that when a light ray enters at a maximum of $R(t_-)$ and exits at a minimum, the total redshift effect is maximised. For such a ray to exist, the relation between the parameters must be
\begin{equation}\label{20}
\omega=\frac{\pi}{2}\frac{2n+1}{r_{\rm s}+d}, \quad\text{with }n\text{ integer less than }\quad \frac{r_{\rm s}+d}{A\pi}-\frac{1}{2}.
\end{equation}
In fig.~\ref{f4} we observe the three terms of the right-hand side of eq.~\eqref{17} plotted (without the square roots) for this case. The peaks of the redshift term, which correspond to precisely the light rays described, reach their highest possible values for the parameters $A$ and $d$ used.

\begin{figure}
	\centering
	\includegraphics[scale=.45]{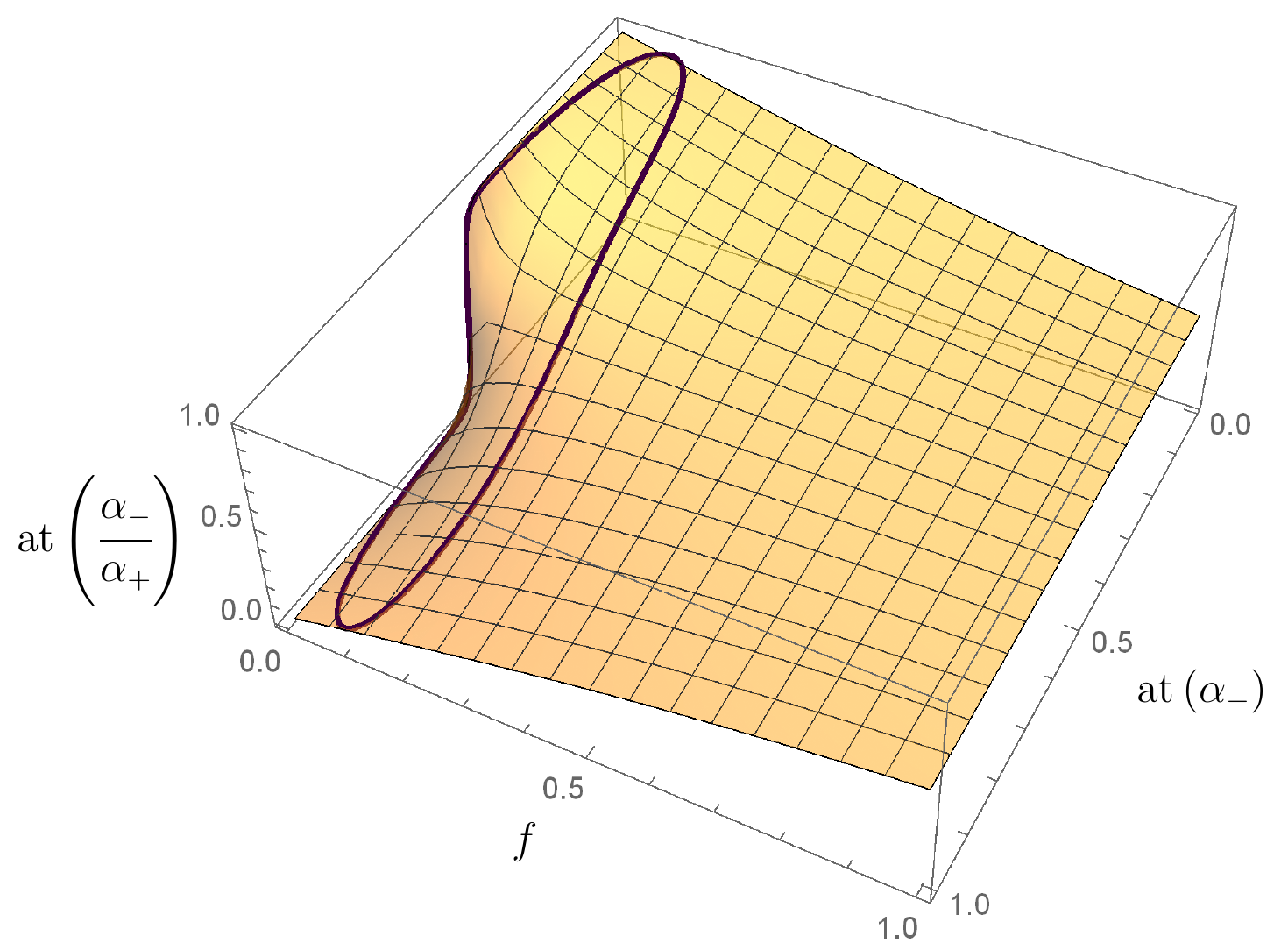}
	\includegraphics[scale=.45]{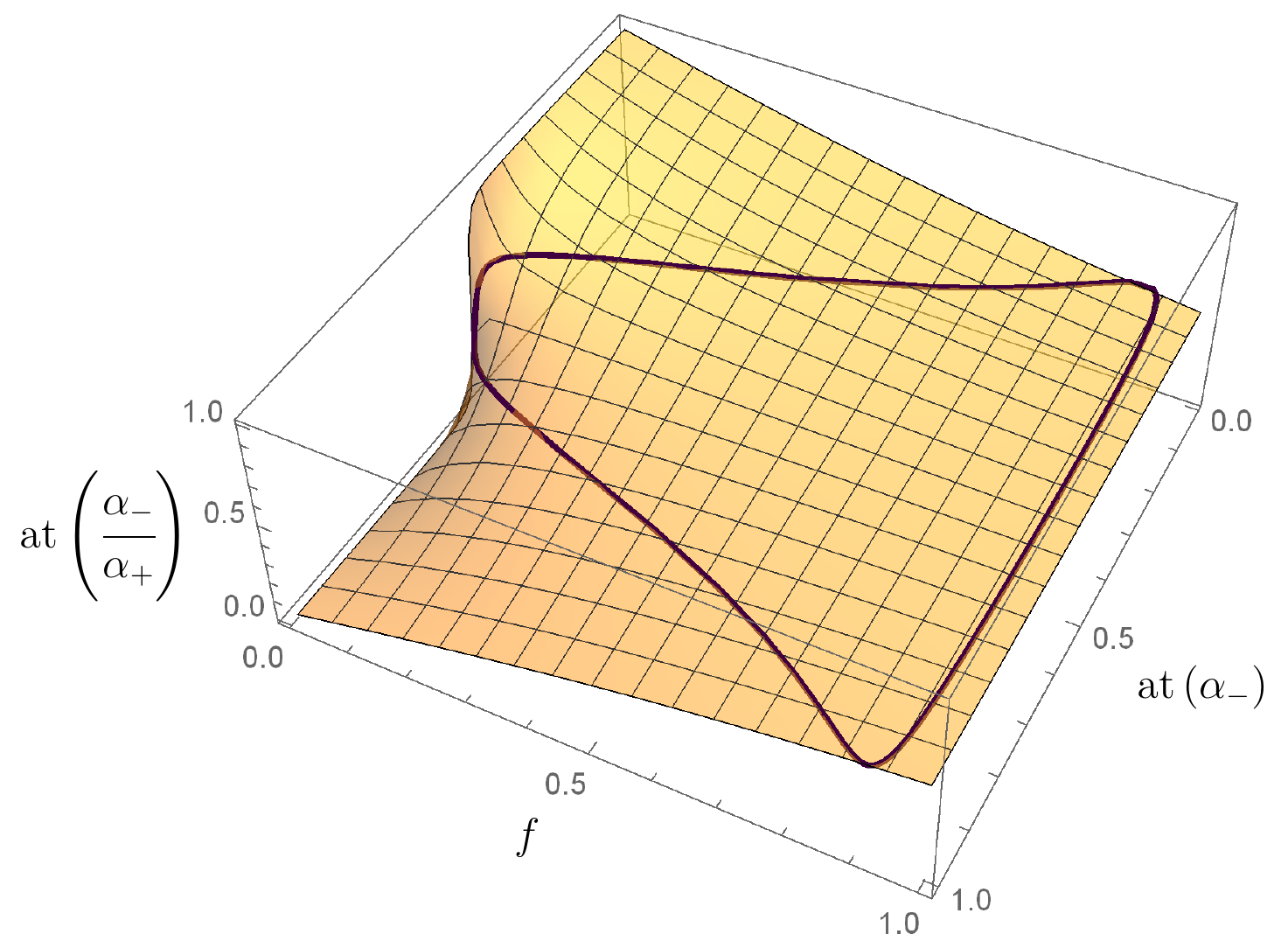}
	\caption{Values of the Doppler term $\alpha_-/\alpha_+$ as a function of $f$ and $\alpha_-$. The axes of $\alpha_-$ and $\alpha_-/\alpha_+$ have been rescaled with a function $\text{at}(x)=\frac{2}{\pi}\tan[-1](x)$ to scale down their whole range into $(0,1)$. The curve drawn on top of the surface on the left represents the values taken during a period of oscillation with parameters $d=0.1$, $A=0.099$ and $\omega=10$, and the curve on the right during an oscillation with parameters $d=10$, $A=9.99$ and $\omega=0.1$. The region with a sharp gradient close to the horizon ($f=0$) is produced around $\alpha_-=1$ ($1/2$ in the graphic), corresponding to the transition from falling inward (during which time $\alpha_-<1$) to going outward (during which $\alpha_->1$). At $f\to1$ the value of the Doppler term tends to 1 smoothly, as the interior and exterior geometries become the same.}
	\label{f5}
\end{figure}

Also in fig.~\ref{f4}, we observe that the individual Doppler terms have distinct maxima. For the in-crossing term the maximum is produced for a ray which enters slightly after the one which maximises redshift (which enters at a maximum of $R$), during the in-fall of the shell. For the out-crossing term it is produced for a ray which exits slightly before redshift maximising one (which exits at a minimum of $R$), so again during an in-fall of the shell. Guided by this result, we can look for the conditions which maximise the individual Doppler terms, and also see whether there is a frequency for the shell at which the two peaks coincide to make a maximum net effect. In fig.~\ref{f5} we can directly see the values which $\alpha_-/\alpha_+$ takes at different redshifts $f$ and velocity parameters $\alpha_-$.

At the very minimum of the oscillation of the shell (the closest point to $r=r_{\rm s}$), $\alpha_-/\alpha_+=1$ and there is no Doppler effect for any redshift $f$. When the shell is moving outward ($\alpha_->1$) but is still close to the horizon ($f\ll1$), from eq. \eqref{7} we get
\begin{equation}
\frac{\alpha_-}{\alpha_+}\simeq\frac{\alpha_-^2}{(1-\alpha_-)^2}f,
\end{equation}
that is, at a constant velocity the Doppler term has a linear dependence on the redshift function, with a slope which grows rapidly as $\alpha_-\to1^+$ and which tends to 1 as $\alpha_-\to\infty$. On the other hand, when the shell is falling in, the function close to the horizon can be expressed as
\begin{equation}
\frac{\alpha_-}{\alpha_+}\simeq\frac{(1-\alpha_-)^2}{f},
\end{equation}
which grows parabolically as $\alpha_-\to0$ (as the in-fall speed increases) and hyperbolically as $f\to0$ (as the formation of the horizon is approached).

With the above equations and fig.~\ref{f5} we can see that the point of the shell trajectory where the Doppler effect reaches a maximum appears in the $\alpha_-<1$ region, and that its precise position is influenced by two factors: at a constant $f$ it is maximum at the highest velocity (lowest $\alpha_-$), increasing parabolically as $\alpha_-$ decreases, while at a constant velocity it is maximum at the lowest $f$, with a hyperbolic divergence at $f=0$. If $\alpha_-$ approaches 1 while $f$ approaches 0, that is, if the shell tends to a full stop just before the formation of the horizon, then, when $f$ is sufficiently small, the hyperbolic divergence dominates over the parabolic tendency to zero and the maximum is reached at a point very close to the minimum value of $f$, just before the region of very large gradient observed in fig.~\ref{f5} is entered. If, on the other hand, the shell oscillations are produced far away from the Schwarzschild radius $r_{\rm s}$, the maximum Doppler effect is reached closer to the point of maximum in-fall velocity (minimum $\alpha_-$).

\begin{figure}
	\centering
	\includegraphics[scale=.59]{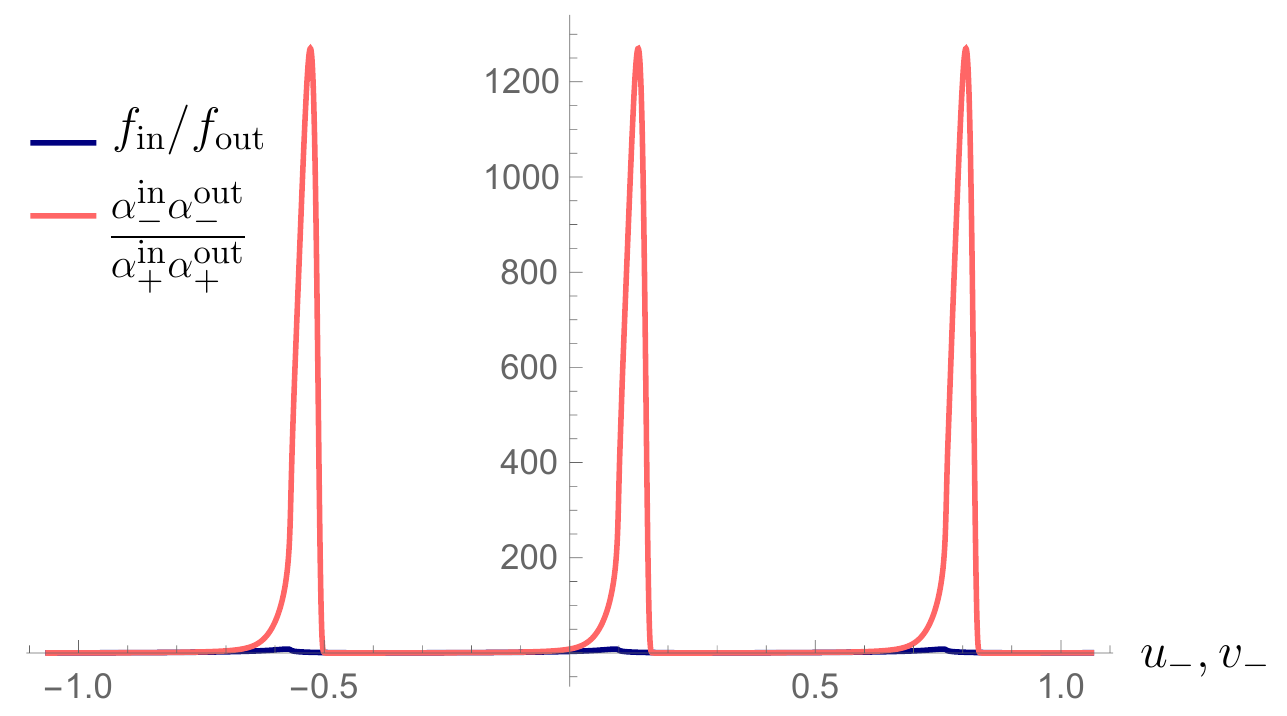}
	\includegraphics[scale=.59]{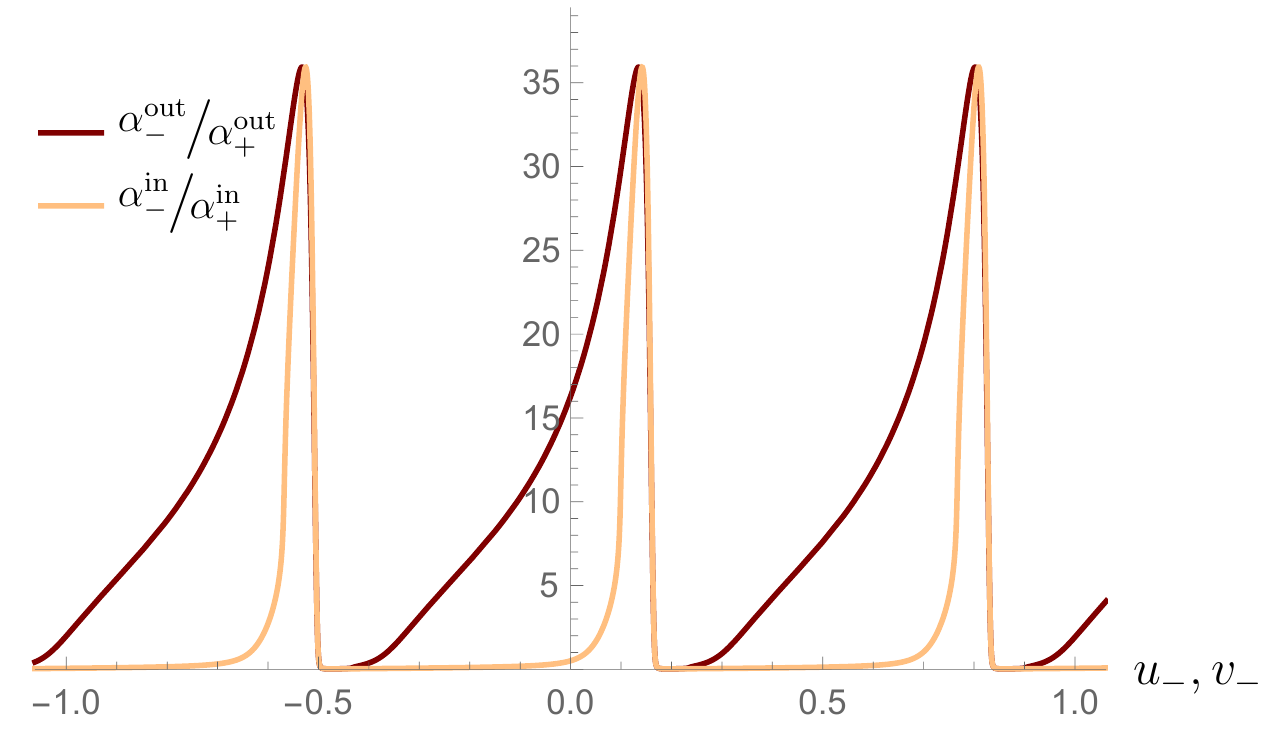}
	\caption{Left: total redshift and Doppler terms plotted separately (the square root of their product gives $du_{\rm out}/dv_{\rm in}$) for an oscillation with parameters $d=0.1$, $A=0.099$ and $\omega\simeq9.42$, given by eq. \eqref{18} with $n=3$. In this case we see the Doppler term clearly dominates. Right: in-crossing and out-crossing Doppler terms plotted separately. We observe the almost-coincidence of the two Doppler peaks, produced for two very close rays passing through the shell slightly before the one which enters and exits at a minimum of the oscillation. This near-coincidence results in the dominance of the net Doppler term in the left graph.}
	\label{f6}
\end{figure}

As an example, in fig.~\ref{f6} we can see the almost-coincidence of the two Doppler peaks (it looks exact in the figure) for an oscillation which bounces at $d-A=10^{-3}r_{\rm s}$, with a frequency $\omega$ which allows rays which enter at a minimum of $R$ to also exit at a minimum. The rays which maximise the in-crossing and out-crossing Doppler effects almost coincide with the ones which cancel out the redshift effect, and even more so with each other. Even when the peaks do not exactly coincide, due to their widths the net Doppler effect given by their product can be very close to its maximum possible value.

To conclude, these resonant cases have allowed us to understand the behaviour of the quotient $g/h$ around its highest values, and relate it to specific dynamical regimes of the shell. As we will see, the observed regions of rapid increase or decrease will have significant influence on the behaviour of semiclassical effects.

\subsection{Semiclassical effects}

So far we have studied the dispersion of light rays (or analogously, of modes of the massless scalar field) which cross the oscillating shell and pass through the interior Minkowski region. From these results we can directly calculate the semiclassical quantities discussed earlier, namely the ETF and the RSET. The behaviour of these quantities will be similar to that of the dispersion functions described above, as the former are constructed simply from derivatives of the latter. The structure of peaks and plateaus for each period of the oscillation will merely become more exaggerated for these new functions. For reference, the structure of the ETF due to a single interval of deceleration during collapse has been previously studied with some detail in \cite{Harada2019}. Our study is based on quite different dynamics, but the results are qualitatively similar.

\begin{figure}
	\centering
	\includegraphics[scale=.6]{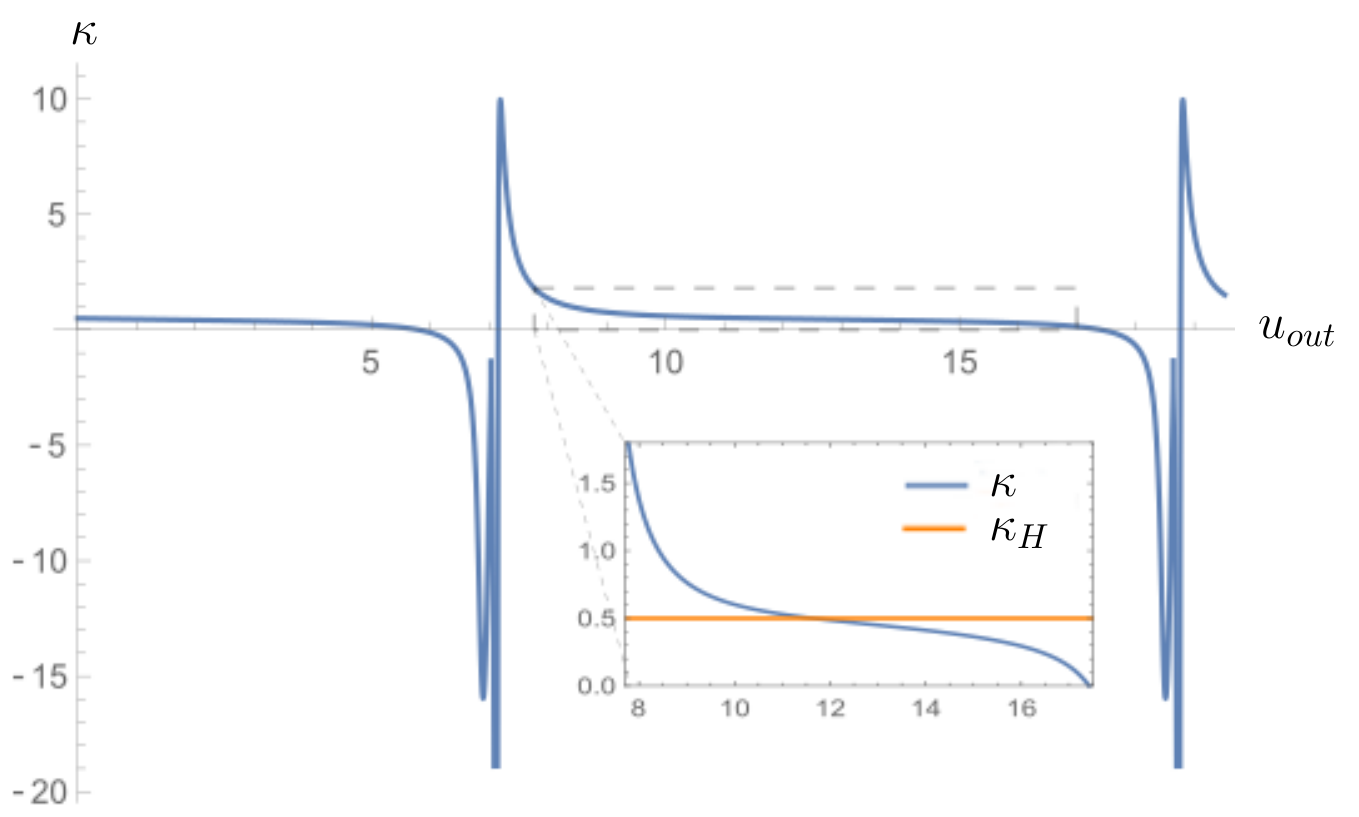}
	\caption{ETF $\kappa_{u_{\rm in}}^{u_{\rm out}}$ produced by an oscillating shell with the same parameters as the ones used for fig.~\ref{f4}, for which the net redshift effect is maximised. The small plot inside the main one is a magnification of the plateau region, along with a comparison with the value $\kappa_{\rm H}$ of the function in the case of Hawking radiation.}
	\label{f7}
\end{figure}

In fig.~\ref{f7} we can see the ETF $\kappa_{u_{\rm in}}^{u_{\rm out}}$, which contains information of the flux of particles seen in the ``in" vacuum state by an inertial observer at future null infinity, calculated with the relation between the ``in" and ``out" coordinates given by the product of the functions plotted in fig.~\ref{f4} through eq. \eqref{17}. As can be guessed by observing the curves in fig.~\ref{f4}, the smaller peaks in $\kappa_{u_{\rm in}}^{u_{\rm out}}$ are produced around the maxima of the Doppler effect contributions. On the other hand, the largest negative and positive peaks are produced in the regions of large gradient on either side of the maximum of the redshift contribution (keep in mind that the horizontal axes of the two plots are rescaled versions of each other). Between each set of peaks there is a region of smoothly decreasing EFT, with values around the Hawking temperature of a BH with the same mass.

\begin{figure}
	\centering
	\includegraphics[scale=.6]{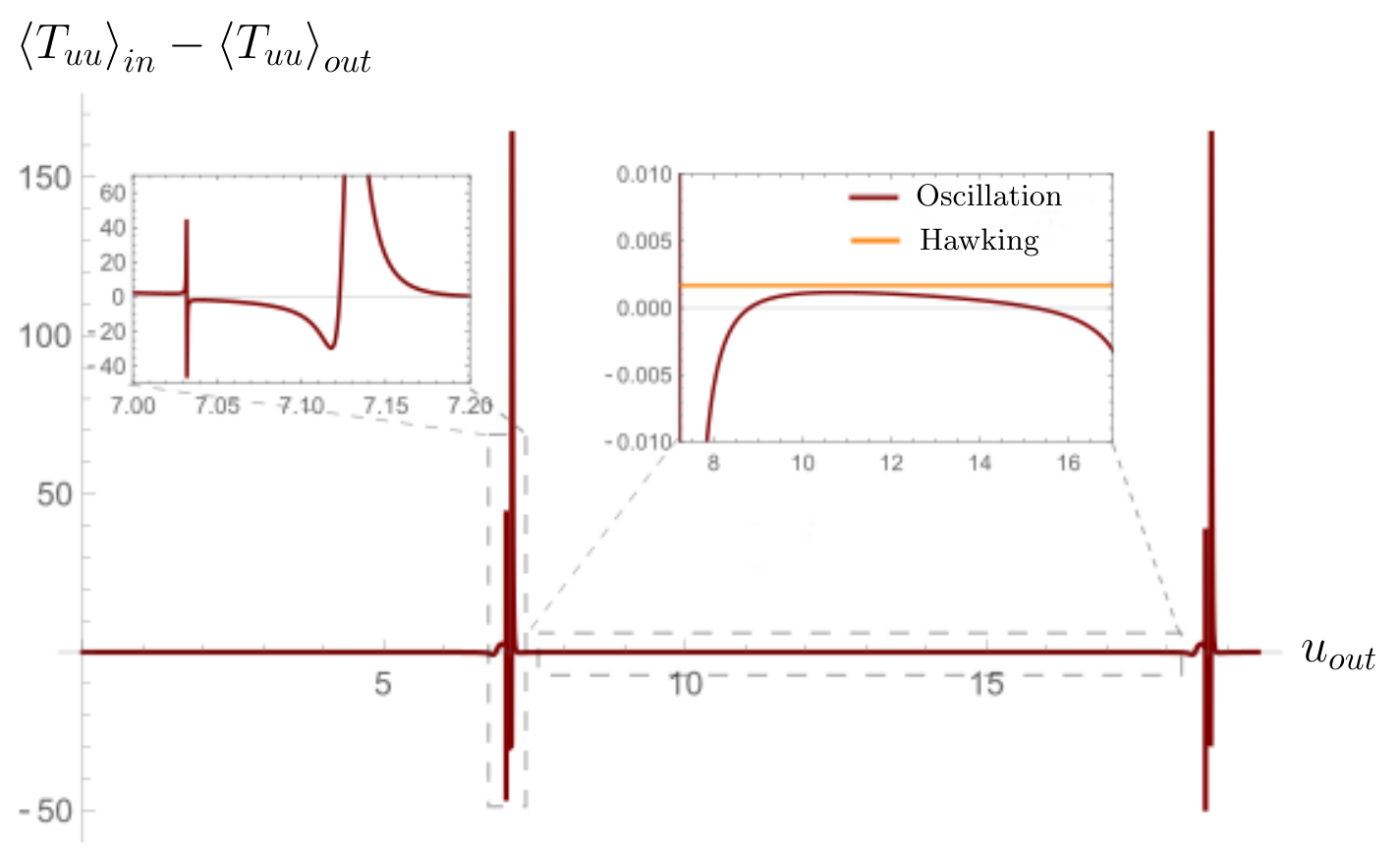}
	\caption{Difference between the $u_{\rm out}u_{\rm out}$ components of the RSET in the ``in" and ``out" vacuum states, corresponding to the outgoing flux of radiation which appears at future null infinity. As in the case of the ETF, we observe periodic peaks, which correspond to the rays which enter at a maximum of the oscillation and exit at a minimum, which maximises the redshift effect, and a more flat intermediate region of values near that of the Hawking radiation flow produced after the formation of a horizon, superimposed in the right zoomed-in rectangle.}
	\label{f8}
\end{figure}

In fig.~\ref{f8} we have plotted the outgoing radiation flux at future null infinity, defined as the difference between $\expval{T_{u_{\rm out}u_{\rm out}}}$ evaluated for the ``in" and ``out" vacuum states. From equations \eqref{27} we see that this quantity depends on $\kappa_{u_{\rm in}}^{u_{\rm out}}$ and its derivative, explaining the somewhat similar, but amplified, characteristics. This quantity alone is representative of the highs and lows of the RSET during the oscillation, since the term which is missing is simply the Boulware vacuum polarisation, which maintains low values in the $u_{\rm out}$ coordinate (outside the horizon it is below the Hawking flux value in fig.~\ref{f8}). It is the $u_{\rm out}$ coordinate itself which tends to become non-regular, leading to a general amplification of both terms (tending to a divergence at the horizon if they do not perfectly compensate each other).

\begin{figure}
	\centering
	\includegraphics[scale=.6]{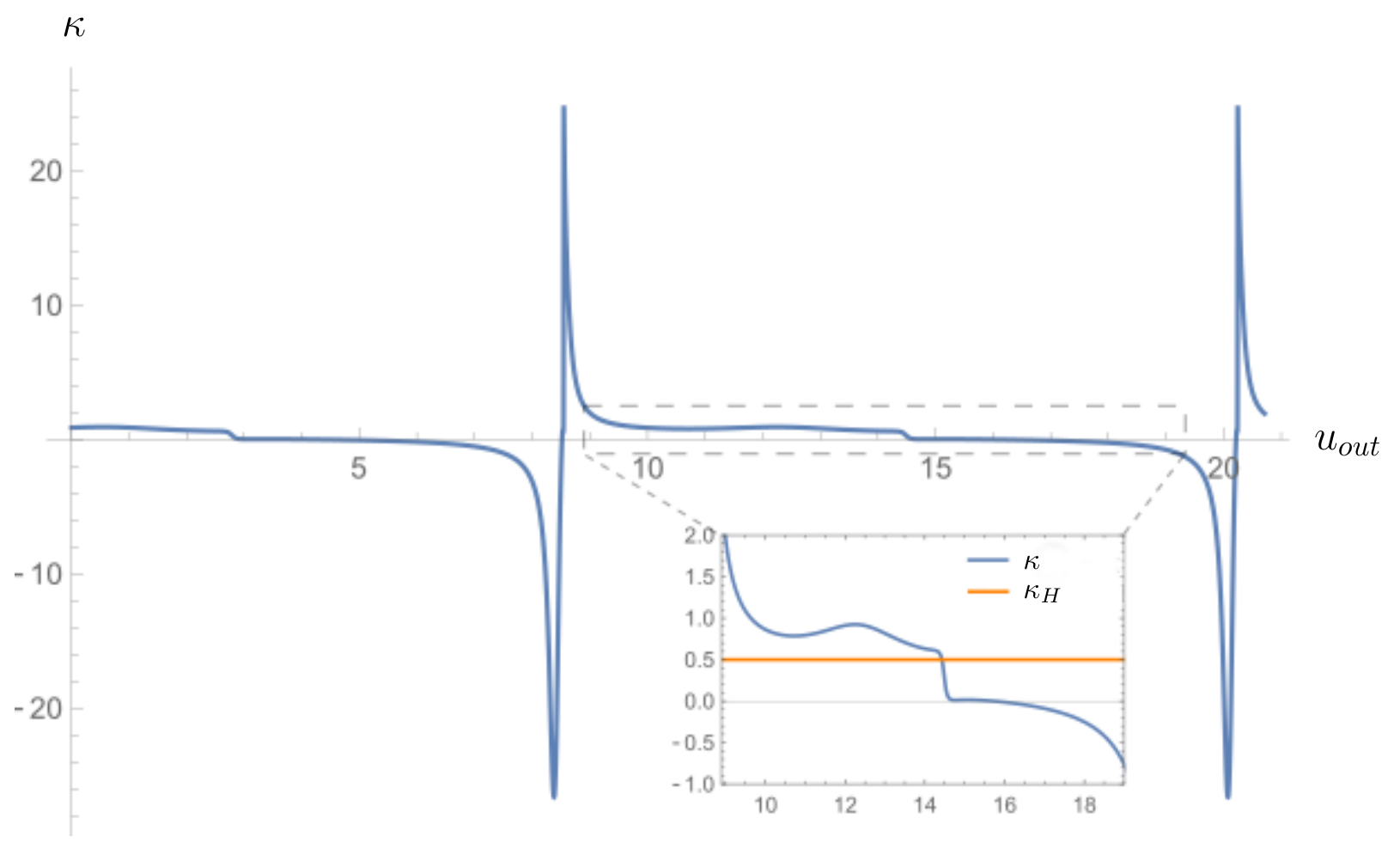}
	\caption{ETF in the outgoing radiation sector, for an oscillation which maximises the net Doppler effect, obtained with the functions plotted in fig.~\ref{f6}.}
	\label{f9}
\end{figure}

In fig.~\ref{f9} we observe the ETF for the oscillation which maximises the net Doppler effect. The two most notable differences with respect to the case which maximises redshift are the somewhat cleaner large peaks, caused by a better overall coincidence in the aspects of the in-crossing and out-crossing effects around the minima of the oscillation, and a less clean intermediate region, caused in turn by a worse coincidence there.

\begin{figure}
	\centering
	\includegraphics[scale=0.6]{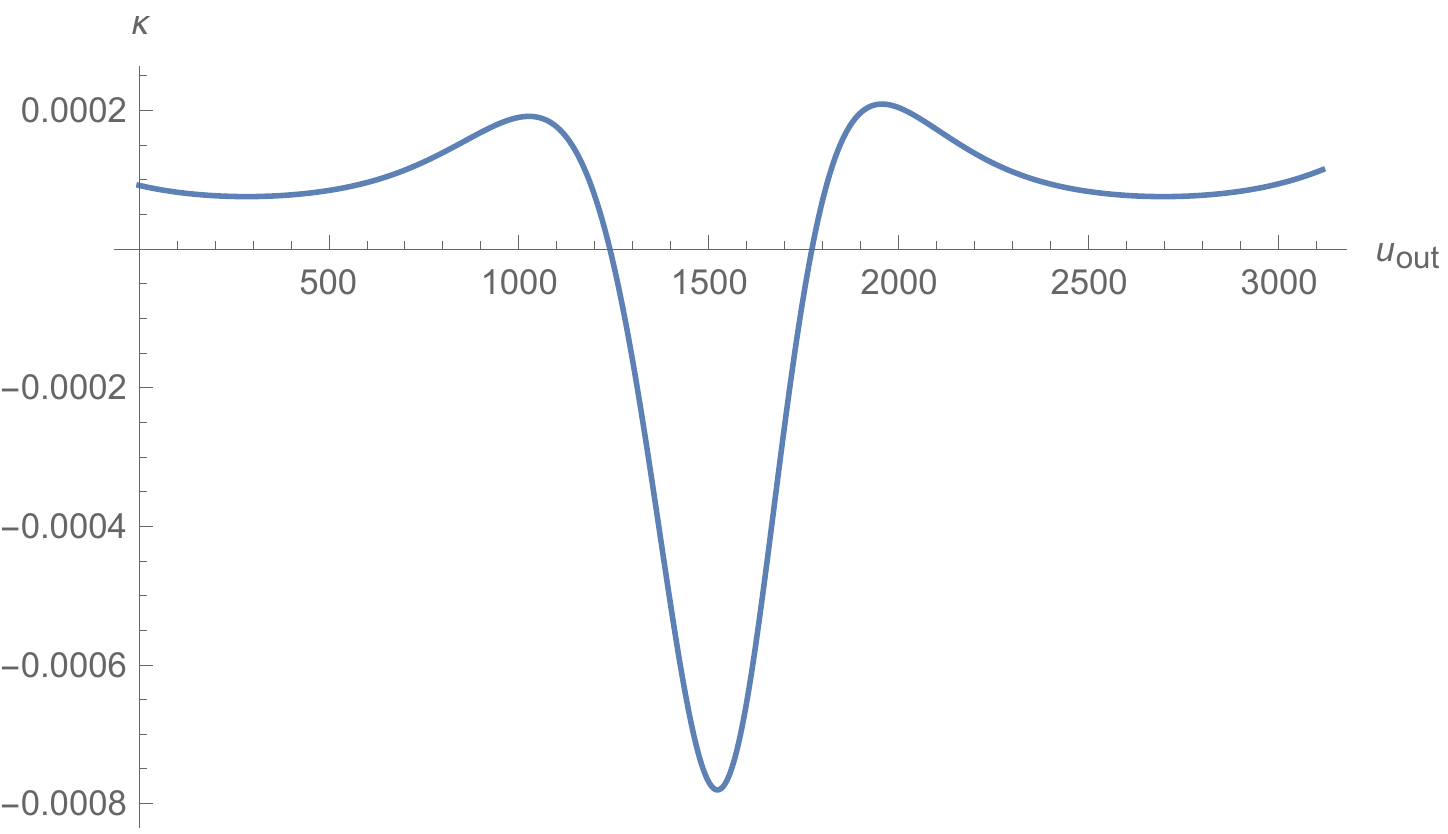}
	\caption{ETF in the outgoing radiation sector, for an oscillation at a low velocity (less then or equal to $0.15\%$ the speed of light) and with a radial proximity to the horizon between $10^{-3}r_{\rm s}$ and $0.2r_{\rm s}$ (the same as in all the cases seen so far). Compared to the cases with higher velocities, we observe a significant decrease of its values and a smoothing of its derivative.}
	\label{f10}
\end{figure}

In order to give a more general picture of the semiclassical effects produced by this type of shell trajectory, we can study the consequence of changing the order of magnitude of each of the oscillation parameters. First, in fig.~\ref{f10} we see the behaviour of the ETF for an oscillation with the same proximity to the horizon (between 0.001 and 0.201 \mbox{times $r_{\rm s}$}) but with a much lower velocity, reaching at most about $0.15\%$ of the speed of light. In this case all semiclassical fluxes are greatly diminished, approaching the static shell limit in which the radiation temperature and flow become zero.

Another possibility is to maintain the maximum proximity to the horizon ($10^{-3}r_{\rm s}$) and the large maximum speed ($\sim99\%$ the speed of light), but to vary the amplitude of the oscillation. Decreasing the amplitude leads to a qualitatively similar result for the ETF: each period contains a cluster of large peaks (larger as the amplitude decreases) surrounded by a region of values close to the Hawking temperature. On the other hand, increasing the amplitude to above $0.1r_{\rm s}$ leads to a general decrease in the values at both the peaks and the intermediate regions. While the shell still gets close to the would-be horizon, the more gradual changes in velocity which come with the larger amplitude make this case more similar to the slow approach toward the Schwarzschild radius which will be discussed in detail in chapter \ref{ch3}. Let us only briefly mention that while the flux at infinity may be small in these cases, the RSET close to $r_{\rm s}$ still tends to become very large.

The last parameter we can vary is the proximity to the horizon. Understandably, if the shell oscillates very far from the horizon, the ETF and its first derivative become very small, even if the maximum velocities are large. On the other hand, if the shell is close to the horizon, around $10^{-3}r_{\rm s}$ or closer at the minimum of the oscillation, and its amplitude is not very large, then the closer it is, the larger the peaks become, but the intermediate region again remains at values around the Hawking temperature on average.

\section{Summary and discussion}

In this chapter we have seen how a thin spherical shell oscillating just above its own Schwarzschild radius can produce bursts of radiation from the quantum vacuum. Depending on the frequency and amplitude of the oscillations and how these resonate with the light-crossing time of the object, the profile of the radiation flux in time can have more pronounced peaks and/or long periods of nearly thermal emission. The bursts of non-thermal emission in fact correspond to the parts of the dynamics in which the shell gets the closest to its Schwarzschild radius, slows down, and bounces back before crossing it; the mismatch between the ETF and the thermal flux necessary to cancel the Boulware divergence in \eqref{27} for these regions already implies large values of the RSET near the shell surface, a fact further exacerbated by the height of the emission peaks.

As far as the emission goes, although for an object of astronomical size these bursts have a significant suppression due to the Planck constant in the RSET (e.g. the flux terms should be multiplied by $l_{\rm p}^2/r_{\rm s}^2$), this may well be compensated by the height of the peaks if the surface gets very close (e.g. a Planck length away) to $r_{\rm s}$ at its minimum radius. While this toy model of a shell with an empty interior is far from a complete model of an ultracompact astrophysical object, it does offer a light-bending behaviour which one expects of such objects and thus a qualitatively robust model for the relation between the ``in" and ``out" sectors.

Perhaps the most striking result is the large gradient region in fig.~\ref{f5}. It clearly points to a set of configurations in which semiclassical corrections can become extremely large due to both Doppler and redshift effects: namely, ones in which the matter surface approaches its Schwarzschild radius at low velocities. We will now explore the magnitude of semiclassical effects in a new set of dynamical scenarios which pass through this region of configuration space, particularly when the shell actually reaches zero redshift and forms a horizon.


\chapter{Horizon formation from slowly collapsing matter}\label{ch2}

The next step in our study is to see exactly what happens when the Schwarzschild radius is crossed at a finite regular time. Particularly, it is a well established fact that the asymptotic ETF is always $1/2r_{\rm s}=1/4M$ in this case (corresponding to the Hawking temperature in natural units~\cite{Hawking1975}), and that this provides an additional term in the RSET with respect to its Boulware vacuum value that precisely regularises the divergence at the horizon \cite{DFU,Fabbri2005}. We will also see this explicitly in the next chapter.

In this chapter we will be interested in semiclassical effects produced in a finite time interval around the formation of a horizon by a shell collapsing \textit{at different velocities lower than the speed of light} at the moment of crossing the horizon. Past studies in this direction, although detailed, have usually involved only a shell collapsing at light-speed (e.g. \cite{SC2014}), justified by the fact that during astrophysical black-hole formation, the velocity of falling matter is expected to be high when crossing the horizon. By contrast, as mentioned earlier, the goal of this part of the thesis is to thoroughly study the semiclassical effects produced in more general dynamical situations.

Since in this case the asymptotic solutions for both the ETF and RSET are known, we will be more interested in short-term dynamical effects. At horizon formation, large values of the RSET are to be expected if the ``in" vacuum approximates the static Boulware vacuum in some way (say, in the case of a very slow collapse). Therefore, it is at the horizon itself where we might expect the most clear estimate of how large semiclassical effects can become. We will thus be interested in obtaining the total values of the RSET components there. To give them a more physical interpretation, we will also calculate the corresponding values of the vacuum energy density and the radial pressure measured by free-falling observers.

\section{Conformal factor at the horizon}

In order to obtain the values of the RSET in the ``in" vacuum, we need to calculate the conformal factor $C(u_{\rm in},v_{\rm in})$ which allows us to write the part of the metric \eqref{geo} restricted to the time-radius subspace in the $u_{\rm in},v_{\rm in}$ null coordinates,
\begin{equation}
	ds^2_{(2)}=-C(u_{\rm in},v_{\rm in})du_{\rm in}dv_{\rm in}.
\end{equation}
Just as a reminder, the ``in" vacuum state of the dimensionally reduced problem is defined by the plane waves in the asymptotic Minkowski region at past null infinity, which is entirely within the asymptotic region of the exterior Schwarzschild geometry if the shell never reaches the speed of light in the past. The ingoing modes, labelled by $v_{in}$, either fall directly into the singularity or are reflected at the origin $r_-=0$ and from there either escape before the formation of the horizon and reach future null infinity, or fall into the singularity. If they reach future null infinity, they can be labelled by the coordinate $u_{\rm out}$, the value of which is a function of the previous label $v_{\rm in}$. Any point in the geometry outside the event horizon (both on the exterior and interior of the shell) can be labelled by a pair $(u_{\rm out}, v_{\rm in})$. In the notation introduced in eqs. \eqref{1} and \eqref{3}, $v_{\rm in}$ is simply $v_+$ and $u_{\rm out}$ is $u_+$. The dispersion of the light rays between past and future null infinity is given by
\begin{equation}
	\frac{du_{\rm out}}{dv_{\rm in}}=\left.\frac{du_+}{du_-}\frac{dv_-}{dv_+}\right|_{v_-=u_-}=\frac{g(u_-)}{h(u_-)},
\end{equation}
where we have made use of the relation $du_-=dv_-$ for the reflection of light rays at the origin, and where $u_-$ is a function of $v_{in}$ through the inverse of the integral of $h(u_-)$. Studying the values of $g(u_-)$ and $h(v_-)$, defined in \eqref{6}, from $-\infty$ until the formation of the horizon for different trajectories of collapse, one can see that $h$ is of order one throughout. On the other hand, $g$ always has a divergence at the horizon since $u_+$ reaches an infinite value while $u_-$ is still finite. The contrast in this behaviour implies that the approximation
\begin{equation}
	\frac{dv_-}{dv_{in}}\simeq 1,
\end{equation}
that is, the approximation of considering our $v_{in}$ coordinate as the Minkowski $v_-$, captures well the magnitude of the physical effects produced in the dynamics around the formation of the horizon. This approximation, apart from simplifying the calculations which follow, also allows us to obtain results while only fixing the trajectory of the shell in an arbitrarily small region around the point of horizon-crossing.

From this point on we will drop the subscripts from the two null coordinates we will use for the most part: $v\equiv v_+$ and $u\equiv u_-$ (we will not use $u_{\rm out}$ since it is divergent at the horizon). Also, we will mostly use the radial coordinate in the exterior region, so $r$ will always refer to $r_+$.

From equations \eqref{3} and \eqref{6} we see that the conformal factor of the dimensionally reduced geometry as a function of $u$ and $v$ is
\begin{equation}
	C(u,v)=|f(r(u,v))|g(u).
\end{equation}
Since we are interested in calculating the RSET at the horizon, where large values might be expected for it, we must evaluate the above quantity and at least its first two derivatives there. A minor inconvenience in that process is the fact that the explicit form of $r(u,v)$ is not generally available, and numerical calculations cannot be relied upon either, since at the horizon $f$ is zero and $g$ diverges. To handle this difficulty, we will use an expansion for $r(u,v)$ around the line corresponding to the horizon, where $u=u_{\rm h}=$ const.,
\begin{equation}\label{8q}
	r(u,v)=q_0(v)+q_1(v)(u-u_{\rm h})+\frac{1}{2}q_2(v)(u-u_{\rm h})^2+\cdots,
\end{equation}
where $q_i$ is the $i$-th derivative of $r$ with respect to $u$ evaluated at $u_{\rm h}$, namely, $q_i=\partial^i r/\partial u^i|_{u=u_{\rm h}}$. In order to calculate the RSET components, we will need up to second derivatives of the conformal factor in $u$. To evaluate them we must use the expansion of $r(u,v)$ in $u$ up to third order, due to the $1/(u-u_{\rm h})$ divergence generally present in $g(u)$. This means that we need only $q_0,q_1,q_2$ and $q_3$.

Let us now see how to calculate these coefficients. The lowest order one $q_0(v)$ is just the value of $r$ at the horizon, namely,
\begin{equation}
	q_0(v)=2M.
\end{equation}
The rest of them can be obtained through the relations
\begin{equation}\label{6q}
	\frac{\partial r}{\partial u}=-\frac{1}{2}g(u)f(r),\qquad \frac{\partial r}{\partial v}=\frac{1}{2}f(r),
\end{equation}
as we show in the following. The first of these equations evaluated at $u_{\rm h}$ gives $q_1(v)$, but its right-hand side is just as difficult to evaluate as the conformal factor itself. However, we can make use of the second equation to write the cross-derivative
\begin{equation}\label{7q}
	\frac{\partial}{\partial{v}}\frac{\partial{r}}{\partial u}=-\frac{1}{2}g(u)f'(r)\frac{\partial r}{\partial v}=-\frac{1}{2}g(u)f'(r)\frac{1}{2}f(r)=\frac{1}{2}f'(r)\frac{\partial r}{\partial u}.
\end{equation}
Taking into account that $f'(r)$ evaluated at the horizon is just $1/r_{\rm s}$, the evaluation of this equation at $u_{\rm h}$ gives us a first order differential equation for $q_1(v)$, namely \mbox{$q_1'(v)=q_1(v)/(2r_{\rm s})$}. Using this method recursively allows us to write analogous equations for all the coefficients $q_i(v)$ in \eqref{8q}. For the ones relevant to our calculation of the RSET we obtain
\begin{equation}\label{17q}
	\begin{split}
		&q_1'(v)=\frac{1}{2r_{\rm s}}q_1(v),\\
		&q_2'(v)=\frac{1}{2r_{\rm s}}q_2(v)-\frac{1}{r_{\rm s}^2}q_1^2(v),\\
		&q_3'(v)=\frac{1}{2r_{\rm s}}q_3(v)-\frac{3}{r_{\rm s}^2}q_2(v)q_1(v)+\frac{3}{r_{\rm s}^3}q_1^3(v).
	\end{split}
\end{equation}
Initial conditions for these equations can be found by fixing the zero of the $v$ coordinate at the point of horizon formation, and considering the relation $r_+=r_-$ at the surface of the shell.

For a shell which crosses the horizon with an approximately constant radial velocity as seen from the inside ($\alpha_-=dv_-/du_-\simeq const.$), from equations \eqref{2} and \eqref{5} we get the relation
\begin{equation}\label{28q}
	r_-\simeq r_{\rm s}+\frac{\alpha_--1}{2}(u-u_{\rm h})
\end{equation}
at the shell surface, which gives us the initial conditions $q_1(0)=(\alpha_--1)/2$, $q_2(0)=0$ and $q_3(0)=0$ (these last two are approximate if $\alpha_-$ is only approximately constant, but the important aspects of our final results do not change if they have different values). We solve the above equations to get
\begin{equation}\label{29q}
	\begin{split}
		&q_1(v)=-\frac{1-\alpha_-}{2}e^{v/2r_{\rm s}},\\
		&q_2(v)=\frac{(1-\alpha_-)^2}{2r_{\rm s}}e^{v/2r_{\rm s}}(1-e^{v/2r_{\rm s}}),\\
		&q_3(v)=-\frac{3(1-\alpha_-)^3}{8r_{\rm s}^2}e^{v/2r_{\rm s}}(1-4e^{v/2r_{\rm s}}+3e^{v/r_{\rm s}}).
	\end{split}
\end{equation}

\section{RSET at the horizon for the ``in" vacuum}

\begin{figure}
	\centering
	\includegraphics[scale=0.7]{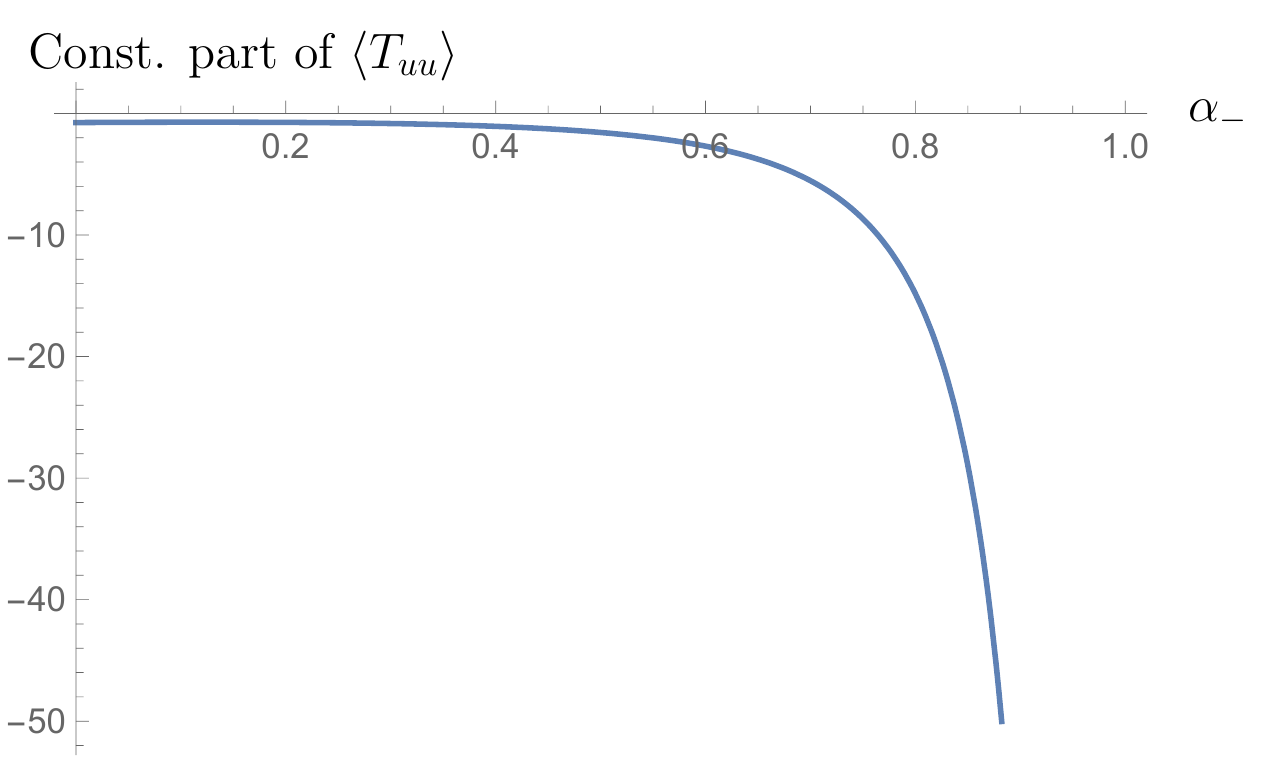}
	\caption{Plot of the constant part in the parenthesis of eq. \eqref{31a} as function of $\alpha_-$ in its domain of possible values. It has negative values throughout and a divergence at $\alpha_-=1$.}
	\label{f17}
\end{figure}

We now have everything prepared to calculate the RSET components at the horizon. Substituting the solutions \eqref{29q} into the series expansion \eqref{8q}, we see how $f$ depends on $u$ and $v$ up to third order in $(u-u_{\rm h})$. As for resolving the dependence of $g$ in $r$ (which appears through $\alpha_+$), we must remember the definition of this function \eqref{6} which tells us that it is evaluated at the shell surface. Therefore, close to the horizon, we can simply use the expression for $r$ given in \eqref{28q}. With these functions we can obtain $C(u,v)$ up to second order in $(u-u_{\rm h})$ [remember $g$ has a leading term $1/(u-u_{\rm h})$],
\begin{equation}
	\begin{split}
		C(u,v)&=(1-\alpha_-)e^{v/2r_{\rm s}}+\left[\left(-\alpha_-^2+\frac{3}{2}\alpha_--1\right)e^{v/2r_{\rm s}}+(1-\alpha_-)^2e^{v/r_{\rm s}}\right]\frac{u-u_{\rm h}}{r_{\rm s}}\\&+\frac{e^{v/2r_{\rm s}}}{8(1-\alpha_-)}\left[3-10\alpha_-+12\alpha_-^2-10\alpha_-^3+3\alpha_-^4\right.\\&\left.-4(1-\alpha_-)^2(3-5\alpha_-+3\alpha_-^2)e^{v/2r_{\rm s}}+9(1-\alpha_-)^4e^{v/r_{\rm s}}\right]\frac{(u-u_{\rm h})^2}{r_{\rm s}^2}+\cdots.
	\end{split}
\end{equation}
Finally, we can use the expression \eqref{e33} to obtain the components of the RSET at the horizon:
\begin{subequations}\label{31q}
	\begin{align}
		\begin{split} \expval{T_{uu}}&=\frac{1}{24\pi r_{\rm s}^2}\left(\frac{-6\alpha_-^4+16\alpha_-^3-27\alpha_-^2-16\alpha_--6}{8(1-\alpha_-)^2}\right.\\&\hspace{35mm}\left.+\frac{\alpha_-}{2}e^{v/2r_{\rm s}}+\frac{3}{4}(1-\alpha_-)^2e^{v/r_{\rm s}}\right), \end{split}\label{31a}\\
		\expval{T_{uv}}&=-\frac{1}{24\pi r_{\rm s}^2}\frac{1-\alpha_-}{2}e^{v/2r_{\rm s}},\\
		\expval{T_{vv}}&=-\frac{1}{24\pi r_{\rm s}^2}\frac{1}{8}.
	\end{align}
\end{subequations}
Their behaviour can be read easily, except perhaps for the first constant term in the parenthesis in $\expval{T_{uu}}$, which has been plotted as a function of $\alpha_-$ in fig.~\ref{f17}. The following observations can be made:
\begin{itemize}
	\item Firstly, the components seem to grow exponentially on the horizon as time passes. This, however, turns out to be a consequence of the coordinate system in which they are expressed. In a system more appropriate for the static Schwarzschild region, say the Eddington-Finkelstein advanced coordinates $(v,r)$, this behaviour is suppressed by factors of $1/C$ arising from the relation $\partial u/\partial r$. A more detailed analysis of the energy density and flux perceived by a free-falling observer will follow shortly.
	\item The second thing one might notice is that $\expval{T_{vv}}$ is constant and therefore completely independent from the dynamics of the collapse. This is a consequence of the fact that we have chosen the Eddington-Finkelstein $v$ coordinate, which is not affected by the interior Minkowski region. We also note that this ingoing flux term is negative, and can thus be identified with the negative energy which drives Hawking evaporation locally at the horizon~\cite{DFU}. In Part \ref{pt2} of this thesis we will explicitly calculate the backreaction on the geometry stemming from this term in a simpler collapse scenario.
	\item Finally, we note that the $\expval{T_{uu}}$ component diverges as $\alpha_-\to 1$, that is, as the collapse becomes slower, approaching the static limit. As we will see, the \mbox{$1/(1-\alpha_-)^n$} terms are then exponentially suppressed in time (in the regular Eddington-Finkelstein coordinates), but they play an important role near the point of horizon formation. The large values of this component for a configuration approaching staticity is reminiscent of the Boulware state divergence of the RSET~\cite{Fabbri2005}, and of a firewall configuration~\cite{Almheiri2013}. The subsequent suppression of these terms can then be seen as a direct consequence of thermalisation.
\end{itemize}

\section{Energy density, flux and pressure perceived by a free-falling observer at the horizon}

Let us consider the four-velocity $w$ of a free-falling observer in the Schwarzschild geometry expressed in $(u,v)$ coordinates, evaluated at the moment of horizon crossing. It has the form
\begin{equation}
	w^\rho=\left(\sqrt{\frac{2}{\beta_0}}\frac{1}{C},\sqrt{\frac{\beta_0}{2}},0,0\right),
\end{equation}
where $\beta_0$ is related to the radius $r_0$ from which the free fall was initiated through
\begin{equation}
	\beta_0=\frac{1}{2}\frac{1}{1-r_{\rm s}/r_0}.
\end{equation}
Let us also introduce the space-like unitary vector perpendicular to this four-velocity and pointing in the outward radial direction,
\begin{equation}
	z^\rho=\left(-\sqrt{\frac{2}{\beta_0}}\frac{1}{C},\sqrt{\frac{\beta_0}{2}},0,0\right).
\end{equation}
We now define the effective energy density $\rho$, flux $\Phi$ and pressure $p$ perceived by this observer as
\begin{subequations}\label{36q}
	\begin{align}
		&\rho\equiv\expval{T_{\mu\nu}}w^\mu w^\nu=\frac{2}{\beta_0 C^2}\expval{T_{uu}}+\frac{1}{C}\expval{T_{uv}}+\frac{\beta_0}{2}\expval{T_{vv}},\\
		&\Phi\equiv-\expval{T_{\mu\nu}}w^\mu z^\nu=\frac{2}{\beta_0 C^2}\expval{T_{uu}}-\frac{\beta_0}{2}\expval{T_{vv}},\\
		&p\equiv\expval{T_{\mu\nu}}z^\mu z^\nu=\frac{2}{\beta_0 C^2}\expval{T_{uu}}-\frac{1}{C}\expval{T_{uv}}+\frac{\beta_0}{2}\expval{T_{vv}}.
	\end{align}
\end{subequations}
As an aside, we note that the conformal factor at the horizon,
\begin{equation}\label{37q}
	C(u_{\rm h},v)=(1-\alpha_-)e^{v/2r_{\rm s}},
\end{equation}
is not equal to 1 when $v=0$, where the geometry must match with the interior Minkowski region, because we are not using the Minkowski $v_-$ coordinate. If we were, we would have to multiply $C$ by $h=dv_+/dv_-$, which at the point of horizon formation has the value $h=1/(1-\alpha_-)$.

\begin{figure}
	\centering
	\includegraphics[scale=0.55]{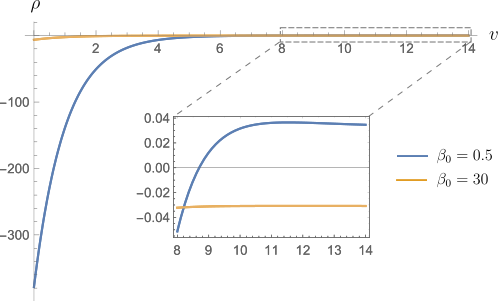}
	\includegraphics[scale=0.55]{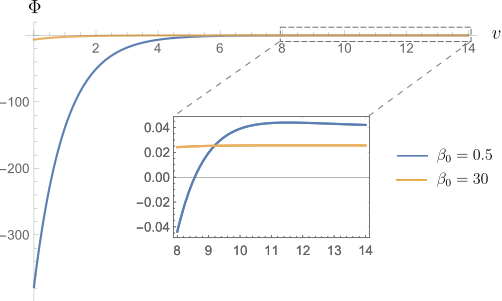}
	\includegraphics[scale=0.55]{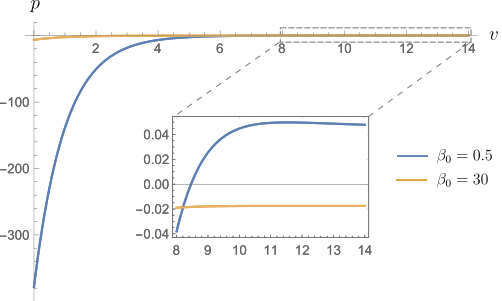}
	\caption{Perceived energy density, flux and pressure at the horizon, as a function of the Eddington-Finkelstein $v$ coordinate, for $\beta_0=0.5$ (free fall from $r_0\to\infty$) and for $\beta_0=30$ (free fall from $r_0\simeq1.017r_{\rm s}$), in the Schwarzschild region of a collapse with parameter $\alpha_-=0.9$ ($v_{\rm therm}\simeq9.2r_{\rm s}$). The point $v=0$ marks the formation of the horizon. Immediately after, we observe these function have very large (negative) values. This is a direct consequence of the proximity of the parameter $\alpha_-$ to 1, as discussed in the text. On the other hand, the asymptotic values are always small (except for observers which start their free fall very close to the horizon, where $\beta\to\infty$). The sign of the asymptotic energy density and pressure depend of the velocity of the observer (they are negative for slower observers), while the outgoing flux is always positive, in accordance with the evaporating black-hole scenario.}
	\label{f18}
\end{figure}

We thus see that the growing exponentials appearing in eqs. \eqref{31q} do not show up in the scalar quantities in \eqref{36q}. In fact, these turn out to have constant, finite asymptotic values that depend on the initial condition $\beta_0$ of the free falling observer
\begin{subequations}
	\begin{align}
		&\rho\xrightarrow[v\to\infty]{} \frac{1}{24\pi r_{\rm s}}\left(-\frac{1}{2}-\frac{\beta_0}{16}+\frac{3}{2\beta_0}\right),\\
		&\Phi\xrightarrow[v\to\infty]{} \frac{1}{24\pi r_{\rm s}}\left(\frac{\beta_0}{16}+\frac{3}{2\beta_0}\right),\\
		&p\xrightarrow[v\to\infty]{} \frac{1}{24\pi r_{\rm s}}\left(\frac{1}{2}-\frac{\beta_0}{16}+\frac{3}{2\beta_0}\right).
	\end{align}
\end{subequations}
When these values are approximately reached, the system can be said to have thermalised, as all other terms are suppressed exponentially. A measure of the time it takes to do so, in the $v$ coordinate, for a slow collapse (when $\alpha_-$ is close to 1) is given by the value
\begin{equation}
	v_{\rm therm}= 4r_{\rm s}\log(\frac{1}{1-\alpha_-}).
\end{equation}
In fig.~\ref{f18} we see plots for $\rho$, $\Phi$, and $p$ for two different values of $\beta_0$, which make the asymptotic values of $\rho$ and $p$ have different signs. Except for the case of extremely small values of $\beta_0$, the asymptotic values of the previous quantities are always negligibly small due to the suppression of the RSET by Planck's constant (which has been omitted in the choice of units). However, this smallness can be compensated during the transient phase of the collapse. Near the point of horizon formation we have $1/(1-\alpha_-)^4$ terms, originating from the $1/(1-\alpha_-)^2$ term in $\expval{T_{uu}}$ in \eqref{31q} and from the $1/C^2$ term evaluated from~\eqref{37q}. These terms can be made arbitrarily large if $\alpha_-$ is very close to 1, compensating the suppression by Planck's constant.

With these results we see that the RSET approaches a physical divergence in the static limit $\alpha_-\to1$. At the limit itself this divergence is hardly surprising, as for a static shell the ``in" vacuum essentially becomes the Boulware one. What is interesting is the fact that this limit can be approached through the single velocity parameter $\alpha_-$ of the shell when it crosses the horizon, without imposing any conditions on its past evolution. This seems to indicate that during the formation of a BH, if by some mechanism the collapse of matter were to be slowed down just before it forms a horizon, its subsequent evolution would become a problem which requires a full semiclassical treatment.

\section{Summary and discussion}

In conclusion, a shell forming a horizon at a very low speed can produce arbitrarily large values of the RSET, tending to a divergence in the static limit. This divergence can be seen as a combination of both a Boulware-state-like behaviour, and the purely dynamical Doppler-like terms discussed in the previous chapter. If backreaction on the geometry is ignored, these large values are quickly suppressed as the BH thermalises, after which the standard evaporative evolution may be expected. However, if backreaction is taken into account sooner, then the dynamics of the collapse itself may be altered before a horizon ever forms.

Combined with the existence of semiclassically sustained static ultracompact objects~\cite{Carballo-Rubio2017,Arrechea2021}, this result strongly suggests that in a more realistic collapse scenario, if the initial conditions are such that the classical matter moves ``slowly" when it is about to cross its own horizon, backreaction may instead lead it to a horizonless equilibrium state. A possibility as to how these initial conditions may come about will be discussed in Part \ref{pt2} of this thesis.


\chapter{Asymptotic horizon formation}\label{ch3}

The final type of geometric configuration we will analyse in this part of the thesis is one in which the formation of a horizon is approached only asymptotically in time, resulting, as we will see, in light-trapping behaviour without the presence of actual trapped surfaces. We find that these geometries cause an emission of thermal Hawking radiation, confirming the fact that no strict trapped surface is required for the geodesic peeling which results in this emission~\cite{Barceloetal2006,BLSV06,BLSV11,Barceloetal2011b}.

In this chapter we will fully explore the properties of these particular geometries. We will work with a more general setup than the thin-shell model of the previous chapters, presenting an exhaustive catalogue of spherically-symmetric geometries with this light-trapping behaviour. Before going into an analysis of their semiclassical properties, we will first present a thorough study of their causal features. We will begin by briefly presenting the geometries in section~\ref{Sec:geometries}, then analysing their distinctive causal structures in section~\ref{Sec:causality}. As we will see, the interesting feature of these spacetimes from a purely geometric perspective is that they contain no trapped surfaces, and yet they form an event horizon. Initially we will once again present them as geometric \textit{ad hoc} constructions, although later in the chapter (in section ~\ref{Sec:energyconditions}) we will analyse in detail whether they could be obtained as solutions of Einstein's equations for some plausible matter content, discussing the energy conditions that can be satisfied around the trapping region while supporting these configurations. Finally, we will look at their semiclassical characteristics. We will show how their thermal emission is generally at a temperature lower than that of the final (asymptotic) horizon's surface gravity, and how this causes the RSET in the vicinity of their would-be horizon to grow exponentially quickly in time.

\section{The geometries}
\label{Sec:geometries}

We start with a generic family of spherically symmetric metrics written in advanced Eddington-Finkelstein coordinates,
\begin{equation}\label{1c}
ds^2=-f\mathop{dv^2}+2g\mathop{dv}\mathop{dr}+r^2\mathop{d\Omega^2},
\end{equation}
where $f$ and $g$ are generally functions of $v$ and $r$, or just of $r$ in static cases. We use the advanced null coordinate $v$ because we are interested only in the future part of the causal structure of these spacetimes (for the past region we assume that they are asymptotically flat, such that we can construct an ``in" vacuum). We will also assume that the geometries are regular at $r=0$. We use this simplifying assumption to avoid causal aspects associated with singularities and concentrate just on those due to the presence of horizons. However, we note that our analysis can be generalised straightforwardly to geometries with singularities, so long as they do not overlap with the region of light-ray trapping that we will study. We restrict our analysis to this local region, which can easily be inserted into a spacetime with a different global structure with much the same consequences.

We will work with two types of geometries, both of which trap outgoing light rays, but are otherwise quite different from one another. Let us present these two cases by first looking at two static configurations, which will later become the asymptotic limit in time of our dynamical models discussed in section~\ref{s3}.

\subsection{Static configurations}
\label{Sec:static}

The two types of configurations which we will study in this chapter originate from a simple consideration. From the line element \eqref{1c} with $f$ and $g$ depending only on $r$, the equation which governs the paths of the outgoing light rays is
\begin{equation}\label{2c}
\frac{dr}{dv}=\frac{1}{2}\frac{f(r)}{g(r)}.
\end{equation}
From this equation, it is apparent that these null trajectories do not distinguish between a situation in which $f$ is zero (as occurs for certain values of the radial coordinate in Schwarzschild, Reissner-Nordström, or similar spacetimes) and one in which $g$ diverges. Needless to say, the two situations are physically quite different in spite of this.

In these static configurations we will assume that in either case the right-hand side of~\eqref{2c} is zero at some radius $r_{\rm h}$ and that it can be expanded in a power series around this point approaching both from the inside \mbox{$r<r_{\rm h}$}, 
\begin{equation}\label{3c}
\frac{1}{2}\frac{f(r)}{g(r)}= k_1(r_{\rm h}-r)+k_2(r_{\rm h}-r)^2+\cdots,
\end{equation}
and from the outside $r\geq r_{\rm h}$,
\begin{equation}\label{4c}
\frac{1}{2}\frac{f(r)}{g(r)}= \tilde{k}_1(r-r_{\rm h})+\tilde{k}_2(r-r_{\rm h})^2+\cdots.
\end{equation}
Since we want the only zero of these expressions to be at $r_{\rm h}$ and we want to avoid creating a trapped region of finite volume, we require that the first non-zero coefficients $k_i$ and $\tilde{k}_j$ of both series be positive. If $g(r)\simeq {\rm const.}$ around $r_{\rm h}$, then we have a BH which allows ingoing causal trajectories across $r_{\rm h}$ but not outgoing ones. On the other hand, if $f(r)\simeq {\rm const.}$, then $g(r)$ diverges as the inverse of a polynomial, which results in the same behaviour for outgoing light rays as before (since $f/g$ is the same), but for ingoing ones there is a difference: they are actually unable to cross the surface $r=r_{\rm h}$ either. This can be deduced from the expressions which describe their paths in this coordinate system, namely the geodesic equations for their radial trajectory $(v(\sigma),r(\sigma))$,
\begin{equation}\label{5c}
v={\rm const},\quad
\ddot{r}=-\frac{\partial_r g(r)}{g(r)}\dot{r}^2\simeq\frac{m}{r-r_{\rm h}}\dot{r}^2,
\end{equation}
where the dot denotes the derivative with respect to the affine parameter $\sigma$, and $m$ is the order of the first non-zero term in the expansion \eqref{4c}. Integrating this equation allows one to see that the affine parameter reaches an infinite value when the ingoing ray gets to $r_{\rm h}$ (e.g. for $m=1$, $r-r_{\rm h}\propto e^{-c\sigma}$, with $c>0$), indicating that, as seen from the outside, this surface is actually an asymptotic region (i.e. a separate future null infinity). The interior region $r\le r_{\rm h}$ is therefore entirely separated from the exterior spacetime.

\begin{figure}
	\centering
	\includegraphics[scale=.6]{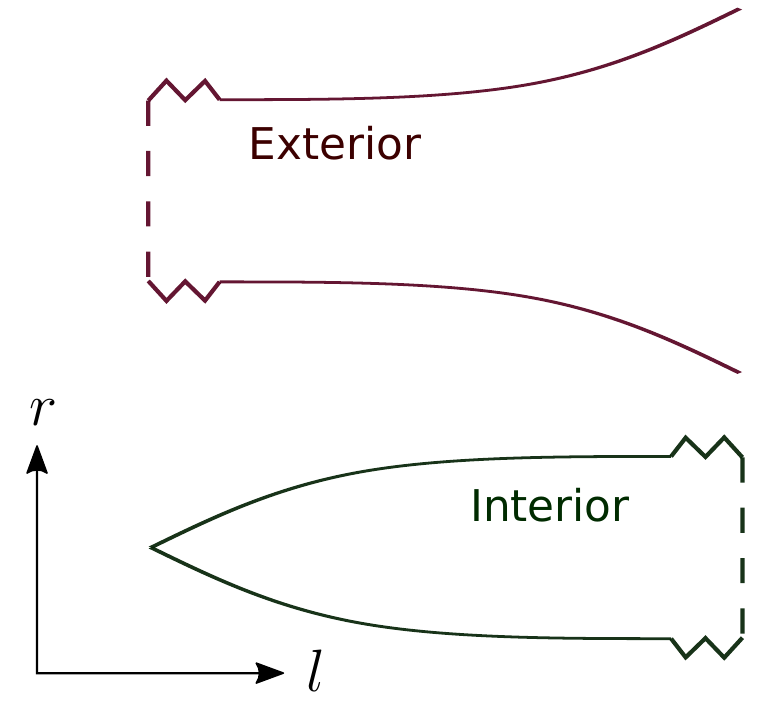}
	\caption{A qualitative representation of the relation between the radial coordinate $r$ and the proper length in the radial direction $l$ for a geometry in which $g$ diverges at some radius $r_{\rm h}$. This divergence corresponds to an infinite stretching of $l$ which completely severs the interior and exterior geometries.}
	\label{f4c}
\end{figure}

To understand this situation better, we remind the reader that there is a more well-known spacetime in which $g$ diverges: that of a traversable wormhole. Particularly, this same configuration would be a standard spherical wormhole~\cite{VisserWH} if the expansion \eqref{4c} had a leading term of order $(r-r_{\rm h})^{k}$, with $0<k<1$, as can be seen by calculating the proper radial length $l$ in slices of constant Schwarzschild time (defined by \mbox{$dt=dv-(g/f)dr$}) and expressing the radial coordinate $r$ as a function of $l$ around $r_{\rm h}$. On the other hand, when the leading order in the series is 1 or greater, as in our working case, the proper length diverges and space becomes infinitely stretched around the neck of the wormhole, akin to an infinite tube. Therefore, the static geometry we are considering here actually consists of two disjoint spacetimes, both having one infinite tubular ending (see fig.~\ref{f4c}).

\subsection{Including time-dependence}\label{s3}

\begin{figure}
	\centering
	\includegraphics[scale=.6]{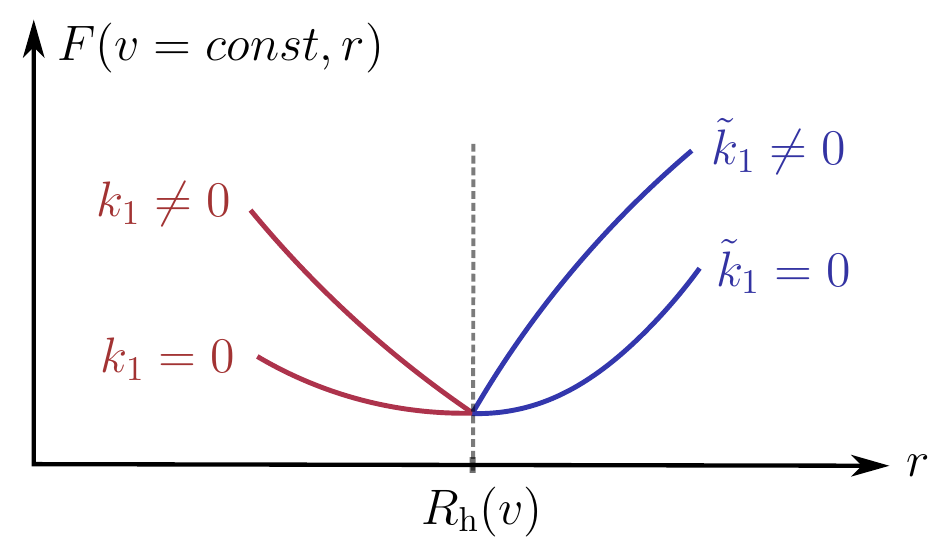}
	\caption{Slice at constant time ($v=const.$) of the generalised redshift function $F(v,r)$ around $R_{\rm h}(v)$. There is a discontinuity in the first derivative of this function at $R_{\rm h}(v)$ if either $k_1\neq 0$ or $\tilde{k}_1\neq 0$, which through the Einstein equations can translate into either a thin shell of matter, or into a discontinuity in pressure, depending on how the two individual degrees of freedom ($f$ and $g$) of the geometry which comprise $F(v,r)$ behave.}
	\label{f2c}
\end{figure}

Having discussed these static spacetimes, we will now include a time-dependence in the metric functions $f$ and $g$ in order to push the formation of the apparent horizon/asymptotic region at $r_{\rm h}$ out to the limit $v\to\infty$. We will start with some definitions. First, we will call the right-hand side of eq. \eqref{2c} the \textit{generalised redshift function} $F$,
\begin{equation}\label{6c}
F(v,r)\equiv\frac{1}{2}\frac{f(v,r)}{g(v,r)}.
\end{equation}
We will assume that this function has a minimum in $r$ at a moving point $R_{\rm h}(v)$, and that it can be approximated by a series expansion on either side,
\begin{equation}\label{7c}
F(v,r)=\Delta(v)+k_1[R_{\rm h}(v)-r]+k_2[R_{\rm h}(v)-r]^2+\cdots
\end{equation}
for $r<R_{\rm h}(v)$, and
\begin{equation}\label{8c}
F(v,r)=\Delta(v)+\tilde{k}_1[r-R_{\rm h}(v)]+\tilde{k}_2[r-R_{\rm h}(v)]^2+\cdots
\end{equation}
for $r\geq R_{\rm h}(v)$ (see fig.~\ref{f2c}). $\Delta(v)$ is a function of $v$ which decreases and tends to zero in the limit $v\to\infty$. The function $R_{\rm h}(v)$ tends to a point $r_{\rm h}$ in the same limit. Our only simplifying assumption will be that the first non-zero coefficients $k_i$ and $\tilde{k}_j$ of the expansion on either side, aside from being positive, are approximately constant at large times (or, equivalently, that they tend to a constant at least as quickly as $R_{\rm h}(v)$).

It is worth mentioning that if either or both $k_1$ or $\tilde{k}_1$ are non-zero, then the function $F$ is continuous but not smooth at $r_{\rm h}$. Through equation \eqref{6c} we see that this translates into a sharp peak in either $f$ or $g$ (or both) in slices of constant $v$. The Einstein tensor for the metric \eqref{1c} only has a second partial derivative of $f$ with respect to $r$ (in its angular components), meaning a peak in $f$ corresponds to a spherical thin shell of matter. If the peak is in $g$, the tensor is only discontinuous, and so are the matter density, flux and stress seen by any observer. We note that the geometry is perfectly regular in spite of this discontinuity, unlike what one might expect in e.g. a static stellar configuration, where a jump in pressure leads to a singularity.

Regardless of the presence of the non-smooth peak (and its corresponding non-zero surface gravity), we will call all geometries in which $F$ has one zero, or an appropriate tendency to produce one zero (see the discussion below), and is positive everywhere else, \textit{extremal} configurations. We use this name because of a shared characteristic they have with the standard extremal BH solutions: the presence of an outer and inner horizon which degenerate to the same radial position. We extend the standard definition of extremality by allowing for a non-zero surface gravity on either side of the horizon, and characterising this horizon purely by its light-ray-peeling properties (such that either $f$ can tend to zero or $g$ to a divergence).

To recover the shell model with a Schwarzschild exterior used in the previous chapters, we would simply have to impose $\Delta(v)=1-2M/R_{\rm h}(v)$ at leading order, with $R_{\rm h}$ then being the shell position, tending to the Schwarzschild radius $2M$ asymptotically. In the interest of generality, in this chapter we will assume that the two functions $\Delta(v)$ and $d_{\rm h}(v)\equiv R_{\rm h}(v)-r_{\rm h}$ decay to zero independently, and that the function $g$ can be different from 1 in the exterior region.

If both $\Delta(v)$ and $d_{\rm h}(v)$ tend to zero sufficiently fast, it turns out that after some point in time some of the light rays which are inside the sphere of radius $r_{\rm h}$ will remain trapped inside, the outermost of which defines the event horizon. This can be seen by analysing the large $v$ limit of the solutions of eq.~\eqref{2c} for $r<R_{\rm h}$. With the series \eqref{7c} we can write this equation as
\begin{equation}\label{9a}
\frac{dr}{dv}=\Delta(v)+k_1[R_{\rm h}(v)-r]+k_2[R_{\rm h}(v)-r]^2+\cdots.
\end{equation}
Let us first see the case in which $k_1\neq 0$. At leading order the equation then becomes
\begin{equation}
\frac{dr}{dv}\simeq\Delta(v)+k_1d_{\rm h}(v)-k_1(r-r_{\rm h}).
\end{equation}
From the functions $\Delta(v)$ and $d_{\rm h}(v)$ on the right-hand side we only need to consider the one which decays more slowly for the asymptotic solution. For example, if the slower of the two decays as $b\mathop{e}^{-\alpha v}$ (with $b$ and $\alpha$ some positive constants), then we can ignore the other one and obtain solutions of the form
\begin{equation}
r-r_{\rm h}\simeq -\frac{b}{\alpha-k_1}\mathop{e^{-\alpha v}}+c\mathop{e^{-k_1 v}},
\end{equation}
where $c$ is an integration constant. There are trapped solutions, which approach $r_{\rm h}$ asymptotically from below, only if $k_1<\alpha$: they correspond to the values $c<0$. On the other hand, if the slower of the two functions $\Delta(v)$ and $d_{\rm h}(v)$ goes to zero more quickly than an exponential, then we again have a solution with a leading-order term $c\mathop{e^{-k_1 v}}$ and corrections which decay much faster, asymptotically recovering the same solutions as above for any value of $k_1$. Finally, if the slower of the two functions goes to zero more slowly than an exponential, e.g. as $1/v^n$, then there are no trapped solutions at all.

Now let us see the case in which $k_1,\dots,k_{m-1}=0$ and $k_m\neq0$. Equation \eqref{9a} becomes
\begin{equation}\label{12c}
\frac{dr}{dv}\simeq\Delta(v)+k_md_{\rm h}(v)^m+k_2(r_{\rm h}-r)^m+\cdots,
\end{equation}
where we have omitted the cross-terms in the leading order. If we assume the $(r_{\rm h}-r)^m$ term dominates the right-hand side, we obtain solutions of the type
\begin{equation}
r-r_{\rm h}\sim -\frac{1}{(v-c)^{\frac{1}{m-1}}},
\end{equation}
where $c$ is again an integration constant, and we have omitted a positive constant multiplying factor. These solutions are consistent with the assumption used above to obtain them (such as ignoring the cross-terms of the $m$-th power) as long as both $\Delta(v)$ and $d_{\rm h}(v)^m$ decay at least as quickly as $1/v^n$, with
\begin{equation}\label{14-2c}
n-1>\frac{1}{m-1}.
\end{equation}
On the other hand, if we assume one of the terms $\Delta(v)$ or $d_{\rm h}(v)^m$ dominates the right-hand side of eq.~\eqref{12c}, then we get a solution of the type
\begin{equation}\label{15c}
r-r_{\rm h}\sim\int^{v}\mathop{dv'}D_{\rm max}(v'),
\end{equation}
where $D_{\rm max}(v')=\max[\Delta(v'),d_{\rm h}(v')^m]$, with the maximum taken at sufficiently large $v$ to be in the asymptotic regime of the two functions. This solution is again only consistent with the assumption for the right-hand side of the differential equation \eqref{12c} if the larger of the two functions decays at least as quickly as $1/v^n$, with $n$ satisfying \eqref{14-2c}.

In summary, light trapping occurs if the following rules are satisfied by $D_{\rm max}(v)$ (or equivalently by both $\Delta(v)$ and $d_{\rm h}(v)^m$):
\begin{itemize}
	\item If $D_{\rm max}\sim 1/v^n$, then light rays are trapped if the power $n$ and the order of the first non-zero coefficient in \eqref{7c}, which we will call $m$, satisfy \eqref{14-2c}.
	\item If $D_{\rm max}\sim e^{-\alpha v}$, then we can have any \mbox{$m\geq 1$}. If $m=1$, there is the additional condition \mbox{$\alpha>k_1$}.
	\item If $D_{\rm max}$ decays more quickly that an exponential, then there are no restrictions to the series~\eqref{7c}.
\end{itemize}

In these cases, the fact that the confined light rays do not reach the exterior future null infinity indicates the presence of an event horizon. This horizon's surface is described by the trajectory of the first trapped light ray, which can be seen to correspond to the solution \eqref{15c}. Any outgoing rays which are outside it reach the asymptotically flat exterior region, and their dispersion is related to the presence and temperature of Hawking radiation. As for those on the inside, they must go to a different asymptotic region. How they end up depends on whether the asymptotic approach to zero in $F$ is due to a zero in $f$ or a divergence in $g$, as we will now see.

\section{Causal structure}\label{Sec:causality}

\subsection{Causal structure for finite $g$}

\begin{figure}
	\centering
	\includegraphics[scale=.55]{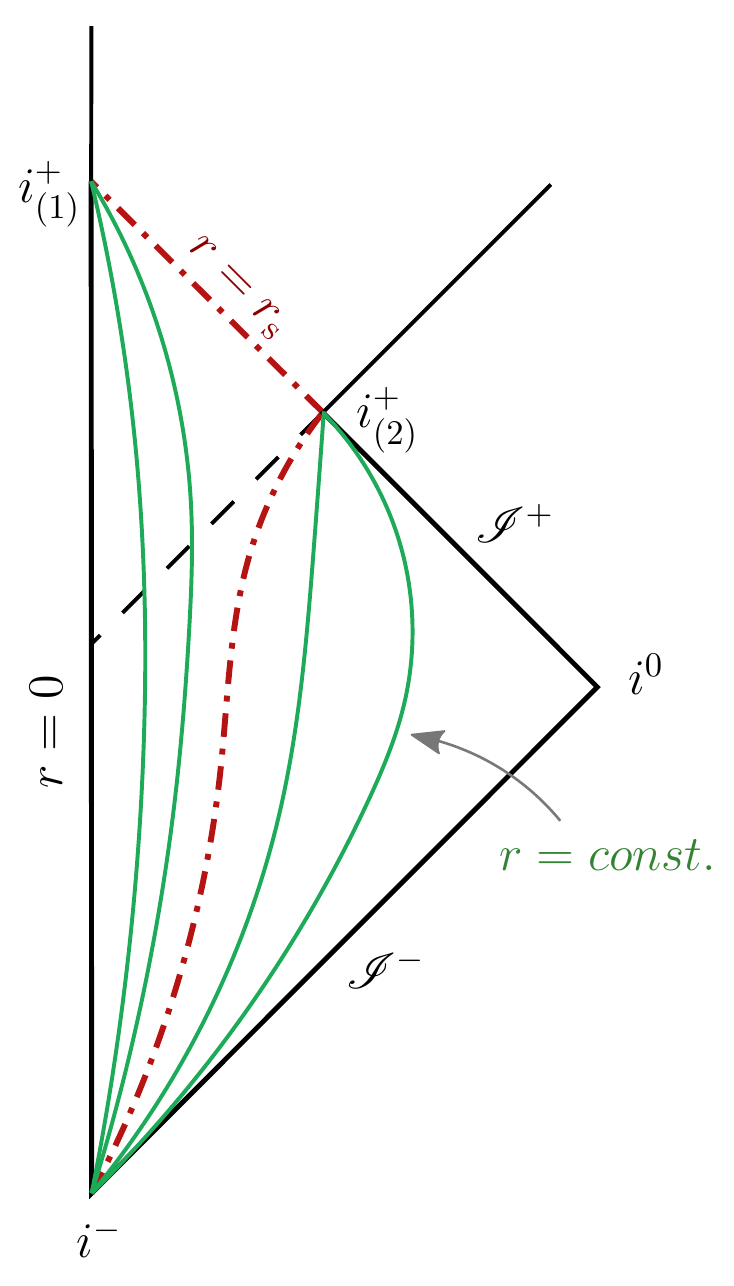}
	\caption{Conformal diagram of the spacetime with $g\simeq1$ and $f$ given by \eqref{7c} satisfying the appropriate conditions for light-ray trapping. The dashed line is the event horizon, corresponding to the first trapped outgoing light ray. The dash-dotted line is the surface $r=r_{\rm h}$, which is described by a timelike curve and the Cauchy horizon. The curves to the left of $r_{\rm s}$ correspond to surfaces of $r={\rm const.}<r_{\rm h}$, while to the right they are $r={\rm const.}>r_{\rm h}$. The lines outside the conformal triangle indicate the need to extend the spacetime.}
	\label{f1c}
\end{figure}

Let us first assume that any light-ray trapping is due to an approach to zero in $f$ of the form \eqref{7c}, and that $g$ remains finite (we will in fact assume $g=1$ for simplicity). The causal structure of the spacetime in this case almost always ends up being the same as that of an extremal (regular) BH, shown in fig.~\ref{f1c}. In the limit $v\to\infty$, the surface $r=r_{\rm h}$ becomes a Cauchy horizon, beyond which the geometry is extendable.

To show this, we can turn to one of the geodesic equations for a radial trajectory $(v(\sigma),r(\sigma))$ in our metric,
\begin{equation}\label{9c}
\ddot{v}+\frac{\partial_rf}{2}\dot{v}^2=0.
\end{equation}
If the first non-zero coefficient in \eqref{7c} is $k_1$, then the solution to this equation close to $R_{\rm h}(v)$ is
\begin{equation}\label{10c}
k_1\dot{v}_0(\sigma-\sigma_0)\simeq 1-e^{-k_1(v-v_0)},
\end{equation}
where the subscript $0$ refers to initial values. Since $k_1$ must be positive, when the affine parameter $\sigma$ reaches the finite value
\begin{equation}
\sigma_{\rm h}=\sigma_0+1/(k_1\dot{v}_0),
\end{equation}
the geodesic has reached the limit $v\to\infty$ and, in the absence of singularities, can be extended past this point. Of course, this is the case only if the geodesic stays close enough to $R_{\rm h}(v)$ so as to keep the approximation \eqref{7c} valid.

As mentioned earlier, if $k_1\neq 0$ then for light rays to be trapped below $r_{\rm h}$ the functions $\Delta(v)$ and $d_{\rm h}(v)$ must both tend to zero at least as quickly as an exponential. For example, if
\begin{equation}
\Delta(v)=e^{-\alpha v},\qquad d_{\rm h}(v)=e^{-\beta v},
\end{equation}
with $\alpha$ and $\beta$ some positive constants, then the solution for the trajectories of outgoing null geodesics is asymptotically
\begin{equation}\label{13c}
r(v)\simeq r_{\rm h}-\frac{1}{\alpha-k_1}e^{-\alpha v}-\frac{k_1r_{\rm h}}{\beta-k_1}e^{-\beta v}+ce^{-k_1 v},
\end{equation}
where $c$ is an integration constant. There are trapped null solutions if $\min(\alpha,\beta)>k_1$ (and the leading order of the solution is only the exponential with the smallest coefficient in absolute value). They correspond to the values $c\le0$ for the integration constant ($c=0$ for the horizon itself). The approximation resulting in eq. \eqref{10c} is valid for these trajectories (since they approach $R_{\rm h}$), and they are therefore extendable past the $v\to\infty$ limit. At this limit they reach $r=r_{\rm h}$, making this surface a Cauchy horizon, as shown in fig.~\ref{f1c}.

As for spacelike and timelike geodesics, the equivalent of eq. \eqref{2c} is
\begin{equation}\label{14c}
\frac{dr}{dv}=F(v,r)\pm\frac{1}{2g(v,r)\dot{v}^2},
\end{equation}
with $+$ for spacelike and $-$ for timelike ones. Since we are interested in the region around $r_{\rm h}$ at large $v$, we only look for geodesics which stick close to this radius asymptotically. Using this as an assumption for the solutions, for $g(v,r)=1$ it is easy to check that with eq. \eqref{10c}, $\dot{v}$ diverges quickly enough for the new term in \eqref{14c} (with respect to the null case) to become negligible at leading order in the asymptotic expansion. Thus, for every null geodesic of the type \eqref{13c} there are also a spacelike and a timelike geodesic with the same approximate expressions. From the signs of the additional term in \eqref{14c} it can be seen that corrections to the leading order would reveal that in terms of radius the spacelike geodesics are actually slightly above the null ones, while the timelike ones are slightly below (moving inside the light-cones). Eq.~\eqref{10c} is also a valid approximation for the affine parameter of these geodesics, meaning that they are also extendable.

The same occurs even when $k_1=0$: geodesics which try to escape from the interior region reach the $v\to\infty$ limit in finite affine parameter. For example, if $k_2\neq 0$ we can solve eq. \eqref{9c} in the vicinity of $r_{\rm h}$ and see that the value this parameter reaches when $v$ diverges is
\begin{equation}
\sigma_{\rm h}=\sigma_0+\frac{1}{\dot{v}_0k_2(r_{\rm h}-r_0)}.
\end{equation}

There are only two exceptions to this scenario of extendable geodesics. The first one is the case in which the function $f$ is constant in $r$, making all coefficients $k_i$ in the expansion \eqref{7c} zero; equation \eqref{9c} then implies $v\propto\sigma$ and there is no Cauchy horizon for trapped geodesics. The second exception is, in a sense, a generalisation of the first: it is the case in which the function $f$ in non-analytical in the $r$ direction about its minimum, and all its derivatives in this direction are zero there. In other words, we can generalise from the case of constant $f$ in $r$ and maintain the non-extendibility by sacrificing the analytic nature of the function. Let us provide an example: suppose we have
\begin{equation}
F(v,r)\simeq e^{-\frac{1}{(r-r_{\rm h})^2}}+\frac{1}{v^n},
\end{equation}
where $r$ and $v$ are expressed in units of some arbitrary length scale. Then the asymptotic solutions for trapped outgoing light rays are
\begin{equation}
r\sim r_{\rm h}-\frac{1}{\sqrt{\log(v-c)}},
\end{equation}
where $c$ is an integration constant and aside from the asymptotic condition $v\gg1$, the range of validity of each solution is $v>c+1$. Along these trajectories eq.~\eqref{9c} becomes
\begin{equation}
\ddot{v}=\frac{2}{(v-c)[\log(v-c)]^{3/2}}\dot{v}^2,
\end{equation}
the asymptotic solution of which is again $v\propto\sigma$ (we have confirmed this both analytically and numerically).

In summary, the requirement on $f$ for the geometry to be non-extendable is that all derivatives at its minimum in the direction of decreasing $r$ be zero, either by making the function constant in $r$ or non-analytical. These cases seem rather unphysical, but they do highlight the fact that the presence of a Cauchy horizon depends entirely on the knowledge of the derivatives of $f$ about a single radial point. It is therefore a case in which an arbitrarily small region of the geometry, the description of which may be expected to change in a complete microscopic theory of gravity, affects our picture of the global causal structure of the spacetime.

\subsection{Causal structure with non-vanishing $f$}

If, on the other hand, light rays are trapped not due to a tendency to zero of $f$ but because of increasing values in $g$ at $R_{\rm h}(v)$, tending to a divergence in the limit $v\to\infty$, the situation is quite different. The geodesic equation relating $v$ to the affine parameter $\sigma$ is in this case
\begin{equation}\label{16c}
\ddot{v}+\frac{\partial_vg}{g}\dot{v}^2=0.
\end{equation}
For simplicity, we will consider $R_{\rm h}(v)\equiv r_{\rm h}$, since not doing so does not lead to any qualitative changes in the causal structure we will obtain (so long as light-ray trapping is maintained). If $k_1\neq 0$, then we can take $\Delta(v)=e^{-\alpha v}$, with $\alpha>k_1$, as it is the slowest allowed approach to zero. The trajectories of trapped outgoing null geodesics are described by \eqref{13c} without the $e^{-\beta v}$ term. Then, for large values of $v$ equation \eqref{16c} takes the form
\begin{equation}
\ddot{v}=-\frac{2\alpha}{k_1|c|}e^{-(\alpha-k_1)v}\dot{v}^2.
\end{equation}
The solution for the affine parameter $\sigma$ is an exponential integral function with argument proportional to $e^{-(\alpha-k_1)v}$, and at large values of $v$ is approximated by the relation
\begin{equation}\label{18c}
\sigma=a_1+a_2v,
\end{equation}
with $a_1$ and $a_2$ being integration constants. Thus in this case the affine parameter of these geodesics reaches infinity at the same time as $v$, meaning that the $r_{\rm h}$ region does not become a Cauchy horizon but a (disjoint) part of future null infinity. If $k_1=0$, there is no change in this behaviour. In fact, eq.~\eqref{18c} is still the approximate solution relating $v$ to the affine parameter for trapped geodesics at large values of $v$ [e.g. if $k_2\neq0$ and we take again $\Delta(v)=e^{-\alpha v}$, the term approximating $v$ on the right-hand side of eq.~\eqref{18c} is in this case $\int_1^v\exp(x^2\exp(-\alpha x))\mathop{dx}$].

\begin{figure}
	\centering
	\includegraphics[scale=.7]{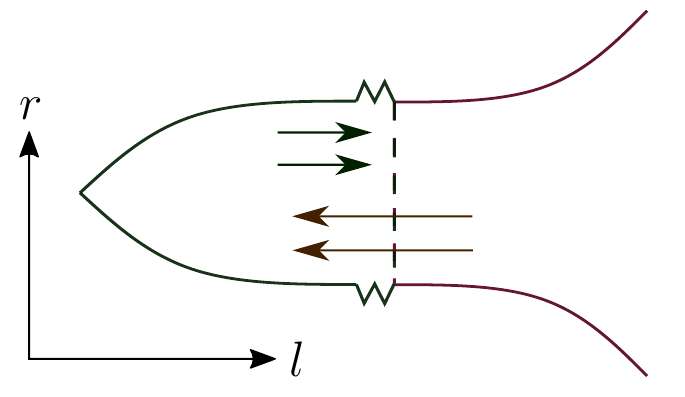}
	\caption{Relation between the radial coordinate $r$ and the proper length in the radial direction $l$ for a geometry in which $g$ tends to a divergence at $r_{\rm h}$. Outgoing light rays become trapped in this infinitely stretching region, while ingoing ones pass right through it.}
	\label{f5c}
\end{figure}

This case of diverging $g$ can be interpreted geometrically from these results for null geodesics: space becomes stretched in the radial direction at $r_{\rm h}$ as the proper radial length $l$ approaches a divergence along with $g$. This stretching is sufficiently quick so as to asymptotically freeze these light rays in their approach toward $r_{\rm h}$ (see fig.~\ref{f5c}). But the key difference with respect to the previous case is that this occurs without the low values of the redshift function $f$, which results in proper time not being slowed down and observers reaching the asymptotic region $v\to \infty$ in infinite time.

As this situation approaches the static case discussed above, one might wonder if ingoing rays would also be affected in a similar manner, becoming unable to cross the $r_{\rm h}$ surface. This turns out not to be the case. The geodesic equation relating the affine parameter to the radial coordinate in $v=\text{const.}$ sections is the same as the second expression in eq. \eqref{5c} in terms of $g$, but in this case, close to $r_{\rm h}$ it takes the form
\begin{equation}
\ddot{r}=-\frac{\partial_rg}{g}\dot{r}^2\simeq\frac{m\tilde{k}_m(r-r_{\rm h})^{m-1}}{\tilde{k}_m(r-r_{\rm h})^m+\Delta(v)}\dot{r}^2,
\end{equation}
where $m$ is again the order of the first non-zero term in the series expansion of $F$, and $\Delta(v)$ is a (small) constant. In contrast to the static case, the right-hand side is not divergent due to the finite $\Delta(v)$ term. Consequently, the affine parameter is finite when crossing $r_{\rm h}$ (e.g. for $m=2$, $r-r_{\rm h}\simeq\sqrt{\Delta(v)}\tan[c_1\sqrt{\Delta(v)}(\sigma-c_2)]$, with $c_1,c_2$ integration constants; crossing occurs at $\sigma=c_2$).

Outgoing light rays become trapped due to the fact that they are moving in a direction in which $g$ increases and $\Delta(v)$ decreases, and actually see space stretching as they go. On the other hand, ingoing rays only see a snapshot of a partially stretched geometry, through which they can easily pass given enough time.

\begin{figure}
	\centering
	\includegraphics[scale=.57]{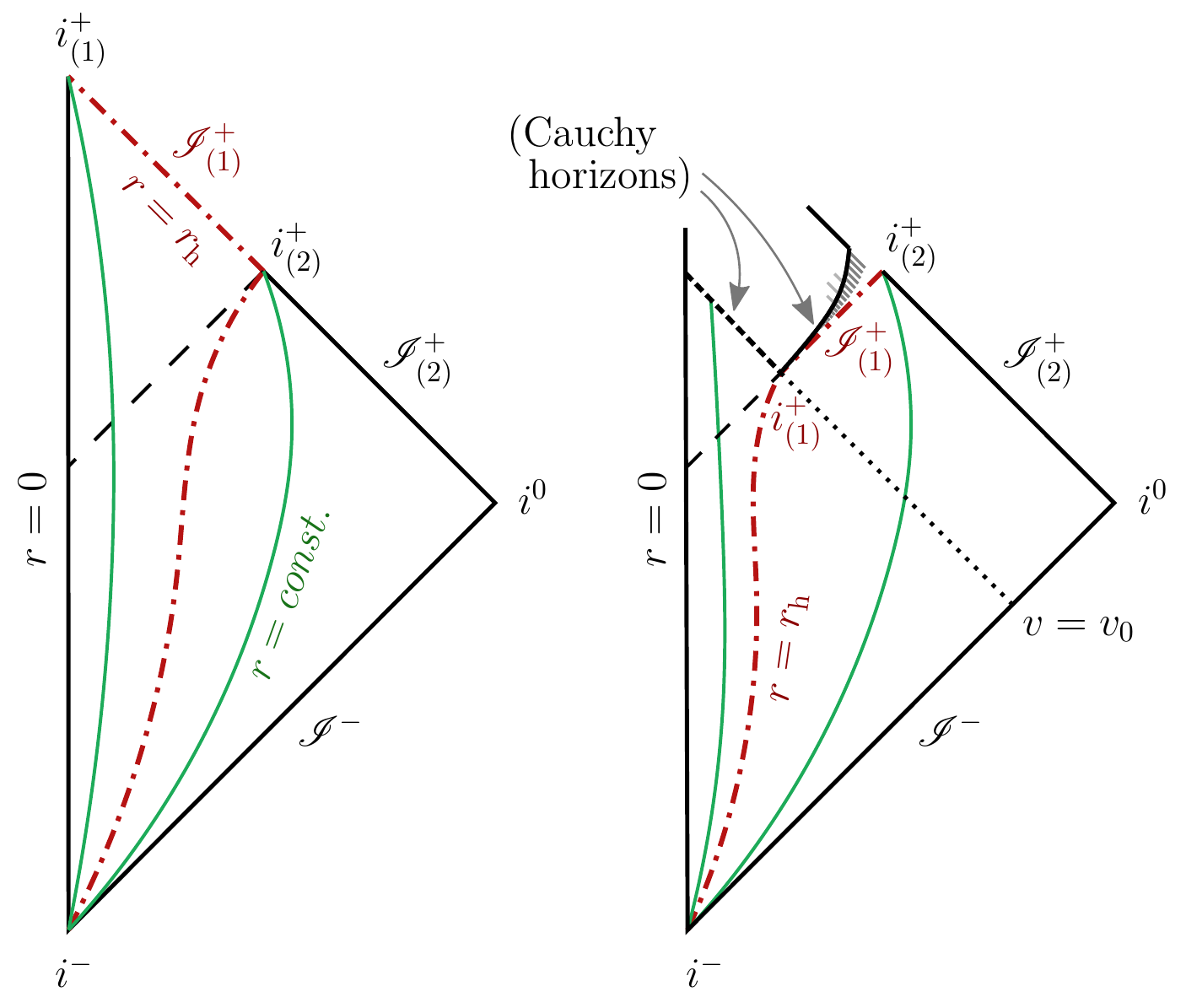}
	\caption{Conformal diagram of the spacetime in which $g$ tends to a divergence. Left: the divergence is reached in infinite time at the surface $r=r_{\rm h}$ (the dash-dotted line), which becomes a separate part of future null infinity for outgoing light rays. This interior null infinity is denoted by $\mathscr{I}_{(1)}^+$, while the exterior one (for escaped light rays) is $\mathscr{I}_{(2)}^+$. Right: the divergence at $r=r_{\rm h}$ is reached at a finite moment $v=v_0$ (and remains thereafter), making $r=r_{\rm h}$ a future null infinity $\mathscr{I}_{(1)}^+$ for ingoing light rays with $v\geq v_0$. The symbols $i^+_{(1)}$ and $i^+_{(2)}$ indicate future timelike infinities for two different sets of observers (for the diagram on the right this is conditional; see discussion below). In general, this latter geometry is extendable past the surface $v=v_0$ for $r<r_{\rm h}$, marked as a Cauchy horizon. The diagram includes what the extension may look like, indicating that to be fixed it requires initial data from another surface, which is effectively another Cauchy horizon in the past of the region.}
	\label{f3c}
\end{figure}

The two future null infinities in this spacetime are separated by an event horizon, as shown in fig.~\ref{f3c} (left). The exterior one we assume is in an asymptotically flat region at $r\to\infty$, while the interior one (at $r\to r_{\rm h}^-$) has a matter content all the way through, which we will briefly analyse in the next section.

\subsection{Diverging $g$ in finite time}

In fig.~\ref{f3c}, the diagram on the right represents a case in which the point in time at which $g$ diverges is brought down to a finite value $v=v_0$ (i.e. $\Delta(v)=0$ for $v\ge v_0$). The spacetime in this case becomes a combination of the static and asymptotically formed cases, and can help shed light on both.

The infinite tube from the static case is now formed dynamically, i.e. space stretches in the radial direction and breaks into two at a point $(v_0,r_{\rm h})$. For $v> v_0$ and $r>r_{\rm h}$ the spacetime is part of the exterior region of the static case, in which the surface $r=r_{\rm h}$ is a future asymptotic region for geodesics which approach it. As for the interior $r<r_{\rm h}$ region, from the moment $v=v_0$ on the evolution is no longer determined by any initial conditions set at any past spacelike 3-surface, making the surface $v=v_0$ for $r<r_{\rm h}$ a Cauchy horizon. The conditions needed to fix a particular extension will generally be determined at a surface which can be thought of as a second Cauchy horizon in the past of the extended region (beyond which the spacetime could again be extended), as shown in fig.~\ref{f3c}.

The only thing left to analyse in order to complete our picture of this geometry is the point $(v_0,r_{\rm h})$. The first thing to note is that there is no curvature singularity there. Considering that it is the point after which $r=r_{\rm h}$ becomes an asymptotic region, one might initially think of it as part of this region. Then all geodesics which approach it would have an affine parameter which tends to infinity there. We can easily check if this is the case with the geodesic equations.

As it turns out, the answer is not that straightforward. Whether this point is part of an asymptotic region for geodesics or not actually depends on how the divergence in $g$ is approached, i.e. how quickly space is stretched. This is encoded in how $\Delta(v)$ reaches zero. The details of this calculation are deferred to appendix \ref{ap1}, but the summary is the following: if we have
\begin{equation}
\Delta(v)\propto(v_0-v)^n,
\end{equation}
then for all geodesics to have their affine parameters diverge when they reach this point, the inequality
\begin{equation}
n-1\geq\frac{1}{m-1}
\end{equation}
must be satisfied, where $m$ is the lesser of the two numbers corresponding to the orders of the first non-zero coefficients $k_i$ and $\tilde{k}_i$ in the expansion of $F$. If this inequality is not satisfied, then some geodesics will reach this point in finite proper time and will be extendable beyond it. For timelike geodesics, the extensions will be into the interior region beyond the $v=v_0$ Cauchy horizon.

On the other hand, in the case in which the divergence in $g$ is reached in infinite time, i.e. the limit $v_0\to\infty$, all geodesics which approach this point have their affine parameter reach an infinite value at the same rate as $v$ [eq. \eqref{18c}]. To understand how the transition from the right to the left diagram in fig.~\ref{f3c} occurs, we can think of the fact that in this limit all ingoing light rays make it through $r_{\rm h}$, are reflected at the origin and then become trapped in an approach toward what is essentially the point $(v_0,r_{\rm h})$ from the inside. The point is then stretched to become a future null infinity region for all these rays, as well as a future timelike infinity for geodesics which may approach it from below [$i^+_{(1)}$] or above [$i^+_{(2)}$], as seen in the left diagram in fig.~\ref{f3c}. These timelike infinities are also the ones for geodesics which approach the asymptotic region with radii $r<r_{\rm h}$ and $r>r_{\rm h}$ respectively.

\section{Energy conditions}\label{Sec:energyconditions}

The physical picture behind the spacetimes of type $f\to 0$ is roughly that of a collapse of matter which slows down progressively, tending to a halt, while also approaching crossing its own gravitational radius. On the other hand, the $g\to\infty$ type seems to describe a stretching of space in the radial direction in a manner similar to cosmological expansion. These are unusual situations, to say the least, so it is interesting to see whether some of them can be associated with the dynamics of classically reasonable matter, i.e. whether their stress-energy tensors can satisfy any of the energy positivity conditions.

We will suppose that $w^\mu$ is any timelike or null vector, and without loss of generality we will suppose its angular component is in the $\theta$ direction, resulting in the inequality
\begin{equation}\label{19c}
-f(w^v)^2+2gw^vw^r+r^2(w^\theta)^2\le 0.
\end{equation}
The only region in which we have needed to fix the spacetime geometry so far is around $R_{\rm h}(v)$, so we will analyse how matter behaves there, using the expansions \eqref{7c} and \eqref{8c}.

Let us again start with the case $g\simeq 1$. To test the weak (and null) energy condition, we contract the Einstein tensor of \eqref{1c} twice with $w^\mu$ and see whether the resulting scalar is positive for all $w^\mu$ satisfying \eqref{19c}. Since we will only analyse this condition in the region where we have fixed the geometry, i.e. around the minimum in $F$, we can omit some terms which will not give leading-order contributions and write
\begin{equation}\label{20c}
G_{\mu\nu}w^\mu w^\nu\simeq\frac{1-r\partial_r f}{r^2}\left[f(w^v)^2-2w^vw^r\right]-\frac{\partial_v f}{r}(w^v)^2+\left(r\partial_r f+\frac{r^2}{2}\partial_r^2f\right)(w^\theta)^2.
\end{equation}
This quantity can be shown to be positive in many cases with some simplifications (we will not attempt to derive the most general conditions for positivity). Particularly, let us assume that $\partial_v f$ is negative [for which $\Delta(v)$ must decrease more slowly than $d_{\rm h}(v)$]. Then the term with this partial derivative will be positive and can safely be ignored. With the inequality \eqref{19c}, the sufficient conditions for the rest of the terms on the right-hand side of \eqref{20c} to be positive turn out to be
\begin{equation}\label{wec}
r_{\rm h}\partial_r f< 1\quad\text{and}\quad \frac{r_{\rm h}^2}{2}\partial_r^2f> -1,
\end{equation}
which can be satisfied or violated with an appropriate choice of coefficients in \eqref{7c} and \eqref{8c}. If they are satisfied, then any timelike observer around this region of ``slowed down gravitational collapse" will see a matter distribution with positive energy density.

With a similar analysis, it can be shown that the strong energy condition \mbox{($R_{\mu\nu}w^\mu w^\nu\ge 0$)} can be satisfied around $r_{\rm h}$ (the asymptotic position of the minimum of $f$) at late times if
\begin{equation}\label{sec}
\partial_v f<0\quad\text{and}\quad\frac{r_{\rm h}^2}{2}\partial_r^2f\ge \max(-r_{\rm h}\partial_rf,-1).
\end{equation}
As for the dominant energy condition (that is, requiring that the momentum flux $-T^\mu_{\hphantom{\mu}\nu}w^\nu$ be causal and future-pointing), it can be satisfied if the weak energy condition is, and
\begin{equation}
(1-r_{\rm h}\partial_rf)^2\ge\left(r_{\rm h}\partial_rf+\frac{r_{\rm h}^2}{2}\partial_r^2f\right)^2,
\end{equation}
which can again be achieved with an appropriate choice of coefficients in \eqref{7c} and \eqref{8c}.

This result implies that it is not necessary to violate energy conditions locally in order to generate the $f\to0$ type geometry, but it does not guarantee that for the whole of our spacetime construction. Indeed, if the interior of this geometry ($r<r_{\rm h}$) is singularity-free, then it resembles that of a regular BH, in which some energy conditions are typically broken around the origin $r=0$~\cite{Ansoldi,Bardeen,BeatoGarcia,Hayward}.

In the local region where energy conditions can be satisfied, it may also be interesting to see what form the energy density and pressure perceived by an observer can take. Let us consider an observer freely falling in the radial direction, who has a four-velocity $w^\mu$ and, for simplicity, at the moment of crossing $r_{\rm h}$ is moving in the $v$-direction with $w^v\simeq 1$. Then, taking again $g\simeq 1$, the energy density seen by this observer when approaching from the outside is approximately
\begin{equation}
\rho\simeq\frac{1}{8\pi G}\left(\frac{1}{r_{\rm h}^2}-\frac{\partial_rf}{r_{\rm h}}\right)
\end{equation}
Note that the condition for $\rho$ to be positive coincides with the first condition in \eqref{wec} (the second condition there is necessary for tangentially moving observers). The radial pressure seen by this observer is
\begin{equation}
p_{\rm r}\simeq -\rho,
\end{equation}
and the tangential pressure is
\begin{equation}
p_\theta\simeq \frac{1}{8\pi G r_{\rm h}^2}\left(r_{\rm h}\partial_rf+\frac{r_{\rm h}^2}{2}\partial_r^2f\right).
\end{equation}
One may note that this expression is positive if the second inequality in \eqref{sec} is satisfied along with the one for $\rho>0$. From all this we see that, although the energy conditions can be satisfied, the corresponding matter content is classically rather strange: it has a pressure which is generally anisotropic, and in some cases the sign of its radial component is opposite to that of its tangential component.

However, if our question is whether these geometries are physically reasonable, even locally, this analysis is incomplete. Due to the causal structure involved, additional considerations must be taken into account. On the one hand, it is well-known that the presence of a Cauchy horizon in a solution of the Einstein equations generally indicates that this solution is unstable under perturbations \cite{PoissonIsrael89}. And even if we ignore the possibility of classical perturbations, if we define a quantum field on this spacetime, an analysis based on semiclassical gravity reveals an even greater instability around Cauchy horizons~\cite{BalbinotPoisson93}. We will explore the potential dynamical effects of these instabilities in Part \ref{pt2} of the thesis.

On the other hand, as we saw in the previous two chapters, the RSET tends to have very large values at and around the outer horizon when the latter forms in dynamical regimes in which the matter content moves ``slowly" enough. The two effects can be distinguished by whether they are primarily related to the accumulation of trapped light rays tending toward the inner horizon, or to the peeling of escaping light rays off of the outer horizon, the latter of which we will analyse in the next section. Either way, a geometry of the type we are studying here will certainly have considerable semiclassical corrections. Therefore, even if it satisfies classical energy conditions, it may not be a self-consistent solution of the semiclassical Einstein equations, due to the negative energies typically present in the RSET~\cite{Visser2}. Conversely, if the stress-energy needed to generate it were not sensible on a purely classical level, this would not be enough to discard it as a solution in semiclassical gravity. A complete analysis of the self-consistency of this type of geometry is, however, beyond the scope of this work.

As for the geometries in which $g$ tends to a divergence, it turns out that they generally violate even the weak energy condition. To see this we can write down
\begin{equation}
\begin{split}
G_{\mu\nu}w^\mu w^\nu&=\left(\frac{1}{r^2}+\cdots\right)\left[f(w^v)^2-2gw^vw^r\right]\\&\quad+\left(\frac{\partial_rg\partial_vg}{g^3}+\cdots\right)r^2(w^\theta)^2+\\&\quad+\frac{2f\partial_vg}{rg^2}(w^v)^2+\frac{2\partial_rg}{rg}(w^r)^2.
\end{split}
\end{equation}
The first three terms can be made positive for all $w^\mu$ with a particular choice of $g$, but with the last term it is no longer possible. Particularly, if the last term is negative (which it is for $r>r_{\rm h}$), then any attempt to compensate it with the other terms fails for some choice of vector $w^\mu$. From a more physical perspective, this implies that observers moving sufficiently fast in the radial direction (which becomes increasingly difficult as space stretches, i.e. it requires them to approach the speed of light) may see a negative energy density content. Thus it appears these spacetimes are not ones we may expect to form from the dynamics of exclusively classical matter.

\section{Hawking temperature}\label{s5}

If a field is quantised on top of these horizon-approaching spacetimes, the magnitude of its RSET can be related to the presence of Hawking radiation and the value of its temperature, as discussed in the \hyperref[chI]{Introduction} [see e.g.~eq.~\eqref{27}] and in ref.~\cite{BBGJ16}. In this section we will present a general method for calculating the asymptotic effective temperature function (ETF) of the Hawking radiation generated by these geometries, while only requiring knowledge of the approximate asymptotic solutions for the trajectories of outgoing null geodesics in a neighbourhood of their event horizons (the ``peeling" of the last escaping light rays). We will then use this to estimate the values of the RSET at and above the asymptotic horizon position $r_{\rm h}$.

To recapitulate, the ETF was introduced in \cite{BLSV11}, and its expression for the outgoing radiation sector is given by
\begin{equation}\label{25c}
\kappa_{u_{\rm in}}^{u_{\rm out}}\equiv-\left.\frac{d^2u_{\rm in}}{du_{\rm out}^2}\right/\frac{du_{\rm in}}{du_{\rm out}},
\end{equation}
where the ``in" and ``out" indices refer to the asymptotically flat regions at past and future null infinities: the coordinates are proportional to the natural Minkowskian coordinates at these regions, and the indices of $\kappa$ refer to the difference between the two natural Minkowskian vacuum states (particularly, how the ``in" region vacuum state is seen as a flux of particles when it evolves and reaches the ``out" region). If this function is approximately constant for a long enough period of time~\cite{BLSV11}, then during this period the geometry will create particles with a Planckian spectrum with temperature $\kappa_{u_{\rm in}}^{u_{\rm out}}/2\pi$ in natural units.

This function depends only on the quotient $f/g$, so the calculation is the same for the two types of geometries we have considered. We will assume $k_1$ and $\tilde{k}_1$ are non-zero, as the case in which either one is zero can be obtained as a limit from the final result. We will also assume that $\Delta(v)$ and $d_{\rm h}(v)$ both decrease as exponentials, since it is the slowest allowed approach to zero for light-ray-trapping to occur in this case, and also because the case of a faster approach can again be obtained from the same result.

To calculate the ETF, we need to obtain the trajectories of outgoing null geodesics in a small region around the spatial minimum of $F$. For this we can make use of the solution \eqref{13c} for $r<R_{\rm h}$ and its analogue with $k_1\to -\tilde{k}_1$ for $r>R_{\rm h}$. We will take the small region $(r_{\rm h}-\epsilon,r_{\rm h}+\tilde{\epsilon})$, with $\epsilon$ and $\tilde{\epsilon}$ arbitrarily small positive constants (with the condition that time has advanced enough for $R_{\rm h}$ to be inside this radial interval). We call $v_\epsilon$ the time at which a particular ray crosses $r_{\rm h}-\epsilon$, $v_{\rm h}$ the time when it crosses $R_{\rm h}$, and $v_{\tilde{\epsilon}}$ the instant it crosses $r_{\rm h}+\tilde{\epsilon}$. For our purposes, the labels $v_\epsilon$ and $v_{\tilde{\epsilon}}$ represent the $u_{\rm in}$ and $u_{\rm out}$ ones respectively.

From the solutions \eqref{13c}, a straightforward calculation leads to the asymptotic (in $v$) result
\begin{equation}\label{26c}
\frac{dv_\epsilon}{dv_{\rm h}}\sim \beta r_{\rm h}e^{-(\beta-k_1)v_{\rm h}}+e^{-(\alpha-k_1)v_{\rm h}},
\end{equation}
where we have omitted a proportionality constant. From here on we must decide which of these two exponentials dominates at large time, i.e. which one decays slower. If $\alpha<\beta$, then the first one dominates, and we also obtain from the exterior solutions the asymptotic relation
\begin{equation}\label{27c}
\frac{dv_{\rm h}}{dv_{\tilde{\epsilon}}}\sim \frac{\tilde{k}_1}{\alpha+\tilde{k}_1}.
\end{equation}
On the other hand, if $\beta<\alpha$, then the second exponential in \eqref{26c} dominates and the result is the same as \eqref{27c}, only substituting $\alpha$ for $\beta$. Defining $\gamma=\min(\alpha,\beta)$ we can proceed with integrating \eqref{27c} in generic terms. Doing so and substituting into \eqref{26c} we obtain the asymptotic relation between the labels
\begin{equation}
\frac{dv_\epsilon}{dv_{\tilde{\epsilon}}}\sim e^{-\tilde{k}_1\frac{\gamma-k_1}{\gamma+\tilde{k}_1}v_{\tilde{\epsilon}}}.
\end{equation}
The ETF is simply minus the coefficient multiplying $v_{\tilde{\epsilon}}$ in the exponential,
\begin{equation}\label{29c}
\kappa_{u_{\rm in}}^{u_{\rm out}}\sim \tilde{k}_1\frac{\gamma-k_1}{\gamma+\tilde{k}_1}.
\end{equation}
If either $\Delta(v)$ or $d_{\rm h}(v)$ decays quicker than an exponential, then the limit $\alpha\to\infty$ or $\beta\to\infty$ can be taken, respectively. Eq. \eqref{29c} still applies if both $\alpha$ and $\beta$ are taken to $\infty$, i.e. if $\gamma\to\infty$, giving simply $\tilde{k}_1$ for the ETF. If the slope on either side of the minimum of $F$ is zero, then the corresponding limits $k_1\to 0$ and $\tilde{k}_1\to 0$ can also be taken, the latter resulting in a zero ETF.

It is worth observing that the (asymptotic) surface gravity of these objects at $r_{\rm h}$ is given by
\begin{equation}
	\kappa=\frac{1}{2}\frac{\partial_r f}{g},
\end{equation}
meaning that when $g$ diverges, the surface gravity always tends to zero. More generally, when $g\not\simeq\text{const.}$, there is no longer a direct relation between the surface gravity and the temperature of Hawking radiation corresponding to the horizon. The latter is instead associated with the slope $\partial_r(f/g)$, i.e. the coefficient $k_1$ of the series \eqref{3c}.

Another interesting observation is that the result in eq.~\eqref{29c} has a clear dependence not only on $\tilde{k}_1$, but also on $k_1$, which is the asymptotic surface gravity of the \textit{inner} horizon. The reason for this is the fact that the light ray which describes the event horizon~\eqref{15c} of these spacetimes travels outward in the $r<r_{\rm h}$ region, and light rays which are slightly above it can still be below $r_{\rm h}$ but end up escaping to infinity. They thus carry information of the inner region $k_1$ out to the asymptotic region. This is also the reason why the ETF is not simply $\tilde{k}_1$ (though it still reduces to this value if the dynamics are too quick for the inner region to have an effect on the outward peeling).

This modification in the peeling of the escaping light rays also has a considerable effect on the RSET. To see this, let us look at the simple case in which the exterior geometry is the Schwarzschild vacuum solution. This can be recovered at leading order if the matter surface has a trajectory $R_{\rm h}(v)$ by imposing
\begin{equation}
\Delta(v)\simeq\frac{R_{\rm h}(v)-r_{\rm s}}{r_{\rm s}},\quad \tilde{k}_1\simeq\frac{1}{2r_{\rm s}},
\end{equation}
with $r_{\rm s}=2M$ the Schwarzschild radius. Then, if one considers eq.~\eqref{27} applied to the Boulware and ``in" (or Unruh) vacua for the exterior region $r>R_{\rm h}(v)$, it can be seen that the Boulware state divergence when $R_{\rm h}(v)\to r_{\rm s}$ can only be avoided if the ETF corresponds exactly to the Hawking temperature value. In other words, when the fraction in eq.~\eqref{29c}, which alters the Hawking temperature given by the surface gravity $\tilde{k}_1$, is different form 1, a Boulware-like divergence is approached in the RSET. Particularly, in the Boulware state the RSET of a BH has a $1/f(r-r_{\rm s})$ divergence (with $f(r)=1-r_{\rm s}/r$ being the Schwarzschild redshift function) \cite{Boulware1975,Visser2}, so if e.g. $R_{\rm h}(v)\simeq r_{\rm s}(1+e^{-\alpha v})$, then there is an exponential growth in the RSET for the exterior region $r>R_{\rm h}(v)$, since
\begin{equation*}
\left.\frac{1}{f}\right|_{\rm surface}\simeq\frac{r_{\rm s}}{R_{\rm h}(v)-r_{\rm s}}\simeq e^{\alpha v}.
\end{equation*}
It is therefore likely that the full dynamics of a spacetime with a transient period akin to these geometries would have considerable deviations from classicality.

\section{Conclusions}
In summary, we have seen that there are spherically symmetric geometries which, through their particular asymptotic evolution in time, can behave like BHs and even have event horizons, without ever having formed any trapped surface. While having this characteristic in common, the family contains various distinct causal structures and different behaviours in terms of energy conditions and production of Hawking radiation. The family of geometries is divided into two categories. The first one is characterised by its similarity with a spacetime in which a standard BH is formed, but with the formation of its first trapped surface pushed to the future asymptotic region. In other words, the strict formation of a trapped surface is replaced by an appropriately quick tendency to its formation, quick enough that although outgoing radial light rays always have a positive expansion, some move out slowly enough to be trapped inside a finite spatial region until the advanced time $v$ reaches infinity. Analysing the causal structure of this first category of geometries, we find that aside from an event horizon (described by the first trapped outgoing light ray) in almost all cases there is also a Cauchy horizon, beyond which the trapped geodesics are extendable, giving the same causal structure near the future horizons as an extremal charged BH.

The second category of geometries in which outgoing light rays are trapped has a very different physical picture behind it. Instead of a decreasing redshift function $f$, what results in the slow-down of the radial escape of the light rays is an actual stretching of space in the radial direction. The proper length becomes vastly greater that the radial length, tending to a divergence in their relation. One can think of it as an attempt at opening a wormhole with an infinitely long neck. To simulate the asymptotic formation of a trapped surface, this divergence only needs to be reached asymptotically as well. Meanwhile, because the stretching increases in the $v$ direction, ingoing geodesics can enter the trapped region after traversing a long, but finite tube-like structure. The difference with the first category of spacetimes is most clearly manifest in the causal structure: outgoing geodesics which are trapped below some finite radius are now \textit{not} extendable beyond the $v\to\infty$ border, i.e. their affine parameter also reaches infinity.

This separation into two categories can be seen as due to the fact that requiring for outgoing null trajectories to be trapped defines only what we call the \textit{generalised redshift function} $F(v,r)$, which amounts to just one of the two degrees of freedom of spherically symmetric geometries. However, the geodesic equations, from which we deduce the causal structure, see both of these degrees of freedom. Thus, different ways of imposing the same behaviour in $F$ result in different behaviours of the geodesic affine parameter.

Having studied the causal structure of these spacetimes, we then looked at the matter content which they require as a source in order to be considered solutions of the Einstein equations. The geometries of the first category can be sustained by a matter content which satisfies any of the energy positivity conditions, that is, at least locally around the point of asymptotic horizon formation, where the geometry is specified. On the other hand, the cases of the second category appear to violate even the weak energy condition. They would thus lose their physical significance in a purely classical theory, but we should keep in mind that the grounds for this study are semiclassical effects in geometries with appropriate null geodesic peeling for non-local quantum effects to manifest. The semiclassical contributions to the stress-energy content are known to violate all energy conditions as well, which calls for a broadening of our physical criteria.

We then looked at the thermal flux of Hawking radiation that these geometries produce. Calculating the ETF, we found that there are three distinct quantities of interest, which degenerate to the same value in the standard Schwarzschild BH formation. First, there is the surface gravity of the horizon in the asymptotic limit (where spacetime tends to staticity). The second is the actual slope of $F$ near the horizon, which governs the peeling of outgoing geodesics in static configurations. The third quantity is the actual ETF. The first two quantities only coincide when $g=1$, such as in the Schwarzschild and Reissner-Nordström BHs. The second and third quantities only differ when the asymptotic horizon formation is slow enough for peeling of the escaping light rays to be affected by the interior region of the geometry, modifying the ETF. This change in the ETF then affects the RSET in a way which triggers an approach toward a Boulware-state-like divergence, suggesting that the source generating such a geometry must be analysed through the semiclassical dynamical equations.


\part{Inner horizon: instabilities and evolution}\label{pt2}

\fancyhead[R]{}
\fancyhead[L]{Part II}

\addtocontents{toc}{\protect\thispagestyle{empty}}

As discussed in the \hyperref[chI]{Introduction}, it is generally accepted that within the semiclassical theory BHs should evaporate. More precisely, this means that their trapped region should gradually reduce in size from the outside-in, eventually revealing what is at their core, where the simplest, spherically symmetric model tells us that there is either a singularity, or some nucleus which can only be described with a complete theory of quantum gravity. However, this picture is missing an essential ingredient, one which is present whenever the BH structure is considered with sufficient generality: the inner horizon, i.e. the inner bound of the trapped region, below which some causal observers are free to stop their descent.

When a trapped region forms during gravitational collapse, it usually has an outer and inner boundary, defined by two apparent horizons. Classically the outer one only moves outwards, doing so as more matter is accreted, while the inner one moves inwards, typically tending to a final position at a finite radius (if e.g. the BH has any angular momentum or electric charge, or forms a regular core by violating the strong energy condition). An inner horizon appears in all realistic BH configurations, and one may therefore wonder whether there is some semiclassical effect associated with its presence, just as the Hawking effect is related to the presence of the outer horizon. Models of evaporation of trapped regions with an inner horizon have indeed been considered, but they take evaporation as an \textit{ad hoc} ingredient, implicitly assuming that it still mainly occurs from the outside, and that the inner horizon just waits around for the outer one to eventually come to it (see e.g. \cite{Hayward,Frolov2017}). On the other hand, calculations of the RSET in the vicinity of static inner horizons have been performed~\cite{Birrell1978,BalbinotPoisson93,Ori2019,Hollands2020a,Hollands2020b,Zilberman2022}, but without analysing backreaction at finite times (usually due to the nature of the stationary backgrounds considered, where the only inner horizon present is also a Cauchy horizon).

In classical general relativity, a well-known result is that long-lived inner horizons lead to the so-called ``mass inflation" instability, wherein small perturbations in the matter content of the geometry result in a highly non-linear response in the increase of curvature~\cite{PoissonIsrael89}. For the charged BH and many other models, this increase in curvature can be related to a growth of the Misner-Sharp mass \cite{Misner1964}, hence the name ``mass inflation". For an initially static geometry with an inner horizon which is then perturbed, the region where mass is ``inflated" begins close to the initial position of this horizon and extends below it. Marking the beginning of this large-curvature region is a shockwave \cite{Marolf2012} located on a null hypersurface which remains in the vicinity of the initial inner horizon position. Meanwhile, the inner horizon moves inside this region along a timelike hypersurface, tending toward the origin. If it reaches the origin before reaching the Cauchy horizon, a spacelike singularity is formed, as observed in the numerical analysis in \cite{Brady1995}, in addition to the null weak singularity at the Cauchy horizon itself \cite{Ori1991,Ori1992,Ori1998,Dafermos2017}. We will discuss the details of this inner structure in the following chapter.

Given this classical instability, one may expect that within semiclassical gravity, if backreaction form the RSET at the inner horizon is taken into account at finite times, an equally non-linear process may take place, possibly with sufficient amplification to lead to significant deviations from classicality in spite of the $\hbar$ suppression. It is the goal of this part of the thesis to explore this possibility. In chapter \ref{ch4} we begin by revisiting the classical problem, using a simple geometric model of a spherically-symmetric BH with an inner horizon to study how perturbations can trigger mass inflation, and how the peculiar causal structure of the resulting spacetime comes about. In chapter \ref{ch5} we then turn to perturbations of semiclassical origin, namely backreaction from the RSET on static and dynamical BH backgrounds. In chapter \ref{ch6} we take a small detour to look at how the tools used in the analysis of BH stability can also be applied to another spacetime with somewhat similar causal features: the superluminal warp drive. Finally, in chapter \ref{ch7} we go back to BHs, and look at what the combination of classical and semiclassical perturbations may lead to for the evolution of the trapped region.

\chapter{Classical inner horizon instability: a model of mass inflation}\label{ch4}

\fancyhead[R]{}
\fancyhead[L]{Part II -- \chaptername\ \thechapter: \leftmark}

In this chapter we will work with a model of a spherical BH undergoing classical mass inflation. The perturbations which trigger the instability will take the form of ingoing and outgoing null thin shells, allowing for an analytical description of the evolution of the geometry through the junction conditions at each of the shell surfaces. We will first look at what the conditions on the infalling mass and the response of the geometry are for the instability to be triggered. Then, we will focus on the evolution of the inner apparent horizon in these spacetimes, analysing the resulting causal structure, and preparing the stage for the combined classical and semiclassical analysis of chapter \ref{ch7}.

\section{Mass inflation with thin shells}

The model we will work with is a spherically symmetric geometry with a line element
\begin{equation}
ds^2=g_{tt}dt^2+g_{rr}dr^2+r^2d\Omega^2,
\end{equation}
where $d\Omega^2$ is the line element of the unit sphere. The simplicity of our construction lies in considering that this geometry is static, i.e. that the metric components $g_{tt}$ and $g_{rr}$ are functions of $r$ only, in patches separated by spherical null shells.

\begin{figure}
	\centering
	\includegraphics[scale=.8]{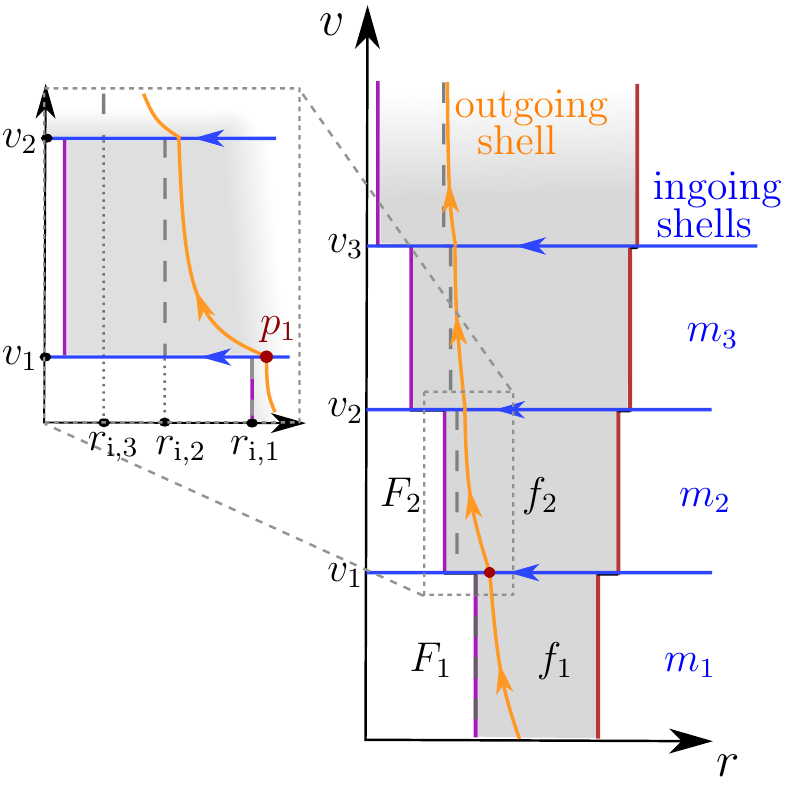}
	\caption{Static BH with outer and inner horizons perturbed by an outgoing shell and a series of ingoing shells with decreasing mass, represented in advanced Eddington-Finkelstein coordinates. The trapped region is shaded in grey. The ingoing shells are located at $v_n$, with $n=1,2,\dots$; $m_n$ refers to the Bondi mass in the asymptotic region before the corresponding ingoing shells (outside the outgoing one); $f_n$ is the $g^{rr}$ metric component in each of these regions down to the outgoing shell, while $F_n$ is the same metric component on the inside of the outgoing shell. $r_{{\rm i},n}$ are the zeros of the $f_n$ functions extended to the region below the outgoing shell (they are the radii which this shell approaches exponentially in $v$ in each patch).}
	\label{f1d}
\end{figure}

To begin with, consider that there are two null shells, one outgoing and one ingoing, which intersect at a point $p_1=(t_1,r_1)$. Continuity of the metric imposes the relation (see \cite{thooft1985,Redmount1985,Barrabes1990})
\begin{equation}\label{intersect}
\left|\frac{f_{2}}{f_{1}}\right|=\left|\frac{F_{2}}{F_{1}}\right|\quad \text{at}\quad p_1,
\end{equation}
where the functions $F$ and $f$ are the $g^{rr}$ component of the metric in different patches, with upper (lower) case letters referring to the interior (exterior) of the outgoing shell and the subscript ``$1$" (``$2$") referring to the interior (exterior) of the ingoing shell, as shown schematically in fig.~\ref{f1d}.

Now consider that this geometry has a trapped region (i.e. a region where $g_{tt}$ becomes positive and $g_{rr}$ negative, which directly relates to the expansion of outgoing null geodesics congruences becoming negative \cite{Hayward1994}). For simplicity, consider also that $-g_{tt}=g^{rr}$ in the static coordinates of each patch in the vicinity of the shells (as would be the case for e.g. the charged BH, and many regular BH models \cite{Hayward,Ansoldi}). We will refer to this metric component as the \textit{redshift function}, this being the previously defined $f$ (and $F$, except for a factor $1/2$).\footnote{Throughout the text, it should be recalled that the properties we require from this function are related to both $g_{tt}$ (which often bears the name redshift function by itself) and $g^{rr}$; the former has more to do with the trajectories of null geodesics, while the latter has to do with the junction conditions which drive the dynamics of the spacetime.} We take the outgoing shell to be travelling inside the trapped region, between the outer and inner horizons. Once we specify some aspects of the initial conditions in the patches of $F_1$, $f_1$, and $f_2$, which amounts to choosing the mass and charge carried by each shell, eq. \eqref{intersect} will tell us how the redshift function in the future of the innermost region of the BH, $F_2$, behaves. We will then repeat the process by adding more ingoing shells, as shown in fig. \ref{f1d}, representing an ingoing perturbation of decreasing amplitude~\cite{Price1972}, and see when and how mass inflation is triggered.

In advanced Eddington-Finkelstein coordinates we can write the line element in the $f_1$ patch as
\begin{equation}\label{geo1}
ds_{1}^2=-f_{1}(r)dv^2+2dvdr+r^2d\Omega^2.
\end{equation}
We note that the same coordinate $v$ can be used for all static representations of the patches external to the outgoing shell, due to the null nature of the ingoing shells (the same applies for the patches on the inside of the outgoing shell, where we will use a null coordinate $V$ for the static description). This outgoing shell is travelling along a null geodesic for both the geometries in the $f_1$ and $F_1$ patches. For now, we will describe its dynamics and its intersections with the ingoing shell(s) in the coordinates of the external $f_1$ patch, while in the next section we will go into more detail regarding its description from the point of view of the interior patches. Being inside the trapped region of \eqref{geo1}, its movement is described by the equation
\begin{equation}
\frac{dr_{\rm shell}}{dv}=\frac{1}{2}f_1.
\end{equation}
Since $f_1$ is negative inside the trapped region, the solution for its radial position $r_{\rm shell}$ decreases in $v$, eventually tending toward the inner gravitational radius of the $f_1$ patch, $r_{\rm i,1}$, where the redshift function can be approximated by
\begin{equation}\label{f1exp}
f_1(r)=-2\kappa_{1}(r-r_{\rm i,1})+\mathcal{O}[(r-r_{\rm i,1})^2],
\end{equation}
with $\kappa_{1}$ being the absolute value of the surface gravity of this inner radius (i.e. $-\frac{1}{2}\partial_rf_1|_{r=r_{\rm i,1}}$). To clarify, $r_{\rm i,1}$ is the position the inner horizon would have if the $f_1$ patch were continued to below the outgoing shell (i.e. the zero of the $f_1(r)$ function when extended to below $r_{\rm shell}$). We note that since below $r_{\rm shell}$ the geometry is described by the $F_1$ function, the position of the actual inner apparent horizon of the geometry is in fact the zero of $F_1$. This position will be denoted by $R_{\rm i,1}$ later on.

The solution which describes the tendency of the shell toward $r_{\rm i,1}$ at sufficiently large values of $v$ can be obtained from the linear order in the above expansion,
\begin{equation}\label{rsell1}
r_{\rm shell}(v)=r_{\rm i,1}+(r_0-r_{\rm i,1})e^{-\kappa_1 v}+\order{e^{-2\kappa_1 v}},
\end{equation}
where $r_0$ is a positive constant representing the position of the shell at $v=0$. Substituting \eqref{rsell1} into \eqref{f1exp}, the redshift function on the shell is then approximated by
\begin{equation}\label{eq:f1_app}
f_{1}|_{\rm shell}=-2\kappa_{1}(r_0-r_{\rm i,1})e^{-\kappa_{1}v}+\order{e^{-2\kappa_1 v}}.
\end{equation}

Now consider that this outgoing shell is intersected by the ingoing one. If the latter is assumed to have positive mass (more generally, satisfies the null energy condition), the inner horizon can only be displaced inward, i.e. $r_{\rm i,1}\ge r_{\rm i,2}$, following a timelike trajectory, in the same way as the outer horizon can only be displaced in an outward (spacelike) direction.\footnote{This is simply an extension of Hayward's theorem for continuous matter \cite{Hayward1994} to the case of shells.} Let us also assume that this mass which the shell carries to the BH is small (compared to the total BH mass), as one may expect from perturbations in an astrophysical scenario. The redshift function after this shell can then be approximated by
\begin{equation}\label{eq:f2_app}
f_{2}|_{\rm shell}=-2\kappa_{2}(r_{\rm shell}-r_{\rm i,2})+\mathcal{O}[(r_{\rm shell}-r_{\rm i,2})^2].
\end{equation}
Particularly, the smallness of the mass carried by the shell is meant to ensure that the order $(r_{\rm shell}-r_{\rm i,2})^2$ remains negligible (in units of the characteristic length scale of the BH, i.e. the initial inner horizon radius), in the same way as the higher order terms in \eqref{f1exp} are, which amounts to a requirement that the displacement $(r_{\rm i,1}-r_{\rm i,2})$ be small (which we will impose explicitly later on). At the intersection point $p_1$, we can insert \eqref{rsell1} for $r_{\rm shell}$ into eq.~\eqref{eq:f2_app} which, together with eq.~\eqref{eq:f1_app}, gives us the expression for one of the two quotients from eq. \eqref{intersect}:
\begin{equation}\label{quotient}
\left.\frac{f_2}{f_1}\right|_{r=r_1}=\frac{\kappa_2}{\kappa_1}+\frac{\kappa_2}{\kappa_1}\frac{r_{\rm i,1}-r_{\rm i,2}}{r_0-r_{\rm i,1}}e^{\kappa_1v_1}\left[1+\order{e^{-\kappa_1v_1}}\right],
\end{equation}
where $v=v_1$ corresponds to the position of the ingoing shell, and is thus also the intersection time. The exponential in this expression, along with eq. \eqref{intersect}, already indicates that the value of redshift function $F_2$ can become much greater than $F_1$, in a manner suggestive of mass inflation. Particularly, one can easily see how this growth of $F_2$ relates to an increase of mass in e.g. the Reissner-Nordström geometry (which we will use as an illustrative example throughout this work), where $F_n=1-2M_n/r+Q_n^2/r^2$. If the redshift function (which is negative, $r_1$ begin inside the trapped region) increases in absolute value, it translates into an increase in the only negative term it contains: the mass of the BH. Assuming this term is already large from previous shell crossings and dominates the behaviour of the redshift function $F_1$ close to the shell, we have the relation
\begin{equation}
F_{1}|_{r=r_1}=-\frac{2M_{1}}{r_1}+\order{M_1^0},
\end{equation}
where $M_1$ is the mass on the inside of the outgoing shell before $v_1$. With the same assumption for the region after the ingoing shell, i.e. that the mass term dominates in the redshift function, we get the inflated mass of the charged BH
\begin{equation}\label{reissner}
M_{2}=M_{1}\left.\frac{f_{2}}{f_{1}}\right|_{r=r_1}+\order{M_1^0}.
\end{equation}
However, as the reader may have already noticed, the exponential in \eqref{quotient} has a prefactor which must also be carefully analysed. For example, if the ingoing shell is considered to have a particular charge and mass which make the displacement of the inner gravitational radius $(r_{\rm i,1}-r_{\rm i,2})$ sufficiently small, the exponential growth could be cancelled. We will discuss this in more detail in the following.

Let us make the perturbation by ingoing shells an iterative process: we represent an ingoing, polynomially decreasing flux of radiation (which is usually the source of mass inflation \cite{PoissonIsrael89,Ori1991} stemming from the decay of perturbations on the geometry \cite{Dafermos2016,Price1972}) with a sequence of ingoing shells of progressively smaller mass,
\begin{equation}\label{deltam}
\delta m_{n}=\frac{a}{v_n^q},
\end{equation}
where $\delta m_{n}$ refers to the change of the exterior mass (i.e. the Bondi mass \cite{Bondi1962} related to past null infinity) produced by a particular ingoing shell located at $v=v_n$, the power $q$ is positive,\footnote{$q$ has a lower bound which depends on the spacing of the shells, needed to guarantee that the total mass thrown into the BH is finite. For example, for a linear distribution of shells in $v$, $q>1$. If the shells become more spread-out, then $q$ can be smaller, while if they become more concentrated it must be larger.} and $a$ is a positive constant with appropriate dimensions. We also impose that there be infinitely many shells and that $\lim_{n\to\infty}v_n=\infty$.

One of the key ingredients necessary for mass inflation is that this increase in the mass (as seen from outside the object) is itself related polynomially to the change in position of the inner gravitational radius $\delta r_{{\rm i},n}$, i.e.
\begin{equation}\label{ingredient1}
\delta r_{{\rm i},n}=-\frac{\beta}{v_n^p},
\end{equation}
where the power $p$ is again positive (though it can be different from $q$), and $\beta$ is again a positive constant (at least asymptotically in $v$). We stress that eq. \eqref{ingredient1} is an \textit{assumption} about how the geometry responds to the ingoing perturbation \eqref{deltam}. For example, in the Reissner-Nordström case, where one has
\begin{equation}\label{rirn}
\delta r_{{\rm i},n}=\frac{1}{\sqrt{m^2-Q^2}}(-r_{{\rm i},n}\delta m_n+Q\delta Q_n),
\end{equation}
the assumption \eqref{ingredient1} restricts the amount of (same sign) charge the ingoing shells can carry. One can easily imagine a case in which the relation between $\delta m_n$ and $\delta Q_n$ is such that e.g. $\delta r_{{\rm i},n}=0$, which would lead to an absence of mass inflation. This is not limited to the Reissner-Nordström case: \eqref{ingredient1} generally avoids the suppression of the exponential in \eqref{quotient} by its prefactor, allowing mass inflation to take place.

This can be seen by looking at the evolution of the redshift function in the interior of the outgoing shell after the $n$th ingoing shell has crossed it, $F_n$. From eqs. \eqref{intersect}, \eqref{quotient}, and \eqref{ingredient1} we get the asymptotic relation at the $n$th intersection point
\begin{equation}\label{massiteration}
\begin{split}
F_n&=F_{n-1}\frac{f_n}{f_{n-1}}\\&=F_{n-1}\frac{\kappa_n}{\kappa_{n-1}}\left(\frac{v_{n-1}}{v_n}\right)^pe^{\kappa_{n-1}\Delta v_n}\left[1+\order{e^{-\kappa_{n-1}\Delta v_n}}\right]\\&=F_{n-1}e^{\kappa\Delta v_n}\left[1+\order{e^{-\kappa\Delta v_n},\frac{\Delta v_n}{v_{n-1}}}\right],
\end{split}
\end{equation}
where $\Delta v_n=v_n-v_{n-1}$ and in the last line we have directly substituted the asymptotic value of the surface gravity of the $f_n$ geometries, $\kappa_n\to\kappa$. We have also used the fact that the differences $v_n-v_{n-1}$ become negligible when compared to the absolute value of $v_n$ asymptotically (for any distribution of infinite shells which covers an infinite range in $v$), and that $e^{\kappa\Delta v_n}>1$, leaving only the dominant contribution as leading order. If we use this relation iteratively from an initial time $v=0$ when the interior redshift function was $F_0<0$, we get the asymptotically exponential increase (of the absolute value of) the redshift function at the shell
\begin{equation}\label{massinfl}
F_n|_{\rm shell}\sim F_0e^{\kappa v_n}.
\end{equation}
The result is independent of the spacing between the ingoing shells: increased spacing only leads to a higher jump in $F_n$ when a shell eventually falls. This tendency continues for as long as more shells are thrown in. Returning to the Reissner-Nordström case, we can once again easily associate this with a proportional increase of the mass term through eq. \eqref{reissner}. In more general geometries, it can be related to an increase of the Misner-Sharp mass given by \cite{Misner1964,Hayward1994}
\begin{equation}\label{Misner-Sharp}
M_{\rm MS}=\frac{1}{2}r(1-F),
\end{equation}
for this interior region. Let us recall that the Misner-Sharp mass provides a quasi-local characterization of gravitational energy in spherical symmetry~\cite{Hayward1994}.

Eq. \eqref{massinfl} captures the main result of mass inflation: the exponential growth of the redshift function (more generally, the $g^{rr}$ metric component) below a certain radius, associated with a corresponding growth of the Misner-Sharp mass. In our model it can be physically interpreted in terms of the exchange of mass between the outgoing and ingoing shells. The outgoing shell, being inside the trapped region, can be seen as having a negative asymptotic mass. At the intersection points, the ingoing shells can therefore take away positive mass by making the negative one of the outgoing shell increasingly more negative. This exchange is mediated by the dynamics of the gravitational field, and in particular the exponential runaway effect is only triggered if the outgoing shell is taken from an initial proximity to the inner gravitational radius of the $f_n$ geometry patches, to subsequently (after the interaction) end up deeper inside the trapped region due to the inward displacement of this inner gravitational radius \eqref{ingredient1}.

It is interesting to note that this process is independent of the particularities of the infalling matter shells, the only requirement being that the infalling perturbations alone induce a polynomially decreasing response in the inner gravitational radius position \eqref{ingredient1}. If the shift in its position were instead to decrease, e.g., exponentially,
\begin{equation}
\delta r_{{\rm i},n}=-\tilde{\beta}e^{-\sigma v},
\end{equation}
where $\tilde{\beta}$ and $\sigma$ are positive constants, then one can observe from \eqref{quotient} and the corresponding equivalent of \eqref{massiteration} that the outcome would depend on the difference between the surface gravity $\kappa$ and the coefficient $\sigma$. Particularly, mass inflation would only take place if $\kappa>\sigma$, as has been shown in the case where $\sigma$ is the surface gravity of a cosmological horizon in asymptotically de Sitter spacetimes, governing the decaying tail of infalling radiation \cite{Mellor1992,Brady1992}.

It is worth noting that this shell-based model has some characteristics in common with the case in which a continuous power-law flux of radiation is present, but there are also some differences. For example, the exponential which appears in the mass inflation of this model is directly related to the exponentially decreasing separation between the outgoing shell and the inner gravitational radius $r_{\rm i}$ (as seen from the outside) in each step represented in fig. \ref{f1d}. Although the average of this distance taken over several steps has the same inverse-polynomial decrease as is expected from a continuous matter case (that is, if $r_{\rm i}$ had a position evolving as a continuous version of \eqref{ingredient1}), taking the limit to a continuous ingoing distribution of matter is not at all straightforward. Although the outgoing shell seems to be a good model for the shockwave which generally appears in this region even with continuous matter \cite{Marolf2012}, the ingoing shells offer qualitatively new features. This may also be inferred by the fact that a generalisation of Ori's model~\cite{Raul2021}, in which the ingoing flux is continuous, leads to a larger variety of asymptotic outcomes, unlike the single exponential behaviour observed here. The question of whether the continuous or the discrete model (or some combination of the two) fits best the behaviour of small (possibly quantised) infalling perturbations may thus turn out to be an important one for a better understanding of the singularity at the Cauchy horizon, though we will not address it here.

\section{Geometry inside the mass-inflated region}\label{secri}

This shell-based construction captures (albeit a simple variant of) the mass inflation process and it can give us some additional insights into the behaviour of the geometry inside the BH at finite times. In particular, to set up our subsequent semiclassical analysis, we want to see how this model answers two questions about the depths of the mass-inflated region. The first one is whether a stream of outgoing radiation below the initial outgoing shockwave may also give rise to an effect similar to mass inflation, which would further increase the rate of growth of mass in the innermost region of the geometry, in the vicinity of the origin. The second one is what particular path the inner apparent horizon actually follows in this whole process, and whether it collapses to the origin to give rise to a spacelike singularity at a finite time $v$, as suggested in the numerical analysis of \cite{Brady1995}.

To answer these questions, we must first get a better understanding of the global structure of the geometry corresponding to the above construction with just a single gravitating outgoing shell. The metrics on the outside and inside of the outgoing shell respectively can be written as
\begin{align}
ds^2&=-f(v,r)dv^2+2dvdr+r^2d\Omega^2,\\
\begin{split}
dS^2&=-F(v,r)dV^2+2dVdr+r^2d\Omega^2\\&=A(v)\left[-A(v)F(v,r)dv^2+2dvdr\right]+r^2 d\Omega^2,
\end{split}\label{geod}
\end{align}
where lower and upper case letters are once again used for quantities in the regions outside and inside the shell, respectively, and $A(v)=dV/dv=f/F|_{\rm shell}$ is a positive function which allows us to switch between the Eddington-Finkelstein coordinates of these two regions, as expressed on the right-hand side of the latter equation.

For simplicity, we will take a smoothed-out average of the metric functions from the previous section (particularly, their dominant behaviour at late times). The redshift function on the shell evaluated on the inside \eqref{massinfl} is then
\begin{equation}\label{rb}
F|_{\rm shell}=-|F_0|e^{\kappa v},
\end{equation}
and, from eq. \eqref{ingredient1} and the shell trajectory in each patch, the redshift function on the outside satisfies
\begin{equation}
f|_{\rm shell}= -\kappa\left[r_{\rm shell}-r_{\rm i}(v)\right]=-\frac{b}{v^p},
\end{equation}
with $b$ a positive constant which depends on the average spacing between ingoing shells (i.e. the average of the quantity $e^{\kappa\Delta v_n}$), into which $\kappa$ (the outside region's inner gravitational radius's surface gravity) and the constant $\beta$ from eq. \eqref{ingredient1} have also been absorbed. The function which relates the outside and inside times, $v$ and $V$, then becomes
\begin{equation}\label{A1}
A(v)=\frac{b}{|F_0|}\frac{e^{-\kappa v}}{v^p}.
\end{equation}

The results we have obtained thus far are valid for any redshift function, but if we want to analyse the deeper regions of the geometry we have to be more specific. Let us therefore first focus on the particular example we have used previously, namely the Reissner-Nordström geometry. If we assume that the ingoing and outgoing shells carry no electrical charge, we can take the evolution of $F$ to be solely due to a change in the mass term. Then, from eqs. \eqref{rb} and \eqref{A1}, and with the Reissner-Nordström redshift function, we get the geometry for the mass-inflated interior region \eqref{geod} written in the $v$ coordinate
\begin{equation}\label{inflRN}
dS^2=\frac{b}{|F_0|v^p}e^{-\kappa v}\left[-\frac{b}{|F_0|v^p}e^{-\kappa v}\left(1-\frac{2M_0e^{\kappa v}}{r}+\frac{Q^2}{r^2}\right)dv^2+2dvdr\right]+r^2d\Omega^2,
\end{equation}
where $M_0$ is a positive constant which represents the initial mass of the BH. Using this metric we see that the equation for outgoing null geodesics (which we can later relate to trajectories of additional gravitating outgoing shells) then takes the form
\begin{equation}\label{outnull}
\frac{dr}{dv}=\frac{b}{|F_0|v^p}\left[-h_0(r)+h_1(r)e^{-\kappa v}\right],
\end{equation}
where $h_0(r)=M_0/r$ and $h_1(r)=(1+Q^2/r^2)/2$ are positive functions. Due to the exponential suppression of the $h_1$ term, it is clear that the term with $h_0$ is dominant on the right-hand side of this equation, except in a progressively smaller region around the origin, where the inner apparent horizon is shrinking toward zero radius. 
In relation to this, we will be able to distinguish between two types of outgoing geodesics in the trapped region: ones whose dynamics is predominantly determined by the $h_0$ term, and ones for which the two terms are comparable. The former exist up to $v\to\infty$ only if $p>1$, which, given eq. \eqref{rirn}, is directly related to the mass thrown into the BH from the outside being finite (for our current example of Reissner-Nordström).\footnote{It is interesting to note that it is quite easy to imagine a geometry in which e.g. due to a relation $\delta r_{\rm i}\propto -(\delta m)^{1/2}$, $p$ ends up being 1 or smaller for a finite accretion of mass. The absence of solutions of the type \eqref{trapped} would then imply a lack of a Cauchy horizon below the outgoing shell, leaving just a trapped region with a tendency toward the formation of a spacelike or null singularity.} If this condition is met and the $h_0$ term continues to dominate, equation \eqref{outnull} generally integrates asymptotically to
\begin{equation}\label{trapped}
r=r_{\rm c}+\frac{h_0(r_{\rm c})}{|F_0|(p-1)}\frac{b}{v^{p-1}}+\mathcal{O}\left[\left(\frac{b}{v^{p-1}}\right)^2\right].
\end{equation}
The integration constant $r_{\rm c}$ represents the final radial position of each light ray when it reaches the Cauchy horizon at $v\to\infty$, which is different for each geodesic depending on initial conditions. In other words, between the outgoing shell and the inner apparent horizon (which is rapidly shrinking toward $r=0$) there are null geodesics which are trapped in a tendency toward a finite radial position which is different from the asymptotic position of this horizon. To see how this journey is perceived from the point of view of an observer in the interior region itself, we can look at the relation between the $v$ and $V$ coordinates, the latter of which, for observers not tending toward the inner apparent horizon, is proportional to the geodesic affine parameter. As we have seen, this relation is given by
\begin{equation}
\frac{dv}{dV}=\left.\frac{F}{f}\right|_{\rm shell}=\frac{1}{A(v)},
\end{equation}
which asymptotically integrates to
\begin{equation}\label{trappedtime}
V=V_{\rm c}-\frac{b}{|F_0|\kappa}\frac{e^{-\kappa v}}{v^p}+\order{\frac{e^{-\kappa v}}{v^{p+1}}}.
\end{equation}
These geodesics therefore reach the Cauchy horizon at a finite time parameter $V=V_{\rm c}$, corresponding to the integration constant of the equation. This gives us an interpretation for the behaviour seen in \eqref{trapped}: outgoing null geodesics are trapped in a tendency toward these different finite radial positions $r_{\rm c}$ because the function $A$ acts to quickly freeze the proper time, and consequently the movement, of observers in this region. Because of this \textit{freezing function} $A$, most of the outgoing radiation in the trapped region would in fact reach the Cauchy horizon before getting close to the inner apparent horizon. This tells us that interactions between the ingoing shells with additional outgoing shells travelling along these geodesics would not have time to produce any sort of amplification of the mass inflation effect, as this would require proximity to the inner gravitational radius $R_{\rm i}$ of this internal region (which in the absence of such shells is in fact the inner apparent horizon position).

The region where eqs. \eqref{trapped} and \eqref{trappedtime} are valid begins on the inside of the outgoing shell and increases in size toward the origin as the mass tends to infinity due to its unbounded growth following eqs.~\eqref{Misner-Sharp} and~\eqref{rb}. The radii $r_{\rm c}$ at which null geodesics freeze can then vary continuously from $r_{\rm i}$ (the inner gravitational radius of the external geometry) to zero, as is observed in the analytical study of the Cauchy horizon in \cite{Ori1998}. However, this does not imply that all outgoing null geodesics are trapped in this way and are unaffected by the shrinking inner apparent horizon. Due to the growing mass, the radial position of this apparent horizon $R_{\rm i}$ can be approximated by a series expansion of the lower of the two roots of the Reissner-Nordström redshift function in $1/M(v)$, revealing its tendency to zero
\begin{equation}\label{ria}
R_{\rm i}(v)=\frac{Q^2}{2M(v)}+\order{\frac{1}{M(v)}}^3=\frac{Q^2}{2M_0}e^{-\kappa v}+\order{e^{-3\kappa v}},
\end{equation}
with a surface gravity (in absolute value)
\begin{equation}\label{RNk1}
K_{\rm i}(v)=\frac{b}{v^p}\frac{2M_0^3}{|F_0|Q^4}e^{2\kappa v}+\order{\frac{1}{v^p}}.
\end{equation}
There are indeed many outgoing null geodesics which are sufficiently close to this horizon to approach it asymptotically, i.e. to tend to zero from above it. These are the second-type geodesics we mentioned before: the ones for which the $h_0$ and $h_1$ terms of eq. \eqref{outnull} are comparable, with the right-hand side of this equation being close to zero, i.e. $r$ being close to $R_{\rm i}$. Their movement can be described by expanding the right-hand side of this equation around $R_{\rm i}$,
\begin{equation}\label{geo-ria}
\frac{dr}{dv}\simeq-K_{\rm i}(v)[r-R_{\rm i}(v)].
\end{equation}
With the rapidly growing value of $K_{\rm i}$, it can be readily checked that this equation has a family of solutions for which $r\to R_{\rm i}$. These geodesics also reach $v=\infty$ with a finite affine parameter, where they converge to $r=0$ along with $R_{\rm i}$, falling into a (strong) curvature singularity.

If an additional mass inflation effect can take place, pushing the position of the inner apparent horizon to below the one given by~\eqref{ria}, it would be triggered by placing an outgoing shell precisely on one of the geodesics described by~\eqref{geo-ria} which tend toward $R_{\rm i}$ from above. To keep the label $R_{\rm i}$ for the zero of the $F$ function~\eqref{ria}, i.e. the inner gravitational radius of the $F$ patch, we will now call the actual inner apparent horizon $\tilde{R}_{\rm i}$. Also, instead of working with the $v$ coordinate, here it will be more convenient to use $V$, with which we can directly apply the relation \eqref{quotient} for the junction conditions on the intersection points with the ingoing shells. The position of the inner gravitational radius $R_{\rm i}$ and its surface gravity (taken as the absolute value of $\partial_rg_{vv}/(2g_{vr})$ in each coordinate system)\footnote{In other words, $K_{\rm i}^V$ is not just $K_{\rm i}$ with a coordinate change, due to the $A(v)$ factor which multiplies the redshift function after the coordinate change, as can be seen in \eqref{inflRN}.} in the $V$ coordinate system $K_{\rm i}^V$ evolve as
\begin{align}
R_{\rm i}&=\xi_1(V_{\rm c}-V)+\mathcal{O}[(V_{\rm c}-V)^3],\\
K_{\rm i}^V&=\frac{\xi_2}{(V_{\rm c}-V)^3}+\order{\frac{1}{V_{\rm c}-V}},
\end{align}
where $\xi_1$ and $\xi_2$ are some positive constants which depend of the asymptotic mass, the charge and the initial conditions. In the shell model we once again consider that these functions actually have discrete jumps at a set of points in $V$, corresponding to the positions of the infalling shells. If we take these shells to be either equispaced in $v$, or at most distributed with a density polynomially dependent in $v$, then their spacing in $V$ decreases as
\begin{equation}
\Delta V_n=V_n-V_{n-1}\sim V_{\rm c}-V_n
\end{equation}
as $V_n$ tends toward the Cauchy horizon $V_{\rm c}$. The jumps in the position of the inner horizon between shells also decrease as
\begin{equation}
\Delta R_{\rm i,n}\sim V_{\rm c}-V_n.
\end{equation}
Eq. \eqref{quotient} can be applied directly here because the outgoing shell has enough time between each iteration to get exponentially closer to $R_{\rm i,n}$, as can be seen by the fact that
\begin{equation*}
e^{-K^V_{\rm i,n}\Delta V_{n}}\sim e^{-1/(V_{\rm c}-V_{n})^2}\to 0, \quad \text{as}\quad V_{n}\to V_{\rm c}.
\end{equation*}
The quotient of surface gravities in \eqref{quotient} once again tends to a constant in the limit of interest, and so does the quotient of the differences between radial positions in front of the exponential.\footnote{This can be seen explicitly by solving the outgoing null geodesic equation between each infalling shell and matching the solutions. The calculation is completely analogous to the one performed in the vicinity of $r_{\rm i}$.} This leaves the redshift function below this new outgoing shell with an increase given by a multiplicative factor
\begin{equation*}
e^{K^V_{\rm i,n}\Delta V_{n}}
\end{equation*}
after each iteration, which diverges as $e^{\xi_2/(V_{\rm c}-V)^2}$ toward the Cauchy horizon. In terms of~$v$, this new mass inflation increases the Misner-Sharp mass in the vicinity of the origin as an \textit{exponential of an exponential}. One can then imagine that each time we repeat this whole process considering the new displacement of the inner horizon caused by this effect, we would get an additional exponential in $v$ to the chain.

\begin{figure}
	\centering
	\includegraphics[scale=1.2]{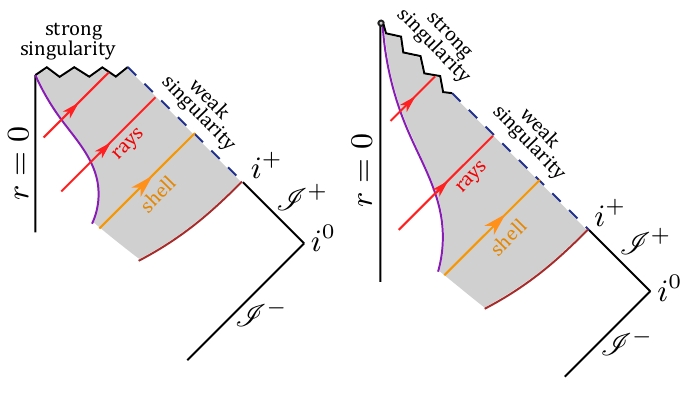}
	\caption{Future part of the causal diagram of the mass inflation geometry. The shaded part is the trapped region and the dashed line is the Cauchy horizon (and a weak singularity). The null shell shown is the upper outgoing one which tends to $r_{\rm i}$ and is responsible for the first mass inflation effect. Left: the inner apparent horizon reaches $r=0$ at finite $v$ and forms a Schwarzschild-type spacelike singularity. Right: the inner horizon only tends to $r=0$ asymptotically in $v$, resulting in a strong null singularity at $v\to\infty$ and $r=0$, at a finite affine distance for geodesics which fall into it.}
	\label{f3d}
\end{figure}

We can now tackle answering the second question posed at the beginning of this section: what is the actual path followed by the inner horizon, and what causal structure does this movement give rise to. It is hardly surprising that in the case of continuous matter the chain effect just described could produce a spacelike singularity at finite $v$, as is observed numerically in \cite{Brady1995} and commented on in later works \cite{Ori1998,Marolf2012}. Even if the exact analytical solution were to only result in a very quick tendency toward the formation of this singularity (i.e. a quick approach of the inner apparent horizon toward the origin, though still asymptotic in $v$), this approach would in fact be so quick that it would likely be numerically indistinguishable from a singularity at a finite time $v$; at any rate, the curvature would become Planckian very fast, making a classical description of the geometry-matter interactions inadequate. However, it is interesting to note that within the classical description, this seemingly small difference between a singularity forming at strictly finite $v$ and having just a tendency toward its formation, however quick, and only forming it at $v\to\infty$, results in two different asymptotic structures, represented in the two diagrams of fig.~\ref{f3d}. In fact, the former case results in the formation of a Schwarzschild-type spacelike singularity, while the latter ends in a (strong) null singularity, both of which are at $r=0$ and are attached to the weak null singularity which spans the Cauchy horizon, where $r$ takes values up to $r_{\rm i}$. To clarify, we use the same criterion for distinguishing strong and weak singularities as the one described in~\cite{Ori1991,EllisSchmidt1977}: both have diverging curvature scalars, but the distortion suffered by an observer of finite size remains bounded when crossing a weak singularity, while it diverges when crossing a strong one. One can readily check that the curvature blow-up of the geometry we use close to the Cauchy horizon [see eq. \eqref{inflRN}] is the same as in refs.~\cite{Ori1991,Brady1995}, $R_{\mu\nu\rho\sigma}R^{\mu\nu\rho\sigma}\sim 48M_0^2r^{-6}e^{2\kappa v}$, leading to the same type of weak singularity there.

Up to here, all we have said regarding these questions has been based on the Reissner-Nordström background. For other geometries besides Reissner-Nordström, the inner structure of the mass-inflated region depends on how the geometry reacts to the increase in mass provided by the infalling null shells, particularly on the trajectory followed by the inner gravitational radius. The relation from the charged BH $\delta r_{\rm i}\propto(-\delta m)$ is also satisfied in the case of a rotating BH, but in regular BH spacetimes this tendency may be modified, depending on how the regularisation of the origin is achieved in the first place. One may expect there to exist trapped geodesics of the type~\eqref{trapped}, but the structure toward the origin may differ, possibly avoiding the formation of a strong singularity altogether by preventing the inner horizon from getting too close to $r=0$.

It is also interesting to note that in general, the explicit divergence of the mass at $v\to\infty$ depends on the ever-smaller ingoing perturbations also continuing up to infinity. Classically it is perfectly natural to consider this to be the case, but one may imagine that a quantum description of the interaction between the BH and infalling matter may have a lower bound on the energy which can actually affect the BH. For example, we can consider that this lower bound is given by the energy of a single photon with a wavelength of the order of the BH external mass $m$ (in geometric units). Then, we can relate this energy to a mass and to a cutoff time $v_{\rm cut}$ through \eqref{deltam}, and for simplicity we can take the polynomial tail in \eqref{ingredient1} to be the same as in \eqref{deltam}, i.e. $p=q$. The result is that by $v_{\rm cut}$ the mass in the interior region would have increased by a factor $e^{(M/l_{\rm P})^{2/p}}$ from just the first exponential mass inflation effect around $r_{\rm i}$, given by eqs. \eqref{massinfl} and \eqref{Misner-Sharp}. Needless to say, the exponent in this number is generally very large. For a solar mass BH and a polynomial decay with $p=12$ (as considered in \cite{PoissonIsrael89}) the mass grows by a factor larger than $e^{10^{6}}$, making a classical description of this region inadequate.

\section{Summary and conclusions}

In this chapter we have constructed a simple geometric model which captures the classical inner horizon instability. It consists of a series of ingoing and outgoing null shells perturbing an otherwise static BH configuration. Matching the mass carried by the ingoing shells with the expected tail of ingoing radiation in astrophysical scenarios~\cite{Price1972,PoissonIsrael89}, we find that mass inflation is triggered if the inner horizon position has a power law relation to the BH mass, akin to the case of charged and spinning BHs. On the other hand, BHs with a particularly stiff inner core, which does not shrink as quickly (or at all) when more mass is introduced, would not experience mass inflation.

For charged BHs, and by extension for rotating ones as well, where an analogous shell-based construction can be made \cite{Barrabes1990}, the instability occurs as long as the perturbations falling in do not carry enough charge or angular momentum to keep the inner horizon still enough, which is unlikely to occur in an astrophysical scenario over long periods of time. The causal structure of these spacetimes is therefore modified: the inner horizon approaches the origin, while outgoing light rays which cross it are frozen in approaches toward finite radial positions slightly below where they enter the trapped region. These rays reach $v\to\infty$ in finite affine parameter, creating a Cauchy horizon there. The curvature divergence at this horizon is weak enough for the metric to be continuously extended beyond it.

For the purposes of the semiclassical analysis of chapter \ref{ch7}, the important part of the evolution is the period between the formation of the BH and the time when its inner apparent horizon reaches a region of Planckian curvature. As it turns out, this initial period may be susceptible to semiclassical modifications, halting the classical tendency toward singularity formation.


\chapter{Semiclassically corrected inner horizon evolution}\label{ch5}

In this chapter we will present a perturbative analysis of backreaction on an inner horizon from the RSET of a quantum scalar field. We will begin by calculating the RSET in the Polyakov approximation on spacetime backgrounds containing dynamically formed BHs. Then we take the RSET as a source of dynamical perturbation and see how the trapped region tends to evolve semiclassically. On the one hand, we recover the evaporative tendency of the outer horizon. On the other, we find a tendency for the inner horizon to move outward, reducing the size of the trapped region from the inside.

\section{Backreaction on a BH with a static inner horizon}\label{static}

We begin by presenting a toy model for the geometry of the formation of a spherical BH with an inner horizon, which captures the main characteristics with sufficient generality, but is simple enough to allow analytical calculations with the semiclassical perturbations caused by the RSET. After its formation, we impose that this BH be classically static (with a timelike Killing vector outside its trapped region). It can be either of a Reissner-Nordström type, with a timelike singularity at the centre, or of a regular type, with a de Sitter or approximately flat core region~\cite{Ansoldi,Simpson2019}. In the interest of avoiding ambiguities in the propagation of the quantum field modes, we will work with the assumption of a regular core, although our results on semiclassical evolution are also applicable to the former case, for a small but finite time period.

\subsection{Geometric model and RSET}

We start with a Minkowski spacetime, which has a line element
\begin{equation}
	ds^2=-dv^2+2dvdr+r^2d\Omega^2
\end{equation}
in advanced Eddington-Finkelstein coordinates. The simplicity of our construction lies in assuming that after a point in time $v_{\rm f}$ the geometry becomes a regular, spherically symmetric static BH,
\begin{equation}\label{metric-vr}
	ds^2=-f(r)dv^2+2dvdr+r^2d\Omega^2.
\end{equation}
The surface which separates the flat and BH regions can be seen as a collapsing null shell, a model often used when calculating semiclassical effects near the outer horizon (see e.g. \cite{SC2014}), though in this case it does not stem from a simple classical solution and is more of a geometric construct. The peeling of null geodesics, and consequently of modes of massless quantum fields, away from the outer horizon makes long-term semiclassical effects exhibit a certain universality there \cite{BLSV11}. Thus a null shell is as good as any model of collapse if we want to study late-time Hawking evaporation. However, as we will see, at the inner horizon null geodesics behave in the opposite way to those at the outer one, i.e. they are accumulated, so the result is the exact opposite: late-time behaviour of semiclassical effects is highly sensitive to the characteristics of the collapse, and to conditions from the past of the spacetime in general. In the face of this complexity, the collapsing shell model can be used as a simple geometric method for obtaining reasonable initial conditions for the quantum modes entering the BH region, without resorting to the numerical computations which would be required in more generic collapse scenarios. This method is also unique in the sense that it makes the quantum modes acquire the least amount of ``noise" from the collapse and serves to isolate the effects coming purely from the quantum field finding itself in the BH geometry. Due to its simplicity, it is also an excellent example in which we can follow the origin and consequences of semiclassical effects by means of analytical expressions, as we will see.

\begin{figure}
	\centering
	\includegraphics[scale=.5]{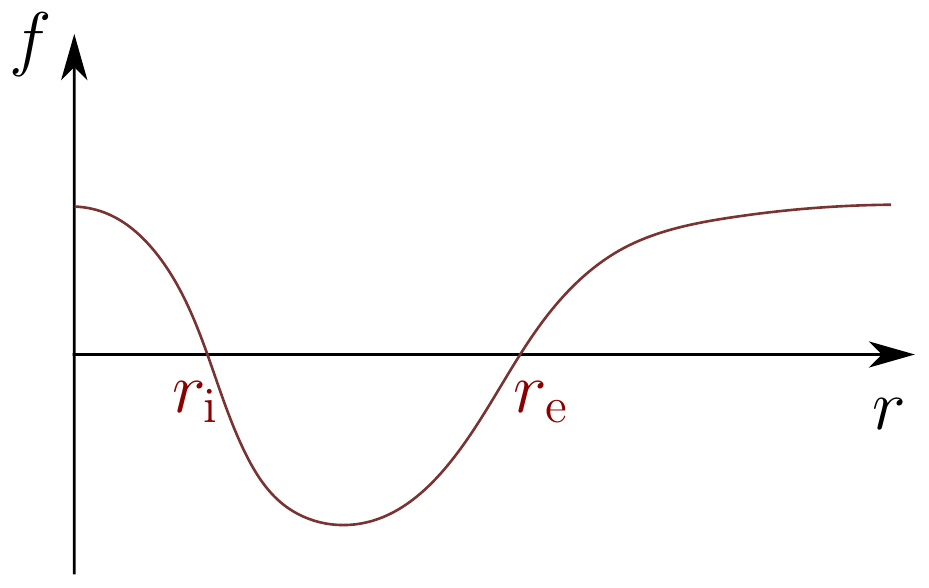}
	\caption{Redshift function $f(r)$ of our schematic regular BH. For the inner horizon $r_{\rm h}=r_{\rm i}$ and for the outer one $r_{\rm h}=r_{\rm e}$.}
	\label{f1e}
\end{figure}

For the redshift function in the BH region $f(r)$ we will use a series expansion around each horizon of the form
\begin{equation}\label{redshift}
	f(r)=2k_1(r-r_{\rm h})+2k_2(r-r_{\rm h})^2+2k_3(r-r_{\rm h})^3+\cdots,
\end{equation}
where $r_{\rm h}$ denotes the position of either the internal or external horizon and $k_{i}$ are constants, $k_1$ corresponding to the surface gravity of the horizon (negative for the internal and positive for the external horizon; note the simplification of the notation with respect to chapter~\ref{ch3}, where we used the absolute values of each surface gravity). Sufficiently close to each horizon, these two expansions are a valid representation of the redshift function, the global structure of which we assume is qualitatively of the form represented in fig. \ref{f1e}. We note that throughout this work, when we construct a series assuming that a quantity with dimensions of length $l$ is ``small", we of course mean comparatively to the scale of the problem, i.e. that the sets of quantities $\{k_1l\}$, $\{k_2l^2,(k_1l)^2\}$, and subsequent orders, are progressively smaller.

\begin{figure}
	\centering
	\includegraphics[scale=.8]{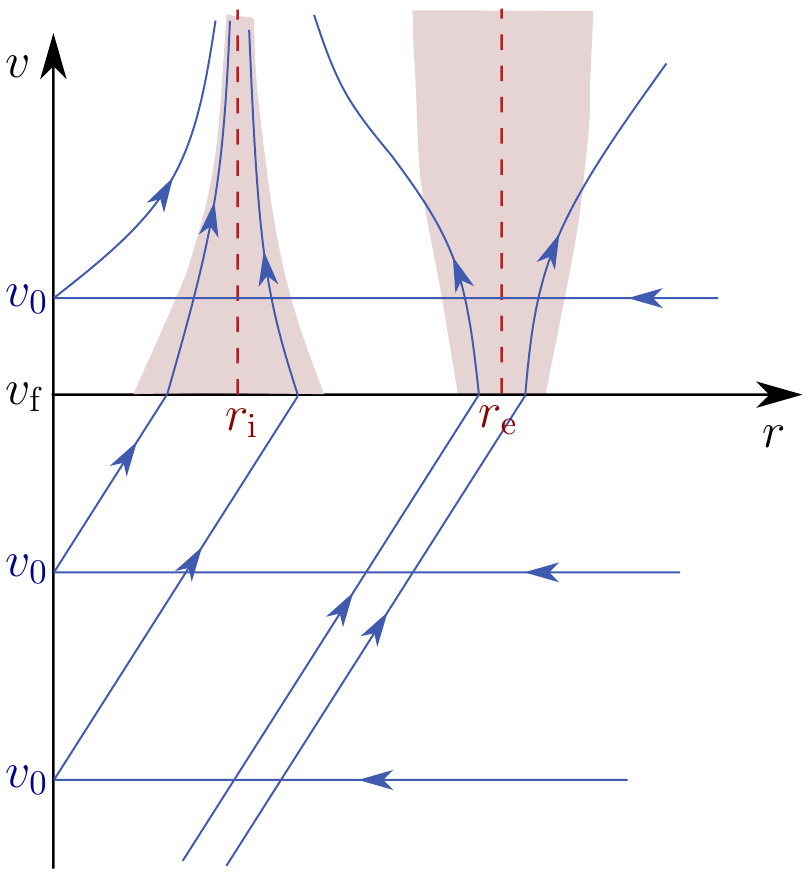}
	\caption{Representation of the trajectories of outgoing radial light rays. The BH geometry begins at $v>v_{\rm f}$. The shaded regions around each horizon are a qualitative representation of the regions where our approximation for the RSET is valid.}
	\label{f2e}
\end{figure}

We are thus treating an arbitrary BH geometry of the form \eqref{metric-vr}. We use only one of the two degrees of freedom of the spherically symmetric spacetime for simplicity, given that the redshift function itself is enough to generate the causal structures we are interested in. The simple dynamical model for the formation of this structure, along with a focus on the areas around each horizon, will make analytical calculations of backreaction tractable. We can now proceed to construct the quantum ``in" vacuum state for this geometry, necessary for calculating the RSET. This vacuum state is defined from the Minkowski region at past null infinity, and its evolution is determined by the evolution of the spacetime. In particular, for the 1+1 dimensional approximation we will be using, its modes can be obtained from the behaviour of the radial null geodesics, represented in fig. \ref{f2e}. The ingoing ones simply satisfy
\begin{equation}
	v=\text{const.}
\end{equation}
For the outgoing ones we must solve the equation
\begin{equation}\label{outnulle}
	\frac{dr}{dv}=\frac{1}{2}f(r).
\end{equation}
For $v<v_{\rm f}$, $f(r)=1$ and the solution is simply
\begin{equation}\label{flatsol}
	r(v)=\frac{1}{2}(v-v_0),
\end{equation}
where $v_0$ is an integration constant, identified as the time at which the light ray passes through the origin $r=0$. The ``in" vacuum is constructed from a pair of null coordinates $(v_{in},u_{in})$, which in this case are in fact $v_{\rm in}=v$ and $u_{\rm in}=v_0$ (i.e. in the $u_{\rm in}=\text{const.}$ outgoing null trajectories the value of $u_{\rm in}$ is the value of $v$ when said trajectories meet the origin). We have conveniently expressed the integration constant in terms of $v_0$, and we need to do the same with all solutions for outgoing light rays from here on, i.e. we need to trace them back to the origin. For the region in which the BH has formed ($v>v_{\rm f}$), we are only interested in analysing the vicinity of each horizon, where \eqref{redshift} is sufficiently accurate. There, the solutions of \eqref{outnulle} can be expressed in a series around $d_{\rm f}=0$, a parameter corresponding to the distance of the null ray from the horizon at $v_{\rm f}$,
\begin{equation}\label{df}
	d_{\rm f}=r(v_{\rm f})-r_{\rm h}=\frac{v_{\rm f}}{2}-\frac{v_0}{2}-r_{\rm h},
\end{equation}
the second equality coming from matching with \eqref{flatsol} at said time. For the purposes of the present calculation, it is sufficient to express the solution of \eqref{outnulle} up to order $d_{\rm f}^3$,
\begin{equation}\label{rv}
	\begin{split}
		r(v)\simeq r_{\rm h}&+e^{k_1(v-v_{\rm f})}d_{\rm f}+\frac{k_2}{k_1}\left[-e^{k_1(v-v_{\rm f})}+e^{2k_1(v-v_{\rm f})}\right]d_{\rm f}^2\\ &+\left[\left(\frac{k_2^2}{k_1^2}-\frac{k_3}{2k_1}\right)e^{k_1(v-v_{\rm f})}-2\frac{k_2^2}{k_1^2}e^{2k_1(v-v_{\rm f})}+\left(\frac{k_2^2}{k_1^2}+\frac{k_3}{2k_1}\right)e^{3k_1(v-v_{\rm f})}\right]d_{\rm f}^3.
	\end{split}
\end{equation}
At the outer horizon, where the surface gravity is positive, the coefficients of this series increase exponentially, making it a bad approximation for any finite $d_{\rm f}$ after sufficient time has passed. The reason for this can be seen in fig. \ref{f2e}: outgoing light rays diverge away from the outer horizon, thus away form where the expansion \eqref{redshift} is valid. The reverse happens at the inner horizon, where the surface gravity is negative, as light rays converge toward this horizon. However, this will translate inversely to the accuracy of the RSET approximation in each region, as this quantity will depend on how the rays evolve \textit{backwards} in time.

To switch from $(v,r)$ to $(u_{in},v_{in})\equiv(u,v)$ coordinates, we use the relation
\begin{equation}
	dr=\left.\frac{\partial r}{\partial v}\right|_{v_0=u}dv+\left.\frac{\partial r}{\partial v_0}\right|_{v_0=u}du,
\end{equation}
and note from \eqref{outnull} that $\partial r/\partial v|_{v_0=u}=f(r)/2$, which means that the term proportional to $dv^2$ in the metric cancels out. We are then left with the line element for the BH region
\begin{equation}
	ds^2=2\left.\frac{\partial r}{\partial v_0}\right|_{v_0=u}dudv+r^2(u,v)d\Omega^2=-C(u,v)dudv+r^2(u,v)d\Omega^2,
\end{equation}
where $C=-2\partial r/\partial v_0|_{v_0=u}$ is the conformal factor of the reduced 1+1 dimensional spacetime. From \eqref{rv} and \eqref{df} we obtain the expansion for this quantity up to order $d_{\rm f}^2$,
\begin{equation}\label{conformal}
	\begin{split}
		C\simeq\; &e^{k_1(v-v_{\rm f})}+2\frac{k_2}{k_1}\left[-e^{k_1(v-v_{\rm f})}+e^{2k_1(v-v_{\rm f})}\right]d_{\rm f}\\&+3\left[\left(\frac{k_2^2}{k_1^2}-\frac{k_3}{2k_1}\right)e^{k_1(v-v_{\rm f})}-2\frac{k_2^2}{k_1^2}e^{2k_1(v-v_{\rm f})}+\left(\frac{k_2^2}{k_1^2}+\frac{k_3}{2k_1}\right)e^{3k_1(v-v_{\rm f})}\right]d_{\rm f}^2.
	\end{split}
\end{equation}

With this we are ready to calculate the components of the RSET in the Polyakov approximation, given in ``in" coordinates by the standard expressions [see eq. \eqref{e33}]
\begin{subequations}\label{RSET}
	\begin{align}
		\expval{T_{uu}}&=\frac{1}{96\pi^2r^2}\left[\frac{\partial_u^2C}{C}-\frac{3}{2}\left(\frac{\partial_uC}{C}\right)^2\right],\\
		\expval{T_{vv}}&=\frac{1}{96\pi^2r^2}\left[\frac{\partial_v^2C}{C}-\frac{3}{2}\left(\frac{\partial_vC}{C}\right)^2\right],\\
		\expval{T_{uv}}&=\frac{1}{96\pi^2r^2}\left[\frac{\partial_uC\partial_vC}{C^2}-\frac{\partial_u\partial_vC}{C}\right].
	\end{align}
\end{subequations}
Calculating these quantities and switching back to $(v,r)$ coordinates we obtain the leading order of the local expressions around each horizon
\begin{subequations}\label{RSETres}
	\begin{align}
		\expval{T_{vv}}&=-\frac{1}{96\pi^2r_{\rm h}^2}\frac{k_1^2}{2}+\order{r-r_{\rm h}},\\ \label{RSETrv}
		\expval{T_{rv}}&=-\frac{1}{96\pi^2r_{\rm h}^2}2k_2+\order{r-r_{\rm h}},\\
		\expval{T_{rr}}&=\frac{1}{96\pi^2r_{\rm h}^2}3\frac{k_3}{k_1}\left[1-e^{-2k_1(v-v_{\rm f})}\right]+\order{r-r_{\rm h}}.
	\end{align}
\end{subequations}
There are several things to note here. The first one is that we can actually obtain one additional order in the expansion of $\expval{T_{rv}}$, and two in the expansion of $\expval{T_{vv}}$, from just the terms in \eqref{conformal}, as only deriving with respect to $u$ reduces its order. This will be useful later in our calculation.

The second thing to note is the exponential in $\expval{T_{rr}}$. Such time-dependent coefficients are also present in higher-order terms in the remaining RSET components, and they affect the distance from $r_{\rm h}$ for which the truncated series expansions are a good approximation. In particular, due to the $C$ in the denominator of \eqref{RSET} and the subsequent change of coordinates, they will be exponentials of positive multiples of $(-k_1v)$. This means that for $k_1>0$ (outer horizon) the coefficients in the series quickly tend to constants, while for $k_1<0$ (inner horizon) they grow exponentially. This is the inverse effect on precision we alluded to earlier, which can be understood easily by observing fig. \ref{f2e}. The light rays converge toward the inner horizon $r_{\rm i}$ so that determining the RSET in a region around it requires past information from larger and larger regions, where \eqref{redshift} is no longer valid. On the other hand, the diverging light rays away from the outer horizon imply that the approximation breaks down only when said rays are sufficiently far away from this horizon for \eqref{redshift} to cease being precise, and not due to incoming information. The latter case can be seen as a manifestation of the universality of quantum effects around the external horizon of a BH at late times, analysed in \cite{BLSV11}, while the former shows the reverse being true for the inner horizon, making its long-term semiclassical dynamics more difficult to pin down. Still, the fact that $\expval{T_{rr}}$ evaluated at the inner horizon itself grows exponentially makes it clear that semiclassical effects are not to be ignored there, as they can overcome their Planck-scale suppression very quickly.

The third thing to note is the negative ingoing flux at both horizons, which can be seen directly in the $\expval{T_{vv}}$ component. This term is clearly non-local in curvature and is there solely due to the presence of each horizon, being determined by their respective surface gravities. At the outer horizon it is this negative flux which drives Hawking evaporation, compensating the positive thermal flux at infinity, as discussed originally in \cite{DFU}. At the inner horizon one may therefore expect that this term would lead to a similar, classically forbidden behaviour of the spacetime, such as increasing the radial position of this horizon and reducing the size of the trapped region from the inside. This indeed seems to occur, as we show below.

Finally, it is worth making some remarks regarding the approximation which we use for the RSET \eqref{RSET}. As discussed in the \hyperref[chI]{Introduction}, one of its most obvious drawbacks is the divergence it generally has at $r=0$. If either horizon in our model were close to the origin (in units of the Planck length), we would have to regularise this tensor (see e.g. \cite{Parentani1994,Arrechea2019}) or use a different approximation. However, we can simply restrict our geometric models to ones in which $r_{\rm h}\gg l_{\rm p}$, which, for BHs, is also a requirement for the semiclassical approximation itself to be valid.\footnote{This assumption is also valid for the dynamical cases studied in the next section, perhaps with the exception of the $n<1$ case of eq. \eqref{a1div} in its final stages, which however occurs only after the bouncing effect we are interested in, and well outside the range of validity of the approximation for the RSET that we use.}

As stated above, the reason why we use the Polyakov approximation is twofold. First, there is as yet no method to compute the exact 3+1 dimensional RSET for the ``in" vacuum with such generality; and second, this approximation seems to be enough to capture horizon-related effects, that is, at least when it comes to Hawking evaporation. However, it is easy to see that the terms local in curvature would differ from the exact RSET by just looking at the trace anomalies in 1+1 and 3+1 dimensions. The former is directly proportional to the $\expval{T_{rv}}$ term \eqref{RSETrv}, which depends only on the coefficient $k_2$ from \eqref{redshift}, while the latter, calculated with the expression given in e.g. \cite{BD}, is at zeroth order in $(r-r_{\rm h})$ a completely different function of the coefficients $k_1$, $k_2$ and $k_3$. We therefore do not exclude the possibility that an exact 3+1 dimensional calculation of backreaction may lead to different dynamics. In what follows it is shown, however, that the first perturbations on the position of either horizon are driven by the non-local flux term in $\expval{T_{vv}}$, making the resulting initial dynamics a robust result whenever such a flux is present.

\subsection{Perturbed Einstein equations}

In order to see the dynamical implications of backreaction near the two horizons, we will perturb the metric (while maintaining spherical symmetry) and source the first order perturbation with the RSET. Without loss of generality, the perturbed metric can be written as
\begin{equation}\label{metric-perturbed}
	ds^2=-\left[f(r)+\delta f(v,r)\right]dv^2+2\left[1+\delta g(v,r)\right]dvdr+r^2d\Omega^2.
\end{equation}
Since we have expanded the RSET in powers of $(r-r_{\rm h})$, we must do the same with these perturbations:
\begin{equation}\label{dfg}
	\begin{split}
		\delta f(v,r)&=\delta f_0(v)+\delta f_1(v)(r-r_{\rm h})+\cdots,\\
		\delta g(v,r)&=\delta g_0(v)+\delta g_1(v)(r-r_{\rm h})+\cdots.
	\end{split}
\end{equation}
It is worth noting that since the RSET we are using is fixed entirely by the background, the matter side of the Einstein equations may also require that a perturbation of the classical matter (the stress-energy tensor of which generates the background) be considered. However, with the series expansion which we are using, we have found that at the order needed to determine $\delta f_0$ and $\delta g_0$ the equations are compatible with this additional perturbation being zero, i.e. the classical matter content retaining its background functional form.

We therefore begin by equating the first order in $\delta f$ and $\delta g$ and zeroth order in $(r-r_{\rm h})$ of the Einstein tensor to the RSET \eqref{RSETres} times $8\pi l_{\rm p}^2$, where $l_{\rm p}$ is the Plank length. From the $vv$ component we obtain
\begin{equation}\label{vv}
	(1-2k_1r_{\rm h})\delta f_0(v)-r_{\rm h}\delta f'_0(v)=-\frac{l_{\rm p}^2}{24\pi}k_1^2.
\end{equation}
Let us first analyse the implications of this equation in the case of the external horizon, with $r_{\rm h}=r_{\rm e}$ and $k_1>0$. If we take the familiar Schwarzschild case, where $2k_1r_{\rm e}=1$, with the initial condition $\delta f(v_{\rm f},r)=0$ the equation simply integrates to
\begin{equation}
	\delta f_0=\frac{l_{\rm p}^2k_1^2}{24\pi r_{\rm e}}(v-v_{\rm f})=\frac{l_{\rm p}^2k_1^3}{12\pi}(v-v_{\rm f}).
\end{equation}
Therefore, the redshift function tends to increase around the horizon, leading to what can be identified as the first stages of Hawking evaporation of the trapped region. This can be seen explicitly by noting from the metric \eqref{metric-perturbed} that the modified equation for the trajectory of outgoing null radial geodesics is
\begin{equation}\label{outnullp}
	\frac{dr}{dv}=\frac{1}{2}\frac{f(r)+\delta f(v,r)}{1+\delta g(v,r)}.
\end{equation}
Equating the right-hand side to zero and substituting the series \eqref{dfg}, we see that the first order change in the radial position of the external horizon is
\begin{equation}\label{linevap}
	r_{\rm e}\quad\to\quad r_{\rm e}-\frac{\delta f_0}{2k_1}=r_{\rm e}-\frac{l_{\rm p}^2k_1^2}{24\pi}(v-v_{\rm f}).
\end{equation}
In other words, the Schwarzschild horizon has a tendency to shrink, albeit slowly.

For a non-Schwarzschild horizon (e.g. Reissner-Nordström, Schwarzschild-dS or -AdS, regular BH models, etc.), the general solution of \eqref{vv} is
\begin{equation}\label{df0}
	\delta f_0=\frac{l_{\rm p}^2k_1^2}{24\pi(2k_1r_{\rm e}-1)}\left[1-e^{-\frac{1}{r_{\rm e}}(2k_1r_{\rm e}-1)(v-v_{\rm f})}\right].
\end{equation}
In the limit $2k_1r_{\rm e}\to 1$ we recover the above Schwarzschild case. For other cases we can understand the initial tendencies by expanding this expression around $v=v_{\rm f}$,
\begin{equation}\label{dfexp}
	\delta f_0=\frac{l_{\rm p}^2k_1^2}{24\pi r_{\rm e}}\left[(v-v_{\rm f})-\frac{2k_1r_{\rm e}-1}{2r_{\rm e}}(v-v_{\rm f})^2+\cdots\right].
\end{equation}
For a horizon with a surface gravity greater than that of a Schwarzschild BH of the same size, i.e. for $k_1>1/(2r_{\rm e})$, we see that at linear order in $(v-v_{\rm f})$ the evaporation tends to be quicker while, when the $(v-v_{\rm f})^2$ term becomes important, it slows down. The opposite is true if $k_1<1/(2r_{\rm e})$, that is, the evaporation begins slower than in Schwarzschild but then tends to quicken. Of course, no definite conclusions can be drawn regarding the long-term evolution of the horizon due to the various approximations involved. Particularly, even if the above exponentials are good approximations initially, the fact that for an external horizon the coefficient $2k_1r_{\rm e}-1$ can change sign as the surface gravity and radius evolve makes it likely that the overall evolution has a kind of intermediate behaviour, more akin to the Schwarzschild case.

To complete our picture of what occurs at the outer horizon, we can look at the rest of the perturbed Einstein equations in search for a solution for $\delta g_0$. Combining the $vr$ component at zero order in $(r-r_{\rm e})$ and the $vv$ component at first order [as we mentioned above, the first and second orders of $\expval{T_{vv}}$ can be obtained easily from \eqref{conformal}], we get
\begin{equation}
	\begin{split}
		(2k_1r_{\rm e}-1)\delta g'_0(v)+\frac{1}{r_{\rm e}}&(1-4k_1r_{\rm e})\delta g_0(v)=\\&\frac{1}{r_{\rm e}}(1+2k_1r_{\rm e}+4k_2r_{\rm e}^2)\delta f_0(v)+\frac{l_{\rm p}^2}{6\pi r_{\rm e}}k_2(1-2k_1r_{\rm e})
	\end{split}
\end{equation}
In the Schwarzschild case, where $k_1=1/(2r_{\rm e})$ and $k_2=-1/(2r_{\rm e}^2)$, this reduces to
\begin{equation}
	\delta g_0=0.
\end{equation}
In more general scenarios, with different relations between $r_{\rm e}$ and the coefficients $k_i$, $\delta g_0$ is a non vanishing function of time which, looking at \eqref{outnullp}, causes a slightly faster or slower (depending on its sign) divergence of null geodesics away from the horizon. From the metric \eqref{metric-perturbed} its effect can also be interpreted physically as a contraction or expansion of space in the radial direction, as discussed in chapter \ref{ch3}.

Having recovered the familiar evaporative tendency of the outer horizon, we can now move on to the analysis of backreaction at the inner horizon. The equations are exactly the same, except that for the sake of notation we switch $r_{\rm e}\to r_{\rm i}$ and we have to keep in mind that $k_1$ is now negative. To see how the position of the inner horizon changes we must look at \eqref{df0}, which we can rewrite in a more convenient manner given the sign of $k_1$ as
\begin{equation}\label{df0i}
	\delta f_0=\frac{l_{\rm p}^2k_1^2}{24\pi(1-2k_1r_{\rm i})}\left[e^{\frac{1}{r_{\rm i}}(1-2k_1r_{\rm i})(v-v_{\rm f})}-1\right].
\end{equation}
We see that, much like before, the redshift function tends to increase, and this leads to a reduction of the size of the trapped region,
\begin{equation}
	r_{\rm i}\to r_{\rm i}+\frac{\delta f_0}{2|k_1|}.
\end{equation}
The series expansion around $v=v_{\rm f}$ is the same as \eqref{dfexp}, meaning that if the absolute value of the surface gravity of the inner horizon is greater than that of the outer horizon (which is usually the case, except in near-extremal configurations), the initial tendency is for the trapped region to begin evaporating more quickly from the inside than from the outside. Additionally, we note that the coefficient multiplying $v$ in the exponential in $\delta f_0$ is positive for any inner horizon, implying the possibility that the exponential behaviour for the evaporation may be a more general characteristic which is maintained beyond the initial tendency. We will analyse this more closely for a particular family of dynamical solutions in the following.

\section{Dynamical horizons and self-consistent solutions}\label{dynamic}

We have seen how the perturbations on the metric caused by the RSET behave around the static inner and outer horizons of a BH. But either due to dynamics in the classical sector, or to semiclassical backreaction itself, the position of these horizons is generally not static. Calculating the RSET on a generic dynamical background can be challenging even in the Polyakov approximation, but we can work around this difficulty in a manner similar to what we employed in the previous section. We will expand the redshift function $f(v,r)$ in a series around a dynamical horizon $r_{\rm h}(v)$ and calculate the RSET in terms of the time-dependent coefficients of this series. This will allow us to again obtain the deviations in the metric $\delta f(v,r)$ and $\delta g(v,r)$ around each horizon. Doing so without completely specifying the dynamics of the background will then allow us to arrive at approximate self-consistent solutions in some particularly simple cases.

\subsection{RSET around dynamical horizons}

To maintain the simplicity of the initial conditions for the quantum modes we had in the previous section, we will once again consider a spacetime model which is flat up to a time $v_{\rm f}$ and then transitions abruptly to a BH geometry. The relation between the parameter $d_{\rm f}$ and the ``in" coordinate $u$, identified with $v_0$, is therefore still given by \eqref{df}, with $r_{\rm h}$ in this case being $r_{\rm h}(v_{\rm f})$.

The BH geometries which we will use for a background are of the form
\begin{equation}
	ds^2=-f(v,r)dv^2+2dvdr+r^2d\Omega^2.
\end{equation}
We note that considering both degrees of freedom of the spherically-symmetric geometry yields an equally straightforward calculation for the RSET, the main complication arising at the stage of resolution of the perturbed Einstein equations. We will use the above form because the simplified cases in which we will be able to obtain approximate self-consistent dynamics also retain it. The expansion we will use for the redshift function is completely analogous to the one in the static case,
\begin{equation}
	f(v,r)=2k_1(v)[r-r_{\rm h}(v)]+2k_2(v)[r-r_{\rm h}(v)]^2+2k_3(v)[r-r_{\rm h}(v)]^3+\cdots,
\end{equation}
where now $k_1,k_2,\dots$ are functions of $v$. Assuming that the quantity $r-r_{\rm h}(v)$ is small, the trajectories of outgoing null radial geodesics can be expanded as
\begin{equation}
	r(v)=r_{\rm h}(v)+p_1(v)+d_{\rm f}e^{\tilde{k}_1(v)}+p_2(v,d_{\rm f})+p_3(v,d_{\rm f})+\cdots,
\end{equation}
where
\begin{align}
	\tilde{k}_1(v)&=\int_{v_{\rm f}}^vk_1(\tilde{v})d\tilde{v},\\
	p_1(v)&=-e^{\tilde{k}_1(v)}\int_{v_{\rm f}}^ve^{-\tilde{k}_1(\tilde{v})}r'_{\rm h}(\tilde{v})d\tilde{v}, \label{p1}\\
	p_2(v,d_{\rm f})&=e^{\tilde{k}_1(v)}\int_{v_{\rm f}}^ve^{-\tilde{k}_1(\tilde{v})}k_2(\tilde{v})\left[p_1(\tilde{v})+d_{\rm f}e^{\tilde{k}_1(\tilde{v})}\right]^2d\tilde{v},\\
	\begin{split}
		p_3(v,d_{\rm f})&=e^{\tilde{k}_1(v)}\int_{v_{\rm f}}^ve^{-\tilde{k}_1(\tilde{v})}\left\{2k_2(\tilde{v})p_2(\tilde{v})\left[p_1(\tilde{v})+d_{\rm f}e^{\tilde{k}_1(\tilde{v})}\right]\right.\\&\hspace{3.6cm}\left.+k_3(\tilde{v})\left[p_1(\tilde{v})+d_{\rm f}e^{\tilde{k}_1(\tilde{v})}\right]^3\right\}d\tilde{v},
	\end{split}
\end{align}
with $r'_{\rm h}(v)=dr_{\rm h}/dv$. Unlike in the static case, this expansion is performed around the first order solution for the separation from the horizon, $r_1(v)\equiv\left[p_1(\tilde{v})+d_{\rm f}e^{\tilde{k}_1(\tilde{v})}\right]$ (in units of the scale of each horizon), making it progressively worse with time no matter how small the initial parameter $d_{\rm f}$ is. In particular, looking at the expression for $p_1$, the larger the rate of change of the horizon position $r'_{\rm h}(v)$ is, the quicker the approximation breaks down. However, this can be delayed if the coefficients $k_i(v)$ with $i\ge 2$ remain small enough compared to powers of $k_1(v)$, as we will consider in one of our simplified models below.\par
With these expressions we can obtain a generalisation of \eqref{RSETres} for dynamical backgrounds, valid for a small but finite time interval. This RSET at zero order in $r_1(v)$ is
\begin{subequations}\label{RSETdyn}
	\begin{align}
		\expval{T_{vv}}&\simeq-\frac{1}{96\pi^2r_{\rm h}(v)^2}\left[\frac{1}{2}k_1(v)^2-k'_1(v)\right],\label{Tvv} \\
		\expval{T_{rv}}&\simeq-\frac{1}{96\pi^2r_{\rm h}(v)^2}2k_2(v),\\
		\begin{split}
			\expval{T_{rr}}&\simeq\frac{1}{96\pi^2r_{\rm h}(v)^2}e^{-2\tilde{k}_1(v)}\left\{\int_{v_{\rm f}}^v\left[8k_2(\tilde{v})e^{\tilde{k}_1(\tilde{v})}\int_{v_{\rm f}}^{\tilde{v}}k_2(\bar{v})e^{\tilde{k}_1(\bar{v})}d\bar{v}+6k_3(\tilde{v})e^{2\tilde{k}_1(\tilde{v})}\right]d\tilde{v}\right.\\ &\hspace{4cm}\left. -6\left[\int_{v_{\rm f}}^vk_2(\tilde{v})e^{\tilde{k}_1(\tilde{v})}d\tilde{v}\right]^2\right\},
		\end{split}
	\end{align}
\end{subequations}
from which we can easily recover \eqref{RSETres} in the static limit (paying attention to the integration limits). As a side note, it is interesting to see that the expression for $\expval{T_{vv}}$ has the same structure as the one found in \cite{BBGJ16} and in \eqref{27} for the difference between the values of the RSET in two different vacuum states, when expressed in terms of the effective temperature function (except for a sign difference between the ETF and $k_1$). This suggests that the flux present in this term is of non-local nature, stemming purely from the light-ray peeling or accumulation around a horizon, and not from curvature.

\subsection{Perturbed equations and Hawking evaporation}

With an expansion analogous to \eqref{dfg} for the metric perturbations, the generalisation of eq. \eqref{vv} is now
\begin{equation}
	\left[1-2r_{\rm h}(v)k_1(v)\right]\delta f_0(v)-r_{\rm h}(v)(\delta f')_0(v)-2k_1r_{\rm h}(v)r'_{\rm h}(v)\delta g_0(v)=-\frac{l_{\rm p}^2}{24\pi}\left[k_1^2(v)-2k'_1(v)\right],
\end{equation}
where $(\delta f')_0=\delta f'_0-\delta f_1r'_{\rm h}$ is the zeroth order of the derivative of $\delta f$ with respect to $v$ (while $\delta f'_0$ is the derivative of the zeroth order; the two only coincide for static backgrounds). We see that the simple decoupling we had for $\delta f_0$ in the static case is not present in general, unless the rate of change of the background is small enough for $r'_{\rm h}(v)\ll k_1(v)r_{\rm h}(v)$ to be satisfied. If the dynamics are induced only by backreaction itself, then $r'_{\rm h}$ and $k'_1$ are initially of order $l_{\rm p}^2$ (in dimensions of the horizon scale), and so are $\delta f$ and $\delta g$, implying that the approximation
\begin{equation}\label{vv-dyn}
	\left[1-2r_{\rm h}(v)k_1(v)\right]\delta f_0(v)-r_{\rm h}(v)\delta f'_0(v)\simeq -\frac{l_{\rm p}^2}{24\pi}k_1^2(v),
\end{equation}
is reasonable. $r'_{\rm h}$ being small also implies that the series expansion itself is accurate for a longer period of time, as can be seem from \eqref{p1}.

Let us once again start by looking at the evolution of an initially static external Schwarzschild horizon. We have the initial condition $2r_{\rm e}(v_{\rm f})k_1(v_{\rm f})=1$, and possible deviations from this equation at later times $v$, once multiplied by $\delta f_0$, are of the same order as the terms we have neglected above, allowing us to now neglect the first term. We also have the relation $\delta f'_0 \simeq -r'_{\rm e}/r_{\rm e}$ for the displacement of the Schwarzschild horizon. Substituting into this equation we get
\begin{equation}\label{hawk}
	r'_{\rm e}\simeq-\frac{l_{\rm p}^2}{96\pi}\frac{1}{r_{\rm e}^2}.
\end{equation}
The solution of this approximate self-consistent equation is simply
\begin{equation}\label{schw}
	r_{\rm e}(v)\simeq \left[r_{\rm e}(v_{\rm f})^3-\frac{l_{\rm p}^2}{32\pi}(v-v_{\rm f})\right].
\end{equation}
To check that we are on the right track, we can see that expanding this solution around $v_{\rm f}$ recovers the linear tendency form \eqref{linevap}. Additionally, we can extrapolate this solution to later times, assuming that \eqref{schw} remains approximately valid at the later stages of evaporation, and obtain the Hawking evaporation time $v_{\rm H}$, defined by $r_{\rm e}(v_{\rm H})=0$,
\begin{equation}\label{hawk-evap}
	v_{\rm H}\simeq 256\pi \frac{M^3}{l_{\rm p}^2}\simeq \left(\frac{M}{M_\odot}\right)^3 10^{73}\,\text{s},
\end{equation}
where $M=r_{\rm e}(v_{\rm f})/2$ is the initial BH mass and $M_\odot$ is the solar mass. The approximations involved in obtaining this extrapolation are equivalent to the quasi-stationary approximation used originally by Hawking to estimate the lifetime of BHs \cite{Hawking1974}. The fact that the Polyakov approximation to the RSET can capture this phenomenon bodes well for the validity of our following calculations.

In other scenarios where the quasi-stationary assumption is valid for a period of time, we can still use eq. \eqref{vv-dyn} to get a first glimpse at self-consistent solutions. We can integrate for $\delta f_0$, obtaining
\begin{equation}\label{df0dyn}
	\delta f_0(v)=e^{\int_{v_{\rm f}}^v[1/r_{\rm h}(\tilde{v})-2k_1(\tilde{v})]d\tilde{v}}\int_{v_{\rm f}}^ve^{-\int_{v_{\rm f}}^{\tilde{v}}[1/r_{\rm h}(\bar{v})-2k_1(\bar{v})]d\bar{v}}\frac{l_{\rm p}^2k_1(\tilde{v})^2}{24\pi r_{\rm h}(\tilde{v})}d\tilde{v}.
\end{equation}
Taking the right-hand side as a function of a slowly-evolving classical background would just give a generalisation of \eqref{dfexp}. Modifying the right-hand side with the backreaction due to $\delta f$ and $\delta g$ would give a more accurate expression of backreaction in a quasi-stationary approximation, the validity of which would have to be checked along the evolution in each case. It is interesting to note that for an inner horizon, where $k_1(v)<0$, unless $k_1(v)$ quickly tends to zero, we once again have the growing exponential we had in \eqref{df0i}, now somewhat hidden in the term given by the lower bound of the middle integral (the one which is outside the exponentials). To see how this exponential behaviour manifests itself in an approximate self-consistent solution, we will now look at a specific set of backgrounds for which we can solve this equation.

\subsection{Inner horizon expansion for a simple background}

Let us consider for the BH region a redshift function which around the inner horizon has the particular form
\begin{equation}\label{bg}
	f(v,r)=1-\frac{\lambda(v)}{2}r-\frac{\alpha(v)}{r},
\end{equation}
with $\lambda(v)$ a positive function and $\alpha(v)$ a function which satisfies the initial condition $\alpha(v_{\rm f})=0$. While $2\lambda\alpha<1$, the position and surface gravity of the inner horizon within this geometry are
\begin{align}
	r_{\rm i}&=\frac{1}{\lambda}(1+\sqrt{1-2\lambda\alpha})=\frac{2}{\lambda}-\alpha-\frac{1}{2}\alpha^2\lambda+\cdots, \label{ri}\\
	k_1&=\frac{1-2\lambda\alpha-\sqrt{1-2\lambda\alpha}}{4\alpha}=-\frac{1}{4}\lambda+\frac{1}{8}\alpha\lambda^2+\cdots, \label{k1dyn}
\end{align}
where the series expansions on the right-hand side are valid while $|\alpha\lambda|\ll 1$.

For $\alpha=0$, the only non-zero term of the RSET on this background would be $\expval{T_{vv}}$ from \eqref{RSETdyn}, and the local approximation for null geodesics involved in obtaining it would actually be exact (i.e. higher order terms in the expansion would be zero) in a finite region around the inner horizon, akin to the left shaded region in fig. \ref{f2e}. There, we would thus have
\begin{equation}\label{Tvvin}
	\expval{T_{vv}}=-\frac{1}{96\pi^2r^2}\left[\frac{1}{2}k_1(v)^2-k'_1(v)\right].
\end{equation}
All this remains approximately true while $\alpha$ is small compared to $\lambda$ in units of $r_{\rm i}$, and we will use this fact to simplify the dynamical perturbation equations in order to obtain an approximate self-consistent solution.

The Einstein tensor of this geometry is
\begin{align}
	G_{vv}&=\frac{\lambda(v)}{r}-\lambda(v)^2+\frac{\lambda'(v)}{2}-\lambda(v)\frac{\alpha(v)}{r^2}+\frac{\alpha'(v)}{r^2},\\
	G_{vr}&=-\frac{\lambda(v)}{r},\\
	G_{\theta\theta}&=\frac{G_{\varphi\varphi}}{{\rm sin}^2\theta}=-\frac{r\lambda(v)}{2}.
\end{align}
We see that $\alpha(v)$ appears only in two terms of the $vv$ component and is divided by $r^2$. We can consider these two terms as the ones sourced by the RSET \eqref{Tvvin}, the rest corresponding to the classical background, which fixes the function $\lambda(v)$ (keep in mind that the classical background is just a toy model used to construct the causal structure we are interested in). Then $\alpha(v)$ becomes the semiclassical perturbation satisfying the equation
\begin{equation}\label{alpha}
	\alpha'(v)-\lambda(v)\alpha(v)=-\frac{l_{\rm p}^2}{24\pi}\left[k_1(v)^2-2k'_1(v)\right].
\end{equation}
Integrating for $\alpha$ we obtain
\begin{equation}\label{alphadyn}
	\alpha(v)=-e^{\int_{v_{\rm f}}^v\lambda(\tilde{v})d\tilde{v}}\int_{v_{\rm f}}^ve^{-\int_{v_{\rm f}}^{\tilde{v}}\lambda(\bar{v})d\bar{v}}\frac{l_{\rm p}^2}{24\pi}\left[k_1(\tilde{v})^2-2k'_1(\tilde{v})\right]d\tilde{v}.
\end{equation}
If we take a background with $\lambda=\text{const.}$ and substitute $k_1$ for its zero-order value from the expansion \eqref{k1dyn}, the integration yields
\begin{equation}\label{alphastat}
	\alpha(v)=-\frac{l_{\rm p}^2\lambda}{192\pi}\left[e^{\lambda(v-v_{\rm f})}-1\right].
\end{equation}
This solution is valid until the initially zero $\alpha$ becomes comparable to $\lambda$ (in units of $r_{\rm i}$), which, as \eqref{df0i} and \eqref{df0dyn} suggested, does not take long due to the growing exponential. Introduced into the geometry, this term behaves like a negative mass, tending to move the inner horizon outward, as can be seen from \eqref{ri}. The exponential growth of this negative mass and the displacement of the inner horizon can be seen as a semiclassical manifestation of the inner horizon instability, with the opposite effect to its classical counterpart.

Given this intriguing tendency to evaporate the trapped region from the inside, it is only natural to ask oneself what may happen if the same behaviour were to continue throughout the evolution of the inner horizon, up until the disappearance of the trapped region. In other words, what would the result be if the driving force of evaporation continued to be the local $\expval{T_{vv}}$ term on the right-hand side of \eqref{alpha}. Although this assumption is less justifiable dynamically than the analogous one used for Hawking evaporation in eq.~\eqref{hawk}, one may think of it as just an extrapolation from the initial tendency. If nothing else, it serves as an example of how the dynamics of the inner horizon can continue with an RSET which continues violating the energy positivity conditions, as it seems likely to do around a horizon, making the geometry evolve in a classically forbidden manner.

To answer this question, we can take into account the change in surface gravity due to the evolution of $\alpha$ on the right-hand side of \eqref{alpha} through \eqref{k1dyn}. Writing $\alpha$ in terms of $k_1$ as
\begin{equation}
	\alpha=\frac{1}{2}\frac{4k_1+\lambda}{(2k_1+\lambda)^2},
\end{equation}
equation \eqref{alpha} becomes
\begin{equation}\label{sc}
	\left[\frac{4k_1}{(\lambda+2k_1)^3}+\frac{l_{\rm p}^2}{12\pi}\right]k'_1+\frac{\lambda'}{2}\frac{\lambda+6k_1}{(\lambda+2k_1)^3}+\frac{\lambda}{2}\frac{\lambda+4k_1}{(\lambda+2k_1)^2}-\frac{l_{\rm p}^2}{24\pi}k_1^2=0.
\end{equation}
This equation contains as solutions the initial behaviours in the backreaction problem given by \eqref{alphadyn} and \eqref{alphastat}, along with their extensions. More generally, it governs the evolution of a geometry whose dynamics is modified by a (generally negative) ingoing flux of energy determined by the surface gravity at its inner apparent horizon through \eqref{Tvvin}.

\begin{figure}
	\centering
	\includegraphics[scale=.45]{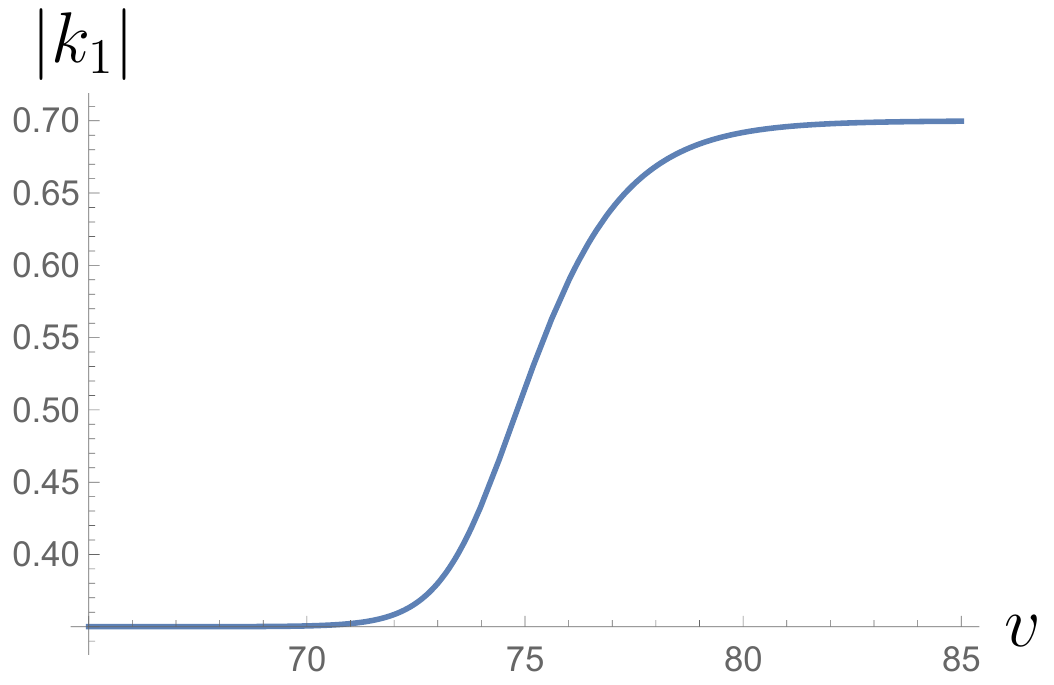}
	\includegraphics[scale=.45]{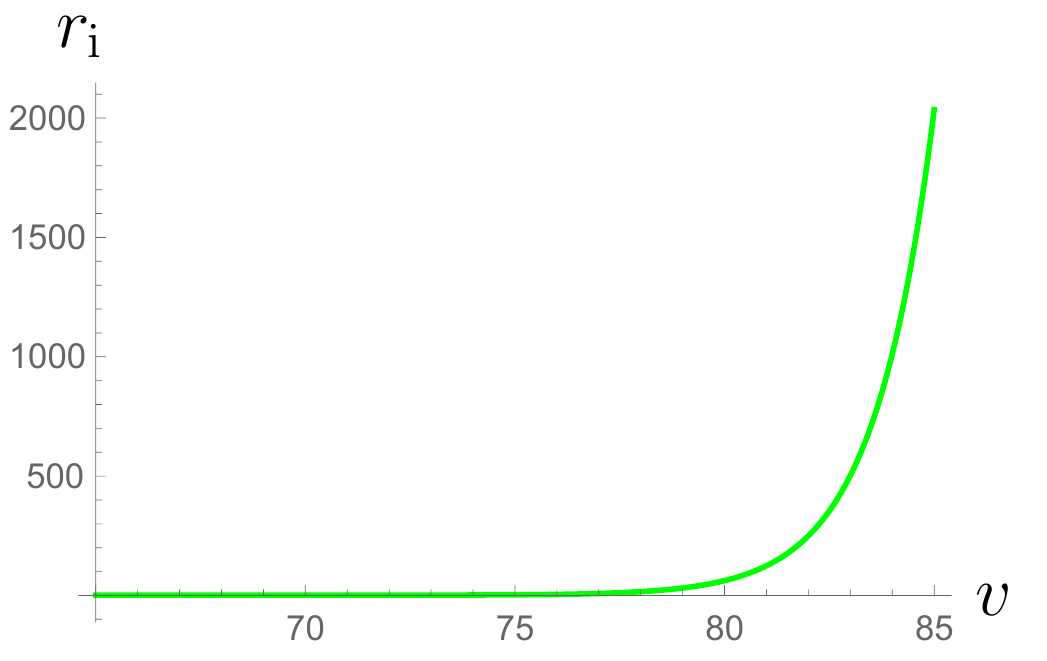}
	\includegraphics[scale=.45]{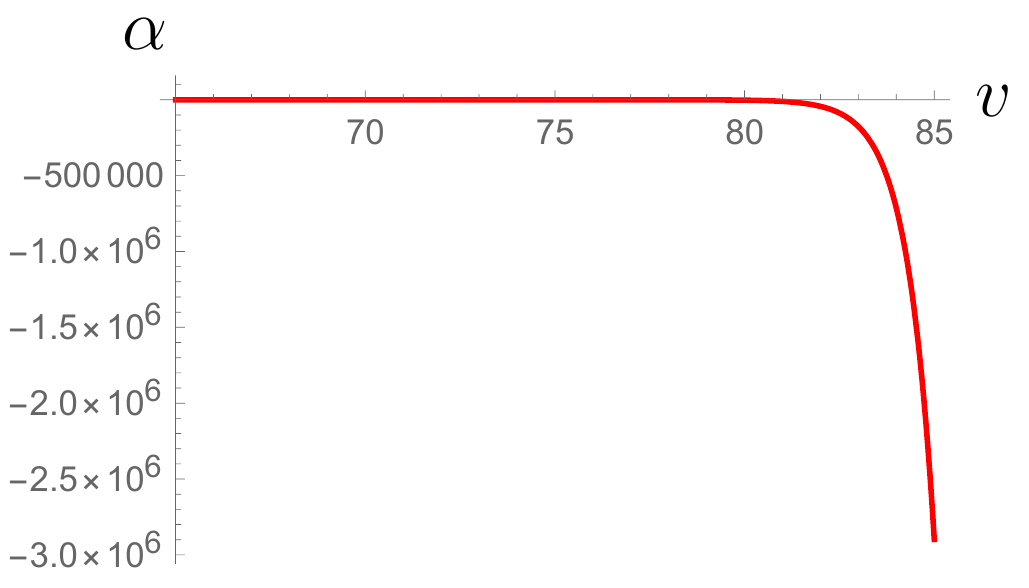}
	\caption{Plots of $|k_1(v)|$, $r_{\rm i}(v)$ and $\alpha(v)$, from left to right, for the evaporation of the inner horizon. We have taken $\lambda=0.7$ and $l_{\rm p}=10^{-8}$ (as a large difference in scales is required but smaller values of $l_{\rm p}$ make numerical evaluation more difficult and lead to no qualitative changes).}
	\label{inner-evap}
\end{figure}

Taking $\lambda$ as a positive constant, all solutions of \eqref{sc} have the same behaviour, shown in fig. \ref{inner-evap}: the decrease in $\alpha$ (also observed perturbatively) initially increases the absolute value of $k_1$ but then makes it tend to a constant (with a value $\lambda/2$). This surface gravity then continues to feed the right-hand side of \eqref{alpha}, extending the exponential behaviour of $\alpha$ indefinitely. The radial position of the inner horizon also continues to increase exponentially.

In summary, the geometry \eqref{bg} fed by the flux \eqref{Tvvin} has an inner horizon which moves outward exponentially quickly. This goes on indefinitely due to the global structure of \eqref{bg}, which does not contain an outer bound for the trapped region (being more akin to an inflationary universe than to a BH). In a more realistic scenario, even if backreaction continues to be governed by a term like \eqref{Tvvin}, we expect such dynamics to end when the trapped region disappears. The main conclusion we can extract from this is that incorporating the modifications of the geometry due to backreaction on the right-hand side of \eqref{alpha} does not tend to decrease the rate of the initially exponential evaporation of the inner horizon.

If we assume that such a behaviour is the dominant factor in the elimination of a trapped region, we can estimate a revised evaporation time for BHs with an inner horizon. Considering a BH of mass $M$, with an outer horizon $r_{\rm e}\sim M$ which evaporates slowly \textit{à la} Hawking, and an inner horizon with an initial position $r_{\rm i,0}$ and surface gravity $k_{1,0}$, the inner horizon would meet the outer one after a time
\begin{equation}
	v_{\rm evap}\simeq \frac{1}{k_{1,0}+(2r_{\rm i,0})^{-1}}\log\frac{M}{l_{\rm p}}\lesssim\frac{M}{M_\odot}\times 10^{-5}\,\text{s},
\end{equation}
where $M_\odot$ is the solar mass, and we have obtained the upper bound on the right-hand side by assuming that the surface gravity at the inner horizon is initially greater than that of the outer horizon, the latter of which we take to be of the order $1/M$ (the logarithmic dependence has been omitted in this bound as for no astrophysically reasonable object would it increase the order of magnitude further). Needless to say, this process is much quicker than the time it would take for a Schwarzschild BH to evaporate from the outside, given by \eqref{hawk-evap}.

Therefore, using the word ``evaporation" to describe the leading effects of semiclassical backreaction on the inner horizon may not be adequate, as the exponential behaviour of its outwards displacement may be better described as an ``inflation" process (akin to the classical mass inflation described in the previous chapter). We therefore dub this phenomenon \textit{inner horizon inflation}.

\subsection{Collapsing matter: singularity or bounce}

So far our results in this section have been a direct generalisation of the perturbation analysis in the previous one. But the treatment on dynamical backgrounds and self-consistent extrapolation allow for a wider range of solutions to be analysed, in particular ones in which the classical backgrounds itself is dynamical. We expect the backreaction of a moving inner horizon to have a similar effect as observed for the initially static background: to push it outward and try to diminish the size of the trapped region. Whether this tendency from backreaction can overcome its Planck-scale suppression and dominate over the dynamics of the classical background is what we will analyse here.

What we will look at is the backreaction problem around the dynamical inner horizon of a gravitational collapse which would classically end in a Schwarzschild-like BH. We construct a geometry around this horizon of the type \eqref{bg} with $\alpha=0$ (classically) and
\begin{equation}\label{a1div}
	\lambda=\frac{\lambda_0^{1-n}}{(v_{\rm s}-v)^n},
\end{equation}
with $v\in(0,v_{\rm s})$, $n>0$ and $\lambda_0$ a constant (with the same dimensions as $\lambda$) which defines the characteristic length scale of the problem. Matching this with a Minkowski region through an ingoing null shell at $v=0$, we get a picture of a collapse in which the inner horizon initially travelled inward at light-speed but then slowed down before continuing to the centre. This is once again a method of simplifying the initial conditions for the quantum modes which enter the BH region by removing their dependence on the details of the collapse in the far past, thus focusing only on the effects caused by these modes entering the vicinity of the inner horizon at the final stages of the collapse. This also goes hand in hand with our approximation \eqref{alpha}.

Introducing these backgrounds into eq. \eqref{alphastat}, we can analyse whether horizon-related semiclassical effects can become relevant to the overall dynamics. What we find is that there are two ways $\alpha$ can become large enough for this to occur. First, if the integral of $\lambda$ diverges, which is the case for $n\ge 1$, then $\alpha$ always diverges as the exponential of this integral, making it clearly dominant over the classical background. Second, regardless of whether the integral of $\lambda$ diverges or not, if the interval $(0,v_{\rm s})$ is large enough, i.e. if the background dynamics is slow enough for a long period of time, then an effect similar to what occurred with a static background may dominate. Then the exponential of the integral of $\lambda$ becomes large enough to overcome the Planck scale suppression, even though it may not tend to a divergence.

\begin{figure}
	\centering
	\includegraphics[scale=.5]{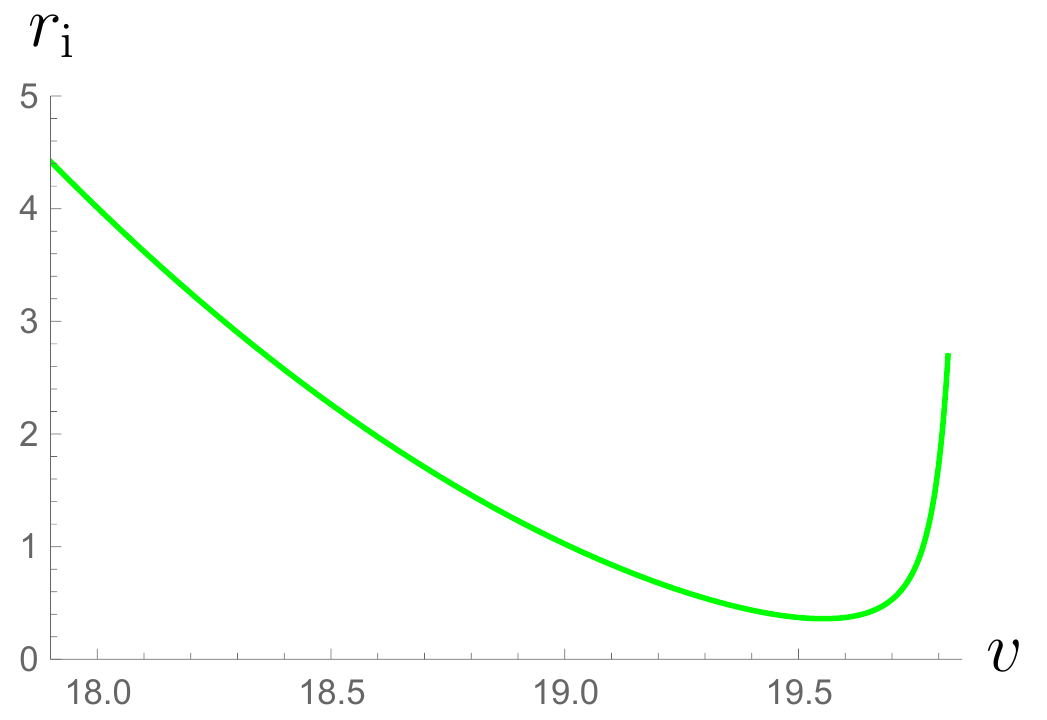}
	\includegraphics[scale=.5]{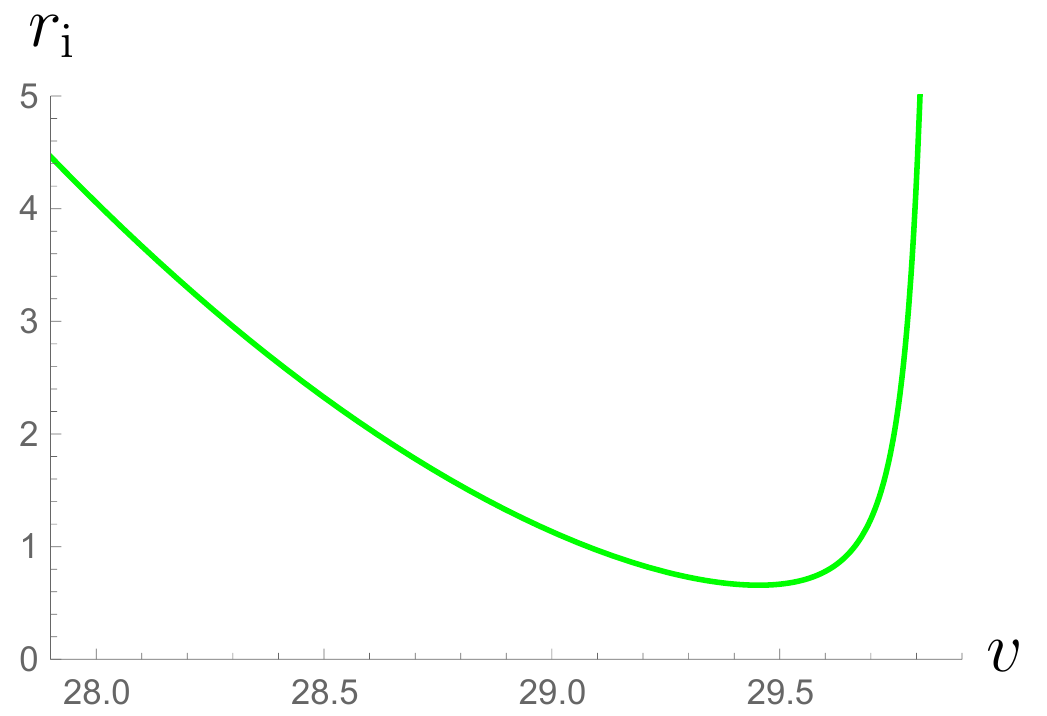}
	\caption{Plots of $r_{\rm i}(v)$ for a background given by \eqref{a1div} with $\lambda_0=1$ and $n=2$, with $v_{\rm s}=20$ on the left and with $v_{\rm s}=30$ on the right (starting from zero perturbation at $v=0$). Units are once again given by $l_{\rm p}=10^{-8}$, for the same computational and qualitative reasons as above.}
	\label{bounce2}
\end{figure}

\begin{figure}
	\centering
	\includegraphics[scale=.5]{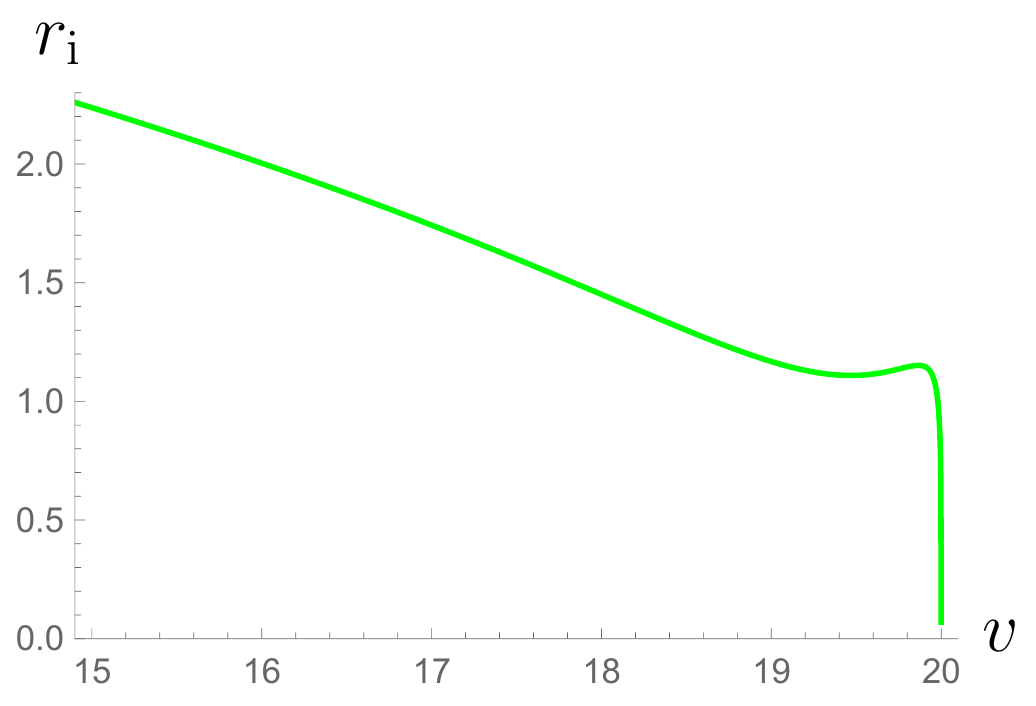}
	\includegraphics[scale=.5]{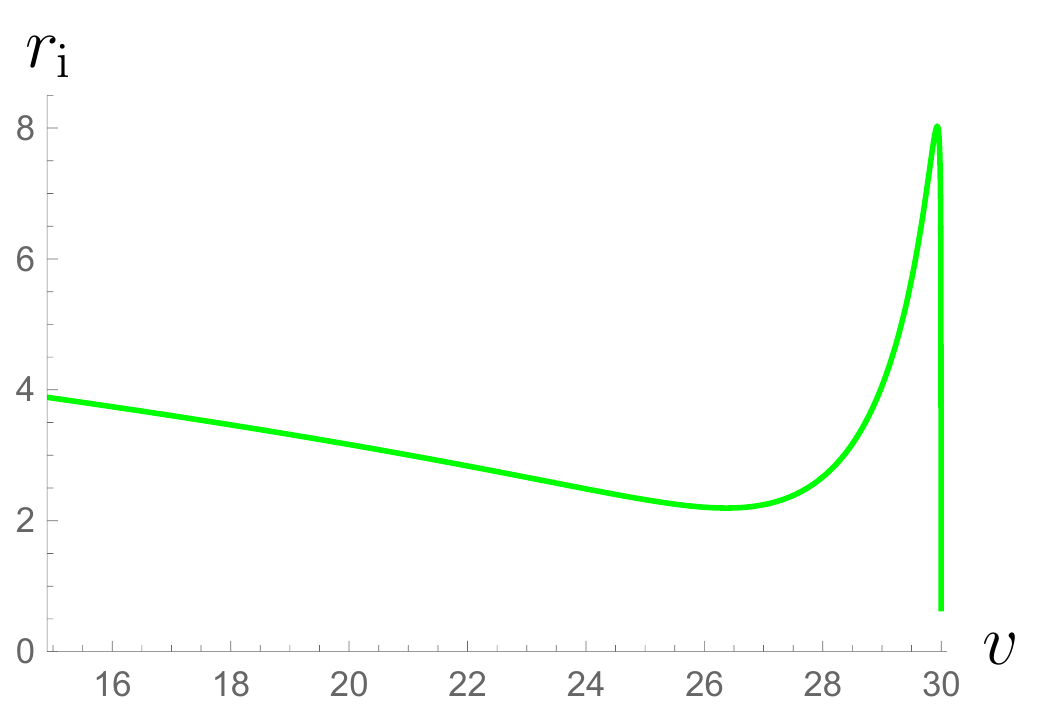}
	\caption{Plots of $r_{\rm i}(v)$ for a background given by \eqref{a1div} with $\lambda_0=1$ and $n=1/2$, with $v_{\rm s}=20$ on the left and with $v_{\rm s}=30$ on the right (starting from zero perturbation at $v=0$). Same units as above.}
	\label{bounce12}
\end{figure}

Indeed, if we integrate \eqref{sc}, which contains these initial tendencies along with their extrapolation to the regime in which semiclassical effects dominate, we get two different types of solutions:
\begin{enumerate}
	\item For $n\ge1$ the semiclassical backreaction always ends up overcoming the contribution of the classical background, resulting in a bounce in the position of the inner horizon, as shown in fig. \ref{bounce2}. We note that the final stages of the Oppenheimer-Snyder collapse correspond to a value $n=1$, as shown in \eqref{OSk}.
	\item For $n<1$, semiclassical backreaction can accumulate and lead to an initial bounce for large enough values of $v_{\rm s}$, as can be seen in fig. \ref{bounce12}. Such a bounce indicates that the collapsing behaviour of the classical matter has been temporarily counteracted, and may subsequently be inverted, making the trapped region disappear completely. However, in our extrapolation, due to the fact that the trapped region is not bounded from above, semiclassical effects eventually lose out and a singularity forms.
\end{enumerate}

We note that the time at which the bounce occurs depends on when this exponential overcomes the Planck length suppression, which generally occurs well before either the surface gravity or the radial position of the inner horizon get close to the Planck scale, as can be checked by inspecting figures \ref{bounce2} and \ref{bounce12}. On the one hand, this allows us to see that we have not obtained an unnaturally large result due to using the Polyakov approximation for the RSET [this already being obvious from the origin of the exponential in e.g. \eqref{alphastat}, which grows even when this horizon is far away from $r=0$]. On the other hand, it is an indication that such dynamics could be accurately described with a semiclassical treatment.

In summary, horizon-related semiclassical effects during gravitational collapse can only be avoided in this model if the classical trajectory of the inner horizon goes to zero quickly (small $v_{\rm s}$, not giving the exponential time to grow) and with a sharp peak at the end in $(v,r)$ coordinates ($n<1$, making the integral of $\lambda$ convergent). Otherwise, in a regime where semiclassical effects are dominated by a term like \eqref{Tvvin}, the collapse will tend to a halt, followed by a quick extinction of the trapped region from the inside, i.e. an inner horizon inflation process.

However, we remind the reader that  the accuracy of our approximation for the RSET \eqref{Tvvin} can break down when $\alpha$ becomes comparable to $\lambda$ at the scale of $r_{\rm i}$. Furthermore, in these dynamical scenarios it may become inaccurate even sooner if the inner horizon reaches a region sufficiently close to the origin to cross paths with light rays which have explored the core of the forming BH, i.e. when it steps out of the left shaded region in fig. \ref{f2e}. Then the precise structure of this core must be specified in order to calculate the RSET. Therefore, although this behaviour is the natural extension of our approximation, we cannot claim with certainty that it represents the complete semiclassical dynamics of gravitational collapse. However, it is a very suggestive possibility.

\section{Conclusions}

In this chapter we produce a bare-bones picture of semiclassical backreaction on black-hole spacetimes which have an inner horizon in addition to an outer one. We construct a simple toy model of a spherically-symmetric geometry in which a regular BH forms, and look at the perturbations caused by the RSET around both the inner and outer apparent horizons. We find that treating these perturbations locally yields analytical results, and we obtain a clear picture of the initial tendencies of this double-horizon structure to evaporate. For the RSET we use a massless scalar field and apply the Polyakov approximation.

At the external horizon, the RSET provides an ingoing flux of negative energy (in accordance with the results of \cite{DFU}). Backreaction from this flux generates a small perturbation which tends to evaporate the horizon. For a Schwarzschild geometry this perturbation initially grows proportionally to the advanced Eddington-Finkelstein time coordinate $v$, with a suppression by a Planck constant, i.e. a slow evaporation. For a background which is not a vacuum spacetime (e.g. Schwarzschild-dS, Schwarzschild-AdS, regular BHs), the modified relation between the radial position of the horizon and its surface gravity results in a somewhat different behaviour for the perturbation: if the surface gravity is larger than it would be for a Schwarzschild BH of the same size, the evaporation is initially quicker than in Schwarzschild, but then has a tendency to slow down, and vice versa if the surface gravity is smaller.

At the inner horizon, the RSET again gives us a negative ingoing flux. The backreaction in this case again results in a reduction of the size of the trapped region, i.e. the inner horizon moves outward. Most importantly, this movement has an overall initial tendency to be much quicker that the evaporation of the outer horizon. This calculation of first order perturbations, if taken as indicative of the qualitative nature of the long-term evolution, strongly suggests a revised picture for evaporation: instead of the outer horizon slowly moving in and eventually revealing the core of the BH, if an inner horizon is present, the trapped region may evaporate more quickly from the inside out. For regular BHs, this coincides with the picture described in \cite{Carballo-Rubio2019,Carballo-Rubio2019b}, which was motivated heuristically by the existence of mass inflation due to classical perturbations \cite{Carballo-Rubio2018b}, although without an explicit discussion of the associated backreaction. Our results here show that the backreaction from semiclassical effects contains the seeds that may lead to a realization of this kind of picture.

In light of these results we extend our background geometries to include dynamical horizons. On the one hand, we do so in order to obtain a better approximation to the complete self-consistent semiclassical solutions which start from a static background. On the other, we are also interested in the backreaction around the dynamical inner horizon in models of spherical BH formation (e.g. Oppenheimer-Snyder collapse \cite{Oppenheimer1939}), where the trapped region first appears close to the eventual outer horizon, and its inner bound quickly moves inward, tending toward the origin (and the formation of a singularity).

Through analysing these additional geometries we indeed obtain approximations for the self-consistent solutions in both static and dynamical backgrounds. Though the range of validity of these approximations is limited, they at least show us the initial tendency of the evolution quite clearly. We find that the semiclassical tendency to inflate the inner horizon remains even in dynamical backgrounds, though whether this can significantly affect the evolution of the geometry varies on a case-by-case basis. Most notably, we find that in many cases in which the background dynamics would make the inner horizon reach the origin (Oppenheimer-Snyder-type collapse), there is a tendency for semiclassical effects to become dominant before this occurs and bounce the horizon back outward. Although the bounce itself occurs in most cases outside the range of validity of the approximation we use for calculating the RSET, the way this result depends on the divergent tendency of the surface gravity is very suggestive of it being a generic property of geometries of this type.

Given that all astrophysical BHs are expected to have an inner horizon, our results at the very least indicate that horizon-related semiclassical effects should never be overlooked when analysing the formation and evolution of these objects.


\chapter{Interlude: blueshift instabilities in the horizon structure of warp drive spacetimes}\label{ch6}

\fancyhead[R]{}
\fancyhead[L]{Part II -- \chaptername\ \thechapter: Blueshift instabilities in warp drive spacetimes}

Before we proceed with our analysis of BH evolution, we will briefly look at a different type of geometry, one whose stability has been questioned on the same grounds as that of a BH with an inner horizon: the Alcubierre warp drive~\cite{Alcubierre1994}. Let us first begin with some introductory remarks regarding this spacetime.

Unlike BHs, the motivation behind its construction is purely geometrical, curving the geometry and bending the lightcones locally in order to produce what appears to be faster-than-light travel for faraway observers. While it brings the idea of superluminal interstellar travel, usually reserved to science fiction, to the realm of physics, its construction in practice requires the creation and manipulation of large quantities of exotic matter~\cite{anti-warp1,Alcubierre2017}. In other words, the stress-energy tensor which generates the warp drive solution of the Einstein equations violates every local energy positivity condition~\cite{Hawking1973}. In fact, this is a manifestation of an even more general restriction: any asymptotically flat configuration which gives rise to apparent superluminal travel seems to require exotic matter~\cite{Olum1998}.

As of yet, there is no experimental evidence of the existence of exotic matter capable of such spacetime distortions. In fact, attempts have been made to provide a geometric interpretation for the absence of gravitating exotic matter through the addition of an underlying causal structure which limits how classical spacetimes can curve~\cite{VisserBassett}. Alternatively, but in the same line of reasoning, this underlying causal structure could be less rigid and allow certain types of emergent warp-drive configurations, though never ones which produce closed timelike curves~\cite{BarceloGerardo2022}. Other arguments suggest that exotic matter cannot be so easily dismissed, and can in fact be engineered at will. One comes from interpreting the Casimir effect in terms of quantum vacuum energy (for an alternative interpretation see~\cite{Jaffe2005}). Indeed, attempts have been made to construct the required negative energy profile present in a warp drive solution through manipulating the boundary conditions of quantum vacuum modes in a Casimir-effect manner \cite{White2021}. There are also other, more robust proposals for methods in which effective exotic matter distributions can be generated, such as the light-matter interaction and the protocol of quantum energy teleportation~\cite{NicoEdu}.

Additionally, effective exotic matter appears naturally in quantum field theory in curved spacetimes, as we have seen in previous chapters. However, given that in this theory it is the curvature of spacetime itself which makes quantum states react to potentially produce negative energies, it is not clear whether such effects would work in favour or against building configurations such as warp drives. The evaporative tendencies of horizons we have seen thus far would in fact suggest the latter possibility, given that the warp drive spacetime has causal aspects which are quite similar to a BH with both an inner and outer horizon, as we will see below. Particularly, the inner-horizon-like front end of the warp bubble can be argued to lead to a similarly quick tendency toward destabilising backreaction.

Among the issues mentioned above, this semiclassical instability is perhaps the most critical roadblock for the feasibility of warp-drive configurations of any size. In~\cite{Hiscock1997} it was established that a 1+1 dimensional warp-drive configuration, corresponding to the central axis of movement of higher-dimensional drives, develops this instability. Then, in \cite{anti-warp2} this analysis was generalised to warp drives formed dynamically from an initially flat spacetime. Calculating the RSET of a quantum scalar field, it was shown that the accumulation of geodesics, and correspondingly of modes of the quantum field, at the front end of the drive leads to an exponential growth in the vacuum energy density, analogous to that found at the inner horizon of BHs \cite{BalbinotPoisson93,Hollands2020a,Ori2019}.\footnote{It is interesting to note that in these works the semiclassical inner horizon instability is quantified not through backreaction analysis, but rather by considering the energy contained in the RSET as seen by inertial observers which approach this horizon. Particularly, the finite value of $\expval{T_{vv}}$ seen in chapter \ref{ch5}, combined with the divergence of $v$ with respect to regular time coordinates at the Cauchy horizon (such as the $V$ coordinate in chapter \ref{ch4}), leads to a divergent physical energy. This can perhaps provide a more intuitive picture of why backreaction seems to act so abruptly on such configurations.} Furthermore, it was shown that this instability survives even in the presence of a modified dispersion relation at high energies~\cite{Coutantetal2012}. Thus, this instability, being caused by the very superluminal movement of the warp-drive with respect to the quantum vacuum, does indeed appear unavoidable.

At least, this is the case in 1+1 dimensions, but whether and to what degree this semiclassical instability is present in more realistic higher-dimensional warp-drive spacetimes, including in the 3+1 dimensions of our universe, has so far remained an open question. Given the lack of spherical symmetry in warp drive geometries, we cannot directly use the Polyakov approximation, and calculating the exact RSET in such spacetimes is both technically and conceptually very challenging (see e.g.~\cite{Zilberman2022}). However, instead of performing such a calculation, we can make use of some important intuitions gained from the 1+1 example, as well as from the calculations in BH spacetimes, to give a rough estimate of the vacuum energy in these warp drives of higher dimensions. Particularly, the exponentially growing accumulation of energy can be related to the presence of surfaces (or points) of infinite blueshift, which are known to cause instabilities even on a classical level (cf. mass inflation instability of BHs analysed in chapter \ref{ch4}). Therefore, a classical analysis of geodesics can likely suffice to identify the regions which may cause such instabilities.

In this chapter we analyse the geodesics of a warp-drive spacetime in 2+1 dimensions, focusing in particular on the vicinity of the walls of the warp bubble, which posses horizon-like properties. Remarkably, we find that for warp bubbles of finite spatial extension there is generally only a single point where infinite blueshift can occur, suggesting that the semiclassical singularity in 2+1 and higher dimensions is far weaker than its 1+1 dimensional counterpart. Particularly, by looking at the geodesics trapped in an approach toward this point, as well as ones which get close to it but end up deflected away, we estimate that the integrated semiclassical energy density around this point should remain bounded in most cases. Furthermore, we show that although changing the shape and trajectory of the warp bubble cannot eliminate this point, it can serve to further disperse the geodesics in its vicinity and, by extension, the semiclassical energy accumulation. Much like how aircraft reduce their air resistance by having particular shapes and adapting to air currents, warp drives must adapt their geometry and dynamics in order for the quantum vacuum to offer as little resistance as possible to their movement.

In section \ref{s2f} we provide a brief introduction to the warp drive and its semiclassical instability in 1+1 dimensions. In section \ref{s3f} we analyse the geodesics in a 2+1 dimensional drive which are relevant for determining its causal structure and its potential instability-inducing points. We estimate the semiclassical energy accumulation at and around these points, and we analyse how this accumulation may be dispersed by changing the shape or trajectory of the warp bubble. In section \ref{s4f} we provide some concluding remarks.

\section{Warp drive and the semiclassical instability}\label{s2f}

The Alcubierre warp drive as an isolated system in an asymptotically flat spacetime has the following metric:
\begin{equation}\label{metricx}
ds^2=-dt^2+[d{\xbb}-{\vbb}(t,{\xbb})dt]^2,
\end{equation}
where ${\xbb}$ represents spatial coordinates and ${\vbb}(t,{\xbb})$ determines the velocity and shape of the warp bubble (both these quantities are defined as Euclidean vectors with as many components as spatial dimensions in the manifold). Flat spacetime is recovered far away from the bubble by imposing that $|{\vbb}|\to 0$ as $|{\xbb}|\to\infty$. We take $\bar{\bm x}_{\text{c}}(t)$ to be the trajectory of the centre of the bubble in this asymptotically-Minkowskian coordinate system, and for convenience we also define the comoving spatial coordinates ${\bm x}={\xbb}-\bar{\bm x}_{\text{c}}(t)$. We can then write the usual definition ${\vbb}(t,{\bm x})=\bar{f}({\bm x}){\bm V}(t)$, where ${\bm V}(t)=d\bar{\bm x}_{\text{c}}(t)/dt$ is the velocity of the bubble and $\bar{f}({\bm x})$ determines its shape. At the centre of the bubble this shape function must satisfy $\bar{f}(0)=1$, and as $|{\bm x}|\to\infty$ it must tend to zero sufficiently quickly. In comoving coordinates the metric can be written as
\begin{equation}\label{metricr}
ds^2=-dt^2+[d{\bm x}+{\bm v}(t,{\bm x})dt]^2,
\end{equation}
where ${\bm v}(t,{\bm x})=f({\bm x}){\bm V}(t)$, with $f=1-\bar{f}$.

\begin{figure}
	\centering
	\includegraphics[scale=0.55]{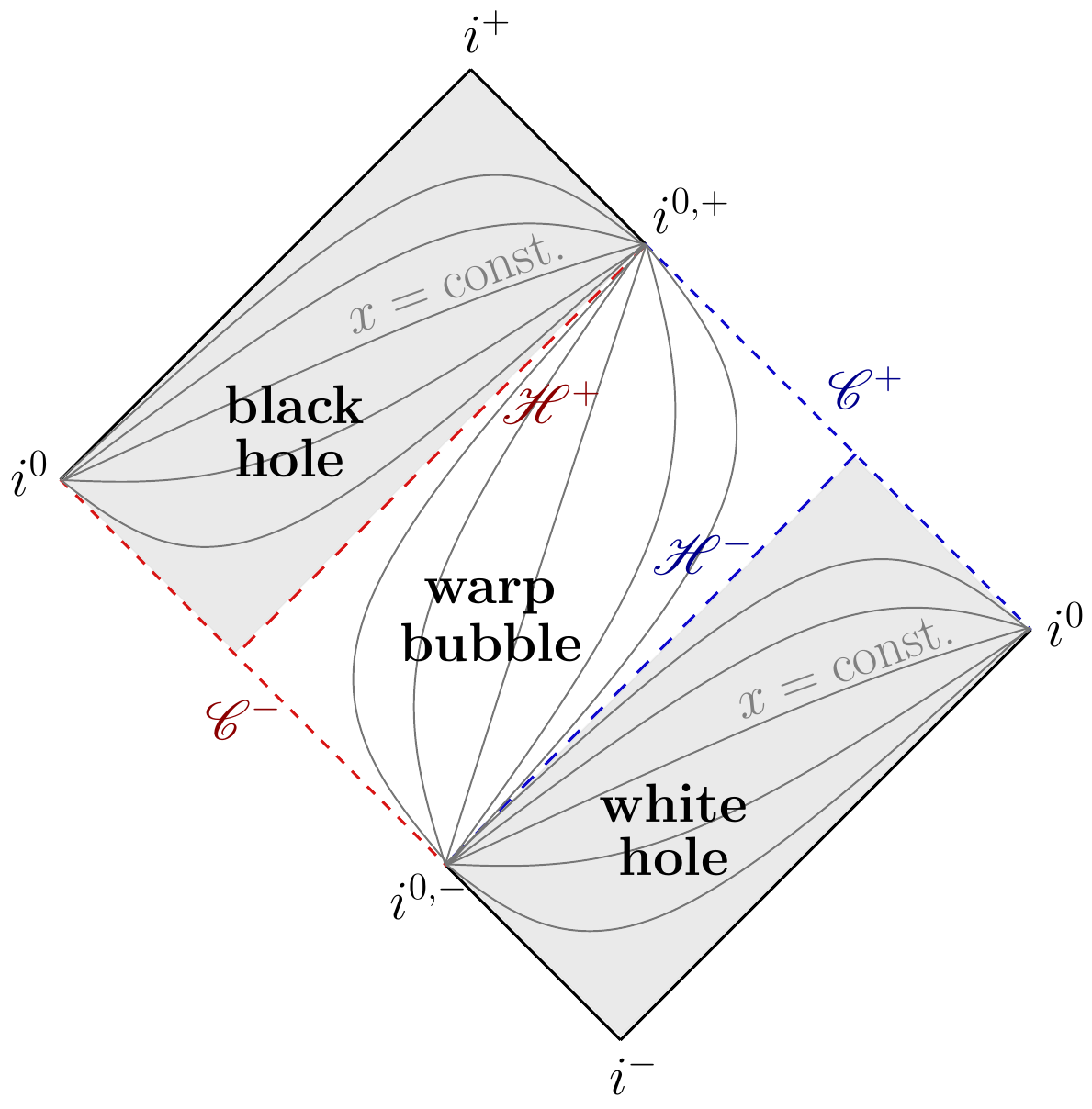}
	\caption{Causal structure of a stationary 1+1 dimensional warp drive spacetime. The warp bubble is located between a white and a BH, separated by the past and future horizons, denoted by $\mathscr{H}^-$ and $\mathscr{H}^+$ respectively. $\mathscr{H}^+$ connects with a Cauchy horizon to the past $\mathscr{C}^-$, and $\mathscr{H}^-$ with a Cauchy horizon to the future $\mathscr{C}^+$. The curved lines correspond to lines of $x=\text{const.}$, which are timelike inside the warp bubble, and spacelike everywhere else.}
	\label{f1f}
\end{figure}

To understand this geometry better, let us start with a 1+1 dimensional stationary (${\bm V}(t)=\text{const.}$) case. In this case, the line element acquires the same form as that of the radial-temporal sector of a BH spacetime written in Painlevé-Gullstrand coordinates~\cite{Martel2001}. Particularly, the front end of the warp drive behaves like a white hole horizon (more precisely, a white hole outer horizon or, equivalently, a BH inner horizon), and the rear end like a BH horizon (time reverse of the former). In comoving coordinates, the inside of the warp bubble appears located between two trapped regions, i.e. between a white and a BH, as shown in fig.~\ref{f1f}.

In \cite{anti-warp2} it was shown that when such a configuration forms from an initially flat region, the presence of any background quantum field, even in vacuum,\footnote{The instability studied here is the same for any Hadamard state.} leads to an exponential growth of energy at the white hole horizon. Particularly, the energy density obtained when contracting the RSET with the velocity $u^\mu$ of a free-falling observer which approaches this horizon grows as
\begin{equation}\label{2div}
\rho=\expval{T_{\mu\nu}}u^\mu u^\nu\sim e^{2\kappa t},
\end{equation}
where $\kappa$ is the surface gravity of the horizon. This result was obtained through a calculation much like the BH inner horizon analysis of the previous chapter. Particularly, if we express the $\expval{T_{vv}}$ component \eqref{RSETrv} in a null coordinate which does not diverge at the Cauchy horizon, it reveals the above exponential tendency. In BHs, the 1+1 dimensional calculation in the radial-temporal sector generalises directly to higher dimensions due to the isotropy of the light-trapping behaviour. In warp drives, however, the generalisation of the horizon structure shown in fig.~\ref{f1f} to higher dimensions is not as straightforward. As it happens, for generic, finite-sized warp bubbles, the dimension of the surface of infinite blueshift does not grow with the dimension of the spacetime, but remains the same as in 1+1. We will now proceed to show this explicitly in 2+1 dimensions.

\section{Geodesics and stability in 2+1 dimensions}\label{s3f}

Let us now turn our attention to the 2+1 dimensional warp drive. We will start with a thorough analysis of a particularly simple, yet quite generic configuration: a stationary warp bubble travelling in a straight line, with a geometry which has a reflection symmetry with respect to a central axis aligned with the direction of motion, as depicted in fig.~\ref{f2f}. The comoving spatial coordinates will be denoted by $\{x,y\}$, where $x$ is taken to be aligned with the direction of motion and $y$ with the direction of symmetry. The line element of the geometry is
\begin{equation}\label{metric2}
ds^2=-dt^2+[dx+v(x,y)dt]^2+dy^2.
\end{equation}
Taking $y=0$ as the position of the central axis, the function $v(x,y)$ has even parity in~$y$. The equations which determine the null geodesics of this spacetime are
\begin{align}
(v^2-1)\dot{t}^2+2v\,\dot{t}\dot{x}+\dot{x}^2+\dot{y}^2&=0,\label{geo0}\\
(v^2-1)\,\dot{t}+v\,\dot{x}&=E,\label{geo1f}\\
\ddot{y}-\partial_y v(v\,\dot{t}^2+\dot{t}\dot{x})&=0\label{geo3},
\end{align}
where $E$ is an integration constant and the dot indicates differentiation with respect to the geodesic affine parameter $\sigma$.

\subsection{Movement on the central axis}

\begin{figure}
	\centering
	\includegraphics[scale=0.8]{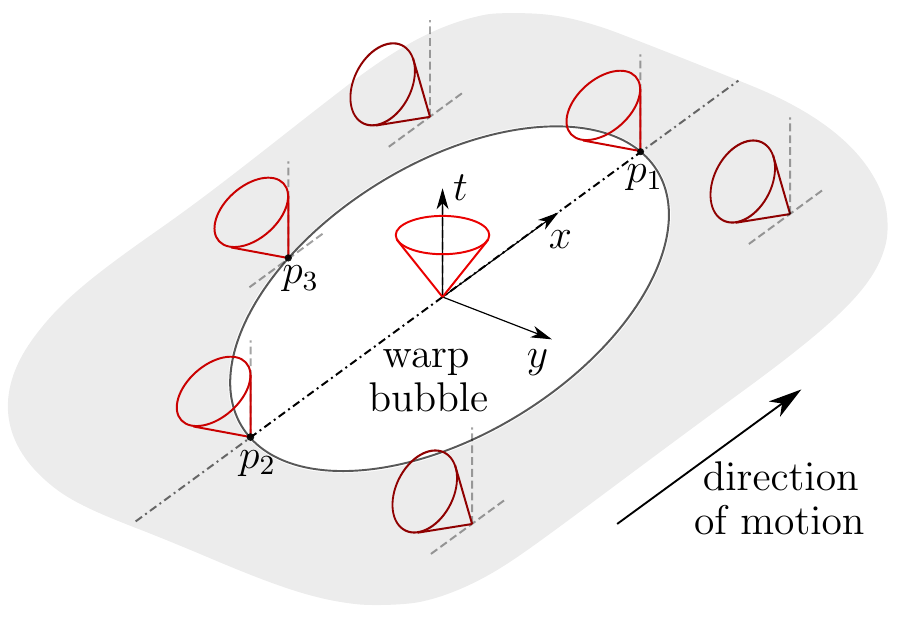}
	\caption{Warp bubble in 2+1 dimensions as seen in comoving coordinates. Light cones show the permitted directions of movement for causal trajectories. The point $p_1$ is the front end of the bubble, which produces a blueshift instability toward the future. Likewise, $p_2$ produces an instability toward the past. $p_3$ represents a generic point of the border of the bubble which does not lie on the central axis.}
	\label{f2f}
\end{figure}

The 1+1 dimensional example presented above corresponds to the movement of geodesics along the central axis. We can recover the causal structure of fig.~\ref{f1f} by integrating the null geodesic equations with the initial conditions $y_0=\dot{y}_0=0$ (which, through Eq. \eqref{geo3} and the fact that $\partial_yv=0$ at $y=0$, implies $y(\sigma)=0$). It is worthwhile to do this explicitly in the vicinity of the edges of the bubble, where $v$ approaches 1. Let $p_1=(x_1,0)$ be the front end of the bubble, as shown in fig.~\ref{f2f}. Near this point we can consider the series expansion in the $x$ direction
\begin{equation}\label{expan1}
v(x,0)=1+\kappa (x-x_1)+\cdots,
\end{equation}
where $\kappa$ is a positive constant (analogous to a horizon surface gravity). At leading order, equations \eqref{geo0} and \eqref{geo1f}, with the expansion \eqref{expan1}, have a family of solutions
\begin{equation}\label{csol}
x-x_1\simeq-E(\sigma-\sigma_1),\qquad
t\simeq-\frac{1}{\kappa}\log|\sigma-\sigma_1|,
\end{equation}
with $\sigma_1$ a constant. We see that these null geodesics reach the limit $t\to\infty$ at a finite~$\sigma$, corresponding to the future Cauchy horizon in fig.~\ref{f1f}. The past Cauchy horizon is obtained analogously by taking the expansion \eqref{expan1} at the rear end of the drive ($p_2$ in fig.~\ref{f2f}), making $\kappa$ negative, and considering ${t\to-\infty}$.

We also note that, given that \eqref{geo0} is quadratic in the coordinate functions, there are in fact two families of solutions of the geodesic equations around $p_1$. Aside from the ones shown above, there are also solutions for which $t$ does not diverge, which can be identified as the ones which cross the black and white hole horizons depicted in fig.~\ref{f1f} (left-moving light rays). Locally, these can be seen as backwards-directed trajectories relative to the bubble.

The blueshift instability is triggered by the rays which take infinite $t$ to reach $p_1$, as can be seen from the fact that the spatial separation between any two distinct lightlike observers which approach this point tends to zero, implying that the wavelength of individual perturbations also goes to zero, i.e. they are infinitely blueshifted. Intuitively, one can then see that backreaction from generic perturbations may well destabilise this configuration even on a classical level. The semiclassical instability is also a direct consequence of this blueshift, as the modes used to define a quantisation become singular when approaching this point. In stationary, eternal BH configurations, it has been shown that the semiclassical tendency toward a singularity at the infinite blueshift surface is in fact stronger than its classical counterpart~\cite{Hollands2020a,Zilberman2022} (as we will see in the next chapter, this appears to also be the case for BHs formed at a finite time).

\subsection{Other unstable points?}

We have shown that the causal structure of the 1+1 case continues to be present in higher dimensions, at least on one axis. However, it is not clear whether other points of infinite blueshift besides $p_1$ (and $p_2$, if we consider past instabilities) are present in other parts of the 2+1 configuration. In fact, for the simplest type of warp bubble, it turns out that there are no other such points, as we will now show.

The points which one can expect to have special causal behaviour are the ones which comprise the rest of the edge of the warp bubble, where $v=1$. Let $p_3$ be one such point (see fig.~\ref{f2f}). At this point, we can use the labels $\kappa=\partial_xv|_{p_3}$ and $\zeta=\partial_yv|_{p_3}$. We define the front and back end of the bubble ($p_1$ and $p_2$ in fig.~\ref{f2f}) as the points where $\zeta=0$, which for the symmetric bubble we are considering lie on its intersection with the symmetry axis $y=0$. These are the points which correspond to the future and past Cauchy horizons shown above. For a smooth convex bubble, $\zeta\neq 0$ at all other points of its frontier.

We are interested in whether there are geodesics for which $t$ diverges at finite $\sigma$ when approaching $p_3$. We can answer this by substituting $v$ and its derivatives for their values on these points in the geodesic equations, and checking whether $t$ can approach infinity while $\sigma$, $x$, and $y$ remain bounded. In Eq. \eqref{geo1f}, the first term on the left-hand side, $(v^2-1)\dot{t}$, can tend either to 0, a constant, or infinity. If it went to infinity, then $\dot{x}$ would also diverge at $p_3$, which, given the analiticity of the geometry, leads to no consistent solutions (one can check this explicitly by taking an arbitrary inverse-polynomial or logarithmic divergence for $\dot{x}$ and checking the requirements imposed on the other derivatives in equations \eqref{geo0} and \eqref{geo3} at $p_3$, arriving at an inconsistency). If this first term of \eqref{geo1f} goes to a constant or to 0, then equations \eqref{geo1f} and \eqref{geo3} become
\begin{equation}
\dot{x}|_{p_3}=\tilde{E}=\text{const.},\qquad \ddot{y}|_{p_3}=\zeta(\dot{t}^2+\tilde{E}\dot{t})|_{p_3}.
\end{equation}
Since the geometry is analytic, the divergence of $t$ implies the divergence of its derivatives. Therefore, $\dot{t}$ and $\dot{t}^2$ would have different rates of divergence, and $\ddot{y}$ would remain finite only if $\zeta=0$. If $\zeta\neq 0$, then $\ddot{y}$ diverges, which does not occur for any consistent solutions at the finite point $p_3$ (this can again be seen by considering the analiticity of the geometry or checking explicitly for such solutions in the geodesic equations).

In other words, the only points on the boundary of the bubble $v=1$ which have the possibility of generating Cauchy horizons are the ones where $\partial_yv=\zeta=0$, i.e. where the derivative of the shape function $v$ in the direction perpendicular to that of motion is zero. Stated as such, this result can be seen to be independent of the particular shape or symmetry of the bubble, as long as the configuration is stationary. For a smooth and convex bubble, this implies that there are strictly only two points of infinite blueshift akin to the ones present in 1+1 dimensions, and time symmetry tells us that only one is unstable toward the future (and the other toward the past).

If the bubble is not convex and additional points of $\partial_y v=0$ are present, then the warp drive could be said to be less ``aerodynamic" in its motion within the quantum vacuum, as it would find further resistance to its stability due to larger energy accumulation. However, as long as such points are isolated from each other, the overall configuration could potentially be stable, as the total amount of accumulated vacuum energy could be finite.

\subsection{Vicinity of the unstable points and vacuum energy divergence}

To find out whether the single-point blueshift instabilities present in 2+1 (and higher) dimensions are actually detrimental to the stability of the whole warp drive configuration, we must estimate the behaviour of the quantum vacuum energy in a small vicinity around these points. An instability is present only when the divergence at these points has a certain ``width", enough to produce a singularity if backreaction is considered. Finding out whether this is the case would generally involve calculating the RSET on this spacetime for a test field in an appropriate vacuum state. However, due to the great technical difficulty involved in such a calculation, we will resort to an estimation based on an extension of the analogy between the movement of geodesics at and around the central axis, and the 1+1 dimensional case.

To set this up, let us begin by considering a solution slightly away from the central axis solutions \eqref{csol}, but still in the vicinity of $p_1$. We now perform the expansion of $v$ around $p_1$ to leading order in both $x$ and $y$,
\begin{equation}\label{shapey}
v(x,y)\simeq 1+\kappa(x-x_1)+\xi y^{2n},
\end{equation}
where $\xi$ is a constant with appropriate inverse-length dimensions, positive if the bubble is convex and negative if it has a concave peak (and zero if it has a finite-sized flat peak), and $n$ a natural number. Larger values of $n$ make the peak more flat in the $y$ direction. We consider the deviation from the solution \eqref{csol} (where $y=0$),
\begin{align}
\delta x(t)&=x(t)-x_1-Ee^{-\kappa t},\label{delx}\\
\delta t(\sigma)&=t(\sigma)+\frac{1}{\kappa}\log|\sigma|,
\end{align}
where $E$ is a constant. It is convenient to rewrite equations \eqref{geo0}, \eqref{geo1f} and \eqref{geo3} at leading order in $\delta x$, $y$ and $\delta t$ (and their derivatives) as
\begin{align}
y''+\kappa y'-2n\xi y^{2n-1}&\simeq 0,\label{devy}\\
\delta x'+\kappa\delta x+\xi y^{2n}+\frac{1}{2}(y')^2&\simeq 0,\label{xprime}\\
\delta\dot{t}+\frac{e^{2\kappa t}}{2\kappa^2 E}(2\kappa\delta x+2\xi y^{2n}+\delta x')\label{dott}&\simeq 0,
\end{align}
where the prime indicates a derivative with respect to~$t$. Eq.~\eqref{devy} can give us a description of the perturbation in $y$, from where we can use Eq. \eqref{xprime} to obtain the perturbation $\delta x$, and Eq. \eqref{dott} to find the modification $\delta t$ to $t(\sigma)$. Particularly, Eq. \eqref{devy} can be solved directly, and the validity of the solution can be checked by making sure that the approximations which lead to \eqref{devy} are accurate, which can be done with the solutions of \eqref{xprime} and \eqref{dott}.

Let us begin by looking at the case of a warp bubble edge with a finite-sized region which is flat in $y$. This, as one might imagine, is not a very ``aerodynamic" shape, as it is not convex. The solutions of \eqref{devy} with $n=0$ tending to this frontal region would be
\begin{equation}\label{flat}
y\simeq c_2+c_3e^{-\kappa t},
\end{equation}
where $c_{2,3}$ are integration constants. The constant $c_2$ is indicative of the fact that $p_1$ is no longer the only point which traps geodesics into a tendency toward a Cauchy horizon. Eq. \eqref{xprime} with $n=0$ furthermore shows us that for these solutions $x$ has the exact same behaviour at large $t$ as it does on the central axis (i.e. $\delta x$ has the same solutions as $x$ in \eqref{csol} and can be absorbed in the integration constants of the latter); and Eq. \eqref{dott} shows the same for $t(\sigma)$, implying that the approximations used to obtain \eqref{devy} are accurate. Therefore, in this case there would be a finite-sized region with a blueshift instability, and we can expect that this configuration would be unstable under both classical and semiclassical perturbations.

If the bubble has, say, a parabolic profile in $y$ (i.e. $n=1$), then the solutions become
\begin{equation}\label{sol}
y\simeq c_2e^{\eta_-t}+c_3e^{-\eta_+t},\qquad
\delta x\simeq c_1\,y^2,
\end{equation}
with $c_1$, $c_2$ and $c_3$ constants, and
\begin{equation}
\eta_\pm=\frac{\kappa}{2}\left(\sqrt{1+8\frac{\xi}{\kappa^2}}\pm 1\right).
\end{equation}
Let us first look at the case of a convex bubble, for which $\xi>0$ and consequently $\eta_\pm>0$. In this case, only initial conditions which give $c_2=0$ correspond to geodesics trapped in a tendency toward the tip of the bubble from outside the axis, since $y\to 0$ as $t$ grows. These fine-tuned geodesics (of measure zero within the total set of solutions) for each value of~$c_3$ represent a separatrix between solutions deflected away (exponentially quickly, while the approximation is valid) to one side ($c_2>0$) and the other ($c_2<0$). As one might expect, the approximations leading to Eq. \eqref{devy} break down quickly when $c_2\neq0$, and become asymptotically exact when $c_2=0$.

\begin{figure}
	\centering
	\includegraphics[scale=0.5]{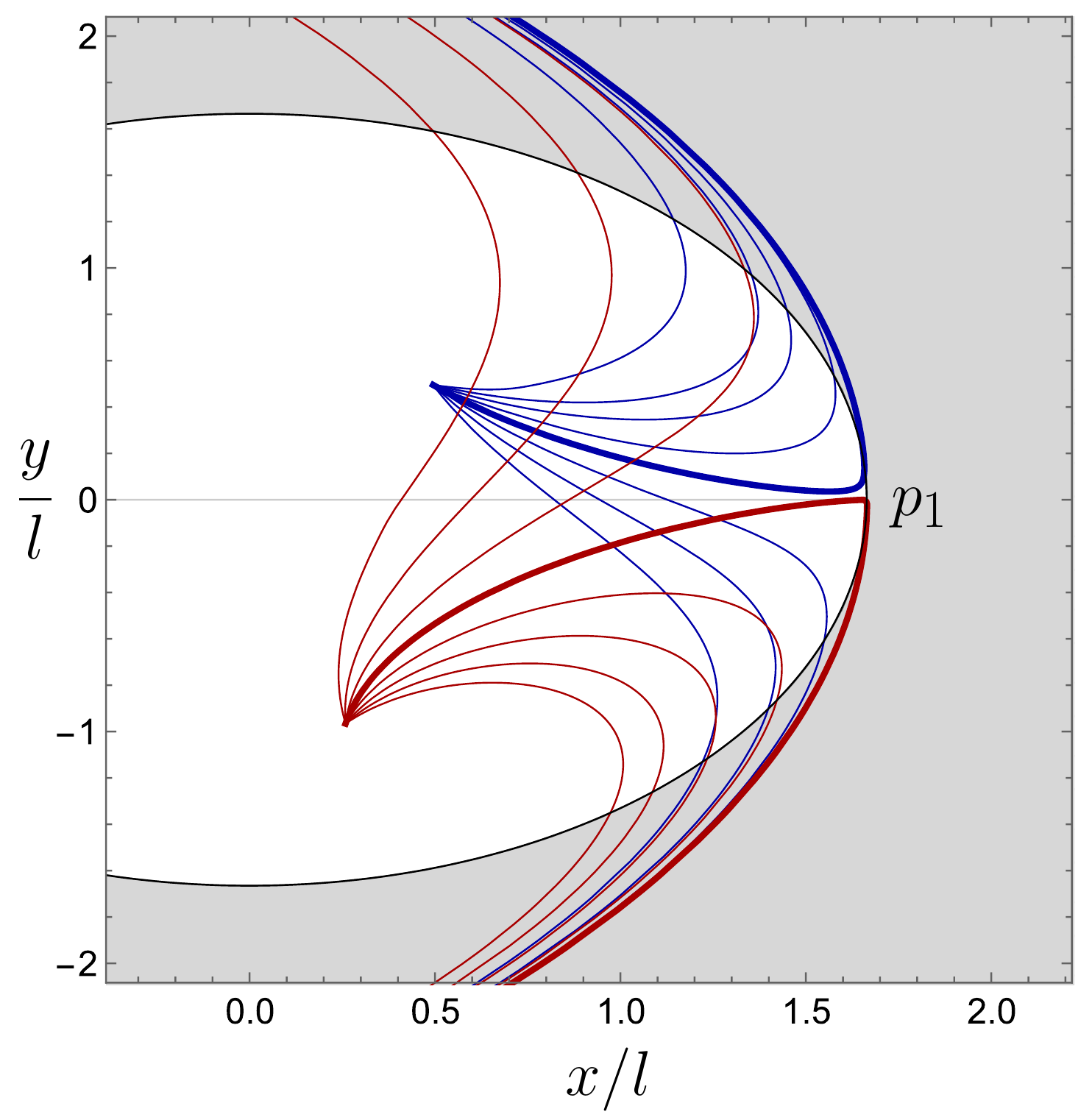}
	\caption{Numerical integration in time of two bundles of null geodesics launched from two different points in the interior of a stationary circular warp bubble moving at twice the speed of light. $l$ is a characteristic length scale of the bubble. The geodesics are launched in the forward direction (with respect to the motion of the drive), with an initial angle dispersion of $\pi/3$ between the first and last of each bundle of curves. The thicker lines of each bundle represent the geodesics which get closest to the point $p_1$, which in a vicinity of this point correspond to curves \eqref{sol} with small values of $c_2$.}
	\label{f3f}
\end{figure}

A further check of this behaviour was performed numerically by directly solving equations \eqref{geo0}, \eqref{geo1f}, and \eqref{geo3} for geodesics launched from within a moving convex warp bubble. The result is shown in fig.~\ref{f3f}, where one observes clearly the general behaviour of light within this geometry. The first thing we note about the geodesics shown is that they follow the restrictions imposed by the light cones represented in fig.~\ref{f2f}: they can only move forward (toward larger values of $x$) while inside the bubble, and when they approach its edges they turn around. The thicker lines of each bundle of geodesics mark two curves which get close to the central axis in the vicinity of the point of runaway blueshift $p_1$. In these curves we see explicitly the behaviour captured in Eq. \eqref{sol} for solutions with small values of $c_2$. They initially approach the axis, until the exponential growth of $e^{\eta_-t}$ overcomes the smallness of $c_2$ and pushes them away. As expected, for geodesics launched from each point, the separatrix ($c_2=0$) between the ones which end up on the left and on the right of $p_1$ turns out to be impossible to capture numerically. This provides further evidence of the fact that, although these solutions end up infinitely blueshifted, they are of measure zero within the whole family of geodesics.

For a bubble with a more flat profile in $y$ at $p_1$ (i.e. $n>1$), there are no analytical solutions to \eqref{devy}, but it can be seen that the source term for the derivatives is smaller and deflection therefore has an initially polynomial (rather than exponential) dependence on $t$. Aside from this, the qualitative behaviour of the geodesics in such a bubble remains the same (this has been checked numerically).

Returning to the $n=1$ parabolic profile, we can make an important observation regarding the deflected geodesics. By taking the geodesic from \eqref{sol} with $c_3=0$ as representative of the generic qualitative behaviour of geodesics with $c_2\neq 0$, we can write its solution in terms of the initial condition $y(0)=y_0$ as
\begin{equation}
y\simeq y_0e^{\eta_-t}.
\end{equation}
We can then define a deflection time $t_{\rm def}$ as the time it takes for the solution to reach a fixed reference point $y_{\rm def}$,
\begin{equation}\label{tdef}
t_{\rm def}=\eta_-^{-1}\log(y_{\rm def}/y_0).
\end{equation}
The value of $y_{\rm def}$ is a characteristic length scale of the geometry which can be defined e.g. as the separation for which the approximation which lead to \eqref{devy} fails. The important part is the dependence on $y_0$, particularly, the logarithmic divergence as $y_0\to 0$.

In 1+1 dimensions, the presence of a Cauchy horizon and the corresponding divergence of $t$ for finite $\sigma$ in \eqref{csol} drives the exponential growth of the energy density \eqref{2div}. In 2+1 and higher dimensions, one may then expect the same type of growth only around points where the null geodesics behave the same way (i.e. as if approaching a Cauchy horizon), which occurs only when they approach the tip of the warp bubble. In other words, the exponential growth of the energy density should only occur at a single point. As to what happens in the vicinity of this point and how this energy accumulation falls off away from it, the geodesics with $c_2\neq 0$ might provide a clue.

Particularly, in the regime $t\ll t_{\rm def}$ these geodesics have a very small deviation from the central solution \eqref{csol}, and one might expect that they bring about a growth similar to~\eqref{2div}, but instead of blowing up to infinity as $t$ grows, tending to a finite cutoff value with a profile given by the logarithm of $y_{\rm def}/y_0$ in \eqref{tdef}, with $y_0$ representing the separation from the central point.

Another argument in favour of this kind of asymptotic density profile can be made by just considering the blueshift of light rays which could be randomly launched in the general direction of the front of the warp bubble. The ones which happen to tend exactly to the tip are the only ones which are trapped and have a divergent tendency in their blueshift. In the rest of the bubble, one may expect that a stationary situation is quickly reached if e.g. the rays are launched at regular intervals. In a given time the same number of rays would enter a given area as the ones which exit it, though the closer this area is to the tip of the bubble, the longer their stay there and the larger their transient blueshift, giving rise to the same logarithmic profile of energy density.

In dimensions larger than 1+1, the ``instability" for a convex warp bubble is therefore just the growth at a single point, and integrating the energy in its vicinity the result would not asymptotically tend to a divergence. Even if the logarithmic profile we obtain for the asymptotic tendency for the semiclassical energy density is not the correct description one would get from calculating the RSET exactly in 2+1 or higher dimensions, at the very least the fact that this profile is related to the accumulation of geodesics is robust. Therefore, additional dispersion of these geodesics would translate into further stabilisation of the semiclassical behaviour, as can be achieved by decreasing $t_{\rm def}$ (e.g. by making the peak of the bubble sharper in $y$, i.e. making $\xi$, and hence $\eta_-$, larger), or by making the trajectory of the drive deviate from the straight line path we have considered here, as we will show numerically below. On the other hand, making $t_{\rm def}$ larger (e.g. by decreasing $\xi$ or increasing $n$) would have the opposite effect and bring the drive closer to instability.

A convex shape with a very sharp peak, which offers the least resistance for travel in the presence of a quantum field, even in vacuum, is reminiscent of the shapes used for supersonic aircraft which minimise the frontal pressure and drag that they experience. By extension of this analogy, one would naturally expect that a warp bubble with a flat or concave peak would experience much more resistance, i.e. a much stronger blueshift instability. Indeed, in the case of a flat peak we found that the solutions which are trapped in a tendency toward a Cauchy horizon are much more abundant \eqref{flat}. For a concave peak with e.g. a locally parabolic profile, the solutions would be the same as \eqref{sol} but, $\xi$ being negative, $\eta_-$ would have a negative real part, making both exponentials decreasing ones. This would be a case in which a divergence of the order of that of a flat peak is concentrated at a single point, making the instability even greater.

\subsection{Numerical analysis of non-stationary configurations: further stabilising the warp drive}

In light of these results, one may wonder how this behaviour generalises to dynamical warp drive spacetimes, i.e. ones in which the warp bubble can change its trajectory and velocity over time. Particularly, we want to see whether the point of divergent blueshift $p_1$ remains, or whether some trajectories for the bubble can ``shake off" the potentially accumulated geodesics around such a point at regular intervals.

There are two types of movement which have the potential to do this: a change in direction, or a temporary reduction of the velocity to a subluminal one. However, we have found through a numerical analysis that neither one of these can fully eliminate the point of asymptotically infinite blueshift (and its corresponding finite-time accumulation of vacuum energy). Nonetheless, they can significantly disperse the geodesics in its vicinity, producing the same effect as making the peak of a straight-line bubble sharper.

\begin{figure}
	\centering
	\includegraphics[scale=0.55]{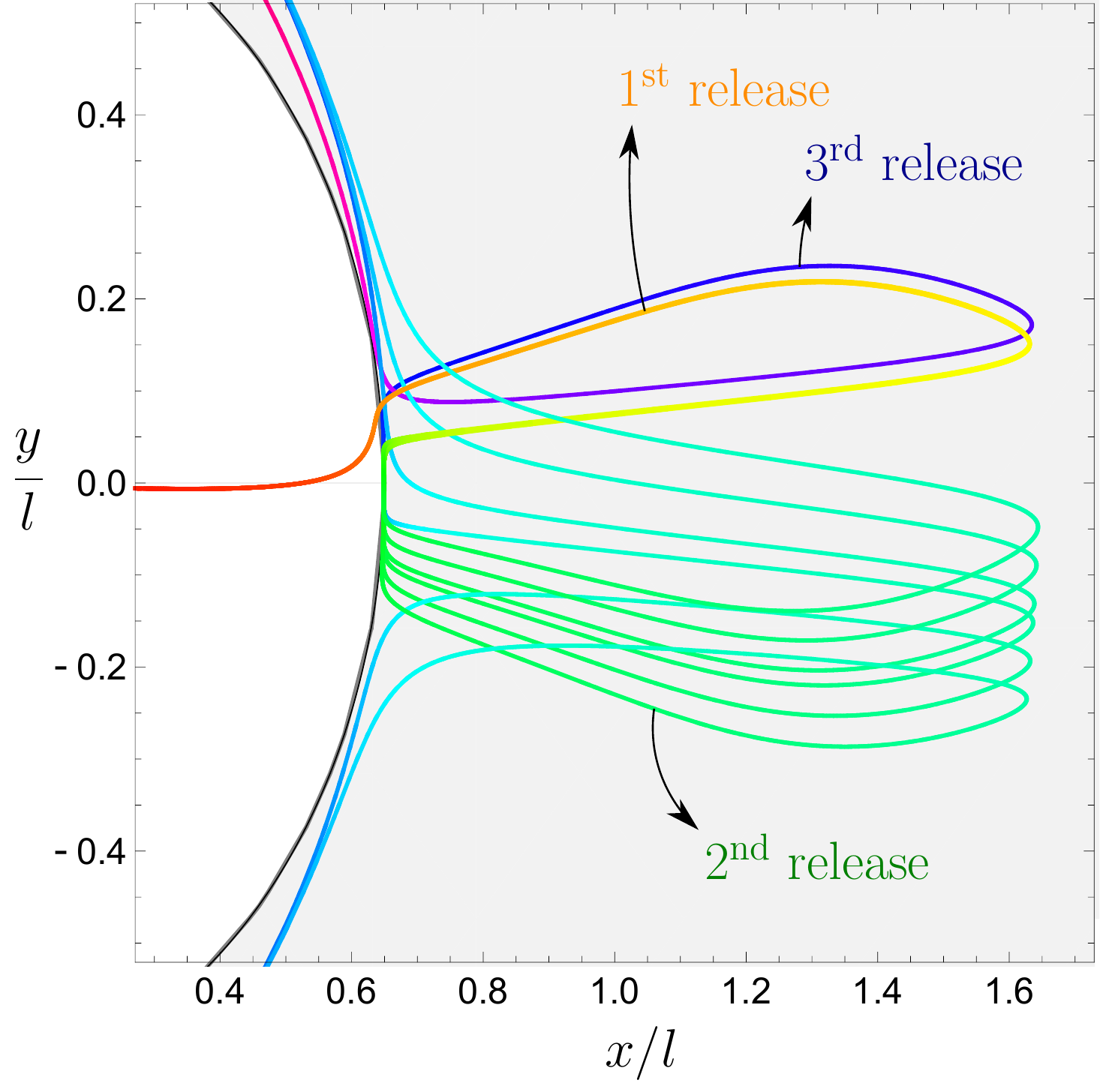}
	\caption{Numerical integration of null geodesics launched from the centre of a circular warp bubble moving in a zig-zag pattern with subluminal and superluminal intervals. The velocity of the drive in the $x$ direction changes between 0.2$c$ and 10$c$ with frequency $2\pi/5$, while in the $y$ direction it changes between 0.8$c$ and -0.8$c$ with frequency $\pi/5$. There are 6 geodesics emerging from the left side of the plot. The angle at which they are emitted only varies by $10^{-4}$ between them, so they initially overlap. Each time the drive becomes subluminal, the geodesics are released forward, only to turn around once it becomes superluminal again. At the second release they become dispersed enough to be visibly separate, while at the third release only one of them remains (the rest having been dispersed to the sides).}
	\label{f4f}
\end{figure}

The fact that the equivalent of $p_1$ cannot be eliminated can be deduced from a simple consideration: if the warp bubble continues to exist indefinitely (and has a well-defined average asymptotic direction), for geodesics launched from each point in its interior there will always be a separatrix between those which end up on one side or the other of its asymptotic trajectory. What can be controlled, however, is the amount of blueshift this separatrix and its adjacent geodesics experience at finite times.

As an example, we present the result from a numerical analysis of a dynamical warp drive configuration which combines the two modifications to the straight-line scenario mentioned above: a change of direction, achieved through a periodically varying velocity in the $y$ direction, and a change to a subluminal velocity in $x$, also performed periodically. Figure \ref{f4f} represents the key features of the behaviour of light rays in such a geometry. Most null geodesics launched from inside the bubble quickly escape to either side of the averaged direction of motion, and never approach the bubble again, akin to those plotted in fig.~\ref{f3f}. The main difference is seen in the rays which are launched approximately in the direction of motion. Particularly, those which remain in the vicinity of the front end of the bubble long enough to catch one of the changes in velocity have the chance to move in the forward direction beyond the confines of the bubble while the drive is subluminal. Then, when the drive becomes superluminal again, the bubble catches up to those rays once again, and they are now deflected to either side of it from the outside. Those which are close to the separatrix between the ones deflected to either side can again remain in the vicinity of the edge of the bubble long enough to catch the next change in velocity and move forward again. This process repeats periodically, and there is once again a particular set of trajectories (which are of measure zero within the total set of null geodesics, and which asymptotically coincide) that define the separatrix between rays deflected to either side of the drive.

Figure \ref{f4f} represents 6 null geodesics which are launched in an approximately forward direction, with an initial angle dispersion of the order of $10^{-4}$ (making them overlap initially, on the left side of the plot). We see that each time the drive becomes subluminal, the geodesics are allowed to leave the bubble in the forward direction (or, as seen from inside the drive, it is the bubble that effectively expands to infinity). Then when it picks up superluminal speed they again turn around. Due to the oscillatory nature of the movement, we can expect that the separatrix also describes a periodic movement in space. In fact, one of the geodesics in fig.~\ref{f4f} is very close to such a behaviour: after the third time it is released in a forward direction (i.e. the third time the bubble becomes subluminal) it nearly follows the same trajectory as after the first time, though the small difference makes it so it is deflected away (toward positive $y$) in the end.

While a convoluted trajectory for the warp drive would have a negative impact on its initial purpose (i.e. shortening travel time), it can, on the other hand, increase its semiclassical stability by reducing the accumulated blueshift around its peak. Constructing an optimal warp bubble shape and trajectory would therefore become a balancing act between having a short travel time and minimising the (possibly already very small) accumulation of unwanted vacuum energy and blueshifted classical perturbations. Of course, this problem would likely have a secondary role when compared to the inevitable engineering difficulties in constructing such configurations to begin with.

\section{Conclusions}\label{s4f}

We have analysed the semiclassical instability present in the Alcubierre warp-drive spacetime through its relation to the behaviour of null geodesics. We have argued that the strong instability found in 1+1 dimensional configurations can actually be tamed in spacetimes of higher dimensions by choosing appropriately the shape and the trajectory of the warp bubble.

First, the warp field should be chosen to have an ``aerodynamic" shape, so as to deflect null geodesics away from its unstable point in the shortest time possible. Second, the trajectory of the drive can be chosen so as to further facilitate this dispersion, particularly with slight changes in its direction of movement (e.g. a small ziz-zag component to the motion), and with alternating intervals of subluminal and superluminal warp field velocities.

Our findings are interesting even from a purely geometrical perspective. In 1+1 dimensions the front wall of a warp drive acts as a pure inner horizon. However, in higher dimensions the warp drive does not have a closed inner horizon (or indeed any closed trapped surfaces). Instead, the warp-drive bubble can be interpreted as an interpolation between an inner horizon point (the front end of the bubble) and an outer horizon point (the back end of the bubble); then, in between we have a causal structure more similar to that of an ergoregion, from which signals can in fact escape. This is the reason why the geodesic accumulation, and the corresponding blueshift instability, is limited to only single points, at least when the shape of the bubble is smooth.

In contrast to BHs with an inner horizon, warp drives could indeed be stable semiclassical configurations, if their shape and trajectory are appropriately chosen, though the mechanism which generates their geometry in the first place may lie outside the semiclassical theory. On the other hand, the less ``aerodynamic" configurations found, such as ones with a flat wall at the front end of the bubble, are an interesting example of the fact that blueshift instabilities (both classical and semiclassical) can appear without the presence closed trapped surfaces.


\chapter{Classical mass inflation vs semiclassical inner horizon inflation}\label{ch7}

\fancyhead[R]{}
\fancyhead[L]{Part II -- \chaptername\ \thechapter: \leftmark}

The main goal of this second part of the thesis is to see whether and how semiclassical physics can have an influence on the evolution of the inner horizon of a BH. In chapter \ref{ch4} we looked at an example of the purely classical evolution of an inner horizon as it undergoes mass inflation. Then, in chapter \ref{ch5} we analysed the semiclassical backreaction around static and dynamical inner horizons in simple geometries which, however, do not incorporate mass inflation. The result of this latter analysis was that the inner horizon tends to be pushed outward due to backreaction from the RSET. The initial tendency for this movement appears to be exponential in time, and extrapolating from it (assuming that the RSET maintains its negative horizon-related flux) we described a potential \emph{inner horizon inflation} process, through which the trapped region is quickly extinguished from the inside out. In this chapter we will extend our semiclassical backreaction analysis to background geometries incorporating the causal properties of mass inflation.

\begin{figure}
	\centering
	\includegraphics[scale=.7]{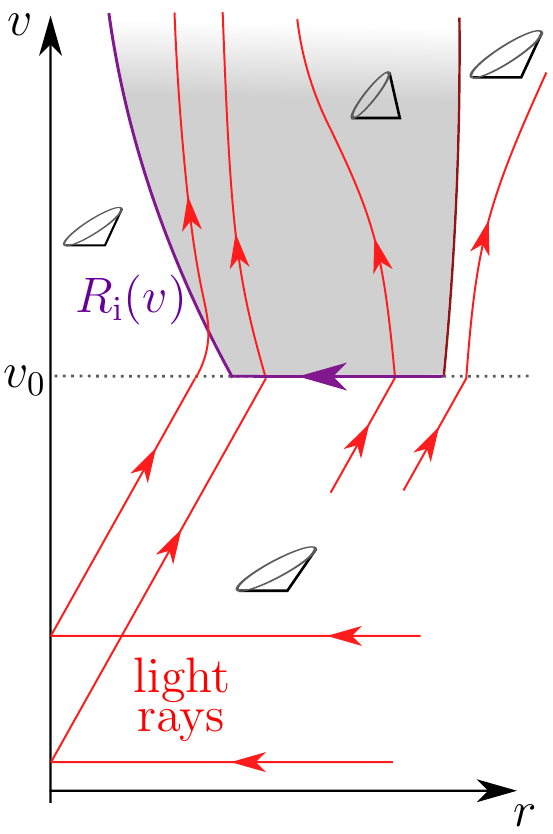}
	\caption{Formation of a BH by a ingoing null shell at $v=v_0$. The ``in" vacuum state is constructed by tracing light rays back to the flat region in the past. Classically, the BH is undergoing mass inflation and the inner horizon is headed toward the origin.}
	\label{fin}
\end{figure}

As in previous chapters, we will use the RSET of a massless scalar field in the Polyakov approximation as the source of backreaction. We construct the ``in" vacuum \cite{Fabbri2005,Frolov2017} of the 1+1 dimensional radial-temporal sector of the spacetime, used to calculate the RSET in this approximation, by following the movement of lightlike geodesics from past null infinity up to the region of interest, which in this case is the vicinity of the inner apparent horizon inside the mass-inflated region of a dynamically formed BH. As discussed in chapter \ref{ch5}, the accumulative effect that the inner horizon has on null geodesics results in a sensitivity of the RSET to the past of a large part of the collapse geometry, unlike what occurs for the outer horizon. We therefore employ the same tactic as in that chapter to simplify the initial conditions of the quantum modes in this region: we consider their propagation as being in a flat spacetime up to an advanced time $v=v_0$, from where it continues inside a BH with an inner horizon, as shown in fig. \ref{fin}. Effectively, this is equivalent to the BH being generated by the collapse of a null shell located at the formation time $v=v_0$, but our motivation for the construction is purely geometrical, to ``clean up" the dependence on the details of the collapse and leave only the part stemming from the quantum modes finding themselves in the trapped region.

In this chapter, we will perform two distinct calculations to estimate semiclassical backreaction in the vicinity of the inner horizon. The first involves a series expansion of the RSET and the metric functions around the point $\{v_0,R_{\rm i0}\}$, where the (timelike part of the) inner horizon $R_{\rm i}$ forms, which allows a simplified term-by-term calculation of the semiclassical perturbations of the metric caused by the RSET. The second is a full self-consistent solution for a classical background of the type \eqref{geod} with a particular family of functions $F(v,r)$, valid for a small but finite time interval after $v_0$. From the latter calculation we find that the initial tendencies seen in the series expansion can lead to very quick accumulative effects which make semiclassical corrections relevant before the spacetime curvature reaches Planckian scales.

\section{Series expansion around horizon formation}

Let us begin by considering the line element
\begin{equation}\label{geoinfl}
ds^2=A(v)\left[-B(v,r)dv^2+2dvdr\right]+r^2d\Omega^2.
\end{equation}
We will use this type of geometry to represent the inner region of a BH undergoing mass inflation, particularly around its inner apparent horizon. In relation to the construction of chapter \ref{ch4}, we are using the geometry obtained with a single outgoing shell, resulting in a single exponential growth of mass, which we expect to be the dominant effect in a transient period between early and very late times, where we place our $v_0$. For the freezing function $A$ we can use the dominant behaviour in $v$ of~\eqref{A1} and set $A(v)=e^{-\kappa v}$, with $\kappa$ a positive constant. Neglecting the $1/v^p$ part of \eqref{A1} amounts to discarding corrections suppressed by an additional $1/v$ factor in the backreaction calculations of this section, which do not have a qualitative influence on our conclusions. For the function $B$, which represents the product of $F$ and $A$ from~\eqref{geod}, we only need to impose that it has a zero at a radius $R_{\rm i}(v)$, corresponding to the inner apparent horizon, with a negative slope in the $\partial_r$ direction.

To calculate the RSET for this spacetime, we first need to construct a quantisation with a vacuum state which is physically adequate for the problem at hand. As previously, we will use the ``in" vacuum state, which is defined as the Minkowski vacuum at the asymptotically flat region of past infinity and its extension to the dynamical region through the propagation of the particle-related modes. As the reader may recall, in the Polyakov approximation we only need to do this in the 1+1 dimensional radial-temporal sector of our geometry, which simplifies the problem greatly, given that 1+1 dimensional spacetimes are conformally flat. Working with a pair of radial null coordinates $\{u,v\}$, the line element \eqref{geoinfl} can generally be written as
\begin{equation}
ds^2=-C(u,v)dudv+r^2d\Omega^2.
\end{equation}
At past null infinity, $C\sim 1$ and Poincaré invariance of the vacuum state in flat spacetime amounts to supertranslation invariance of the ``in" state \cite{Hawking1975}. Following the construction of chapter \ref{ch4}, the $v$ coordinate used in \eqref{geoinfl} is in fact one of these ``in" coordinates, while $u$ must be obtained through its relation to $r$.

The conformal factor of the radial-temporal sector is given by
\begin{equation}\label{conformalg}
C(u,v)=-2A(v)\frac{\partial r(u,v)}{\partial u}.
\end{equation}
The components of the RSET in the Polyakov approximation for the vacuum state selected by the coordinate system $\{u,v\}$ are then given in terms of $C$ through eqs. \eqref{RSET}. Like in chapter \ref{ch5}, we will only use this RSET in a small vicinity around the inner horizon, and only while this horizon is considerably farther away from the origin than a Planck length (i.e. early enough in the evolution of the mass inflation background), so that the Polyakov approximation can be qualitatively accurate. To calculate the conformal factor~\eqref{conformalg}, we need to obtain the function $r(u,v)$ from the solutions of radial null geodesics. The ingoing geodesics are just $v=\text{const.}$, while the outgoing ones are solutions to
\begin{equation}\label{geodesic}
\frac{dr}{dv}=\frac{1}{2}B(v,r).
\end{equation}

At this point we must either specify the function $B$, or try to see what general conclusions could be obtained from just the mere fact that there is an inner horizon in this structure, i.e. that $B$ has a zero at some $R_{\rm i}(v)$ with a negative slope. In the next section we will specify some functions $B$ which can simplify our analysis while still reproducing the causal properties of mass inflation, but for now we will maintain generality and perform a perturbative analysis. Particularly, we will consider a generic expansion of the function $B$ around the point at which this horizon forms $\{v_0,R_{\rm i0}\}$,
\begin{align}
\begin{split}
\frac{1}{2}B(v,r)&=k_1(v)(r-R_{\rm i}(v))+k_2(v)(r-R_{\rm i}(v))^2+\\&\quad+k_3(v)(r-R_{\rm i}(v))^3+\cdots,\end{split}\label{redshiftseries}\\
R_{\rm i}(v)&=R_{\rm i0}+R_{\rm i1}v+R_{\rm i2}v^2+R_{\rm i3}v^3+\cdots,\\
k_n(v)&=k_{n0}+k_{n1}v+k_{n2}v^2+k_{n3}v^3+\cdots,\label{kn}
\end{align}
with $n=1,2,\dots$, and where for simplicity we have set $v_0=0$. The only conditions we impose on these series is that $R_{\rm i0}>0$ and $k_{10}<0$ (this being the inner horizon). The smallness of the terms in the expansions of quantities with (inverse) length dimensions, here and throughout this section, can be measured in terms of their respective initial values at $v=0$, or in units of the characteristic initial scale $R_{\rm i0}$.

For the solution of \eqref{geodesic} we consider the series expansion
\begin{equation}\label{null}
r(v)=r_0+r_1v+r_2v^2+r_3v^3+\cdots
\end{equation}
Substituting this expression into \eqref{geodesic}, we obtain the coefficients of the null trajectories~\eqref{null} in terms of derivatives of $B$ [i.e. the coefficients of its expansion \eqref{redshiftseries}-\eqref{kn}] and a free parameter fixed by an initial condition. We will use $d_0=r_0-R_{\rm i0}$ as this parameter. Tracing back the null trajectories through the Minkowski region $v<0$ (see fig.~\ref{fin}) we find that our missing ``in" coordinate is $u=-2d_0$ (up to a constant which fixes the origin of $u$, taken as zero).

Constructing $r(u,v)$ in this manner, we calculate the conformal factor \eqref{conformalg} and then the RSET components~\eqref{RSET} in the ``in" coordinate system. Switching them back to the Eddington-Finkelstein system, at zeroth order in the series expansion, they are
\begin{subequations}\label{RSETmi}
	\begin{align}
	\begin{split}
	\expval{T_{vv}}&=\frac{l_{\rm P}^2}{96\pi^2R_{\rm i0}^2}\left(-\frac{1}{2}k_{10}^2+k_{11}-2k_{20}R_{\rm i1}-\right.\\&\qquad\qquad\qquad\left.-\frac{1}{2}\kappa^2+k_{10}\kappa\right)+\order{v,d},\end{split}\\
	\expval{T_{rr}}&=\order{v,d},\\
	\expval{T_{vr}}&=-\frac{2l_{\rm P}^2k_{20}}{96\pi^2R_{\rm i0}^2}+\order{v,d},
	\end{align}
\end{subequations}
where $d=r-R_{\rm i0}$ [here $\order{d}$ can be though of as $\order{d_0}$ or $\order{u}$, as the difference between them is $\order{v}$]. Note that if we take \eqref{RSETdyn} and incorporate the expansions in the $v$ direction, we recover the above expressions with $\kappa=0$. The introduction of a non-trivial freezing function $A$ is what makes the calculation in this chapter more involved, requiring the additional series expansion in time (or the specification of a particularly simple function $B$, as in the next section) to obtain a meaningful result.

We now consider the semiclassical Einstein equations
\begin{equation}\label{semiclg}
G_{\mu\nu}+\delta G_{\mu\nu}=T_{\mu\nu}^{\rm class}+\expval{T_{\mu\nu}},
\end{equation}
where $\delta G_{\mu\nu}$ is a perturbation to the background Einstein tensor and $T_{\mu\nu}^{\rm class}$ is the classical matter content sourcing the zeroth order background. In general, the perturbation caused by the RSET would also affect $T_{\mu\nu}^{\rm class}$ through its dependence on the metric. However, since we want to remain as agnostic as possible about this classical matter, we consider that $\delta T_{\mu\nu}^{\rm class}=0$ at zeroth order in the series expansion (in its functional form in Eddington-Finkelstein coordinates), i.e. that
\begin{equation}\label{einsteinpert}
\delta G_{\mu\nu}=\expval{T_{\mu\nu}}+\order{v,d}.
\end{equation}
This simplifying assumption serves two purposes: on the one hand, it allows us to continue to work in purely geometric terms, with as few ingredients in the dynamics as possible. On the other hand, it exemplifies well what backreaction from the RSET can look like in its purest form, where the potentially negative-energy terms in this tensor directly source~$\delta G_{\mu\nu}$. Technically, considering $\delta T_{\mu\nu}^{\rm class}=0$ would over-determine the system of equations, but it turns out to work consistently up to second order in our series expansion, allowing us this first geometric glimpse into backreaction.

Using the expansion of the function $B$ and its coefficients, the Einstein tensor of our generic background is
\begin{subequations}\label{einstein}
	\begin{align}
	G_{vv}&=\frac{2k_{10}R_{\rm i1}}{R_{\rm i0}}+\order{v,d},\\
	G_{rr}&=\order{v,d},\\
	G_{vr}&=\frac{2k_{10}R_{\rm i0}-1}{R_{\rm i0}^2}+\order{v,d}\\
	G_{\theta\theta}&=2R_{\rm i0}(k_{10}+k_{20}R_{\rm i 0}).
	\end{align}
\end{subequations}
with $G_{\phi\phi}=\sin^2(\theta)G_{\theta\theta}$. To construct $\delta G_{\mu\nu}$ we can consider perturbations to the coefficients in~\eqref{einstein}, e.g. $k_{10}\to k_{10}+\delta k_{10}$. Eq.~\eqref{einsteinpert} allows us to fix one of the three coefficients present in the tensor components \eqref{einstein} to its classical value as an initial condition, and we choose the initial position of the inner horizon $R_{\rm i0}$. Then, perturbing the surface gravity $k_{10}$, the initial time derivative of the inner horizon trajectory $R_{\rm i1}$, and the second spatial derivative of the redshift function $k_{20}$, we obtain the leading order of the perturbed tensor
\begin{subequations}\label{einsteinpr}
	\begin{align}
		\delta G_{vv}&=2\frac{\delta k_{10} R_{\rm i1}+k_{10}\delta R_{\rm i1}}{R_{\rm i0}}+\order{v,d},\\
		\delta G_{rr}&=\order{v,d},\\
		\delta G_{vr}&=\frac{2\delta k_{10}R_{\rm i0}-1}{R_{\rm i0}^2}+\order{v,d}\\
		\delta G_{\theta\theta}&=2R_{\rm i0}(\delta k_{10}+\delta k_{20}R_{\rm i 0}).
	\end{align}
\end{subequations}
As per eq.~\eqref{einsteinpert}, we equate these components to the RSET generated by the background \eqref{RSETmi}, keeping in mind that $\expval{T_{\theta\theta}}=\expval{T_{\phi\phi}}=0$ in the Polyakov approximation, thus implying $\delta G_{\theta\theta}=0$. From the other two equations we obtain two key relations,
\begin{align}
\delta k_{10}&=-\frac{l_{\rm P}^2}{12\pi}\frac{k_{20}}{R_{\rm i0}},\\
\delta R_{\rm i1}&=\frac{l_{\rm P}^2}{48\pi}\left(-\frac{k_{10}}{R_{\rm i0}}-\frac{\kappa^2}{R_{\rm i0}k_{10}}+\frac{1}{2}\frac{\kappa}{R_{\rm i0}}+\frac{2k_{11}}{R_{\rm i0}k_{10}}\right).\label{dr}
\end{align}
On the one hand, we can see that the semiclassical contribution to the modification of the surface gravity can be either positive or negative, depending on the sign of the background coefficient $k_{20}$. This initial semiclassical contribution very much depends on the details of the initial background geometry. On the other hand, the modification of the derivative of the inner horizon trajectory is almost always positive, implying a decrease in the rate at which it moves inward. This is a first indication of the regularising tendency which semiclassical corrections can add to the inner horizon dynamics in these spacetimes. Particularly, it can be seen from the fact that $k_{10}<0$, $\kappa>0$, $R_{\rm i0}>0$, and the assumption that $k_{11}<0$, i.e. that the background surface gravity tends to become increasingly more negative, which is certainly the case in mass inflation, as can be seen from e.g. eq. \eqref{RNk1}. Interestingly, the further along an evolution of the type \eqref{ria} we set our initial conditions for semiclassical backreaction, the smaller the background value of $R_{\rm i1}$ would be (approaching zero as $v\to\infty$) and the larger $\delta R_{\rm i1}$ would be in comparison. If the background surface gravity  $k_{10}$ has a divergent behaviour akin to \eqref{RNk1}, then the growth of $\delta R_{\rm i1}$ as we take our initial radial position $R_{\rm i0}$ to 0 cannot be said to be a consequence of the unphysical divergent $1/r^2$ factor present in the RSET approximation. More generally, if $k_{10}$ diverges at least as strongly as $1/R_{\rm i0}$, then a regularised version of this perturbation $R_{\rm i0}^2\delta R_{\rm i1}$ remains at least finite, as opposed to the background $R_{\rm i1}$ which is expected to approach zero unless a spacelike singularity forms at finite $v$.

Using these initial tendencies as an estimate of the magnitude of this effect later on in the evolution, one is led to the hypothesis that semiclassical backreaction will always become dominant at some point before a singularity is formed, perhaps leading instead to a non-singular future. The only caveat might appear when interpreting the semiclassical effects as only the first set of corrections towards a quantum gravity theory. Then, if the corrections occurred only when curvature becomes Planckian, i.e. when the background values in \eqref{dr} overcome the suppression by the $l_{\rm P}^2$ factor, one could argue that these semiclassical effects would have been already superseded by other effects of unknown nature. However, as we will see in the following example, a full time evolution can lead to a very different result, in which semiclassical corrections have a much quicker accumulative effect.

\section{Time-integrable example}

In order to see what the semiclassical evolution of the inner horizon could look like beyond the initial tendencies calculated above, we can use a particular family of geometries for the classical background which simplify our semiclassical analysis greatly. Particularly, we will use geometries which, in a vicinity around the inner horizon, take the form \eqref{geoinfl} with
\begin{equation}\label{redshiftcl}
B(v,r)=e^{-\kappa v}-\frac{1}{2}\lambda(v)r,
\end{equation}
where $\lambda(v)$ is a positive, but otherwise arbitrary function. This type of geometry, along with the below perturbation \eqref{redshiftg}, is a straightforward generalisation of \eqref{bg} from chapter \ref{ch5}. The inner horizon described by this geometry,
\begin{equation}
R_{\rm i}(v)=2\frac{e^{-\kappa v}}{\lambda(v)},
\end{equation}
moves toward the origin as long as $\lambda$ does not decrease faster than $e^{-\kappa v}$. The relation between this position and its surface gravity is not quite the same as in e.g. the Reissner-Nordström case (where for $R_{\rm i}\sim e^{-\kappa v}$, the surface gravity has an increase with a rate $e^{2\kappa v}$, as seen in chapter \ref{ch4}), but a growing surface gravity can still be replicated by choosing a $\lambda$ which increases in time.

The RSET in the Polyakov approximation corresponding to this classical background geometry has a single non-zero component: the ingoing flux
\begin{equation}\label{RSETg2}
\expval{T_{vv}}=\frac{l_{\rm P}^2}{96\pi^2r^2}\left[-\frac{1}{4}\lambda'-\frac{1}{32}\lambda^2-\frac{\kappa}{4}\lambda-\frac{\kappa^2}{2}\right],
\end{equation}
which is negative as long as $\lambda'\geq0$ (this being a reasonable requirement for a mass inflation background, i.e. that the surface gravity of the inner horizon does not decrease).

The main motivation for using these geometries is that, much like the dynamical examples of chapter \ref{ch5}, backreaction from the RSET around the inner horizon has the effect of changing the $B$ function to
\begin{equation}\label{redshiftg}
B(v,r)=e^{-\kappa v}-\frac{1}{2}\lambda(v)r+\delta B(v,r),
\end{equation}
where the first two terms are just its background form, and the perturbation $\delta B$, obtained from the semiclassical Einstein equations, will have a particularly simple form in terms of its dependence in $r$,
\begin{equation}\label{Br}
\delta B(v,r)=-\frac{\alpha(v)}{r}
\end{equation}
(the minus sign serves to make an analogy with a mass term, as discussed below). This can be readily checked by calculating the Einstein tensor with \eqref{redshiftg}, which has the non-zero components
\begin{subequations}
	\begin{align}
	G_{vv}&=\frac{\lambda'}{2}-\frac{\lambda^2}{2}+\frac{e^{-\kappa v}\lambda}{r}+\frac{\kappa\lambda}{2}+\frac{1}{r^2}\left[\alpha'-(\lambda-\kappa)\alpha\right],\label{Gvv}\\
	G_{vr}&=-\frac{\lambda}{r},\\
	G_{\theta\theta}&=\frac{G_{\phi\phi}}{\sin^2\theta}=-\frac{r\lambda e^{\kappa v}}{2},
	\end{align}
\end{subequations}
where primes denote derivatives with respect to $v$. We see that $\alpha$ appears only in $G_{vv}$, and that the terms which contain it can be directly equated to the RSET flux \eqref{RSETg2}, given that they have the same dependence in $r$ [the form of $\delta B$ in \eqref{Br} was obtained by requiring this]. Here we are once again assuming that the classical part of the equations, i.e. the rest of the terms of $G_{\mu\nu}$ and their source $T_{\mu\nu}^{\rm class}$, remain unchanged. Analogously to the series expansion calculation above (and the one in chapter \ref{ch5}), our motivation for doing this is to introduce as little information about the classical matter content as possible, while also getting a cleaner backreaction problem for the Einstein tensor sourced solely by the RSET (in this case it is the particular form of the background which allows us to do this without over-determining the system of equations). As long as this perturbation of non-zero $\alpha$ is negligible for the calculation of the RSET itself, and while \eqref{RSETg2} is accurate (which is the case for a finite time interval after $v_0$, which is smaller for faster dynamics of the background), this gives an approximate self-consistent solution to the semiclassical Einstein equations. In the language of eq.~\eqref{einsteinpert}, we are once again considering a $\delta G_{\mu\nu}$ which is equated to the Polyakov RSET. The only non-zero component of this perturbation tensor is thus $\delta G_{vv}$, sourced by \eqref{RSETg2}, corresponding to the terms involving $\alpha$ in \eqref{Gvv}, while all other $\delta G_{\mu\nu}$ are zero.

Note that the function $\alpha$ in $B$ is analogous to the $M_0$ term in \eqref{inflRN}, which is constant in the product of $A$ and $F$ used to construct $B$ (up to the decaying inverse polynomial $1/v^p$ terms, which we have neglected in our construction here), and represents the classically exponentially growing mass. Thus, if $\alpha$ grows, toward either positive of negative values, it could be taken as an indication of the semiclassical backreaction tending to become the dominant source of dynamics. We will now see that $\alpha$ in fact becomes negative, and in many cases its absolute value tends to grow exponentially quickly.

Equating \eqref{RSETg2} to the terms containing $\alpha$ in \eqref{Gvv}, leaving the remaining terms as fixed by the background, the evolution of this semiclassically sourced $\alpha$ is given by the equation
\begin{equation}\label{self-consistent}
\alpha'(v)-\eta_1(v)\alpha(v)=\frac{l_{\rm P}^2}{48\pi}\eta_2(v)+\order{l_{\rm P}^4},
\end{equation}
where
\begin{equation}\label{eta2}
\eta_1=\lambda-\kappa,\quad \eta_2=-\lambda'-\frac{1}{8}\lambda^2-\kappa\lambda-2\kappa^2
\end{equation}
are two functions determined by the choice of background. The general solution of this equation is
\begin{equation}
\alpha(v)=\frac{l_{\rm P}^2}{48\pi}e^{\int^v\eta_1(\tilde{v})d\tilde{v}}\left[c_1+\int^ve^{-\int^{\tilde{v}}\eta_1(\bar{v})d\bar{v}}\eta_2(\tilde{v})d\tilde{v}\right],
\end{equation}
where $c_1$ is the integration constant which can be fixed by initial conditions.

Let us first look at the simple case in which $\lambda$ is constant, which represents an inner horizon shrinking in radius proportionally to $e^{-\kappa v}$ while maintaining a constant surface gravity. Here we already see the main difference from the situations studied in chapter~\ref{ch5}. Depending on whether $\kappa$ or $\lambda$ is larger, $\eta_1$ can be either a positive or negative constant, while $\eta_2$ is always a negative constant. With the condition of the semiclassical perturbation being initially zero, $\alpha(0)=0$, we get the solution
\begin{equation}
\alpha(v)=\frac{l_{\rm P}^2}{48\pi}\frac{\eta_2}{\eta_1}[e^{\eta_1v}-1].
\end{equation}
Whether $\eta_1$ is positive or negative, $\alpha$ evolves toward negative values. In the former case its absolute value grows exponentially quickly, while in the latter it tends to a constant. There is also the particular case in which $\lambda=\kappa$, for which, since $\eta_1=0$, the solution is
\begin{equation}
\alpha(v)=\frac{l_{\rm P}^2}{48\pi}\eta_2v,
\end{equation}
with $\eta_2$ being once again negative. In all these cases, the (increasingly) negative values of $\alpha$ tend to push the inner horizon outward. Expanding the radial position of this horizon around $\alpha=0$,
\begin{equation}
R_{\rm i}=2\frac{e^{-\kappa v}}{\lambda}-\alpha e^{\kappa v}+\cdots,
\end{equation}
we see that once $\alpha$ acquires a non-vanishing value, even in the case where it tends to a constant, this radius quickly acquires non-perturbative corrections. The inner horizon thus begins to move outward, which, although in apparent violation of causal evolution, is hardly surprising considering that the source given by the RSET \eqref{RSETg2} is an ingoing flux of negative energy.

For more general backgrounds given by different functions $\lambda(v)$, we can see from equations \eqref{self-consistent} and \eqref{eta2} that starting from $\alpha(0)=0$, $\alpha(v)$ will tend to decrease and the inner horizon will tend to move outward unless $\lambda(v)$ decreases sufficiently quickly for $\eta_2$ to become positive. For example, if $\lambda$ tends to zero asymptotically in $v$ and its tendency is quicker than $1/v$ (but slower than $e^{-\kappa v}$, so the inner horizon does not move outward classically), and if $\kappa$ is initially negligible in \eqref{eta2}, there can be a period of time in which $\alpha$ increases toward positive values. However, except in these specific scenarios, $\eta_2$ will generally be negative and $\alpha$ will acquire negative values, making the semiclassical movement of the inner horizon an outward one.

Therefore, the results obtained in chapter \ref{ch5} appear to still hold in most of these mass inflation geometries. In other words, while the flux \eqref{RSETg2} dominates the RSET, backreaction tends to push the inner horizon outward. However, it is worth reminding the reader that the conditions for which we have been able to show that this flux is dominant only hold true for a short period of time after the formation of the BH, given by the time it takes for outgoing null geodesics which come from outside the region where~\eqref{redshiftg} is accurate (with $\delta B$ sufficiently small) to intersect the inner horizon. The result for the movement of the inner horizon is therefore only accurate as an initial tendency. Still, the fact that the RSET is likely to keep violating energy positivity conditions even in the later parts of the evolution makes the possibility that the trapped region continue to evaporate from the inside a likely one.

\begin{figure}
	\centering
	\includegraphics[scale=.7]{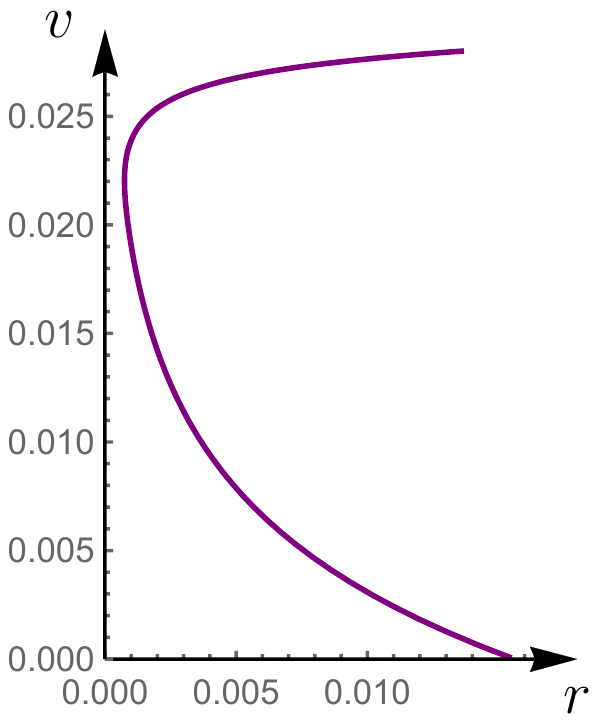}
	\caption{Trajectory of the inner apparent horizon in an extrapolated semiclassical solution, with a classical background which simulates the behaviour of the interior of a mass-inflated charged BH. Length units are taken in terms of the exterior mass $M_0$, the charge is taken to be $Q=M_0/2$ and the Planck length is set as $10^{-5}M_0$ (as a large difference in orders of magnitude is necessary, but a smaller value only increases the computational difficulty while giving no qualitative difference). Note that the bounce occurs at $r\simeq 10^{-3}M_0$, which illustrates that semiclassical effects can become dominant without the radius of the inner horizon becoming Planckian}.
	\label{f4g}
\end{figure}

Assuming that a term like \eqref{RSETg2} continues to dominate the RSET even after backreaction has become significant, we can extrapolate the movement of the inner horizon further along the evaporation process. For example, for a classical background in which the surface gravity increases exponentially as in the Reissner-Nordström case, the extrapolated self-consistent solution can be observed in fig. \ref{f4g}. The inner horizon initially moves inward while the classical background still dominates, but when the semiclassical perturbation has enough time to accumulate it produces an outward bounce and a rapid \textit{inflationary} extinction of the trapped region from the inside.

\section{Conclusions and discussion}\label{secconcl}

In this chapter we have extended our analysis of backreaction on BH inner horizons from chapter \ref{ch5} to background geometries which incorporate the causal features of a classical mass inflation process, as analysed in chapter \ref{ch4}. The initial formation of the BH, which defines the ``in" vacuum state, is modelled the same way as in chapter \ref{ch5}: an ingoing radial null surface separates a past Minkowski region from a future BH. The BH region is taken to initially be in the early stages of mass inflation, where the inner horizon is approaching the origin as a decaying exponential, with a timescale $\kappa$ which is typically related to the initial surface gravity of this horizon before mass inflation began, as seen in chapter \ref{ch4}.

First, using series expansions in the radial and temporal direction, we have looked at the initial perturbative effects the RSET has on the background geometry. We have found that the tendency for the negative ingoing flux of the RSET to push the inner horizon outward is still present. Due to the Planck scale suppression, this tendency is initially very small (for BHs at astrophysical scales). However, this does not preclude the possibility that the exponential accumulation observed over time in the geometries of chapter \ref{ch5} be present once again. Indeed, following the steps in that chapter, we found a particularly simple family of geometries in which backreaction at finite times can be analysed.

The result we found is that semiclassical backreaction once again has a tendency to source a negative mass term, which now turns out to grow more quickly than the positive mass in classical mass inflation. We are thus lead to the conclusion that an inner horizon may well undergo an outward inflation, even in the presence of mass-inflation-inducing classical perturbations, as an extrapolation of our results shows in figure \ref{f4g}.

With this we conclude our analytical analysis of semiclassical inner horizon evolution. The next steps in this study will require numerical computations of self-consistent solutions with particular classical matter sources, which we leave for future work.


\setcounter{secnumdepth}{-1}
\chapter{Conclusions: the evolution of black holes in semiclassical gravity}\label{chC}

\fancyhead[R]{}
\fancyhead[L]{Conclusions}

\begin{quote}
	\textit{Lying on his back, he looked up into the high, cloudless sky . ``Do I not know," thought he, ``that that is infinity of space, and not a vault of blue stretching above me? But, however I strain my sight, I can see only a vaulted dome; and, in spite of my knowledge of infinite space, I have more satisfaction in looking at it as a blue, vaulted dome, than when I try to look beyond."}
	
	\raggedleft - Lev Tolstoy, \textit{Ana Karenina} (1878).
\end{quote}

\vspace*{8mm}

Modern physics lies at the interface between two spheres of reality. The first is the material world, which we can observe and manipulate, and which we aim to describe. The second is the mathematical and logical constructs of our minds, which allow us to extrapolate, generalise, and ultimately to understand. From observation and experiment, we find logical patterns in the behaviour of matter, which we then aim to distil into mathematically concise theories. The full descriptive capability of these theories then typically goes beyond the observations and intuitions which resulted in their conception, leading to predictions of entirely new and as yet unobserved aspects of the physical world.

In their inception, black holes (BHs) were one such prediction. They resulted from an extrapolation of a very particular solution of the Einstein field equations with a high degree of symmetry, which was not expected to be realisable in genuine physical scenarios, at least initially~\cite{Schwarzschild1916,Thorne1970}. However, subsequent theoretical developments showed that classical matter does indeed form trapped regions and singularities under generic circumstances~\cite{Penrose1965,Hawking1973}. This then brought the spotlight to the question of where exactly the limits of applicability of the classical theory lie. Singularities themselves are certainly considered a step too far, but what then is to be thought of the long-lived trapped regions which bend causality in a singularity-directed manner?

If we take the standard approach and assume that corrections to the classical gravitational picture \textit{only} occur when curvature reaches the Planck scale, then long-lived BH trapped regions may well be allowed to exist in our universe. However, quantum field theory in curved spacetimes and semiclassical gravity show that not all quantum corrections to the geometry need to be triggered by local curvature-related effects. Indeed, non-local effects seem to also be present, and they turn out to be of particular importance precisely when causality is bent to the point of forming trapped regions. Most importantly, the backreaction effect they have on the dynamics of the geometry seems to always head in the direction of undoing these causal knots and returning spacetime to normal. Aside from being highly suggestive that quantum regularising tendencies would affect not only singularities, but trapped regions themselves, these semiclassical results already suffice to challenge the classical BH paradigm.

The first result in this direction was that of Hawking evaporation~\cite{Hawking1975}: the peeling of light rays off of the outer horizon of a dynamically formed BH has an effect on quantum fields (in the appropriate ``in" quantisation) which causes an outgoing flux of positive energy to appear far away from the BH, and a compensatory ingoing flux of negative energy to appear near the BH itself~\cite{DFU}, encoded in the renormalised stress-energy tensor (RSET) of the fields. The BH thus tends to lose mass and its horizon area tends to shrink.

The effect on the outer horizon is even more drastic for static BH spacetimes. If one quantises a field on top of a static BH in a manner which respects the time-reversal symmetry and asymptotic flatness, then one ends up with the Boulware state~\cite{Boulware1975,Fabbri2005}, in which the RSET has a divergence at the horizon. A non-perturbative extrapolation of backreaction from this RSET (in the absence of other sources of gravity) tends to eliminate the trapped region altogether, modifying the geometry toward a wormhole configuration~\cite{Fabbri2005b,Arrechea2019}.

While eternal BHs are not expected to exist in our universe (even less so given this result), this state of quantum matter does have potential physical significance in the case of static horizonless stellar configurations, particularly in the limit in which their compactness tends to that of a BH. Such configurations, even when formed dynamically, have a vacuum which relaxes to a Boulware-like state in their exterior, making the RSET around (and usually also below) their surface have extremely large values. And once again, backreaction from the RSET in these stellar objects seems to pull the configuration away from trapped region formation, allowing objects to remain in equilibrium with a compactness arbitrarily close to that of a BH~\cite{Carballo-Rubio2017,Arrechea2021}.

The study of these objects is of paramount importance due to the fact that, if they turn out to be stable, they could potentially mimic astrophysical BHs in observations of both electromagnetic and gravitational nature. However, the precise mechanism behind their hypothetical formation is as yet unknown. One essential ingredient to understand this mechanism is the relation between the ``in" vacuum of gravitational collapse, which when nearing the compactness of a BH is usually expected to have negligible energy contributions~\cite{DFU,Parentani1994,Barcelo2008,Unruh2018}, and the final static vacuum, which has an extremely large semiclassical energy content. In \hyperref[pt1]{part I} of this thesis we explored this relation with a series of \textit{ad hoc} dynamical, spherically-symmetric geometric constructions which connect the collapse and static scenarios in various ways, focusing in particular on the magnitude of the semiclassical effects they produce, as encoded in the RSET and effective temperature function.

In chapter \ref{ch1} we begun with a simple model which captures a large variety of dynamical behaviours: an oscillating spherical shell of matter which periodically approaches the formation of a horizon but bounces back just before it is formed. At each bounce, where matter slows down its motion, we indeed found large values of the RSET. We analysed this by looking at the bursts of radiation at infinity, obtained from the effective temperature function. These bursts in turn also guarantee the presence of large values of the RSET close to the matter surface. Curiously, between each bounce we also found a period of emission of nearly thermal Hawking-like radiation. Aside from having large values of the RSET and thus significant semiclassical effects in its motion, an object with this type of dynamics (e.g. a perturbed ultracompact horizonless object) would also produce a non-negligible emission of radiation which would likely act to reduce its overall mass and size, much like is what expected to occur for BHs.

In chapter \ref{ch2} we again used a thin-shell model, and analysed the case of a trajectory which forms a horizon in finite time. The parameter we were interested in was the speed at which the shell crosses its Schwarzschild radius. As a function of this parameter, we calculated the values of the RSET at the horizon, and the corresponding energy density, flux and pressure perceived by free-falling observers. We saw that these physical quantities can become arbitrarily large (and also stay large for longer) the lower this speed parameter is, approaching a divergence in the static limit.

To further explore the low-velocity regime in gravitational collapse, in chapter \ref{ch3} we analysed surface trajectories which approach the horizon radius monotonously, but reach it only asymptotically. In this case we observed that semiclassical effects are very sensitive to the structure of the geometry close to the surface, and for completeness we went beyond the thin-shell approximation and used an arbitrary spherically-symmetric geometry near horizon formation (in the sense of light-ray trapping). With minimal assumptions, we showed that this type of dynamics results in the emission of thermal Hawking radiation, with values of the effective temperature function and RSET at late times depending only on a few characteristics of the geometry through one of its degrees of freedom, which we called the generalised redshift function: particularly, the speed at which its minimum approaches zero (i.e. the speed at which the formation of an apparent horizon is approached) and its spatial derivatives on both sides of this minimum (a generalisation of the notion of surface gravity). Depending on these quantities, the dynamical ``in" vacuum can behave as in the usual case of black-hole formation in finite time, or it can become similar to the static Boulware vacuum (generally at lower speeds of approach). In the latter case, while thermal emission is still possible, it is at a temperature below that of Hawking evaporation (related to the generalised surface gravity), and the RSET thus acquires very large values around the horizon radius, tending to a divergence asymptotically in time. As a side note, we also found that this asymptotic tendency toward the formation of an apparent horizon, if maintained indefinitely, actually leads to the formation of an event horizon, despite the absence of trapped surfaces. The causal structure of these spacetimes and the energy conditions they can satisfy were also analysed.

A clear-cut result from all the above situations is that semiclassical backreaction on the geometry (through the RSET) is indeed a necessary ingredient in analysing any geometry in which matter happens to be contracting or expanding at very low velocities (much lower than the speed of light) when it is so compact that it approaches crossing its own gravitational radius and forming a trapped surface. As a purely kinematic exercise, our analysis shows the richness of the situations around the threshold of horizon formation. Beyond that, if a scenario in which matter actually enters such low-velocity regimes, it is possible that backreaction may tend to further halt the collapse, and one may expect that such dynamics would connect with the static semiclassical stellar ultracompact equilibrium solutions~\cite{Carballo-Rubio2017,Arrechea2021} as a final state.

The remaining question is how these low-velocity initial conditions could be achieved in a region where the gravitational pull is so strong that a trapped region is nearly formed. In standard astrophysical collapse, matter is expected to be moving at high speeds when it reaches this level of compactness, making it likely that semiclassical effects are insufficient to prevent the initial formation of a trapped region~\cite{DFU,Parentani1994,Barcelo2008,Unruh2018}. Since it is generally believed that once a trapped region forms, a region of Planckian curvature is also inevitable, one may decide to relegate the issue of the subsequent evolution of the BH to a matter only solvable within a full theory of quantum gravity. Indeed, different approaches to quantum gravity have attempted to give an effective description of what this evolution may look like, often involving a bounce which eliminates the trapped region from the inside~\cite{Haggard2015,Husain2021,Husain2022}. However, whether curvature does become Planckian before deviations from classicality take place is not as clear as one may assume. To see whether it is indeed the case, we must turn to an analysis of the evolution of the interior of BHs once they are formed, both in classical and in semiclassical gravity.

The most important feature of the interior of BHs for semiclassical physics is the inner apparent horizon. While classical analyses of the mass inflation instability which takes place around this horizon are quite thorough~\cite{PoissonIsrael89,Ori1991,Ori1998,Ori1992,Marolf2012}, the same cannot be said for past semiclassical ones. Works which have performed analyses in this direction have mostly focused on calculating the RSET in background geometries of eternal BH solutions, looking at the vicinity of a Cauchy horizon rather than a dynamically formed inner horizon. A divergent behaviour found at these Cauchy horizons is then typically used to argue that strong cosmic censorship could be saved (from its classical problem of extensions past the weak null singularity) through semiclassical backreaction, as generic initial conditions would seemingly change the Cauchy horizon into a strong singularity. Less often, it has been indeed appreciated~\cite{BalbinotPoisson93,Ori2019,Zilberman2022} that semiclassical backreaction might lead to ``defocusing" or ``expansion" of the Cauchy horizon to large radii due to the addition of negative mass. However, how this effect translates to dynamically formed trapped regions at finite times has not been addressed. Crucially, the idea that trapped regions have a finite lifetime has not been incorporated into these analyses. In \hyperref[pt2]{part II} of this thesis we have attempted to address this issue.

To first understand the evolution of the classical background, in chapter \ref{ch4} we used a simple shell-based construction to analyse classical perturbations and the mass inflation instability~\cite{PoissonIsrael89}. This lead to interesting results in its own right: on the one had, mass inflation being triggered depends strongly on the mass to charge (and, by extension, angular momentum) ratio of the infalling matter. In more general BH constructions with an inner horizon (e.g. singularity-free BHs~\cite{Ansoldi,Hayward}) the necessary condition for this instability appears to be even more strongly model dependent, as it comes down to how the infalling perturbations affect the inner horizon position in time. Particularly, horizons which are harder to displace (e.g. a regularised core with a size which is not shrunk with the addition of more mass) may not trigger the instability.

Then, in chapter \ref{ch5} we performed a calculation of semiclassical backreaction on a static (or nearly static) inner horizon, absent of mass inflation. We showed that backreaction from the RSET has a tendency to induce an inflationary instability, wherein this horizon is displaced in an outward direction in an exponential manner, reminiscent of (a negative version of) the classical mass inflation effect. We found this tendency both perturbatively for quite generic inner horizons, and in an approximate full solution to the semiclassical equations. We then extended the analysis to a family of dynamical background geometries which have a tendency to form a spacelike singularity in finite time, akin to the Oppenheimer-Snyder model, and once again found that backreaction pushes the inner horizon outward. Whether the classical or semiclassical tendency came out on top in these models turned out to depend on the speed at which the inner horizon initially tends toward the origin. Generally, given enough time, the semiclassical effect seems to always end up being dominant, suggesting that such a collapse (particularly, one in which the inner horizon classically describes a timelike surface which approaches the origin) may rather result in a semiclassical bounce. Extrapolating further, the trapped region can be assumed to have a very short lifetime, namely of the order of the mass of the BH (in natural units), as opposed to typical results of quantum bounces which place the time scale at the order of the mass squared (which is typically longer than the age of the universe).

In chapter \ref{ch6} we applied the semiclassical inner horizon instability analysis to another type of geometry: the Alcubierre warp drive~\cite{Alcubierre1994}. The motivation behind this is the fact that a 1+1 dimensional warp bubble configuration has very similar causal characteristics to a BH with an inner and outer horizon, which has previously been used to argue against the stability of such spacetimes in the presence of quantum fields~\cite{Hiscock1997,anti-warp2}. Given the inner horizon inflation effect found in chapter \ref{ch5}, this would become an even more compelling argument. However, our analysis here found that this horizon structure of warp drives is not generalised straightforwardly to higher dimensions. Particularly, in 2+1 and higher dimensions, we found that the horizon-like surfaces are not retained, but are rather reduced to single isolated points, and no closed trapped surfaces are present. We then analysed the deflection of null geodesics around the inner-horizon-like points to estimate the degree of instability, finding that it can be negligible with appropriate geometric shapes for the warp bubble.

Then, in chapter \ref{ch7} we went back to the semiclassical BH analysis and extended it to backgrounds which, in the absence of other regularising mechanisms, are undergoing classical mass inflation. We first performed a perturbative backreaction analysis, which by itself already suggested that semiclassical effects always become dominant at some point in the evolution, as they depend on the surface gravity of the inner horizon, which grows exponentially during mass inflation. We then obtained a simple set of self-consistent solutions which, in the absence of the RSET source, can reproduce the structure of a mass inflation geometry, but in a complete semiclassical treatment reveal a very different behaviour. They present a tendency for the semiclassically induced outward push on the inner horizon to again accumulate exponentially quickly, leading to an overall evolution similar to the cases studied in chapter \ref{ch5}. Crucially, the inner horizon inflation effect can potentially take place before curvature becomes Planckian, ensuring the validity of the semiclassical approximation.

\begin{figure}[t]
	\centering
	\includegraphics[scale=1.1]{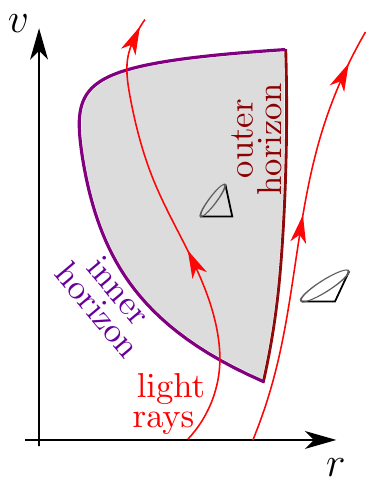}
	\caption{Qualitative picture of the extrapolated evolution of a trapped region in semiclassical gravity.}
	\label{f5g}
\end{figure}

With the addition of this final analysis, we conclude that the semiclassical evolution of BHs could have one of three outcomes. First, we must admit that the standard picture remains a possibility. Classical mass inflation may continue to dominate the dynamics around the inner horizon in later parts of the evolution, where the approximations we have used to estimate semiclassical backreaction cease to be accurate. In this case Hawking evaporation of the outer horizon would dominate the first part of the semiclassical evolution, up to the point at which the mass-inflated region (the upper part of which need not be close to the origin) is revealed to the external universe, where a more detailed analysis would be necessary. The physics of the inner horizon plays a secondary role in this picture until very late times.

However, our present results point to two alternative possibilities which could be realized in a fully self-consistent semiclassical evolution. The first one is that of the inner horizon moving outward due to backreaction from the RSET, but only up to the point at which it meets the outer one, leaving an extremal BH remnant. The second alternative, represented in fig.~\ref{f5g}, is one in which the trapped region disappears completely. The matter which then escapes the BH would likely recollapse, but the dissipation that this mechanism (or several iterations thereof) would produce could bring matter to the slow-moving initial conditions discussed in \hyperref[pt1]{part I}, which may be necessary for the formation of semiclassically sustained horizonless BH mimickers \cite{Visser2008,Barcelo2019,Carballo-Rubio2017,Arrechea2021}. Any trapped regions which may be subsequently formed in perturbations of such objects \cite{Guo2022} would then be just as short-lived, due to the same mechanism.

It is important to note again that the semiclassical inner and outer horizon effects we have analysed here can occur not due to the spacetime curvature becoming Planckian, as one may typically expect from quantum corrections, but rather due to the causal structure around the horizons themselves, and its effect of the modes of quantum fields in these spacetimes. The curvature remaining well below the Planck scale could in fact be taken as a precondition for the semiclassical description itself to be fully reliable. The fact that, where backreaction has been analysed, the tendency of semiclassical corrections is to eliminate trapped regions is a fascinating result. Whether this points to a general rule for quantum corrections, or just one applicable in these particular scenarios, is a question which requires further theoretical scrutiny, especially given the potential relation with observational signals.

Overall, this thesis shows the importance that horizon-related (rather than curvature-related) effects can have in the deviations from the classical BH formation and evolution picture on short timescales. To fully investigate the viability of the semiclassical scenarios presented above, more work is obviously required, such as a full numerical analysis of gravitational collapse including the RSET in the Polyakov or other approximations (with a background containing a timelike inner horizon, as opposed to the spacelike one of ref.~\cite{Parentani1994}). However, our analyses already show that these not much contemplated possibilities deserve full attention.


\newpage

\setcounter{secnumdepth}{2}

\appendix

\fancyhead[R]{}
\fancyhead[L]{Appendix \thechapter: \leftmark}

\chapter{RSET in 1+1 dimensions}\label{RSETcalc}

\section{Geodesic point-splitting}

The parallel transport of a vector $a^\mu_1$, initially given at a point $x^\rho$, along the curve defined by $x^\rho(\epsilon)\equiv x^\rho_\epsilon$, where $\epsilon$ is a real parameter, satisfies the equation
\begin{equation}\label{e15}
	\frac{da^\mu(\epsilon)}{d\epsilon}+\Gamma^\mu_{\nu\rho}(x_\epsilon)a^\nu(\epsilon)\frac{dx^\rho_\epsilon}{d\epsilon}=0.
\end{equation}
If the curve $x^\rho_\epsilon$ is a geodesic, its tangent vector $t^\rho(\epsilon)=dx^\rho_\epsilon/d\epsilon$ satisfies this equation,
\begin{equation}
	\frac{dt^\mu(\epsilon)}{d\epsilon}+\Gamma^\mu_{\nu\rho}(x_\epsilon)t^\nu(\epsilon)t^\rho(\epsilon)=0.
\end{equation}
The parameter $\epsilon$ for the initial point can be fixed to zero ($a^\rho(0)=a_1^\rho$), and each component of the transported vectors can be expressed as a power series around its initial value:
\begin{equation}
	\begin{split}
		a^\mu(\epsilon)&=a^\mu_1+\epsilon a^\mu_2+\frac{1}{2}\epsilon^2a^\mu_3+\cdots,\\
		x^\mu_\epsilon&=x^\mu+\epsilon t^\mu_1+\frac{1}{2}\epsilon^2t^\mu_2+\cdots.
	\end{split}
\end{equation}
Given that $t^\mu_n$, with $n=1,2,\dots$, have been defined as the $n$-th derivatives of $x^\mu_\epsilon$ evaluated at the initial point, the series expansion for the Christoffel symbols is
\begin{equation}
	\Gamma^\mu_{\nu\rho}(x_\epsilon)=\Gamma^\mu_{\nu\rho}+\epsilon\Gamma^\mu_{\nu\rho,\alpha}t^\alpha_1+\frac{1}{2}\epsilon^2(\Gamma^\mu_{\nu\rho,\alpha\beta}t^\alpha_1t^\beta_1+\Gamma^\mu_{\nu\rho,\alpha}t^\alpha_2)+\cdots,
\end{equation}
where the quantities in the coefficients on the right-hand side are evaluated at $\epsilon=0$. Substituting these expressions into \cref{e15}, the total coefficients of each power of $\epsilon$ must be zero, and a relation between the values of the different order derivatives at $\epsilon=0$ of the vector components is obtained. In the two dimensional case discussed thus far, the expressions one obtains are
\begin{equation}\label{e19}
	\begin{split}
		&a_2^u+\frac{C_u}{C}a_1^ut_1^u=0,\\
		&a_3^u+C\left[\frac{t_1^u}{C^3}(CC_{uu}-3C_u^2)-\frac{t^v_1}{4}R\right]a_1^ut_1^u=0,\dots,
	\end{split}
\end{equation}
where the derivative of order $n$ is obtained as a function of the 0th derivatives (the initial values, $a^\rho_1$ and $t^\rho_1$) by recursive substitution. The result for the $a^v_n$ derivatives has the same form as \cref{e19}, interchanging the $u$'s and $v$'s in indices and derivatives.

If the action of parallel transport of a vector by an amount $\epsilon$ is expressed as the result of a tensor acting on the initial value, $a^\mu(\epsilon)=e^\mu_\rho(\epsilon)a^\rho_1$, the fact that the transported component $a^u(\epsilon)$ only depends on the initial value in the same direction $a^u_1$ (which algebraically is a consequence of the fact that in this coordinate system the only non-zero Christoffel symbols are those with all three indices equal) means that this tensor is diagonal. From \cref{e19}, its non-zero terms can be expressed as the series
\begin{equation}\label{e20}
	\begin{split}
		U_\epsilon\equiv e^u_u(\epsilon)&=1-\frac{C_u}{C}t_1^u\epsilon+\frac{1}{2}Ct_1^u\left[\frac{t_1^u}{C^3}(3C_u^2-CC_{uu})+\frac{t^v_1}{4}R\right]\epsilon^2+\cdots,\\
		V_\epsilon\equiv e^v_v(\epsilon)&=1-\frac{C_v}{C}t_1^v\epsilon+\frac{1}{2}Ct_1^v\left[\frac{t_1^v}{C^3}(3C_v^2-CC_{vv})+\frac{t^u_1}{4}R\right]\epsilon^2+\cdots.
	\end{split}
\end{equation}\par
The above equations were obtained for parallel transport of the contravariant components of a vector. For covariant components an expression analogous to \cref{e15} is obtained from the intrinsic form of this equation ($\nabla_{t(\epsilon)}a=0$), namely
\begin{equation}
	\frac{da_\mu(\epsilon)}{d\epsilon}-\Gamma^\nu_\mu\rho(x_\epsilon)a_\nu(\epsilon)t^\rho(\epsilon)=0.
\end{equation}
With the non-zero Christoffel symbols, the equations for both covariant components in null coordinates only differ from the case of their contravariant counterparts by a minus sign in the second term, which would translate into a sign change of the terms with odd powers of $\epsilon$ in \cref{e20}. The same change is produced by $\epsilon\to-\epsilon$. Therefore, defining the new tensor components analogous to \eqref{e20} by $a_u(\epsilon)=\tilde{U}_\epsilon a_{1\,u}$ and $a_v(\epsilon)=\tilde{V}_\epsilon a_{1\,v}$, these can be obtained by
\begin{equation}\label{e22}
	\tilde{U}_\epsilon=U_{-\epsilon};\;\tilde{V}_\epsilon=V_{-\epsilon}.
\end{equation}

\section{Renormalisation}

As discussed earlier, the divergence of the stress-energy tensor vacuum expectation value can be obtained from taking the coincidence limit of the (otherwise convergent) two-point function $\bra{0}\nabla_\mu\hat{\phi}(x^\rho)\nabla'_{\nu'}\hat{\phi}(x'^\rho)\ket{0}$. In this case, at the coincidence limit this function actually coincides with the stress-energy tensor, as can be seen from eq. \eqref{SET} and the metric \eqref{metric2d}. To see the divergent term explicitly as a function of the distance between the points before taking the limit, the operators for parallel transport of covariant vectors \eqref{e22} can be applied to the covariant derivatives symmetrically along opposite directions of the geodesic. The non-zero components become
\begin{equation}
	\begin{split}
		\expval{T_{uu}}_\epsilon\equiv&\bra{0}\tilde{U}_\epsilon\nabla_{u_\epsilon}\hat{\phi}(x_\epsilon)\tilde{U}_{-\epsilon}\nabla_{u_{-\epsilon}}\hat{\phi}(x_{-\epsilon})\ket{0}=\frac{1}{4\pi}U_\epsilon U_{-\epsilon}\sum_\omega\omega e^{i\omega\Delta u},\\
		\expval{T_{vv}}_\epsilon\equiv&\bra{0}\tilde{V}_\epsilon\nabla_{v_\epsilon}\hat{\phi}(x_\epsilon)\tilde{V}_{-\epsilon}\nabla_{v_{-\epsilon}}\hat{\phi}(x_{-\epsilon})\ket{0}=\frac{1}{4\pi}V_\epsilon V_{-\epsilon}\sum_\omega\omega e^{i\omega\Delta v}.
	\end{split}
\end{equation}
For a field with no boundary conditions the sum in $\omega$ becomes an integral in $\omega\in(0,\infty)$, which can be evaluated by an analytic continuation of the coordinates to the complex plane, $\Delta u\to\Delta u+i\delta$, with $\delta>0$,
\begin{equation}
	\int_{0}^{\infty}\mathop{d\omega}\omega e^{-\omega\delta}e^{i\omega\Delta u}=-\frac{1}{(\Delta u+i\delta)^2}.
\end{equation}
The result of the integral for $\delta=0$ is then defined as $-1/(\Delta u)^2$. For the $uu$ component, the series expansion in $\epsilon$ becomes
\begin{equation}
	\begin{split}
		&\expval{T_{uu}}_\epsilon=-\frac{1}{4\pi}\left(1-\frac{C_u}{C}t_1^u\epsilon+\frac{1}{2}Ct_1^uH\epsilon^2+\cdots\right)·\\&\qquad\qquad\qquad\cdot\left(1+\frac{C_u}{C}t_1^u\epsilon+\frac{1}{2}Ct_1^uH\epsilon^2+\cdots\right)·\\
		&\qquad\qquad\qquad\cdot\left[\left(u+\epsilon t_1^u-\frac{1}{2}\frac{C_u}{c}(t_1^u)^2\epsilon^2+\frac{1}{6}CH(t_1^u)^2\epsilon^3+\cdots\right)\right.\\&\qquad\qquad\qquad-\left.\left(u-\epsilon t_1^u-\frac{1}{2}\frac{C_u}{c}(t_1^u)^2\epsilon^2-\frac{1}{6}CH(t_1^u)^2\epsilon^3+\cdots\right)\right]^{-2},
	\end{split}
\end{equation}
where $H=\left[t_1^u(3C_u^2-CC_{uu})/C^3+\frac{t^v_1}{4}R\right]$. With the series $1/(1-x)=1+x+\cdots$ for the denominator, the two-point tensor component becomes
\begin{equation}
	\begin{split}
		\expval{T_{uu}}_\epsilon&=-\frac{1}{16\pi(t_1^u)^2\epsilon^2}+\frac{1}{16\pi}\left[-\frac{C_u^2}{C^2}+\frac{2}{3}\frac{C_{uu}}{C}-\frac{1}{6}\frac{t_1^v}{t_1^u}R\right]+\order{\epsilon^2}\\
		&=-\frac{1}{16\pi(t_1^u)^2\epsilon^2}-\frac{1}{12\pi}\sqrt{C}\partial_u^2\frac{1}{\sqrt{C}}-\frac{1}{96\pi}\frac{t_1^v}{t_1^u}R+\order{\epsilon^2}.
	\end{split}
\end{equation}
The result for $\expval{T_{vv}}_\epsilon$ can again be obtained by switching the $u$'s and $v$'s, while $\expval{T_{uv}}$ remains zero.

The goal of renormalisation is to covariantly remove the terms which produce divergences in these expectation values, and be left with a quantity which can be used as the source in the semiclassical Einstein equations~\cite{Wald1995}. With the scalar
\begin{equation}
	\Sigma=g_{\mu\nu}t^\mu t^\nu=Ct^ut^v=g^{\mu\nu}t_\mu t_\nu=\frac{4}{C}t_ut_v
\end{equation}
the vector components in terms with $1/\epsilon^2$ can be expressed covariantly as
\begin{equation}
	\text{diag}\left(\frac{1}{(t^u_1)^2},\frac{1}{(t^v_1)^2}\right)=\frac{4}{\Sigma^2}\text{diag}\left(t_{1\,u}^2,t_{1\,v}^2\right)=\left(\frac{4}{\Sigma^2}t_\mu t_\nu-\frac{2}{\Sigma}g_{\mu\nu}\right).
\end{equation}
The terms multiplying $R$ can be written as this same tensor multiplied by $\Sigma$. The remaining terms are independent of the point-splitting direction and cannot be written in terms of the metric or curvature tensors. Therefore, a new tensor quantity needs to be defined from its value in null coordinates:
\begin{equation}
	\Theta_{\mu\nu}=-\frac{1}{12\pi}\sqrt{C}\,\text{diag}\left(\partial_u^2\frac{1}{\sqrt{C}},\partial_v^2\frac{1}{\sqrt{C}}\right).
\end{equation}
The result for the two-point stress-energy tensor then takes the form
\begin{equation}\label{e32}
	\expval{T_{\mu\nu}}_\epsilon=\frac{1}{8\pi}\left[\frac{1}{\Sigma\epsilon^2}+\frac{1}{6}R\right]\left(g_{\mu\nu}-\frac{2}{\Sigma}t_\mu t_\nu\right)+\Theta_{\mu\nu}+\order{\epsilon^2}.
\end{equation}

The removal of the divergent term in the limit $\epsilon\to0$ is the first priority of this renormalisation procedure. The term proportional to $t_\mu t_\nu/\Sigma$ should also be removed, firstly, because it makes no sense for the value in the single-point limit to depend on the direction of point-splitting; secondly, because for a generic vector field $t_\mu$ keeping this term is (generally) incompatible with the local conservation of the resulting RSET. Removing these two terms and taking the $\epsilon\to0$ limit, the RSET of the massless scalar field in 1+1 dimensions becomes
\begin{equation}\label{e33a}
	\expval{T_{\mu\nu}}=\frac{1}{48\pi}Rg_{\mu\nu}+\Theta_{\mu\nu}.
\end{equation}
This quantity is regular, and satisfies certain conditions~\cite{Wald1995} which one might expect from a physically reasonable RSET, such as local conservation, $\nabla^\mu\expval{T_{\mu\nu}}=0$, and being zero for the Minkowski quantisations in flat spacetime ($C=$ const.), making this renormalisation compatible with the standard normal ordering in that case.

However, it is worth noting that there is a certain degree of arbitrariness in the particular subtraction chosen. Strictly speaking, at the level of the semiclassical equations of motion, the only requirement for the RSET is conservation. Thus, any conserved geometric counter-terms can be added or subtracted as part of the renormalisation procedure, e.g. $\Lambda g_{\mu\nu}$, with $\Lambda=$ const. In higher spacetime dimensions, the ambiguities in the renormalisation of the stress-energy tensor become even more clear~\cite{Brown1986,Davies1977,Taylor2021}. This is not to say that no robust results can be obtained from the RSET; on the contrary, certain particle creation effects, such as Hawking radiation, are captured well by the RSET independently of how the ambiguities are fixed. For the purposes of this work, we can safely assume that \eqref{e33a} is the correct expression for the RSET in 1+1 dimensions, as the horizon-related effects we are interested in are non-local in curvature, and are thus unaffected by the ambiguities. Also, the fact that the RSET is not given by geometric quantities that can be reduced to the metric and curvature tensors is precisely why it can become large in regions of low curvature (as we will see explicitly).

\section{Trace anomaly}

One interesting observation about the RSET result \eqref{e33a} is that it is not traceless, unlike its classical counterpart. The stress-energy tensor of a field with conformally invariant dynamics is always traceless~\cite{Hawking1973}, and ideally one would want this symmetry to be retained in the quantised theory. Indeed, up until \cref{e32} the tensor we work with is traceless, but after renormalisation this ceases to be the case. Particularly, it is not the removal of the divergent term, but rather the $t_\mu t_\nu/\Sigma$ term which breaks this property, bringing into question whether there might be a different subtraction which preserves it. However, it turns out that conservation of this RSET is incompatible with tracelessness. Recovering a zero trace would require the reintroduction of a term with the form $Rt_\mu t_\nu/\Sigma$, and to keep conservation a normalized vector field $b^\mu=t^\mu/\sqrt{\Sigma}$ must be found such that
\begin{equation}\label{e35}
	\nabla_\mu(Rb^\mu b^\nu)=0.
\end{equation}
Since in two dimensions this vector field has only one degree of freedom, it cannot be made to satisfy both equations in \eqref{e35} for a generic function $C$.

The nonzero trace is therefore an inevitable result for the RSET. And this is not restricted to the 1+1 dimensional case either: an ``anomalous" trace appears in the RSETs of all conformally invariant field theories in any number of spacetime dimensions, with all known renormalisation procedures~\cite{Wald1978,BD,Fabbri2005}.

\chapter{Oppenheimer-Snyder collapse}\label{OSmodel}

The Oppenheimer-Snyder model is constructed by matching a section of a closed Friedman-Robertson-Walker universe, representing the interior of the dust cloud, with a patch of Schwarzschild spacetime, describing the vacuum exterior. The metric for the interior section is then most conveniently expressed in cosmological coordinates,
\begin{equation}
	ds^2=a^2(\tau)\left(-d\tau^2+d\chi+{\rm sin}^2\chi d\Omega^2\right),
\end{equation}
where $a(\tau)$ is the conformal factor, which has the form
\begin{equation}
	a(\tau)=\frac{a_0}{2}\left(1+{\rm cos}\,\tau\right).
\end{equation}
The collapse starts at $\tau=0$ and ends at $\tau=\pi$. The coordinate $\chi$ goes between $0$ and $\chi_0<\pi/2$. The two constants $a_0$ and $\chi_0$ are related to the initial conditions through
\begin{equation}
	a_0=\sqrt{\frac{r_0^3}{2M}},\qquad {\rm sin}^2\chi_0=\frac{2M}{r_0},
\end{equation}
where $r_0$ is the initial radius of the ball and $M$ is its mass.

Generally, when a Schwarzschild BH forms from gravitational collapse, the trapped region is formed before the singularity and has both an outer and inner apparent horizons. The outer horizon is either stationary or moves outward until all the collapsing matter has crossed it. The inner horizon moves inward and reaches the origin as the singularity is formed. The causal diagram in figure \ref{fig23} illustrates this (see \cite{Bengtsson2013} for a more detailed discussion).

\begin{figure}
	\centering
	\includegraphics[scale=.5]{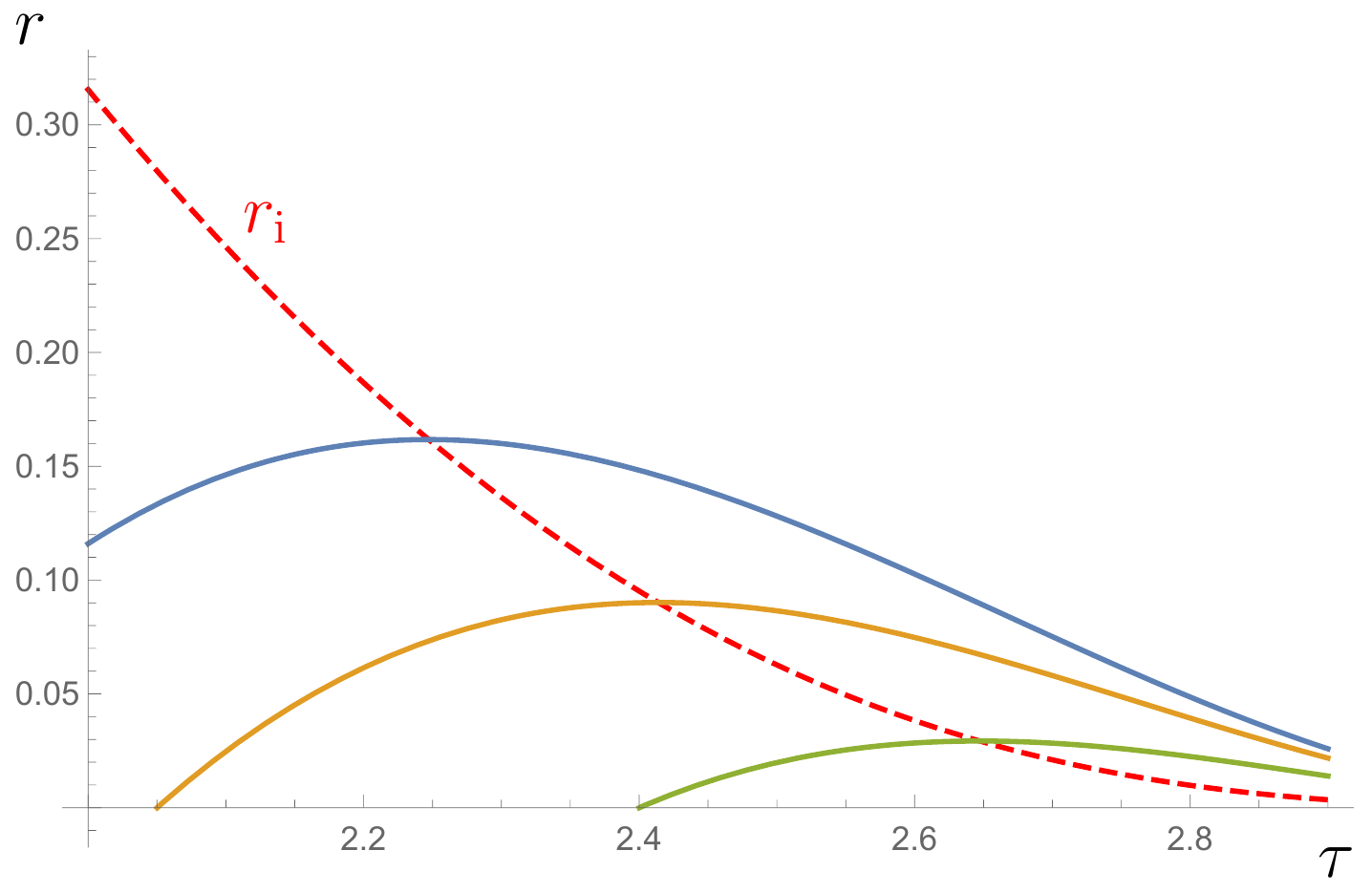}
	\caption{Outgoing null trajectories in the Oppenheimer-Snyder interior region. The dashed line represents the dynamical inner horizon, beyond which even these null rays start moving inward.}
	\label{fig22}
\end{figure}

Now, it is clear that in this interior region causal trajectories are seemingly allowed to move outward throughout the whole collapse, even when the surface has already crossed the Schwarzschild radius of the external geometry. However, when considering whether this movement is actually in the direction of increasing radius, we must take into account that it is relative to the collapsing matter distribution. The radius, as defined by the surface area of spheres in sections of constant $\tau$ and $\chi$, is given by
\begin{equation}
	r(\tau,\chi)=a(\tau){\rm sin}\,\chi.
\end{equation}
If we consider the quickest causal outward movement, i.e. outgoing radial null geodesics, given by $\tau=\chi+U$ with $U$ a constant in the range $(0,\pi)$, their radial positions are given by
\begin{equation}\label{OSnull}
	r=\frac{a_0}{2}(1+{\rm cos}\,\tau){\rm sin}(\tau-U),
\end{equation}
with $\tau$ ranging between $U$ and either $U+\chi_0$, if the latter is less than $\pi$ (in which case the light ray escapes from the surface into the Schwarzschild region), or up to $\pi$ if the opposite inequality is satisfied (in which case the ray remains in the interior until it falls into the singularity). What we are interested in is the inner apparent horizon, where outgoing light rays switch from going in a direction of increasing $r$ to one of decreasing $r$. In terms of the parameter $\tau$ this can be simply obtained by looking for the spot where the derivative of \eqref{OSnull} with respect to it becomes zero. The trajectory of the inner horizon is thus described by the timelike curve
\begin{equation}\label{OSri}
	r_{\rm i}(\tau)=\frac{a_0}{2}(1+{\rm cos}\,\tau){\rm sin}\left(\frac{\pi-\tau}{2}\right).
\end{equation}
This relation is only valid when $r_{\rm i}<r_{\rm s}$, with $r_{\rm s}(\tau)=a(\tau){\rm sin}\,\chi_0$ being the position of the surface, which implies that it is valid for $\tau\in(\pi-2\chi_0,\pi)$. At the lower bound of this interval we have $r_{\rm i}=r_{\rm s}=2M$, i.e. the inner horizon is formed at the same moment as the outer horizon, it stays within the matter distribution and reaches the origin when the singularity is formed. Figure \ref{fig22} shows part of the trajectory of the inner horizon $r_{\rm i}(\tau)$, superposed with the trajectories of a few outgoing light rays. If we were to also draw the ingoing light rays ($\tau+\chi=V=$ const.), we would see that the inner horizon lies inside the local light cones at each point, i.e. this horizon describes a timelike surface, implying that its presence is independent of the chosen spacetime slicing.

We thus see an example of the fact that gravitational collapse in almost all its forms is accompanied by an inner horizon, however briefly. But even this brief existence may create a sufficient environment for the non-local terms of the RSET, usually magnified at horizons, to manifest. This may be so here especially due to the fact that the time scale in which such effects become important is usually related to the surface gravity of the horizon $k_1$ (a term we use to refer to the radial slope of the redshift function, generalised from its definition in the presence of Killing vector fields), and this surface gravity tends to a divergence as the singularity is approached, as we will see. It will therefore be particularly useful for our later analysis to look at how $r_{\rm i}$ approaches the origin, and how $k_1$ diverges there. Expanding \eqref{OSri} around the singularity ($\tau=\pi$) we get the leading order term
\begin{equation}
	r_{\rm i}=\frac{a_0}{8}(\pi-\tau)^3+\cdots.
\end{equation}
In most of this work, we use advanced Eddington-Finkelstein coordinates to describe BH geometries, so we note that this expansion in these coordinates becomes
\begin{equation}\label{OSv}
	r_{\rm i}\propto(v_{\rm s}-v)^{3/4}+\cdots,
\end{equation}
where $v_{\rm s}$ is the instant of singularity formation. To see how the surface gravity diverges, first we need to define this quantity more precisely. Following the procedure from our previous work \cite{Barcelo2020}, we will use the \textit{generalised redshift function} $F(v,r)$, which goes on the right-hand side of the equation for outgoing null radial geodesics,
\begin{equation}
	\frac{dr}{dv}=F(v,r).
\end{equation}
We define the surface gravity at a dynamical horizon as the absolute value of the slope in the radial direction of $F(v,r)$ at the horizon.\footnote{In the static case this definition differs slightly from what is usually referred to as surface gravity \cite{Wald1984}, coinciding only for metrics in which $F=-g_{vv}/2$, as explained in chapter \ref{ch3}.} With this definition we obtain the divergent expression for the surface gravity
\begin{equation}\label{OSk}
	k_1\propto\frac{1}{r_{\rm i}\,\text{sin}\,\tau}\propto\frac{1}{v_{\rm s}-v}.
\end{equation}
It is fairly easy to understand the origin of this expression: the divergence would go as $1/r_{\rm i}$ if the profile of $F$ in the $r$ direction were a straight line starting from a fixed (positive) point at the origin and with decreasing slope; in the Oppenheimer-Snyder case there is an additional diverging factor $1/\text{sin}\,\tau$, i.e. the slope becomes more vertical more quickly, due to the geometry satisfying certain smoothness conditions at the origin (before the singularity forms).

This model serves as a good example of the behaviour of the inner horizon and its surface gravity in gravitational collapse which results in the formation of a Schwarzschild BH. In more general scenarios of collapse the inner horizon may not reach the origin, instead halting a finite distance away. This is the case in the formation of a charged BH, as we saw earlier, and also occurs when rotating and regular BHs form. Classically the dynamics of this horizon are restricted to either moving inward or halting, as moving back out would require the violation of causality, which in turn requires a matter source of negative energy density (i.e. violating the null energy condition) \cite{Hayward1994}. Semiclassically, however, there are no such restrictions, as we know well from Hawking evaporation.

\chapter{Geodesics approaching the point of divergent radial stretching}\label{ap1}
If the function $g(v,r)$ of the metric \eqref{1c} diverges at a point $(v_0,r_h)$, close to this point we can quite generally assume it has the form
\begin{equation}\label{a1}
g=\frac{1}{a(v_0-v)^n+\tilde{k}_m(r-r_h)^m},
\end{equation}
for which we have also assumed that we are approaching from a smaller $v$ and a larger $r$, with $a$ and $\tilde{k}_m$ being positive constants. What we want to find out is, depending on the values of $n$ and $m$, whether there are geodesics which approach this divergent point, and if there are, whether they take a finite of infinite proper time to reach it.\par
The easiest way to obtain an answer is to assume we already have it, and then check if it is true. In other words, let us first assume that there are timelike geodesics which reach this point at a finite affine parameter $\sigma_0$ as
\begin{align}\label{a2}
v-v_0&=-\beta(\sigma_0-\sigma)^q+\cdots,\\ \label{a3}
r-r_{\rm h}&=\alpha(\sigma_0-\sigma)^p+\cdots,
\end{align}
with $\beta$, $\alpha$, $p$ and $q$ positive constants. The geodesic equations these trajectories must satisfy are
\begin{equation}\label{a4}
\ddot{v}=-\frac{\partial_vg}{g}\dot{v}^2,\qquad \ddot{r}=-\frac{\partial_rg}{g}\dot{r}^2+\frac{f}{g}\ddot{v},
\end{equation}
where we have assumed $f$ is constant. Plugging the expressions \eqref{a2} and \eqref{a3} into these equations, we get the following results for the leading order
\begin{align}\label{a6}
q&=\frac{2}{1-n+n/m},\qquad \\ \label{a7}
p=\frac{n}{m}q&=\frac{2}{1-m+m/n},\\ \label{a8}
\frac{\alpha^m}{\beta^n}&=\frac{a}{\tilde{k}_m}\frac{mn-m+n}{mn+m-n}.
\end{align}
There is a single degree of freedom left in the proportionality coefficients, meaning we have found a whole uniparametric family of solutions. An important point is that these solutions are valid representations of geodesics which reach the point of divergent $g$ only if $q$ and $p$ are positive, which implies the restriction
\begin{equation}\label{a9}
n-1<\frac{1}{m-1}
\end{equation}
for the geometry. The smaller the exponents $n$ and $m$, the quicker the divergence is approached, so this inequality can be interpreted as the fact that geodesics only take a finite time to reach the point if the divergence is generated suddenly enough.\par
On the other hand, if we assume the geodesics take an infinite time to reach the point, say as
\begin{align}\label{a10}
v-v_0&=-\frac{\beta}{\sigma^q}+\cdots,\\ \label{a11}
r-r_{\rm h}&=\frac{\alpha}{\sigma^p}+\cdots,
\end{align}
then the opposite inequality,
\begin{equation}
n-1>\frac{1}{m-1},
\end{equation}
must be satisfied, i.e. the divergence of $g$ must be reached slowly enough. Equations \eqref{a6} and \eqref{a7} now hold with a change of sign of the rhs, and eq. \eqref{a8} holds as such.\par
We may then ask whether such geodesics exist for a geometry which precisely satisfies
\begin{equation}
n-1=\frac{1}{m-1}.
\end{equation}
They do, and they take the form
\begin{align}
v-v_0&=-\beta\mathop{e^{-q\sigma}}+\cdots,\\
r-r_{\rm h}&=\alpha\mathop{e^{-p\sigma}}+\cdots,
\end{align}
i.e. they also take infinite proper time to reach the point but they have a different approach. In this case the restrictions on the coefficients imposed by the geodesic equations are
\begin{equation}
\frac{p}{q}=n-1,\qquad \frac{\alpha^m}{\beta^n}=\frac{a}{\tilde{k}_m}(n-1).
\end{equation}\par
So far we have only considered timelike geodesics which fall into $(v_0,r_{\rm h})$ from larger radii. If we also consider ones which may approach this point from the inside, we obtain some additional solutions. Assuming the point is reached in finite proper time, i.e. taking eqs. \eqref{a2} and \eqref{a3}, the latter with a change of sign for the approach from the inside, we get on the one hand solutions which again satisfy eqs. \eqref{a6}, \eqref{a7} and \eqref{a8} (with $\tilde{k}_m\to k_m$, as we are now on the inside), and on the other we obtain some independent additional solutions which satisfy
\begin{align}\label{a18}
p=\frac{1+n}{1-n},&\qquad q=\frac{1}{1-n},\\ \label{a20}
\frac{\alpha}{\beta^{n-1}}&=\frac{f}{2}\frac{a}{1+n}.
\end{align}
The restriction on the geometry for these solutions to exist is simply
\begin{equation}
n<1.
\end{equation}
This kind of additional solutions also exist if we assume an approach in infinite proper time using eqs. \eqref{a10} and \eqref{a11}, the latter again with a change of sign. They satisfy eqs. \eqref{a18} with a change of sign on the rhs, and eq. \eqref{a20} changing the power of $\beta$ from $n-1$ to $n+1$. The geometries on which these solutions exist only need to satisfy
\begin{equation}
n>1.
\end{equation}\par
The conclusion is that if the geometry is given by \eqref{a1} and satisfies
\begin{equation}
n-1\ge\frac{1}{m-1},
\end{equation}
then all geodesics which approach $(v_0,r_{\rm h})$ have their affine parameter tending to infinity. If the opposite relation is satisfied, but $n>1$, then depending on their approach some geodesics will reach this point in finite affine parameter, and some others in infinite. We also remind the reader that throughout the main text we assumed $m\ge1$, which is required for light-ray trapping if the approach toward the divergence in $g$ occurs in infinite advanced time $v$. If we want to relax this restriction in the finite-time diverging case, then the solutions obtained at the beginning of this appendix for an approach in finite proper time \eqref{a2}, \eqref{a3} only exist if the additional restriction $m>1-1/(n+1)$ is satisfied. Also, the ingoing null geodesic which reaches this point does so in finite affine parameter if $m<1$, whereas it always did so in infinite time (just as in the static case) when $m\ge1$.

\newpage
\fancyhead[R]{}
\fancyhead[L]{\leftmark}

\nocite{*}
\bibliography{Bibliografia}
\bibliographystyle{ieeetr}

\end{document}